\PassOptionsToPackage{unicode}{hyperref}
\PassOptionsToPackage{hyphens}{url}
\PassOptionsToPackage{dvipsnames,svgnames,x11names}{xcolor}
\documentclass[
  12pt]{article}

\usepackage{amssymb}
\usepackage{mathrsfs}
\usepackage{iftex}

\usepackage{dsfont} 
\usepackage{graphicx,psfrag,epsf}
\usepackage{enumerate}
\usepackage{array}
\usepackage{url} 
\usepackage{amsmath}
\usepackage{tcolorbox}
\usepackage{subcaption}
\usepackage{bm}
\usepackage{hyperref}
\usepackage{arydshln}
\usepackage[ruled,vlined,linesnumbered]{algorithm2e}
\usepackage{enumitem}
\usepackage{caption}
\usepackage[]{natbib}
\usepackage{xr}

\newtheorem{assumption}{Assumption}

\newcommand{\chengyu}{\color{magenta}}

\newcommand{\Trc}{\text{Trace}}
\newcommand{\Fn}{\mathrm{F}}
\newcommand{\init}{\mathrm{init}}
\newcommand{\Mff}{n^{-1}\bZ^\T\bZ}
\newcommand{\Mffh}{n^{-1}\hat\bZ^\T\hat\bZ}

\newcommand{\Rdis}{R_{nq}}
\newcommand{\rdis}{r_{nq}}

\let\hat\widehat
\let\tilde\widetilde

\newcommand{\ba}{\bm{a}}
\newcommand{\bb}{\bm{b}}

\newcommand{\be}{\bm{e}}

\newcommand{\bh}{\bm{h}}

\newcommand{\bn}{\bm{n}}

\newcommand{\bs}{\bm{s}}

\newcommand{\bu}{\bm{u}}
\newcommand{\bv}{\bm{v}}

\newcommand{\bx}{\bm{x}}
\newcommand{\by}{\bm{y}}
\newcommand{\bz}{\bm{z}}

\newcommand{\Ab}{\mathbf{A}}
\newcommand{\Bb}{\mathbf{B}}

\newcommand{\bA}{\bm{A}}
\newcommand{\bB}{\bm{B}}

\newcommand{\bD}{\bm{D}}
\newcommand{\bE}{\bm{E}}

\newcommand{\bG}{\bm{G}}
\newcommand{\bH}{\bm{H}}
\newcommand{\bI}{\bm{I}}

\newcommand{\bK}{\bm{K}}
\newcommand{\bL}{\bm{L}}
\newcommand{\bM}{\bm{M}}
\newcommand{\bN}{\bm{N}}
\newcommand{\bO}{\bm{O}}
\newcommand{\bP}{\bm{P}}
\newcommand{\bQ}{\bm{Q}}
\newcommand{\bR}{\bm{R}}
\newcommand{\bS}{\bm{S}}

\newcommand{\bU}{\bm{U}}
\newcommand{\bV}{\bm{V}}

\newcommand{\bX}{\bm{X}}
\newcommand{\bY}{\bm{Y}}
\newcommand{\bZ}{\bm{Z}}

\newcommand{\cC}{\mathcal{C}}
\newcommand{\cD}{\mathcal{D}}

\newcommand{\cI}{\mathcal{I}}

\newcommand{\cL}{\mathcal{L}}

\newcommand{\cN}{\mathcal{N}}

\newcommand{\cP}{\mathcal{P}}

\newcommand{\cR}{\mathcal{R}}
\newcommand{\cS}{{\mathcal{S}}}
\newcommand{\cT}{{\mathcal{T}}}

\newcommand{\EE}{\mathbb{E}}

\newcommand{\PP}{\mathbb{P}}

\newcommand{\RR}{\mathbb{R}}

\newcommand{\bbeta}{\bm{\beta}}

\newcommand{\btheta}{\bm{\theta}}

\newcommand{\bphi}{\bm{\phi}}

\newcommand{\bDelta}{\bm{\Delta}}

\newcommand{\bSigma}{\bm{\Sigma}}
\newcommand{\bUpsilon}{\bm{\Upsilon}}
\newcommand{\bPhi}{\bm{\Phi}}
\newcommand{\bPsi}{\bm{\Psi}}
\newcommand{\bOmega}{\bm{\Omega}}

\newcommand{\argmin}{\mathop{\mathrm{argmin}}}

\newcommand*{\zero}{{\bm 0}}
\newcommand*{\one}{{\bm 1}}

\def\T{{ \intercal }}

\ifx\BlackBox\undefined
\newcommand{\BlackBox}{\rule{1.5ex}{1.5ex}}  
\fi

\ifx\QED\undefined
\def\QED{~\rule[-1pt]{5pt}{5pt}\par\medskip}
\fi

\ifx\proof\undefined
\newenvironment{proof}{\par\noindent{\bf Proof\ }}{\hfill\BlackBox\\[2mm]}
\fi

\ifx\theorem\undefined
\newtheorem{theorem}{Theorem}
\fi
\ifx\example\undefined
\newtheorem{example}{Example}
\fi
\ifx\property\undefined
\newtheorem{property}{Property}
\fi
\ifx\lemma\undefined
\newtheorem{lemma}{Lemma}
\fi
\ifx\proposition\undefined
\newtheorem{proposition}{Proposition}
\fi
\ifx\remark\undefined
\newtheorem{remark}{Remark}
\fi
\ifx\corollary\undefined
\newtheorem{corollary}{Corollary}
\fi
\ifx\definition\undefined
\newtheorem{definition}{Definition}
\fi
\ifx\conjecture\undefined
\newtheorem{conjecture}{Conjecture}
\fi
\ifx\fact\undefined
\newtheorem{fact}{Fact}
\fi
\ifx\claim\undefined
\newtheorem{claim}{Claim}
\fi
\ifx\cond\undefined
\newtheorem{cond}{Condition}
\fi

\ifPDFTeX
  \usepackage[T1]{fontenc}
  \usepackage[utf8]{inputenc}
  \usepackage{textcomp} 
\else 
  \usepackage{fontspec}
  \defaultfontfeatures{Scale=MatchLowercase}
  \defaultfontfeatures[\rmfamily]{Ligatures=TeX,Scale=1}
\fi
\usepackage{lmodern}
\ifPDFTeX\else  
\fi
\IfFileExists{upquote.sty}{\usepackage{upquote}}{}
\IfFileExists{microtype.sty}{
  \usepackage[]{microtype}
  \UseMicrotypeSet[protrusion]{basicmath} 
}{}
\makeatletter
\@ifundefined{KOMAClassName}{
  \IfFileExists{parskip.sty}{%
    \usepackage{parskip}
  }{
    \setlength{\parindent}{2pt}
    \setlength{\parskip}{6pt plus 2pt minus 1pt}}
}{
  \KOMAoptions{parskip=half}}
\makeatother
\usepackage{xcolor}
\makeatletter
\ifx\paragraph\undefined\else
  \let\oldparagraph\paragraph
  \renewcommand{\paragraph}{
    \@ifstar
      \xxxParagraphStar
      \xxxParagraphNoStar
  }
  \newcommand{\xxxParagraphStar}[1]{\oldparagraph*{#1}\mbox{}}
  \newcommand{\xxxParagraphNoStar}[1]{\oldparagraph{#1}\mbox{}}
\fi
\ifx\subparagraph\undefined\else
  \let\oldsubparagraph\subparagraph
  \renewcommand{\subparagraph}{
    \@ifstar
      \xxxSubParagraphStar
      \xxxSubParagraphNoStar
  }
  \newcommand{\xxxSubParagraphStar}[1]{\oldsubparagraph*{#1}\mbox{}}
  \newcommand{\xxxSubParagraphNoStar}[1]{\oldsubparagraph{#1}\mbox{}}
\fi
\makeatother

\usepackage{longtable,booktabs,array}
\usepackage{calc} 
\usepackage{etoolbox}
\makeatletter
\patchcmd\longtable{\par}{\if@noskipsec\mbox{}\fi\par}{}{}
\makeatother
\IfFileExists{footnotehyper.sty}{\usepackage{footnotehyper}}{\usepackage{footnote}}
\makesavenoteenv{longtable}
\usepackage{graphicx}
\makeatletter
\def\maxwidth{\ifdim\Gin@nat@width>\linewidth\linewidth\else\Gin@nat@width\fi}
\def\maxheight{\ifdim\Gin@nat@height>\textheight\textheight\else\Gin@nat@height\fi}
\makeatother
\setkeys{Gin}{width=\maxwidth,height=\maxheight,keepaspectratio}
\makeatletter
\def\fps@figure{htbp}
\makeatother

\makeatletter
\@ifpackageloaded{caption}{}{\usepackage{caption}}
\AtBeginDocument{%
\ifdefined\contentsname
  \renewcommand*\contentsname{Table of contents}
\else
  \newcommand\contentsname{Table of contents}
\fi
\ifdefined\listfigurename
  \renewcommand*\listfigurename{List of Figures}
\else
  \newcommand\listfigurename{List of Figures}
\fi
\ifdefined\listtablename
  \renewcommand*\listtablename{List of Tables}
\else
  \newcommand\listtablename{List of Tables}
\fi
\ifdefined\figurename
  \renewcommand*\figurename{Figure}
\else
  \newcommand\figurename{Figure}
\fi
\ifdefined\tablename
  \renewcommand*\tablename{Table}
\else
  \newcommand\tablename{Table}
\fi
}
\@ifpackageloaded{float}{}{\usepackage{float}}
\floatstyle{ruled}
\@ifundefined{c@chapter}{\newfloat{codelisting}{h}{lop}}{\newfloat{codelisting}{h}{lop}[chapter]}
\floatname{codelisting}{Listing}

\makeatother
\makeatletter
\@ifpackageloaded{caption}{}{\usepackage{caption}}
\@ifpackageloaded{subcaption}{}{\usepackage{subcaption}}
\makeatother

\IfFileExists{xurl.sty}{\usepackage{xurl}}{} 
\hypersetup{
  pdftitle={Beyond Vintage Rotation: Bias-Free Sparse Representation Learning with Oracle Inference},
  pdfauthor={Chengyu Cui; Yunxiao Chen; Jing Ouyang; Gongjun Xu},
  pdfkeywords={Latent embedding; Rotation method; Folded concave loss; Non-convex optimisation; Local quadratic approximation.},
  colorlinks=true,
  linkcolor={blue},
  filecolor={Maroon},
  citecolor={Blue},
  urlcolor={Blue},
  pdfcreator={LaTeX}}

\begin{document}
\addtocontents{toc}{\protect\setcounter{tocdepth}{-1}}

\def\spacingset#1{\renewcommand{\baselinestretch}%
{#1}\small\normalsize} \spacingset{1}

\title{Beyond Vintage Rotation: Bias-Free Sparse Representation Learning with Oracle Inference}
\author{Chengyu Cui$^1$, Yunxiao Chen$^2$, Jing Ouyang$^3$, and Gongjun Xu$^1$\\
\small 1. Department of Statistics, University of Michigan, Ann Arbor, Michigan, USA\\
\small 2. Department of Statistics, London School of Economics and Political Science, London, UK\\
\small 3. Faculty of Business and Economics, University of Hong Kong, Hong Kong
}
\date{}
\maketitle

\spacingset{1.2}
\begin{abstract}
Learning low-dimensional latent representations is a central topic in statistics and machine learning, and rotation methods have long been used to obtain sparse and interpretable representations. Despite nearly a century of widespread use across many fields, rigorous guarantees for valid inference for the learned representation remain lacking. In this paper, we identify a surprisingly prevalent phenomenon that suggests a reason for this gap: for a broad class of vintage rotations, the resulting estimators exhibit a non-estimable bias. Because this bias is independent of the data, it fundamentally precludes the development of valid inferential procedures, including the construction of confidence intervals and hypothesis testing.
To address this challenge, we propose a novel bias-free rotation method within a general representation learning framework based on latent variables.
We establish an oracle inference property for the learned sparse representations: the estimators achieve the same asymptotic variance as in the ideal setting where the latent variables are observed. To bridge the gap between theory and computation, we develop an efficient computational framework and prove that its output estimators retain the same oracle property. Our results provide a rigorous inference procedure for the rotated estimators, yielding statistically valid and interpretable representation learning.
\end{abstract}


\spacingset{1.19}
\newpage

\section{Introduction}\label{sec_intro}
Learning latent representations is an important topic in statistical and machine learning. It aims to explain the complex dependence structures in high-dimensional observations via a low-dimensional latent representation/embedding. 
In statistical terms, given an observed response vector $\bY\in\RR^{q}$, we posit a low-dimensional latent variable (or embedding) $\bz = (z_1,\dots,z_r)^\T\in\RR^{r}$. 
The conditional distribution of $\bY$ given $\bz$ is specified as
\begin{equation}
    \bY\mid\bz \;\sim\; \cP(\cdot\mid\lambda(\bz)),\label{eq_general}
\end{equation}
where $\cP(\cdot\mid\lambda)$ denotes a family of distributions indexed by $\lambda$, and $\lambda(\cdot)$ is a representation map that carries the latent variable $\bz$ to the index. In this paper, we consider a linear map $\lambda(\bz) = \bA\bz$ for some representation matrix $\bA\in\RR^{q\times r}$, with each component of $\bA\bz$ governing the conditional distribution of the corresponding component of $\bY$. This framework is appealing for its interpretability: for $j\in\{1,\dots,q\}$ and $l\in\{1,\dots,r\}$, the $(j,l)$-th entry of $\bA$ quantifies the influence of the latent variable $z_{l}$ on the $j$-th observed variable.

The framework~\eqref{eq_general} accommodates a broad class of models used for representation learning. One popular example is the linear latent variable model $\bY = \bA\bz + \bE$, where $\bE$ denotes additive noise. This specification covers, among others, linear factor models~\citep{bai2003inferential,bai2012statistical}, {independent component analysis~\citep{hyvarinen2001independent},} stochastic block models~\citep{abbe2018community,mao2021estimating}, and latent class models~\citep{hagenaars2002applied,lyu2025degree}. Beyond the linear setting, the framework also covers generalised latent variable models~\citep{skrondal2004generalized,bartholomew2011latent} and nonlinear factor models~\citep{fernandez2016individual,chen2020structured,chen2021nonlinear,wang2022maximum,chen2025item}, where suitable link functions accommodate binary, ordinal, count, or compositional data. More broadly, our framework~\eqref{eq_general} accommodates representation learning approaches that do not specify the parametric form of $\cP(\cdot|\cdot)$ but focus on optimising certain loss functions~\citep{udell2016generalized,chi2019nonconvex}. See Examples~\ref{example_1}--\ref{example_3} in Section~\ref{sec_setup} for details.

It is often of interest to learn a sparse representation matrix $\bA$. This idea can be traced back to \citet{thurstone1947vectors} in psychological measurement, where it is argued that a ``scientifically meaningful'' latent structure should exhibit a sparse pattern in the representation map. Specifically, in \eqref{eq_general} with $\lambda(\bz) = \bA\bz$, sparsity in $\bA$ implies that many observed responses depend on only a small subset of the latent variables. This enhances the interpretability of the latent structure, as the resulting mapping clarifies the relationship between latent variables and groups of observed responses, making the latent variables easier to interpret and associate with concrete scientific meanings.
This conceptually simple principle has been widely adopted to learn interpretable latent structures from data, with applications across psychology, education, genetics, and social sciences~\citep{andersonintroduction,mulaik2009foundations,bartholomew2011latent,bengio2013representation,kline2023principles}.

To learn a sparse representation matrix $\bA$, rotation methods are widely used. They select among equivalent representations the one that minimises a certain rotation criterion. In particular, one starts with an initial estimate $\hat\bA$, and then searches over rotated versions $\hat\bA\bG$ to find the minimiser of a function $Q(\bA):\RR^{q\times r}\to \RR$:
\begin{equation}{
    \bA^{\natural} = \argmin_{\bA \in\{ \hat\bA\bG:\bG\in\Xi\}}Q(\bA),\label{eq_rotation_intro}}
\end{equation}
where $\Xi\subset\RR^{r\times r}$ denotes the set of admissible rotation matrices (see Section~\ref{sec_setup_rotate} for details). 
The function $Q(\bA)$ is referred to as {\it rotation criteria} and is designed to measure the sparsity/complexity of $\bA$, so that the solution {$\bA^{\natural}$} in \eqref{eq_rotation_intro} is as sparse as possible~\citep[see][for an overview of rotation methods]{browne2001overview}, and we refer to the rotation methods surveyed there as \emph{vintage} to emphasise their early development and longstanding use in applied work. 
Despite their popularity in applications, the statistical properties of these rotation methods remain poorly understood, aside from very recent results regarding the consistency of point estimation under varimax~\citep{Rohe2020VintageFA,bing2023optimal}. We attribute this gap to a prevalent issue of vintage rotation methods: they introduce a non-estimable and non-negligible bias in {$\bA^{\natural}$}. This deterministic bias makes it difficult to develop statistical guarantees for estimators obtained from vintage rotation methods, especially for valid statistical inference, as we explain below.



\subsection{Limitations of Vintage Rotations}\label{sec_sub_limitation}

\begin{figure}
    \centering
    \includegraphics[width=0.9\linewidth]{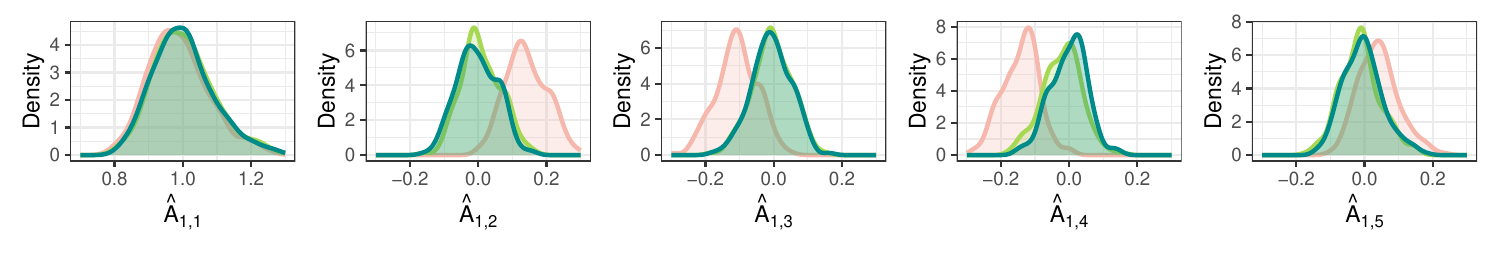}\includegraphics[width=0.1\linewidth]{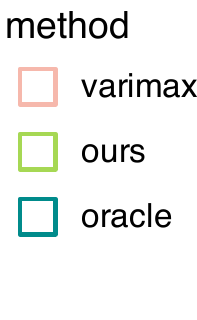}
        \caption{\small Histogram of estimates for the first row of $\bA^*$, given as $(1,0,0,0,0)$, from different methods, with each component shown in a separate panel. The estimates for the remaining rows exhibit a similar pattern, as reported in Section~\ref{subsec_simu}. Here, ``varimax'' denotes the varimax estimator; ``ours'' denotes the Folomin estimator with the minimax concave penalty (MCP) loss (see details in Section~\ref{sec_main}); and ``oracle'' denotes the estimator computed with the latent variables observed. The full simulation study is given in Section~\ref{subsec_simu}.} 
    \label{fig_oracle}
\end{figure}

Many smooth vintage rotation methods induce an unestimable and non-negligible bias. Specifically, under a broad range of settings, these rotation criteria do not admit the true sparse matrix $\bA^*$ as the optimiser. In such cases, even if \eqref{eq_rotation_intro} is initialised at $\bA^*$, the rotated solution {$\bA^{\natural}$} differs from $\bA^*$, i.e., {$\bA^{\natural} \neq \bA^*$}. The discrepancy {$\bA^{\natural} - \bA^*$} is not estimable from the data as it depends only on the underlying $\bA^*$.
Moreover, the discrepancy can be non-negligible in the sense that its magnitude can be comparable to the sample-level estimation error. 
Figure~\ref{fig_oracle} illustrates this bias via a numerical study: although varimax provides a fairly accurate point estimate, it exhibits a clear bias, whereas our method yields an accurate, bias-free estimate and matches the oracle estimation error obtained when the latent variables are assumed to be {\it known}.




This bias makes it particularly challenging to establish statistical guarantees for the rotation methods. For the widely used varimax rotation, only very recently have \citet{Rohe2020VintageFA} and \citet{bing2023optimal} established consistency by showing that the bias vanishes to $0$ as $q$ grows under certain distributional assumptions on $\bA^*$ (see Remark~\ref{remark_bias_magni} for further discussions). These results, however, guarantee point estimation only. Valid inference, including the construction of confidence intervals and hypothesis testing, is generally precluded because the bias is deterministic and can be of the same order as the estimation error. 
Moreover, due to the deterministic nature of the bias, it cannot be corrected by any data-driven debiasing procedure with the point estimation, for varimax or other vintage rotation criteria.

{ Another outstanding limitation is that statistical guarantees for rotation methods allowing correlated latent variables remain limited.}
While varimax assumes uncorrelated latent variables, correlated ones are frequently encountered and preferred in many applications~\citep[e.g.][]{thurstone1947vectors,kline2023principles,cui2025identifiability}. However, allowing correlations among latent variables introduces a major challenge of latent variable collapse, in which the latent variables can become highly collinear after rotation~\citep{browne2001overview}. This potential for collinearity leads to numerical instability in computing the rotation and greatly complicates theoretical analysis.
Due to these challenges, to our knowledge, no existing method can provably recover sparse representations when latent variables are correlated.


\subsection{Our Contributions}

In this paper, we address the challenges by proposing a novel rotation method, {FO}lded LOss MINimisation (Folomin), which enables bias-free representation learning with correlated latent variables. We refer to the resulting estimator as the Folomin estimator and establish both theoretical and computational guarantees. Our contributions are summarised as follows.


\begin{enumerate}[label=(\alph*), leftmargin=0.5cm]
\item\; Our first contribution is to characterise the bias inherent in a broad class of rotation methods and to develop a novel, bias-free Folomin rotation. We show that any rotation method with smooth and symmetric criteria can incur an unestimable bias when recovering many practically relevant sparse representations. To address this issue, we introduce a general class of folded loss functions with a non-smooth peak at zero, which includes the folded concave penalties popularly used in high-dimensional regression~\citep{fan2001variable,zhang2010nearly,fan2011nonconcave,shen2012likelihood}.
We further show that, with the proposed loss, our method is bias-free in the sense that it can exactly recover the true representation matrix at a population level for many sparse structures common in applications.

    \item\; Our second contribution is to establish the oracle inference properties for the Folomin estimator. We say that an estimator of $\bA$ has the oracle inference property if its asymptotic distribution coincides with that of the oracle estimator that would be available if the latent variables were observed, thereby enabling valid statistical inference on $\bA$.
    We also derive an analogous oracle inference property for the latent variables, which are treated as model parameters in our setting. To the best of our knowledge, this is the first work to resolve the problem of enabling valid statistical inference for the sparse matrix $\bA$ under a general setup, in contrast to existing results that only allow inference identified up to an unknown rotation~\citep{bai2003inferential,fan2016projected,chen2019inference,wang2022maximum}. Our results enable practically meaningful and scientifically interpretable inference, supporting downstream tasks such as simultaneous inference on multiple entries and hypothesis tests comparing alternative sparse structures.
    \item\; The third contribution concerns computation.  Computing the Folomin estimator requires solving a non-convex, nonlinear, and non-smooth optimisation problem. To ensure that the estimator is numerically attainable and that our theory applies to algorithm output, we develop an efficient computational framework that yields a computable approximation to the Folomin estimator with the same oracle inference property. Specifically, to address the non-convexity and avoid latent variable collapse, we first construct a consistent initialiser to localise the analysis. 
    We then develop a local quadratic approximation (LQA) scheme to handle the non-smooth objective with nonlinear constraints. We show that a single LQA update produces an estimator with the oracle property, providing the computational guarantee for the proposed method. 
    
\end{enumerate}

 The rest of the article is organised as follows. Section~\ref{sec_setup} introduces the representation learning setup and formalises the bias of vintage rotation methods. Section~\ref{sec_main} presents the proposed Folomin estimator. Section~\ref{sec_oracle_property} establishes its oracle inference property, and Section~\ref{sec_comp_guarantee} develops an efficient computational framework with provable guarantees for the algorithm output. Section~\ref{sec_empirical} presents simulation studies and an analysis of a personality dataset. Section~\ref{sec_conclude} concludes. Additional numerical studies and proofs are provided in the Supplementary Material.

\noindent{\bf Notation.} 
For any integer $N$, let $[N] = \{1, \dots, N\}$. For any $a,b\in\RR$, let $a\vee b = \max(a,b)$ and $a\wedge b = \min(a,b)$. 
For $\bx = (x_1,\dots,x_n)^\T\in\mathbb{R}^n$, let $\|\bx\| := (\sum_{i=1}^nx_i^2)^{1/2}$ and $\|\bx\|_{\infty} := \max_{1\le i\le n}|x_i|$. For $\bx \in \mathbb{R}^n$ and $\cS \subseteq [n]$, let $\bx_{\cS}\in\RR^{|\cS|}$ denote the sub-vector indexed by $\cS$, with $|\cS|$ denoting the cardinality. For $\bM = (M_{ij})_{n\times m}$ and index sets $\cS_1\subseteq[n], \cS_2 \subseteq [m]$, let $\bM_{\cS_1,\cS_2}\in\RR^{|\cS_1|\times|\cS_2|}$ denote the corresponding sub-matrix. We write $\bM_{,\cS_2}$ when $\cS_1=[n]$ and $\bM_{\cS_1,}$ when $\cS_2=[m]$. We use the Frobenius norm $\| \bM\|_{\Fn} := (\sum_{i=1}^n\sum_{j=1}^mM_{ij}^2)^{1/2}$, the operator norm $\|\bM\| := \sup_{\|\bx\|=1}\|\bM\bx\|$, and the two-to-infinity norm $\| \bM\|_{2\to\infty} := \sup_{\|\bx\| = 1} \|\bM\bx\|_{\infty}$. For a square matrix $\bM\in \mathbb{R}^{n \times n}$, denote $\lambda_{i}(\bM)$ as the $i$-th largest eigenvalue of $\bM$. We write $\bI_n$ for the $n\times n$ identity matrix and $\zero_n$ for the zero vector in $\RR^n$.

\section{Problem Setup}\label{sec_setup}
We formalise the representation learning framework in~\eqref{eq_general}. Suppose there are $n$ subjects, and for each $i \in [n]$, let $\bz_i=(z_{i1},\dots,z_{ir})^\T$ denote the latent variable associated with subject $i$ with the matrix form $\bZ = (\bz_1,\cdots,\bz_n)^\T$. For the representation matrix $\bA\in\RR^{q\times r}$, we write $\bA = (\ba_1,\cdots,\ba_q)^\T$ with $\ba_j = (a_{j1},\dots,a_{jr})^\T$ for each $j\in[q]$. For each $i \in [n]$, we observe a $q$-dimensional response vector $\bY_i = (Y_{i1},\dots,Y_{iq})^\T$. 
In \eqref{eq_general}, the distribution of each $Y_{ij}$ depends on the $j$-th component of $\bA\bz_i$, denoted by $\theta_{ij} := \ba_j^\T\bz_i$. Accordingly, we posit that $Y_{ij}$ follows a distribution $\cP_{\theta_{ij}}$ from some parametric family $\{\cP_{\theta} : \theta \in \Theta\}$ indexed by scalar $\theta_{ij}$. The family may also depend on $j$ (for example, $\{\cP^{(j)}_{\theta}\}$), but we suppress this dependence in the notation for simplicity. Given $\{\theta_{ij}\}_{i\in[n],j\in[q]}$, we assume that $\{Y_{ij}\}_{i\in[n],j\in[q]}$ are independent.

We consider a general risk function $\ell:\RR\times\RR\to\RR$ instead of assuming a specific form for $\cP_{\theta}$. We refer to $\ell$ as the risk function, reserving the term loss function exclusively for describing the rotation criteria. For each response $Y_{ij}$, $\ell(\theta;Y_{ij})$ measures the fit of the parameter $\theta$ to the observation $Y_{ij}$. For notational convenience, we write $\ell_{ij}(\theta) = \ell(\theta;Y_{ij})$. The true parameter is then defined as a minimiser of the population risk as
\begin{equation*}
    (\bZ^*,\bA^*) ~\in ~\mathop{\arg\min}_{\bZ\in\RR^{n\times r},\bA \in \RR^{q\times r}}\bar\cL(\bZ,\bA)~ =~ \mathop{\arg\min}_{\bZ\in\RR^{n\times r},\bA \in \RR^{q\times r}}\EE_{(\bZ^*,\bA^*)}\big[\sum_{i=1}^n\sum_{j=1}^q\ell_{ij}(\ba_j^\T\bz_i)\big],
\end{equation*}
where $\bar\cL(\bZ,\bA)$ denotes the population risk, and the expectation is taken with respect to the joint distribution of $\{Y_{ij}\}_{i\in[n],j\in[q]}$. We treat $\bZ$ and $\bA$ as model parameters. 
Without loss of generality, we impose that $\bZ^*$ has normalised columns for identifiability, i.e., $\mathrm{diag}(n^{-1}\bZ^*{}^\T\bZ^*) = \bI_r$, where $\mathrm{diag}(\bM)$ denotes the diagonal matrix formed from the diagonal entries of $\bM$. The true representation matrix $\bA^*$ is assumed to be sparse, with a formal definition given in Section~\ref{sec_sparse_def}. The number of latent dimensions $r$ is assumed to be known and fixed, and $\bZ^*$ and $\bA^*$ are full column-rank. In practice, $r$ can be selected using information criterion-based methods~\citep{bai2002determining,chen2022determining}. Several examples of this formulation are given below.

\begin{example}\label{example_1}
    Linear latent variable models are one widely used special case of the considered framework~\citep{andersonintroduction,bai2003inferential,izenman2008modern}. In this setup, each response satisfies $Y_{ij} = \theta_{ij} + \epsilon_{ij}$ where $\epsilon_{ij}$ is additive noise. The associated risk function is often taken to be the least squares risk: $\ell_{ij}(\theta) = (\theta - Y_{ij})^2$~\citep{joreskog1972factor}.
\end{example}
\begin{example}\label{example_gfm}
    The generalised latent variable model extends the linear latent variable model to handle diverse response types within the generalised linear model (GLM) framework~\citep{skrondal2004generalized,bartholomew2011latent}. In this setup, each response $Y_{ij}$ is assumed to follow an exponential family distribution with conditional density $p(Y_{ij};\theta_{ij}) = g(\theta_{ij}|Y_{ij})$ for some prespecified link function $g(\cdot|\cdot)$ chosen to match the data type (e.g., logistic for binary data or Poisson for count data). The risk function is often taken as the negative log-likelihood: $\ell_{ij}(\theta) = -\log p(Y_{ij};\theta)$.
\end{example}
\begin{example}\label{example_3}
Our framework also accommodates risk functions beyond likelihood-based specifications. For instance, the generalised low-rank models are often formulated by minimising certain risk functions without committing to a fully specified distribution~\citep{udell2016generalized}. Common choices include the $\ell_1$ loss~\citep{candes2011robust} and loss in the exponential family~\citep{collins2001generalization}.
Similar loss-based formulations are also common in the low-rank factorisation literature~\citep{srebro2004maximum,chen2015fast,chi2019nonconvex}. 
\end{example}

\begin{remark}
In some applications~\citep{reckase2009,bai2012statistical,chen2019joint,Rohe2020VintageFA}, it is common to include a row-specific intercept $\beta_{0j}$, specifying $\theta_{ij}$ as $\theta_{ij} = \ba_j^\T\bz_i + \beta_{0j}$.  In our setup, this can be realised by setting $z_{i1}=1$ for all $i\in[n]$, in which case $a_{j1}$ plays the role of the intercept. Introducing these intercepts does not affect the analysis of the rotation. For ease of presentation, we therefore focus on the no-intercept formulation. In some settings, a subject-specific intercept $\beta_{i0}$ is also included, yielding $\theta_{ij} = \ba_j^\T\bz_i + \beta_{0j}+\beta_{i0}$, and the same discussion applies. 

\end{remark}

\subsection{Sparse Representation}\label{sec_sparse_def}
In this paper, we introduce a practically motivated sparsity notion that is typically satisfied in real-world settings and consistent with existing assumptions in the literature.
Following the literature~\citep{jennrich2006rotation,bing2020adaptive,Liu2023RotationTS}, we introduce the notion of {\it simple} rows in $\bA$. For each $l\in[r]$, define the simple rows associated with dimension $l$ as $\cS_l(\bA) = \{j\in[q]:\ba_{j} = \nu_j\be_l\text{ for some }\nu_j\neq 0\}$, where $\be_l\in\RR^{r}$ is the $l$-th standard basis vector { and scalar $\nu_j$ may vary across rows}. Let $\cS(\bA) = \cup_{l\in[r]}\cS_l(\bA)$ denote the collection of all simple rows. 
To quantify angular similarity among nonzero rows, for $\epsilon\ge 0$ and $\ba\in\RR^r\setminus\{\zero_r\}$, define the $\epsilon$ cone neighbourhood (index set) at $\ba$ as \begin{equation*}\cC_{\epsilon}(\ba;\bA) ~=~ \big\{j\in[q]:\|\ba_{j}\|\neq 0\text{ and }|\cos\angle(\ba,\ba_{j})|\ge 1-\epsilon\big\},\end{equation*} 
where $\cos\angle(\bx,\by) =\bx^\T\by/(\|\bx\|\|\by\|) $. For a simple row $j\in\cS_l(\bA)$, we have $\cC_{0}(\ba_j;\bA) = \cS_l(\bA)$. If $\ba = \zero_r$, we set $\cC_{\epsilon}(\zero_r;\bA) = \varnothing$ for all $\epsilon\ge0$. We introduce the following definition.
\begin{definition}\label{def_sparsity}\it We say $\bA$ is $(\lambda,\epsilon)$-sparse if it satisfies $ \min\{|a_{jl}|:a_{jl} \neq 0,j\in[q],l\in[r]\}\ge \lambda$ and 
\begin{equation}\max_{j\in[q]\setminus\cS(\bA)}\big|\cC_{\epsilon}(\ba_j;\bA)\big|~<~ \min_{j\in\cS(\bA)}\big|\cC_{0}(\ba_j;\bA)\big| ~=~  \min_{k\in[r]}\big|\cS_k(\bA)\big|.\label{eq_angular_sep}
\end{equation}
In addition, the row norms are uniformly bounded: $\|\bA\|_{2\to\infty}\le M$ for some constant $M>0$.
\end{definition} 

The notion of $(\lambda,\epsilon)$-sparsity requires that all nonzero entries of $\bA$ have magnitude at least $\lambda$, and that there are enough simple rows for each dimension so that, for some $\epsilon>0$, $\min_{l\in[r]}|\cS_l(\bA)|$ exceeds the size of the $\epsilon$ cone neighbourhood at any non-simple row. Both $\lambda$ and $\epsilon$ may vary~with~$q$. {One notable feature of this sparsity definition is that it accommodates heterogeneous simple rows. That is, simple rows are allowed to have different magnitudes in their nonzero entries, similar to the multiple parallel-row structure considered in \citet{bing2023detecting}.} 

The presence of simple rows is a natural and commonly adopted assumption in latent variable models~\citep{jennrich2006rotation,trendafilov2014simple,xu2017,bing2020adaptive}. A popular example is the \emph{perfectly simple structure}, where all $q$ rows are simple, i.e., $\cS(\bA)=[q]$~\citep{thurstone1947vectors}. 
Similar assumptions appear across diverse fields, including the anchor word condition in topic models~\citep{donoho2003does}, the completeness assumption in cognitive diagnosis models~\citep{chen2015statistical,gu2023joint}, and the pure node assumption in mixed membership stochastic block models~\citep{mao2021estimating,jin2024mixed}.


Geometrically, the dominance of simple rows in Definition~\ref{def_sparsity} can be interpreted as ensuring the identification of $r$ unique ``radial streaks'' as the axes~\citep{thurstone1947vectors,Rohe2020VintageFA,bing2023detecting}. When $\bA$ is $(\lambda,\epsilon)$-sparse, its rows concentrate around $r$ dominant directions in $\RR^r$, namely the directions spanned by the simple rows indexed by $\cS(\bA) = \cup_{l\in[r]}\cS_l(\bA)$, thereby forming $r$ distinct streaks in this space. 
Consequently, among all alternative representation matrices in $\{\bA\bG:\bG\text{ is invertible}\}$, $\bA$ is uniquely identified as the desired sparse representation that achieves the required dominance of simple rows, up to a column-wise scaling and permutation. Thus, the $(\lambda,\epsilon)$-sparsity can be regarded simply as an axis identification condition, which is practically mild and aligns with the prevailing emphasis on simple-row structure in the literature~\citep{harman1976modern,mulaik2009foundations}.

\begin{remark}\label{remark_distributional}
Our definition of $(\lambda,\epsilon)$-sparsity can be satisfied with certain distributional assumptions on $\bA$. 
Specifically, in Section~\ref{supp_sec_distributional} of the Supplementary Material, we show that, if the entries of $\bA$ are i.i.d. from a distribution that places probability at least $1/2$ on zero and satisfies some mild regularity conditions, for some $\epsilon>0$ and $\lambda>0$, $\bA$ is $(\lambda,\epsilon)$-sparse with high probability.
    
\end{remark}

\subsection{Bias in Existing Rotation Methods}\label{sec_setup_rotate}

Now, we characterise the non-estimable bias in smooth vintage rotation methods. To emphasise its deterministic and data-independent nature, we first consider an ideal setting where the population risk $\bar\cL(\bZ,\bA)$ is known, and we obtain its minimiser as $(\bar\bZ,\bar\bA)$. 
Note that $\bar\cL(\bZ,\bA) $ can be written as $ \sum_{i\in[n]}\sum_{j\in[q]}\EE_{\theta^*}[\ell_{ij}(\ba_j^\T\bz_i)]$ with each expectation taken with respect to $Y_{ij}$.
Under mild regularity conditions (see Assumption~\ref{assump_limit} in Section~\ref{sec_oracle_property}), for $i\in[n]$ and $j\in[q]$, the minimiser of each $\EE_{\theta^*}[\ell_{ij}(\theta)]$ is unique and equals $\theta_{ij}^* := \ba_j^*{}^\T\bz_i^*$, and thus $\bar\bZ\bar\bA^\T = \bZ^*(\bA^*)^\T$. For instance, when $\ell_{ij}(\theta) = -\log p(Y_{ij};\theta)$ as in Example~\ref{example_gfm}, this uniqueness follows from the standard Kullback--Leibler divergence argument~\citep{van2000asymptotic}. Then, the minimiser $(\bar\bZ,\bar\bA)$ equals $(\bZ^*,\bA^*)$ up to an invertible transformation: 
\begin{equation}
   (\bZ^*, \bA^*) ~=~ \big(\bar\bZ\bar\bG^\T,\bar\bA\bar\bG^{-1}\big)\text{, for some invertible }\bar\bG\in\RR^{r\times r} .
   \label{eq_start_rotation}
\end{equation}
{ We impose no constraint on $\bar\bA$ and, for notational simplicity, adopt the normalisation $n^{-1}\bar\bZ^\T\bar\bZ=\bI_r$.} 
 Given this minimiser $(\bar\bZ,\bar\bA)$, a rotation method with criterion $Q(\cdot)$ computes 
\begin{equation}{
    \bG^{\rm opt} ~=~ \argmin_{\bG\in\Xi}Q(\bar\bA\bG^{-1}),\label{eq_general_rotation}}
\end{equation} and takes {$(\bar\bZ(\bG^{\rm opt})^\T,\bar\bA(\bG^{\rm opt})^{-1})$} as the solution. The admissible rotation set $\Xi$ is usually specified in two ways, leading to {\it oblique} and {\it orthogonal} rotations. In oblique rotations, we take $\Xi=\{\bG:\mathrm{diag}(\bG\bG^\T)= \bI_r\}$, so that the rotated latent variables {$\bar\bZ(\bG^{\rm opt})^\T$} remain normalised but may be correlated. Specifically, the diagonal entries of {$n^{-1}\{\bar\bZ(\bG^{\rm opt})^\T\}^\T\bar\bZ(\bG^{\rm opt})^\T = \bG^{\rm opt}(\bG^{\rm opt})^\T$} equal $1$ and its off-diagonal entries are unconstrained.
In orthogonal rotations, we take $\Xi=\{\bG:\bG\bG^\T = \bI_r\}$, in which case the rotated latent variables satisfy {$n^{-1} \{\bar\bZ(\bG^{\rm opt})^\T\}^\T\bar\bZ(\bG^{\rm opt})^\T = \bI_r$}, and hence are constrained to be uncorrelated. In orthogonal rotations, the true latent variables are assumed to be orthogonal, i.e., $n^{-1}(\bZ^*)^\T\bZ^* = \bI_r$. { The rotation method can be extended without normalisation $n^{-1}\bar\bZ^\T\bar\bZ=\bI_r$ by replacing the oblique feasible set with $\{\bG:{\rm diag}\{\bG(n^{-1}\bar\bZ^\T\bar\bZ)\bG^{\T}\} = \bI_r\}$ and orthogonal feasible set with $\{\bG:\bG(n^{-1}\bar\bZ^\T\bar\bZ)\bG^{\T}= \bI_r\}$ for the corresponding rotation. }

To recover $(\bZ^*,\bA^*)$ via the rotation method, it is natural to require that $\bar\bG$ is at least a local optimiser of \eqref{eq_general_rotation}. Otherwise, even if {$\bG^{\rm opt}$} is chosen as a local optimiser of \eqref{eq_general_rotation}, we would have {$\bA^* = \bar\bA\bar\bG^{-1}\neq \bar\bA(\bG^{\rm opt})^{-1}$}, which induces bias in the recovered sparse representation matrix. We formalise this with the following definition.

\begin{definition}\label{def_local_id}
We say a rotation method is rotationally bias-free if, there exist some $\lambda_0 > 0$ and $\epsilon_0 > 0$ such that, for any $(\lambda_0,\epsilon_0)$-sparse $\bA^*$, given $(\bar\bZ,\bar\bA)$ in \eqref{eq_start_rotation}, $\bar\bG$ is a strict local optimum of \eqref{eq_general_rotation}. Otherwise, the method is rotationally biased.
\end{definition}
We require this optimality to hold for the $(\lambda_0,\epsilon_0)$-sparse family for one pair $(\lambda_0,\epsilon_0)$, so that the rotation method is guaranteed to recover at least this class of sparse matrices.
For a rotationally bias-free rotation method, it yields $\bar\bG$ as a strict local optimum in \eqref{eq_general_rotation}, so $(\bZ^*,\bA^*)$ can be exactly recovered, either by starting with a proper initialisation close to $\bar\bG$ or by enumerating all local minima of \eqref{eq_general_rotation}.
In contrast, for a rotationally biased rotation method, there exists a $(\lambda_0,\epsilon_0)$-sparse $\bA^*$ such that every local optimum it attains differs from $\bA^*$ by a nonzero deterministic bias, which depends on the underlying structure of $\bA^*$. The deterministic nature of this bias makes it non-estimable from the data and therefore cannot be corrected.


\begin{remark}
\it We do not require global optimality in Definition~\ref{def_local_id}. First, the $(\lambda_0,\epsilon_0)$-sparse matrix of interest is defined in terms of the presence of simple rows, rather than as the global minimiser of any particular rotation criterion. This notion is both more natural and more consistent with the literature, with a clearer geometric interpretation.
Practically, the optimisation problem~\eqref{eq_general_rotation} can be highly non-convex with nonlinear constraints, making its global minimiser computationally intractable in general. By contrast, local optima are often attainable with suitable initialisation, as in the computational framework developed later in Section~\ref{sec_comp_guarantee}.
\end{remark}



Next, we show that a general class of vintage rotation methods is rotationally biased.
\begin{proposition}[{Bias in Vintage Rotations}]\it\label{prop_anti_smooth}
    Consider an orthogonal/oblique rotation method with criterion $Q(\cdot)$. If $(i)$ $Q(\cdot)$ is entry-wise differentiable, i.e., $\partial_{jl}Q(\bA) := \partial_{a_{jl}}Q(\bA)$ exists for any $j\in[q]$ and $l\in[r]$ and $(ii)$ $\partial_{jl}Q(\bA)\big|_{a_{jl} = 0} = 0$, then the method is rotationally biased.  
    \end{proposition}
The requirement in the proposition holds for most existing vintage rotation methods, which typically involve smooth and entry-wise symmetric criteria~\citep{kaiser1958varimax,yates1987multivariate,browne2001overview}.
Indeed, these requirements follow naturally from the smoothness and symmetry. Specifically, these criteria usually treat positive and negative entries of the same magnitude equally: if $\bA^{\prime}$ is obtained from $\bA$ by flipping the sign of the $(j,l)$-th entry, then $Q(\bA)=Q(\bA^{\prime})$. When $Q(\cdot)$ is smooth (entry-wise differentiable), this symmetry implies $\partial_{jl}Q(\bA)|_{a_{jl}=0}=0$.

{ The intrinsic bias of vintage rotation methods arises from their flatness at zero entries. To see this, suppose $\bar\bA=\bA^*$ in \eqref{eq_general_rotation} and $\bA^*$ is $(\lambda,\epsilon)$-sparse. When $\bG$ is perturbed slightly away from $\bI_r$, the zero entries of $\bA^*$ contribute little to the first-order change of the criterion because $\partial_{jl}Q(\bA)$ vanishes, or is very small, near $a_{jl}^*=0$. The direction of the rotation is therefore mainly determined by the nonzero entries, which may favor a non-identity rotation unless their contributions cancel under special symmetry or balanced sparsity patterns. Moreover, the magnitude of this bias depends on the sparsity structure and may scale with $q$. However, for the bias to be negligible for asymptotic distributions, it would typically need to be $o(n^{-1/2})$, which is too restrictive and is generally not guaranteed.
 We provide the following example for the popular varimax criterion as an illustration, and further examine other vintage rotation methods through the numerical study in Section~\ref{supp_sec_bias} of the Supplementary Material.}



\begin{example}\it\label{prop_anti_varimax}
    The varimax rotation \citep{kaiser1958varimax} is an orthogonal rotation with criterion:
\begin{equation*}
{    Q_{\rm varimax}(\bA) = -q^{-1}\sum_{l=1}^r\sum_{j=1}^q\Big\{a_{jl}^4 - \big(q^{-1}\sum_{t=1}^qa_{tl}^2\big)^2\Big\}}.
\end{equation*}
It is easy to verify that $Q_{\rm varimax}(\cdot)$ satisfies the two requirements in Proposition~\ref{prop_anti_smooth}, and thus it is rotationally biased. In particular, consider $\bA^0 = (\bA_1^\T,\cdots,\bA_r^\T,\bA_{r+1}^\T)^\T$, with $\bA_{l} := \one_{q_l}\otimes \be_l^\T$ for $l\in[r]$ where $\one_{q_l}$ denotes the all-ones vector in $\RR^{q_l}$ and $\otimes$ denotes the Kronecker product, and $\bA_{r+1} := (\nu_{1},\dots,\nu_{r})$ where for some $k,l\in[r]$, $\nu_l \nu_k \neq 0$ and $\nu_{k}^2 - \nu_{l}^2 \neq (q_k - q_l)/(q-1)$. Then $\bI_r$ is not a stationary point of  {$Q_{\rm varimax}(\bA^0\bG)$} within $\{\bG\bG^\T = \bI_r\}$. { In practice, varimax is often applied after normalising the rows of the loading matrix to have unit $\ell_2$ norm~\citep{kaiser1958varimax}. Our construction also covers this normalised version by further requiring $\|\bA_{r+1}\|_2=1$, and thereby every row of $\bA^0$ exactly has unit $\ell_2$ norm. The same argument shows that the normalised procedure remains to be rotationally biased.
}

\end{example}

\begin{remark}\label{remark_bias_magni}
    \it The magnitude of the bias in the varimax rotation can be comparable to, or even larger than, the estimation error and is therefore non-negligible for valid inference. Specifically, \citet{Rohe2020VintageFA} established in Lemma G.7 that, under their distributional assumptions on $\bA^*$, the minimiser $\argmin_{\bA\in\{\bA^*\bG:\bG\bG^\T=\bI_r\}}Q(\bA)$ converges to $\bA^*$ at a rate of $O_p(q^{-1/4})$, which is substantially slower than the typical estimation error for $\bA^*$, of order $O_p(n^{-1/2} + q^{-1})$~\citep{cui2025identifiability}. 
    The magnitude and exact form of the bias depend on the structure of $\bA^*$ and the rotation criterion applied, and can vary substantially across settings. Section~\ref{subsec_simu} provides a numerical illustration for varimax, demonstrating that while the point estimation is accurate under the presence of this bias, it fundamentally precludes valid inference.


    
    
\end{remark}

\begin{remark}
    We remark that, besides the rotation methods, the literature also considers adding penalties to the empirical risk to encourage sparse estimation of the representation matrix~\citep{trendafilov2014simple,jin2018approximated}. However, compared with rotation methods, penalising the empirical risk can affect model fit and, more importantly, introduce additional bias in estimating $\bA$. As a result, inference typically requires careful de-biasing procedures~\citep{jin2018approximated}.

\end{remark}

\section{Proposed Method}\label{sec_main}

We introduce the proposed Folomin method. We consider the following rotation criterion:
\begin{equation}Q_{\rho_{\gamma}}(\bA) ~= ~\sum_{j=1}^q\sum_{l=1}^r\rho_{\gamma}(a_{jl}).\label{eq_solo}\end{equation} 
Here, $\rho_{\gamma}(\cdot)$ belongs to a class of folded concave losses satisfying the following conditions. 
\begin{cond}\label{assump_concave_function}
\spacingset{1.6} For some positive constants $a_0$, $a_1$, $a_2$, and $a_3$ with $a_1\le a_3$, the following hold   
    \begin{enumerate}[label=$(\roman*)$,leftmargin=0.6cm]
    \item\label{item_subdif}~ $\rho_{\gamma}(t)$ is symmetric around $0$ and is differentiable almost everywhere; 
   \item\label{item_nonsmooth}~  $\rho_{\gamma}^{\prime}(0+) = a_0\gamma$ for some $a_0>0$, 
    and in $(0,a_1\gamma]$, $\rho_{\gamma}^{\prime}(t)$ exists and is $a_2\gamma$-Lipschitz ;
    \item\label{item_concave}~ $\rho_{\gamma}(t)$ is nondecreasing and concave in $(0,\infty)$, with $\rho_{\gamma}^{\prime}(t) = 0$ for $t\in[a_3\gamma,\infty)$.
    \end{enumerate}
\end{cond}
 
Condition~\ref{assump_concave_function}\ref{item_subdif} preserves symmetry around zero and requires differentiability except at countably many points. Condition~\ref{assump_concave_function}\ref{item_nonsmooth} imposes a non-vanishing right derivative $\rho_{\gamma}^{\prime}(0+)$ and regularises the derivative in a neighbourhood of zero. This yields a non-smooth peak near zero. Condition~\ref{assump_concave_function}\ref{item_concave} formalises concavity and requires that the derivative vanishes beyond a certain threshold, so that changes in large entries in $\bA$ have a negligible impact on the value of the criterion. 
Popular examples include smoothly clipped absolute deviation (SCAD)~\citep{fan2001variable}, minimax concave penalty (MCP)~\citep{zhang2010nearly}, and truncated $\ell_1$ loss~\citep{shen2012likelihood}. 

Condition~\ref{assump_concave_function} ensures that $\rho_{\gamma}(\cdot)$ exhibits a non-smooth peak at zero, which can address the bias inherent in smooth rotation criteria. In particular, we show that for any $\rho_{\gamma}(\cdot)$ satisfying Condition~\ref{assump_concave_function}, both orthogonal and oblique rotations with criterion $Q_{\rho_{\gamma}}(\cdot)$ in \eqref{eq_solo} are rotationally bias-free.
\begin{proposition}[{Bias-free Rotation}]\it\label{prop_idtfblty}
Given any $\lambda_0>0$ and $\epsilon_0>0$, for  $0<\gamma\le\lambda_0/(a_3+1)$, rotation method with $Q_{\rho_{\gamma}}(\cdot)$ is rotationally bias-free under the following settings.

 \noindent{{\bf Orthogonal Setting.} Let $\bA^*$ be $(\lambda_0,\epsilon_0)$-sparse and $\bZ^*$ satisfy $n^{-1}\bZ^*{}^\T\bZ^* = \bI_r$. 
     Then for any $c \le \min(\gamma/M,1/2)$ with $M$ specified in Definition~\ref{def_sparsity}, it holds that
      \begin{equation}
          \bI_r ~= ~\mathop{\arg\min}_{\bG:\|\bG - \bI_r\|\le c,\; \bG\bG^\T = \bI_r}Q_{\rho_{\gamma}}(\bA^*\bG^{-1}).\label{eq_population_solo_ortho}
      \end{equation}
      This implies the orthogonal rotation method with $Q_{\rho_{\gamma}}(\cdot)$ is rotationally bias-free.}
      
\noindent {\bf Oblique Setting.} Let $\bA^*$ be $(\lambda_0,\epsilon_0)$-sparse and $\bZ^*$ satisfy $\mathrm{diag}(n^{-1}\bZ^*{}^\T\bZ^*) = \bI_r$. Define
      \begin{equation}
         \Xi(c,\bZ)~ := ~\Big\{\bG\in\RR^{r\times r}:\|\bG - \bI_r\|\le c\,\text{ and }\, \mathrm{diag}\big\{\bG(\Mff)\bG^\T\big\} = \bI_r\Big\}.\label{eq_true_local_region}
      \end{equation}
     Then for any $\gamma\le \lambda_0/(a_3+1)$ and any $c \le \min(\gamma/M,1)/2$ with $M$ specified in Definition~\ref{def_sparsity},
      \begin{equation}
          \bI_r ~= ~\mathop{\arg\min}_{\bG\in\Xi(c,\bZ^*)}Q_{\rho_{\gamma}}(\bA^*\bG^{-1}).\label{eq_population_solo}
      \end{equation}
      This implies that $\bI_r$ is a strict local minimiser, and thus the oblique rotation method with $Q_{\rho_{\gamma}}(\cdot)$ is rotationally bias-free. 

\end{proposition}
Proposition~\ref{prop_idtfblty} applies to any objective of the form \eqref{eq_solo} with $\rho_{\gamma}(\cdot)$ satisfying Condition~\ref{assump_concave_function}, ensuring that $Q_{\rho_{\gamma}}(\cdot)$ identifies $\bA^*$ as the unique optimiser over the equivalence class $\{\bA^*\bG^{-1}:\bG\in\Xi(c,\bZ^*)\}$. The condition $(a_3+1)\gamma\le \lambda_0 = \min\big\{|a_{jl}^*|:a_{jl}^*\neq 0, j\in[q],l\in[r]\big\}$ imposes a lower bound on the signal strength in $\bA$, analogous to those studies of folded concave losses in regression problems~\citep{fan2011nonconcave,fan2014strong}. 
In \eqref{eq_population_solo}, we focus on a local region $\Xi(c,\bZ)$ within the feasible set of oblique rotations. This restriction helps avoid the difficulties of characterising the global, non-convex landscape of $Q_{\rho_\gamma}(\cdot)$ and prevents degeneracies in the rotated latent variables, that is, the collapse of $\bZ$ due to highly collinear columns \citep{browne2001overview}.
  This motivates the Folomin framework, where we begin with a suitable initial estimator and then optimise $Q_{\rho_\gamma}(\cdot)$ over a local region specified by this initial estimator, as we describe next. 



We introduce the Folomin framework under the oblique setting, which is technically more challenging and includes the orthogonal setting as a special case.
We start with the following empirical risk minimiser:
\begin{equation}
        (\hat\bZ,\hat\bA)~\in~\argmin_{(\bZ,\bA)\in\Theta^*(\epsilon_{nq})}\cL(\bZ,\bA) ~=~ \argmin_{(\bZ,\bA)\in\Theta^*(\epsilon_{nq})} \sum_{i=1}^n\sum_{j=1}^q\ell_{ij}(\ba_j^\T\bz_i),\label{eq_theory_init}
    \end{equation}where $\Theta^*(\epsilon_{nq})$ is a local region of the true parameters $(\bZ^*,\bA^*)$, defined by 
    \begin{equation*}
    \Theta^*(\epsilon) := \Big\{(\bZ,\bA):n^{-1/2}\|\bZ - \bZ^*\|_{\Fn} + q^{-1/2}\|\bA - \bA^*\|_{\Fn}\le \epsilon, \|\bZ\|_{2\to\infty}\le M,\|\bA\|_{2\to\infty}\le M\Big\},
\end{equation*} and $\epsilon_{nq}$ is some sequence converging to zero as $n,q\to\infty$.
Given $(\hat\bZ,\hat\bA)$, we then solve an empirical counterpart of~\eqref{eq_population_solo} as
\begin{equation}
        \hat\bG ~=~ \argmin_{\bG\in \Xi(\epsilon_{nq}^{\prime}, \hat\bZ)}Q_{\rho_{\gamma}}\left(\hat\bA\bG^{-1}\right),\label{eq_theory_solo}\end{equation}
where $\epsilon_{nq}^{\prime}$ is another sequence converging to zero as $n,q\to\infty$ and $\Xi(\epsilon_{nq}^{\prime},\hat\bZ)$ is defined in~\eqref{eq_true_local_region}. This folded loss minimisation problem yields our final Folomin estimator $(\hat\bZ\hat\bG^\T,\hat\bA\hat\bG^{-1})$.

Compared with \eqref{eq_population_solo}, analysing the minimiser of \eqref{eq_theory_solo} presents several non-trivial challenges. First, the noise from all $nq$ observations enters both the objective $Q_{\rho_{\gamma}}(\cdot)$ through $\hat\bA$ and the feasible set $\Xi(c,\hat\bZ)$ through $\hat\bZ$. Since $(\hat\bZ,\hat\bA)$ is obtained under a general nonlinear model, the relationship between the estimator $(\hat\bZ,\hat\bA)$ and the data must be carefully tracked in the analysis.
Moreover, the objective $Q_{\rho_{\gamma}}(\cdot)$ is non-convex, nonlinear, and non-smooth, and the feasible set $\Xi(c,\bZ)$ is itself defined by nonlinear constraints, making it hard to characterise even local minima.
From a computational perspective, both optimisation problems are challenging. For \eqref{eq_theory_init}, while related estimators are studied in the literature~\citep{chen2019joint,chen2021nonlinear,wang2022maximum}, it remains an open problem to compute a solution guaranteed to lie sufficiently close to the true parameters $(\bZ^*,\bA^*)$. For \eqref{eq_theory_solo}, the combination of a non-convex, non-smooth objective and nonlinear constraints makes direct optimisation numerically costly and unstable. Efficient algorithms with reliable and computationally attainable solutions are highly desirable for practice.


In the subsequent sections, we address these challenges by establishing the oracle inference property of the Folomin estimator and by developing a computational framework with provable guarantees that yields an estimator achieving the same oracle property.
The oracle estimator of $\bA^*$ is defined as $\bA^{\mathrm{oracle}} = \argmin_{\bA}\sum_{j=1}^q\sum_{i=1}^n\ell_{ij}(\ba_j^\T\bz_i^*)$, obtained when the true latent variables $\bZ^*$ are assumed to be {\it known}. Similarly, $\bZ^{\mathrm{oracle}} = \argmin_{\bZ}\sum_{i=1}^n\sum_{j=1}^q\ell_{ij}(\ba_j^*{}^\T\bz_i)$. 

We provide an informal summary of our main results below.
\begin{enumerate}[label=(\arabic*), leftmargin = 0.8cm]
    \item \;(Oracle inference; Section~\ref{sec_oracle_property})
    We show that, as $n,q\to\infty$, each row of $\hat\bZ\hat\bG^\T$ and $\hat\bA\hat\bG^{-1}$ has the same limiting distribution as the corresponding row of the oracle estimator $\bZ^{\mathrm{oracle}}$ and $\bA^{\mathrm{oracle}}$, respectively. Moreover, dependence across different rows of $\hat\bZ\hat\bG^\T$ and $\hat\bA\hat\bG^{-1}$ is asymptotically negligible. 
    \item \;(Computational guarantee; Section~\ref{sec_comp_guarantee}) We establish that Algorithm~\ref{alg_init} in Section~\ref{sec_alg_init} produces a consistent initial estimator $(\hat\bZ_{\init},\hat\bA_{\init})$ as one solution of \eqref{eq_theory_init} satisfying $(\hat\bZ_{\init},\hat\bA_{\init})\in\Theta^*\big((n\wedge q)^{-c}\big)$ for some $0<c< 1/2$.
    Furthermore, given $(\hat\bZ_{\init},\hat\bA_{\init})$ from Algorithm~\ref{alg_init}, we show that the solution to \eqref{eq_theory_solo} can be well approximated by the one-step update of Algorithm~\ref{alg_lqa} in Section~\ref{sec_alg_lqa}, denoted by $\hat\bG^{(1)}$, where we prove that the resulting estimator $(\hat\bZ_{\init}(\hat\bG^{(1)})^{\T},\hat\bA_{\init}(\hat\bG^{(1)})^{-1})$ has the same oracle inference  property as $(\hat\bZ\hat\bG^\T,\hat\bA\hat\bG^{-1})$.
\end{enumerate}

Our results are of both theoretical and practical interest. From the theoretical perspective, we develop the first distributional theory for estimators obtained from rotation methods under the general representation learning framework introduced in Section~\ref{sec_setup}. More importantly, this theory applies to a broad class of sparse representation matrices, characterised by the $(\lambda,\epsilon)$-sparsity condition that is straightforward to interpret and easy to satisfy in practice. In contrast, existing work~\citep{Liu2023RotationTS,cape2024varimax} focuses on linear models and provides results only under perfect or nearly perfect simple structure, a stronger assumption that is often violated in practice. 
On the computational side, we derive a numerically attainable estimator and establish that it enjoys the same oracle property as the Folomin estimator, closing the gap between the asymptotic theory and its computational implementation. Moreover, we show that the rows of $\hat\bA\hat\bG^{-1}$ are independent asymptotically, which facilitates various downstream inference tasks, such as testing sparsity structure in $\bA$~\citep{brown2015confirmatory}, and comparing latent variable distributions across subgroups~\citep{Putnick2016MeasurementIC}.


\begin{remark}\it
 \citet{jennrich2006rotation} studied rotation with concave component losses and provided empirical evidence that such criteria can outperform many popular methods. However, the primary focus of \citet{jennrich2006rotation} was on identifiability issues in low-dimensional settings with $q$ fixed and noiseless settings. A more compelling and important question is how to conduct valid statistical inference, which motivates the theory developed in this paper.
\end{remark}

\begin{remark}
    In this paper, we focus on the folded concave loss functions. \citet{Liu2023RotationTS} studied the $\ell_1$ loss and demonstrated its performance via empirical studies. However, their analysis is restricted to a perfect simple structure, and in Section~\ref{supp_sec_l1_discussion} of the Supplementary Material, we show that the rotation method with the $\ell_1$ loss is rotationally biased.
    
\end{remark}





\begin{remark}\it
The folded concave functions have also been studied in the high-dimensional regression literature~\citep{fan2001variable,zhang2010nearly,fan2014strong}, where they act as regularisers. In contrast, in our settings, these folded concave functions serve directly as the loss of the optimisation problem. More importantly, the resulting optimisation problem \eqref{eq_theory_solo} depends implicitly on the data through the empirical risk minimiser $(\hat\bZ,\hat\bA)$: the objective involves $\hat\bA$ and the feasible set is defined by highly nonlinear constraints involving $\hat\bZ$. Consequently, both the theoretical analysis and the computation are fundamentally different and require new technical arguments.



\end{remark}



\section{Oracle Inference Properties}\label{sec_oracle_property}


To derive our theoretical results, we impose the following regularity conditions. 


\begin{assumption}[{Parameter Regularity}]\label{assump_psd_covariance}\it \begin{enumerate}[label=$(\roman*)$, leftmargin=0.8cm] \item\;$\bA^*$ is $(\lambda_0,\epsilon_0)$-sparse for some $\lambda_0>0$ and $\epsilon_0>0$;
\item\;$\|\bZ^*\|_{2\to\infty}\le M$ for $M$ specified in Definition~\ref{def_sparsity}, and $\mathrm{diag}(n^{-1}\bZ^*{}^\T\bZ^*) = \bI_r$;
    \item\;$\bSigma_z^* = \lim_{n \rightarrow \infty}n^{-1}{\bZ^*}^\T\bZ^* $ and $\bSigma_{a}^* = \lim_{q \rightarrow \infty}q^{-1}{\bA^*}^\T\bA^* $ exist and are positive definite.
\end{enumerate}
\end{assumption}

\begin{assumption}[{Scaling}]\it\label{assump_scaling}
    As $n,q\to\infty$, $\delta_{nq} := \log(n\vee q)/\sqrt{n\wedge q}\to 0$ as $n,q\to\infty$.
\end{assumption}
\begin{assumption}[{Smoothness}]\label{assump_limit}\it
The risk function $\ell_{ij}(\theta)$ is three times differentiable to $ \theta$, with the first, second, and third order derivatives denoted by $\ell_{ij}^{\prime}( \theta)$, $\ell_{ij}^{\prime\prime}( \theta)$, and $\ell_{ij}^{\prime\prime\prime}( \theta)$, respectively. For $i\in[n]$ and $j\in[q]$, $\ell_{ij}^{\prime} ( \theta_{ij}^*)$ is mean zero and sub-exponential with the sub-exponential norm {$\|\ell_{ij}^{\prime} ( \theta_{ij}^*)\|_{\psi_1} \le C$ for some constant $C>0$}. 
 Within a compact set of $\theta$, there exist $0<b_L\le b_U$ such that $b_L\le \ell_{ij}^{\prime\prime}(\theta)\le b_U$ and $|\ell_{ij}^{\prime\prime\prime}(\theta)| \leqslant b_U$. 
    
\end{assumption}
\begin{assumption}[{Information Limits}]\label{assump_clt}
    For $j\in[q]$, $\bPhi_{j}^* := \lim\limits_{n\to\infty}n^{-1}\sum_{i=1}^n\ell_{ij}^{\prime\prime}(\theta_{ij}^*)\bz_i^*(\bz_i^*)^\T$ exists and is positive definite, and $\bOmega_{j,a}^* = \lim\limits_{n\to\infty}n^{-1}\sum_{i=1}^n\EE[\ell_{ij}^{\prime}(\theta_{ij}^*)^2]\bz_i^*(\bz_i^*)^\T$ exists. For $i\in[n]$, $\bPsi_i^*:=\lim\limits_{q\to\infty}q^{-1}\sum_{j=1}^q\ell_{ij}^{\prime\prime}(\theta_{ij}^*)\ba_j^*(\ba_j^*)^\T$ exists and is positive definite, and $\bOmega_{i,z}^* := \lim\limits_{q\to\infty}q^{-1}\sum_{j=1}^q\EE[\ell_{ij}^{\prime}(\theta_{ij}^*)^2]\ba_j^*(\ba_j^*)^\T$ exists.
\end{assumption}
Assumption~\ref{assump_psd_covariance} imposes regularity conditions on the true parameters $(\bZ^*,\bA^*)$. Parts $(i)$ and $(ii)$ follow directly from our model setup, and part $(iii)$ is standard in the literature \citep{bai2003inferential,bai2012statistical,fan2016projected,wang2022maximum}. Assumption~\ref{assump_scaling} imposes a mild scaling condition, which is readily satisfied when $\log n\ll \sqrt q$ and $\log q\ll\sqrt{ n}$. Assumption~\ref{assump_limit} imposes standard smoothness conditions on the risk functions $\ell_{ij}(\cdot)$, which are satisfied by many commonly used choices, including the squared risk and negative log-likelihood function from linear, logistic, and Poisson models.  {Assumption~\ref{assump_clt} is a common regularity condition to ensure that $(\bZ^{\mathrm{oracle}},\bA^{\mathrm{oracle}})$ admit well-defined limiting information matrices and asymptotic variances. They are used to establish the asymptotic normality result, where the limiting covariance takes the usual sandwich form.}

For each $l\in[r]$, let $\cT_l ^*: = \{j:a_{jl}^* = 0\}$ and 
\begin{equation*}
\sigma_q^{-1} ~:=\;\; \min_{l\in[r]}\lambda_{r-1}\Big\{q^{-1}(\bA^*_{\cT_l^*,})^\T\bA^*_{\cT_l^*,}\Big\}.\end{equation*}
The quantity $\sigma_q$ measures the effective strength of the zero pattern in $\bA^*$. 
If $\sigma_q$ is large, then either $\cT_l^*$ contains few indices (few zeros) or the columns of $\bA^*_{\cT_l^*,-l}$ are nearly collinear. In either case, the sparsity structure in $\bA^*$ becomes less informative and therefore harder to recover.
Compared with the $(\lambda,\epsilon)$-sparsity notion in Definition~\ref{def_sparsity}, which serves as a certain identification condition, $\sigma_q$ is a technical measure of the sparsity in $\bA^*$.  Next, we show that the Folomin estimator achieves the oracle inference property over a broad range of sparsity regimes. 

\vspace{-0.2in}


\begin{theorem}[Oracle Inference]\it\label{thm_oracle}
    Suppose Assumptions~\ref{assump_psd_covariance}--\ref{assump_clt} hold. Fix $\gamma \le \lambda_0/(a_3+1)$, and choose $\epsilon_{nq}$ in \eqref{eq_theory_init} and $\epsilon_{nq}^{\prime}$ in \eqref{eq_theory_solo} such that, as $n,q\to\infty$,
    \begin{equation}\delta_{nq} = o(\epsilon_{nq}),\quad \epsilon_{nq} = o(\epsilon_{nq}^{\prime}), \quad  \epsilon_{nq}^{\prime}= o\big(\min\{\gamma, (n\wedge q)^{-c}, \sigma_q^{-1}\}\big)\label{eq_epsilon_scaling} \end{equation}
     for any constant $0<c<1/2$. Then for $(\hat\bZ,\hat\bA)$ obtained in \eqref{eq_theory_init} and $\hat\bG$ obtained in \eqref{eq_theory_solo}, we have the following asymptotic expansions
    \begin{equation}\label{eq_oracle_expand}
        \hat\bA\hat\bG^{-1} - \bA^* ~= ~ \bN_a + \bR_{a}\quad\text{ and }\quad\hat\bZ \hat\bG^{\T} - \bZ^* ~ = ~ \bN_z + \bR_z,\end{equation}with $\|\bR_{a}\|_{2\to\infty} = O_p(\sigma_q(n\wedge q)^{-1+\varepsilon})$ and $\|\bR_z\|_{2\to\infty} = O_p(\sigma_q(n\wedge q)^{-1+\varepsilon})$ for any constant $\varepsilon>0$. 
        Write $\sqrt{n}\bN_a = (\bn_1^a,\dots,\bn_q^a)^\T$ and $\sqrt{q}\bN_z = (\bn_1^z,\dots,\bn_n^z)^\T$, and then for each $j\in[q]$ and $i\in[n]$,
        \begin{equation}\label{eq_oracle_asymp_dis}
            \begin{aligned}
                \bn_j^a = \sqrt{n}\big\{\sum_{i=1}^n\ell_{ij}^{\prime\prime}(\theta_{ij}^*)\bz_i^*\bz_i^*{}^\T\big\}^{-1}\sum_{i=1}^n\ell_{ij}^{\prime}(\theta_{ij}^*)\bz_i^*\overset{{ \rm d}}{\to}\cN\big(\zero_r,(\bPhi_{j}^*)^{-1}\bOmega_{j,a}^*(\bPhi_{j}^*)^{-1}\big)\text{ as }n\to\infty,\\
                \bn_i^z=\sqrt{q}\big\{\sum_{j=1}^q\ell_{ij}^{\prime\prime}(\theta_{ij}^*)\ba_j^*\ba_j^*{}^\T\big\}^{-1}\sum_{j=1}^q\ell_{ij}^{\prime}(\theta_{ij}^*)\ba_j^*\overset{{ \rm d}}{\to}\cN\big(\zero_r,(\bPsi_{i}^*)^{-1}\bOmega_{i,z}^*(\bPsi_{i}^*)^{-1}\big)\text{ as }q\to\infty,
            \end{aligned}
        \end{equation}
        and $\{\bn_j^a,\bn_i^z\}_{i\in[n],j\in[q]}$ are pairwise asymptotically independent as $n,q\to\infty$. Notably, $\bn_{j}^a$ and $\bn_i^z$ are asymptotically equivalent in distribution to the $j$-th row of $\sqrt{n}(\bA^{\mathrm{oracle}}-\bA^*)$ and the $i$-th row of $\sqrt{q}(\bZ^{\mathrm{oracle}} - \bZ^*)$, respectively.
    Furthermore, $(\bPhi_{j}^*)^{-1}\bOmega_{j,a}^*(\bPhi_{j}^*)^{-1}$ and $(\bPsi_{i}^*)^{-1}\bOmega_{i,z}^*(\bPsi_{i}^*)^{-1}$ admit consistent plug-in estimators, $\hat\bPhi_j^{-1}\hat\bOmega_{j,a}\hat\bPhi_j^{-1}$ and $\hat\bPsi_i^{-1}\hat\bOmega_{i,z}\hat\bPsi_i^{-1}$, obtained by replacing $(\bZ^*,\bA^*)$ with their consistent estimator $(\hat\bZ\hat\bG^\T,\hat\bA\hat\bG^{-1})$; and \eqref{eq_oracle_asymp_dis} holds with $(\bPhi_{j}^*)^{-1}\bOmega_{j,a}^*(\bPhi_{j}^*)^{-1}$ and $(\bPsi_{i}^*)^{-1}\bOmega_{i,z}^*(\bPsi_{i}^*)^{-1}$ replaced by $\hat\bPhi_j^{-1}\hat\bOmega_{j,a}\hat\bPhi_j^{-1}$ and $\hat\bPsi_i^{-1}\hat\bOmega_{i,z}\hat\bPsi_i^{-1}$, respectively.
\end{theorem}
For the residual terms $\bR_a$ and $\bR_z$ to be negligible relative to the asymptotically normal terms, we require $\sigma_q = o( n^{-1/2}q^{1-\varepsilon} \wedge n^{1-\varepsilon}q^{-1/2})$ for any constant $\varepsilon>0$.
When $\sigma_q$ is of constant order, this reduces to $n = o(q^{2(1 - \varepsilon)})$ and $q = o(n^{2(1 - \varepsilon)})$, in line with conditions commonly imposed in the literature~\citep{bai2003inferential,wang2022maximum,cui2025identifiability}. When $n$ and $q$ are of the same order, the requirement allows $\sigma_q^{-1}\approx q^{-1/2+\varepsilon}$ for some constant $\varepsilon>0$, thereby accommodating a relatively dense structure in $\bA^*$. 
Under the scaling conditions, each row of $\hat\bA\hat\bG^{-1} - \bA^*$ has the same asymptotic distribution as that of $\bN_a$, which coincides with that of $\bA^{\mathrm{oracle}} - \bA^*$ by standard $M$-estimation theory~\citep{van2000asymptotic}; an analogous statement holds for $\hat\bZ\hat\bG^\T-\bZ^*$.

In the literature, asymptotic distributions for the latent variables and representation matrix have mostly been developed either under an oracle rotation matrix that depends on the true parameters~\citep{bai2003inferential,chen2019inference,zhang2022heteroskedastic} or under mathematically convenient identifiability conditions such as {$q^{-1}\bA^\T\bA$ is a diagonal matrix}~\citep{anderson1956statistical,bai2012statistical}.  
As a result, the associated inferential theory often has limited practical relevance and scientific interpretability. A few studies derive asymptotic distributions under more practical identifiability conditions~\citep{bai2012statistical,cui2025identifiability}, but they require additional constraints in $\bA^*$ that are {prespecified}, and the estimators under these settings do not achieve the oracle property. In contrast, our results show that, with $\hat\bG$ obtained from~\eqref{eq_theory_solo}, the rotated estimator $\hat\bA\hat\bG^{-1}$ can recover a sparse $\bA^*$ with the oracle inference property.

Establishing Theorem~\ref{thm_oracle} is technically involved. As a first step, we derive a sharp asymptotic expansion that holds for any empirical risk minimiser $(\hat\bZ,\hat\bA)$ in~\eqref{eq_theory_init}, expressed after alignment by the oracle rotation $\hat\bG^*$ defined in~\eqref{eq_opt_align} of the Supplementary Material (see Lemma~\ref{lemma_asymp_start} therein). Developed within the considered latent representation learning framework, this expansion consists of an asymptotically normal leading term and a remainder term for which we obtain sharp uniform error bounds. The expansion 
is the key tool to characterise the dependence of $(\hat\bZ,\hat\bA)$ on the responses $Y_{ij}$s and to analyse~\eqref{eq_theory_solo}. 
To handle the non-convex, nonlinear, and non-smooth objective $Q_{\rho_\gamma}(\hat\bA\bG^{-1})$, we partition the feasible set of~\eqref{eq_theory_solo} into a sequence of nested rings centred at $\hat\bG^*$. Using the developed asymptotic expansion, we establish sharp concentration inequalities showing that, on each ring, $Q_{\rho_\gamma}(\hat\bA\bG^{-1})$ attains a smaller value in the inner region. This ultimately locates the minimiser of $Q_{\rho_\gamma}(\hat\bA\bG^{-1})$ in a sufficiently small neighbourhood of $\hat\bG^*$, which then leads to the asymptotic results in Theorem~\ref{thm_oracle}
(See Lemma~\ref{lemma_oracle_folomin_solu} in the Supplementary Material for more details).


\section{Computational Guarantee}\label{sec_comp_guarantee}

 In Section~\ref{sec_oracle_property}, we established the oracle property for $(\hat\bZ\hat\bG^\T,\hat\bA\hat\bG^{-1})$, with $(\hat\bZ,\hat\bA)$ obtained from \eqref{eq_theory_init} and $\hat\bG$ from \eqref{eq_theory_solo}.
As discussed in Section~\ref{sec_main}, directly solving problems~\eqref{eq_theory_init} and~\eqref{eq_theory_solo} can be computationally challenging. In this section, we develop two computationally efficient algorithms to address these challenges, presented in Sections~\ref{sec_alg_init} and~\ref{sec_alg_lqa}, respectively.



\subsection{Initialisation}\label{sec_alg_init}
We aim to find $(\hat\bZ_{\init},\hat\bA_{\init})$ as a solution of \eqref{eq_theory_init} within $\Theta^*(\epsilon_{nq})$ for $\epsilon_{nq}$ required by Theorem~\ref{thm_oracle}. 
Motivated by the $\epsilon$ cone neighbourhood defined in Section~\ref{sec_sparse_def}, we first introduce a quantity that is invariant under the transformation $(\bZ,\bA)\mapsto (\bZ\bG^\T,\bA\bG^{-1})$ for any invertible $\bG$:
\[
\cD_{(j_1,j_2)}^{\delta}(\bZ,\bA)
=\begin{cases}
\cos\angle(\bZ\ba_{j_1},\bZ\ba_{j_2}), & n^{-1}\|\bZ\ba_{j_1}\|\,\|\bZ\ba_{j_2}\|>\delta,\\
0, & \text{otherwise},
\end{cases}
\]
with some prespecified $\delta>0$.
Its invariance implies that $\cD_{(j_1,j_2)}^{\delta}(\bZ^*,\bA^*)$ can be well approximated by $\cD_{(j_1,j_2)}^{\delta}(\hat\bZ^0,\hat\bA^0)$ for any estimator $(\hat\bZ^0,\hat\bA^0)$ if the product $\hat\bZ^0\hat\bA^0{}^\T$ is close to $\bZ^*\bA^*{}^\T$, without requiring $\hat\bZ^0$ and $\hat\bA^0$ to be individually consistent.
This quantity is informative for detecting simple rows. Specifically, for any $l\in[r]$ and any $j_1,j_2\in\cS_l(\bA^*)$, for sufficiently small $\delta$, we have $\cD_{(j_1,j_2)}^{\delta}(\bZ^*,\bA^*) = 1$. In contrast, if $j_1$ and $j_2$ do not belong to the same $\cS_l(\bA^*)$, we can show $\cD_{(j_1,j_2)}^{\delta}(\bZ^*,\bA^*) < 1$ unless $\cos\angle(\ba_{j_1},\ba_{j_2}) = 1$, the occurrence of which is restricted when $\bA^*$ is $(\lambda,\epsilon)$-sparse by Definition~\ref{def_sparsity}. Motivated by this observation, we identify simple rows $\cS_k(\bA^*)$ by examining $\cD_{(j_1,j_2)}^{\delta}(\hat\bZ^0,\hat\bA^0)$ for some suitable $(\hat\bZ^0,\hat\bA^0)$, and then use the simple rows to construct the initialisation $(\hat\bZ_{\init},\hat\bA_{\init})$. The procedure is summarised in Algorithm~\ref{alg_init}.

The algorithm starts with the following minimiser
\begin{equation}
    (\hat{\bZ}^0,\hat{\bA}^{0}) = \mathop{\arg\min}_{\substack{\|\bZ\|_{2\to\infty}\le M,\;\|\bA\|_{2\to\infty}\le M,\\n^{-1}\bZ^\T\bZ = \bI_r,\;q^{-1}\bA^\T\bA = \mathrm{diag}(q^{-1}\bA^\T\bA)}}\cL(\bZ,\bA).\label{eq_init_of_init_est}
\end{equation}
Here, $M$ is as specified in Definition~\ref{def_sparsity}, and the constraints of $n^{-1}\bZ^\T\bZ = \bI_r$ and $q^{-1}\bA^\T\bA$ being diagonal are imposed primarily for mathematical convenience and to facilitate the theoretical development. Similar constrained optimisation problems have been studied in the literature~\citep{bai2012statistical,chen2019joint,wang2022maximum,cui2025identifiability}, and we therefore adopt $(\hat\bZ^0,\hat\bA^0)$ as our starting point. To solve \eqref{eq_init_of_init_est}, we provide a gradient descent scheme and establish that it converges linearly to $(\hat{\bZ}^0,\hat{\bA}^{0})$.
Implementation details and theory are provided in Section~\ref{supp_sec_gda} of the Supplementary Material.
{\begin{algorithm}[ht]

\caption{Initialisation}
\label{alg_init}

\KwIn{$(\hat\bZ^0,\hat\bA^0)$ from \eqref{eq_init_of_init_est}; thresholds $\delta,\delta^{\prime}>0$;}

{For each $j\in[q]$,
  construct $\hat\cI_j = \big\{j^{\prime}\in[q]:  \cD_{(j, j^{\prime})}^{\delta}(\hat\bZ^0,\hat\bA^0) > 1-\delta^{\prime}\big\}$;
}

Select the largest $r$ non-overlapping sets in $\{\hat\cI_j\}_{j\in[q]}$ and denote them by $\hat\cS_{1},\hat\cS_{2},\dots,\hat\cS_{r}$\;

{For $k\in[r]$, set $\hat\cS_{-k}:=\cup_{l\neq k}\hat\cS_l$ and compute $\hat\bv_k$ as the $r$-th right singular vector of $\hat\bA_{\hat\cS_{-k},}^0$;
}

Set $\tilde\bG = (\hat\bv_1,\dots,\hat\bv_{r})$ and $\hat\bG^0 = \big\{\mathrm{diag}\big(\tilde\bG^{-1}\tilde\bG^{-\T}\big)\big\}^{-1/2}\tilde\bG^{-1}$;

\KwOut{$(\hat\bZ_{\init}, \hat\bA_{\init}) = \big(\hat\bZ^0(\hat\bG^0)^{\T}, \hat\bA^0(\hat\bG^0)^{-1}\big)$.}
\end{algorithm}}


Starting from this initial value $(\hat\bZ^0,\hat\bA^0)$, Step 1 constructs, for each $j\in[q]$, an index set $\hat\cI_j$ consisting of all $j^{\prime}\in[q]$ such that $\cD_{(j, j^{\prime})}^{\delta}(\hat\bZ^0,\hat\bA^0)>1-\delta^{\prime}$, where $\delta$ and $\delta^{\prime}$ are two thresholds. With suitable choices of $\delta$ and $\delta^{\prime}$, $\{\hat\cI_j\}_{j\in[q]}$ serves as a good approximation to $\cC_{\delta}(\ba_j^*,\bA^*)$ defined in Section~\ref{sec_sparse_def}. We describe a data-adaptive calibration of $\delta$ and $\delta^{\prime}$ in Section~\ref{supp_subsec_tuning_para} of the Supplementary Material.
By the definition of $(\lambda,\epsilon)$-sparsity, when $\delta$ is sufficiently small, the $r$ largest non-overlapping sets in $\{\cC_{\delta}(\ba^*_j;\bA^*)\}_{j\in[q]}$ are precisely $\{\cS_l(\bA^*)\}_{l\in[r]}$. Consequently, the selected sets $\{\hat\cS_l\}_{l\in[r]}$ in Step 2 approximate $\{\cS_l(\bA^*)\}_{l\in[r]}$. In Steps 3 and 4, we use the estimated simple row sets to construct a rotation matrix that aligns $\hat\bA^0$ with the simple structure and rescale the rotation with a diagonal matrix to normalise the columns of the latent variables.

{Algorithm~\ref{alg_init} is related to the simple-row recovery procedures in \citet{bing2020adaptive,bing2023detecting}. In particular, \citet{bing2020adaptive} consider a linear latent variable model and use entry-wise information in $n^{-1}\bY^\T\bY$ to identify simple rows. \citet{bing2023detecting} further study a multiple parallel-row structure, where the number of parallel row classes may exceed $r$ and the parallel rows may have heterogeneous signal strengths.
Our setting is closely related to the latter, under which one may identify more than $r$ parallel directions from noisy estimators of $\bZ$ and $\bA$. Equation~\eqref{eq_angular_sep} in the $(\lambda,\epsilon)$-sparsity condition therefore serves as a key separation assumption ensuring that the $r$ major directions are distinguishable. Our setting, however, differs from these works in that the entrywise information in $n^{-1}\bY^\T\bY$ is no longer directly applicable because the current model may be nonlinear. We thus introduce a new measure $D_{(j,j')}^{\delta}(\hat\bZ^0,\hat\bA^0)$ constructed from  $(\hat\bZ^0,\hat\bA^0)$ to select simple rows. The following result shows that the algorithm yields a consistent initial estimator.
}

\begin{theorem}[{Consistent Initialisation}]
    \label{thm_init}\it Suppose Assumptions~\ref{assump_psd_covariance}--\ref{assump_limit} hold. Choose thresholds $\delta,\delta^{\prime}$ such that, as $n,q\to\infty$,  
    \begin{center}$\delta_{nq} = o(\delta\wedge \delta^{\prime}),\quad \delta^{\prime}=o(\epsilon_0),\quad \delta\le \lambda_0^2\lambda_{r}(\bSigma_z^*)/16,$\end{center}
    with $\lambda_0$, $\epsilon_0$ and $\bSigma_z^*$ defined in Assumption~\ref{assump_psd_covariance}, and $\delta_{nq}$ defined in Assumption~\ref{assump_scaling}. Let $\tilde{\sigma}_q^{-1} := |\cS(\bA^*)|^{-1}\min_{l\in[r]}\sum_{j\in\cS_l(\bA^*)}(a_{jl}^*)^2$ and assume that $\tilde\sigma_q\delta_{nq}=O((n\wedge q)^{-c})$ for some constant $0<c<1/2$. Then $(\hat\bZ_{\init}, \hat\bA_{\init})$ from Algorithm~\ref{alg_init} satisfies
    \begin{align*}
        n^{-1/2}\|\hat\bZ_{\init} - \bZ^*\|_{\Fn} + q^{-1/2}\|\hat\bA_{\init} - \bA^*\|_{\Fn}= O_p(\tilde{\sigma}_q\delta_{nq}) = O_p\big((n\wedge q)^{-c}\big);\\
        \|\hat\bZ_{\init} - \bZ^*\|_{2\to\infty} + \|\hat\bA_{\init} - \bA^*\|_{2\to\infty}= O_p(\tilde{\sigma}_q\delta_{nq}) = O_p\big((n\wedge q)^{-c}\big).
    \end{align*} 
\end{theorem}
Theorem~\ref{thm_init} establishes consistency of $(\hat\bZ_{\init},\hat\bA_{\init})$ produced by Algorithm~\ref{alg_init}. As is standard under sign indeterminacy, this consistency holds up to column-wise sign changes, which we suppress for notational simplicity. The quantity $\tilde\sigma_q$ captures the relative signal strength contributed by the simple rows across the $r$ dimensions. { This quantity reflects that we do not require the simple rows to have the same magnitude, rather $\tilde\sigma_q$ represents an average quantity which is large when each $l\in[r]$ has proportionally many simple rows and the corresponding signal strength $\nu_l$ is not too small.} The condition $\tilde\sigma_q =O\big( (n\wedge q)^{-c}\delta_{nq}^{-1}\big) = O\big((n\wedge q)^{\frac{1}{2}-c}\log^{-1}(n\vee q)\big)$ for some constant $c>0$ allows a broad range of simple row patterns and is typically mild in applications.
Moreover, when $\tilde\sigma_q$ is of constant order, the convergence rate of $(\hat\bZ_{\init},\hat\bA_{\init})$ is $O_p(n^{-1/2}+q^{-1/2})$ up to some small logarithmic factors, which matches that of the oracle estimator $(\bZ^{\mathrm{oracle}},\bA^{\mathrm{oracle}})$, where $n^{-1/2}\|\bZ^{\mathrm{oracle}} - \bZ^*\|_{\Fn} + q^{-1/2}\|\bA^{\mathrm{oracle}} - \bA^*\|_{\Fn} = O_p(n^{-1/2}+q^{-1/2})$ by the standard $M$-estimation theory~\citep{van2000asymptotic}.



\subsection{Local Quadratic Approximation}\label{sec_alg_lqa}

Next, we develop a local quadratic approximation (LQA) algorithm for solving \eqref{eq_theory_solo}. The LQA scheme is widely used in high-dimensional regression with folded concave losses~\citep[e.g.][]{fan2001variable,hunter2005variable}. We adapt it to our setting with several tailored designs. Specifically, the algorithm iteratively optimises a local quadratic surrogate of $Q_{\rho_{\gamma}}(\hat\bA\bG^{-1})$ and handles the constraints in $\Xi(\epsilon_{nq}^{\prime},\hat\bZ)$ via a relaxation followed by a normalisation, as summarised in Algorithm~\ref{alg_lqa}. 
In Step 4, we approximate $Q_{\rho_{\gamma}}(\hat\bA\bG^{-1})$ via the quadratic surrogate obtained from the LQA weights $w_{jk}^{(t)}$ from Step 3, yielding a weighted least-squares type subproblem that can be efficiently solved. The parameter $\eta$ is a prespecified parameter introduced to prevent unbounded weights. We set $\rho^{\prime}_{\gamma}(0) = \rho^{\prime}_{\gamma}(0+)$ and $\rho_{\gamma}'(x)=0$ at any point $x$ where $\rho_{\gamma}(\cdot)$ is not differentiable. To handle the constraint $\bH\in \Xi(\epsilon_{nq}^{\prime},\hat\bZ)$ in the original problem~\eqref{eq_theory_solo}, we work with the relaxed feasible region $\{\bH:\|\bH - \bI_r\|\le R,\mathrm{diag}( \bH^\T\bH) = \bI_r\} $ for some prespecified constant $0<R<1$, followed by a normalisation step (Step 5) to obtain the rotation $\hat\bG^{(t)}$. The bound $\|\bH-\bI_r\|\le R$ is included only to prevent numerical issues, and any fixed choice of $R$ suffices. Discussion on the tuning parameters used in Algorithm~\ref{alg_lqa} is left to Section~\ref{supp_subsec_tuning_para_2} of the Supplementary Material. 

\begin{algorithm}[h]
\caption{The local quadratic approximation (LQA) algorithm}
\label{alg_lqa}

\KwIn{$(\hat\bZ,\hat\bA)$ from \eqref{eq_theory_init}; folded concave loss $\rho_{\gamma}(\cdot)$; radius $R$; approximate constant $\eta$; number of iterations $T$;}

Set $\hat\bG^{(0)} = \bI_r$ and $(\hat\bZ^{(0)}, \hat\bA^{(0)}) = (\hat\bZ,\hat\bA)$\;

\For{$t = 0,1,\dots, T-1$}{

  \tcp{\textbf{Calculate weights}}
  For $j\in[q],\, l\in[r]$, set{
    \[
      w_{jl}^{(t)}=\big(|\hat\bA^{(t)}_{j,l}|^2 + \eta^2\big)^{-1/2}\rho_{\gamma}^{\prime}\big(|\hat\bA^{(t)}_{j,l}|\big);
    \]
  }\vspace{-0.6cm}

  \tcp{\textbf{optimise}}\vspace{-0.6cm}
  \[
    \hat\bH^{(t+1)}
    = \mathop{\arg\min}_{\bH:\|\bH - \bI_r\|\le R,\ \mathrm{diag}(\bH^\T\bH)=\bI_r}
    \sum_{j=1}^q\sum_{l=1}^r
    w_{jl}^{(t)}
    \left\{\hat\bA^{(t)}_{j,}\bH_{,l}\right\}^2;
  \]\vspace{-0.4cm}
  
  \tcp{\textbf{Normalise}}\vspace{-0.6cm}
  \[
    \hat\bG^{(t+1)}
    =
    \Big(
      \mathrm{diag}\big[
        \{\hat\bH^{(t+1)}\}^{-1}
        \big\{n^{-1}\hat\bZ^{(t)}{}^\T\hat\bZ^{(t)}\big\}
        \{\hat\bH^{(t+1)}\}^{-\T}
      \big]
    \Big)^{-1/2}
    (\hat\bH^{(t+1)})^{-1};
  \]\vspace{-0.4cm}

  \tcp{\textbf{Update}}
  Set $(\hat\bZ^{(t+1)},\hat\bA^{(t+1)}) =
  \big(\hat\bZ^{(t)}(\hat\bG^{(t+1)})^\T,\ \hat\bA^{(t)}(\hat\bG^{(t+1)})^{-1}\big)$\;
}
\KwOut{$\hat\bG^{(T)}$ and the final estimator $(\hat\bZ^{(T)},\hat\bA^{(T)})$.}
\end{algorithm}

Next, we show that one iteration of Algorithm~\ref{alg_lqa} yields estimator $\hat\bG^{(1)}$ that closely approximates the solution to \eqref{eq_theory_solo}, and therefore $(\hat\bZ^{(1)},\hat\bA^{(1)})$ also enjoy the oracle inference property.
\begin{theorem}[{ One-step Oracle Inference}]\label{thm_lqa}\it
    Suppose Assumptions~\ref{assump_psd_covariance}--\ref{assump_clt} hold. Fix $\gamma(a_3+1) \le \lambda_0$ and assume that  $\rho_{\gamma}(\cdot)$ is differentiable over $(0,\infty)$. Suppose $\epsilon_{nq}$ and $\epsilon_{nq}^{\prime}$ satisfy \eqref{eq_epsilon_scaling} in Theorem~\ref{thm_oracle}, $\eta \ge \epsilon_{nq}$, and $R$ is any constant in $(0,1)$. 
    Then for $\hat\bG^{(1)}$ obtained after one-step update of Algorithm~\ref{alg_lqa} and $\hat\bG$ from \eqref{eq_theory_solo},  we have
     $\|\hat\bG^{(1)} - \hat\bG\| = O_p(\sigma_q(n\wedge q)^{-1+\varepsilon})$ for any constant $\varepsilon>0$. Therefore, for $(\hat\bZ^{(1)},\hat\bA^{(1)})$ obtained after one-step update of Algorithm~\ref{alg_lqa}, Theorem~\ref{thm_oracle} holds when replacing $(\hat\bZ\hat\bG^\T, \hat\bA\hat\bG^{-1})$ with $(\hat\bZ^{(1)},\hat\bA^{(1)})$.
\end{theorem}
Theorems~\ref{thm_init} and~\ref{thm_lqa} together provide the computational guarantee for the proposed Folomin estimator. In particular, if starting from the consistent initialiser $(\hat\bZ_{\init},\hat\bA_{\init})$ produced by Algorithm~\ref{alg_init}, which satisfies $(\hat\bZ_{\init},\hat\bA_{\init})\in\Theta^*(\epsilon_{nq})$ by Theorem~\ref{thm_init} (with $\epsilon_{nq}$ as required), then Algorithm~\ref{alg_lqa} yields a one-step refined estimator $(\hat\bZ_{\init}(\hat\bG^{(1)})^\T,\hat\bA_{\init}(\hat\bG^{(1)})^{-1})$, which attains the oracle inference property by Theorem~\ref{thm_lqa}.
Importantly, one iteration of Algorithm~\ref{alg_lqa} suffices for valid statistical inference, providing an efficient method for solving the highly nonlinear constrained problem~\eqref{eq_theory_solo}. This one-step update argument aligns with prior analyses of optimisation problems concerning folded concave losses~\citep{zou2008one,fan2014strong}. In practice, performing a few additional iterations can further improve the solution. Moreover, Theorem~\ref{thm_lqa} provides a general guarantee for Algorithm~\ref{alg_lqa}, showing that it can refine any estimator from \eqref{eq_theory_init} into one with the oracle inference property, provided that $\epsilon_{nq}$ satisfies the theorem's condition. This allows Algorithm~\ref{alg_lqa} to be combined with alternative initialisation procedures beyond Algorithm~\ref{alg_init}, as long as they provide a consistent starting point at the required rate.


\begin{remark}
    Theoretically, Algorithm~\ref{alg_lqa} requires $\rho_{\gamma}(\cdot)$ to be differentiable on $(0,\infty)$, which accommodates the SCAD and MCP. In practice, however, we can also employ the truncated $\ell_1$ loss by fixing the value of its derivative at discontinuity. Since this point is encountered with zero probability, the local quadratic approximation remains practically applicable and produces reliable estimates with $\rho_{\gamma}(\cdot)$ specified as the truncated $\ell_1$ loss.
\end{remark}

\section{Empirical Studies}\label{sec_empirical}

This section presents empirical studies to assess our findings. Specifically, Section~\ref{subsec_simu} reports simulation studies that illustrate the bias in popular vintage methods and evaluate the performance of the proposed method. Section~\ref{subsec_real_data} analyses a personality assessment dataset using our method. 
{ The implementation of the proposed method and the code for reproducing the empirical results are available at the GitHub repository: \href{https://github.com/chengyu06/Folomin.git}{https://github.com/chengyu06/Folomin.git}.}

\subsection{Simulation Studies}\label{subsec_simu}
{\small \spacingset{1}
\begin{figure}[h]
    \centering
    \includegraphics[width=6in]{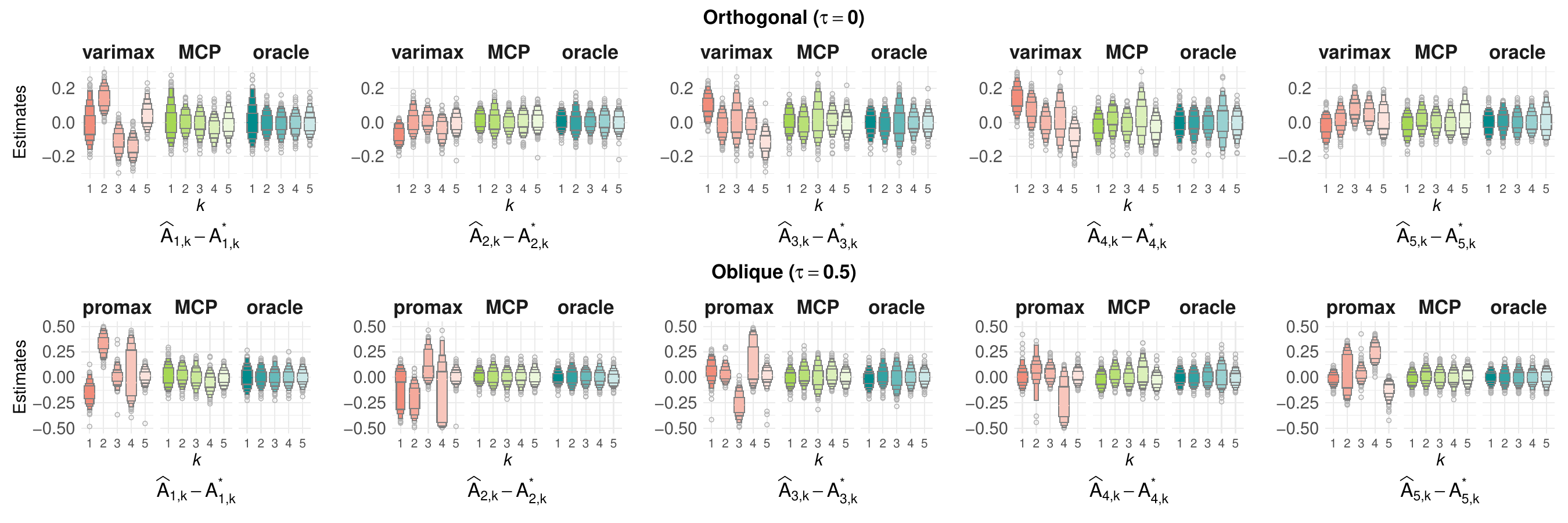}
    \caption{Bias of the estimation for $\bA^*_{1:5,}$ under $n=2000$, $q=2000$, $\lambda = 0.1$ and different $\tau$. Each panel displays the distribution of $\hat\bA_{j,} - \bA^*_{j,}$ for $j=1,2,\dots,5$, with methods indicated above (varimax/promax, Folomin with MCP, and the oracle benchmark).} 
    \label{fig_loading_bias}
\end{figure}
\begin{figure}[h]
    \centering
    \includegraphics[width=5.5in]{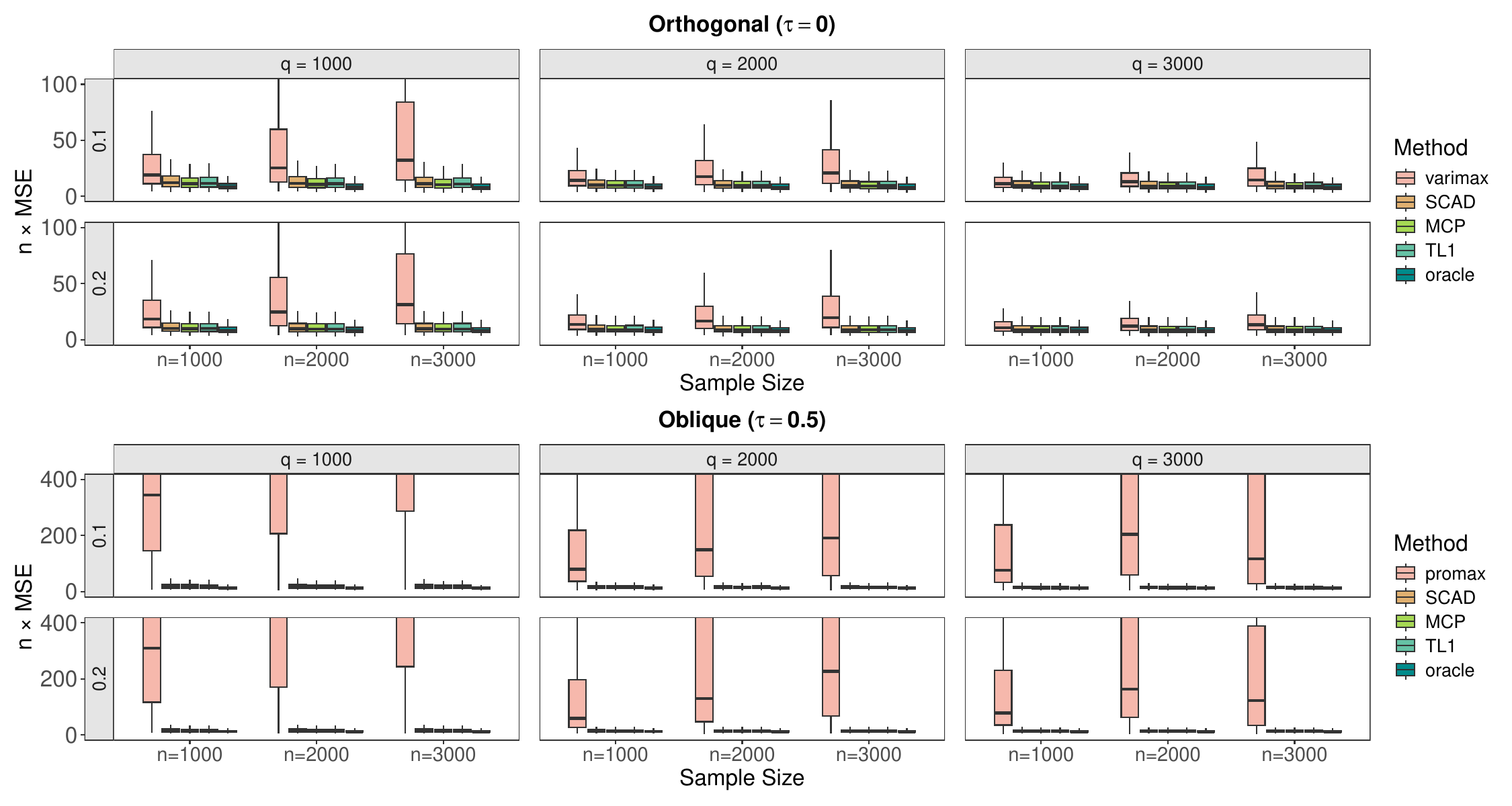}\\
    \caption{Scaled mean squared errors of estimation for $\bA^*$, defined entry-wise as $n\times \sum_{t=1}^{200}(\hat \bA_{j,l}^{(t)} - \bA_{j,l}^*)^2/200$ over 200 replications, where $\hat\bA^{(t)}$ is the estimator produced by different methods at replication $t$, across different settings of $n$, $q$, $\lambda$, and $\tau$.} 
    \label{fig_loadings_mse}
\end{figure}
}

We start by describing the data-generating mechanism. For the representation matrix $\bA\in\RR^{q\times r}$, we construct it to be $(\lambda,\epsilon)$-sparse by generating $\lfloor 0.1q\rfloor$ simple rows, allocating $\lfloor 0.1q/r\rfloor$ simple rows to each of the $r$ dimensions. The nonzero entries in the simple rows are i.i.d. from ${\rm Unif}(1,2)$. The remaining entries are drawn independently from a truncated Gaussian distribution ${\rm Trun}_{(\lambda,2.5)}\cN(0,1)$ with $\lambda$ being the minimal signal strength. Here, $\mathrm{Trun}_{(a,b)}(x)$ denotes truncation in magnitude, namely, $\mathrm{Trun}_{(a,b)}(x) = {\rm sgn}(x)\min(|x|,b)1_{(|x|\ge a)}$ with ${\rm sgn}(x) = 1_{(x> 0)} - 1_{(x<0)}$. We generate the latent variables from $\cN(\zero_r,\bSigma_{\tau})$ and rescale them so that $\mathrm{diag}(n^{-1}\bZ^\T\bZ) = \bI_r$. The covariance matrix $\bSigma
 _\tau\in\RR^{r\times r}$ has the $(l,h)$-th entry set as $\tau^{|l-h|}$ for $l,h\in[r]$, where $\tau$ controls the correlation in the latent variables. Given $(\bZ,\bA)$, we generate responses according to $Y_{ij}\sim \mathrm{Bernoulli}(\mathrm{logit}^{-1}(\ba_j^\T\bz_i)) $ with $\mathrm{logit}^{-1}(x) = \exp(x)/\{1+\exp(x)\}$. We use the risk function $\ell_{ij}(\theta) = -Y_{ij}\theta + \log\{1+\exp(\theta)\}$. 

The manipulated settings include $n\in\{1000,2000,3000\},q \in\{1000,2000,3000\}$, and minimal signal level $\lambda\in\{0.1, 0.2\}$. The number of dimensions is fixed at $r=5$. 
We consider an orthogonal setting ($\tau=0$) and an oblique setting ($\tau=0.5$). In the orthogonal setting, we further normalise the latent variables so $n^{-1}\bZ^\T\bZ = \bI_r$. As baselines, we use varimax for orthogonal rotation and promax \citep{browne2001overview} for oblique rotation, as two popular vintage methods in practice. 
 For our method, we consider three folded concave losses: SCAD, MCP, and truncated $\ell_1$ loss. In the orthogonal setting, we apply the proposed Folomin method as in the correlated case, without imposing the constraint $n^{-1}\bZ^\T\bZ=\bI_r$. 
We also compute the oracle estimation results for $\bA^*$ and $\bZ^*$, obtained as if $\bZ^*$ and $\bA^*$ were known, respectively.
We run 200 replications and apply the computational framework proposed in Section~\ref{sec_comp_guarantee} to obtain the estimators. We summarise the estimation results of $\bA$ in Figures~\ref{fig_loading_bias} and~\ref{fig_loadings_mse}. The results for estimating $\bZ$ are left to Section~\ref{supp_sec_simu_1} of the Supplementary Material.

In Figure~\ref{fig_loading_bias}, we report the estimation bias for the first 5 rows of $\bA^*$, which are fixed at $\bI_5$, under $(n,q,\lambda) = (2000,2000,0.1)$, as an illustrative example (Figure~\ref{fig_oracle} in Section~\ref{sec_sub_limitation} presents the result for the first row). The results under other settings are similar and provided in Section~\ref{supp_sec_simu_1} of the Supplementary Material. Figure~\ref{fig_loading_bias} shows that varimax yields relatively good point estimates, yet still exhibits noticeable bias in many entries. For the promax method, the bias is similarly evident. For both vintage methods, the bias is entry dependent and comparable in magnitude to the estimation error.
By contrast, our method, illustrated here using the MCP, performs well in both settings, exhibiting negligible bias and achieving estimation error comparable to the oracle benchmark. Figure~\ref{fig_loadings_mse} further presents the scaled mean squared errors. As $n$ and $q$ grow, the Folomin estimators, across all three folded concave losses considered, exhibit decreasing scaled mean squared errors and approach the oracle benchmark. In contrast, the vintage estimators exhibit a non-decreasing gap relative to the oracle setting. Their scaled mean squared errors increase with $n$ and decrease with $q$. This pattern suggests that a substantial portion of the error is driven by the bias component that depends only on the dimension and the structure of $\bA^*$ and is independent of the samples. This trend is consistent with existing results~\citep{Rohe2020VintageFA}; see Remark~\ref{remark_bias_magni}.

{\spacingset{1}
\begin{figure}[h]
    \centering
    \includegraphics[width=5.5in]{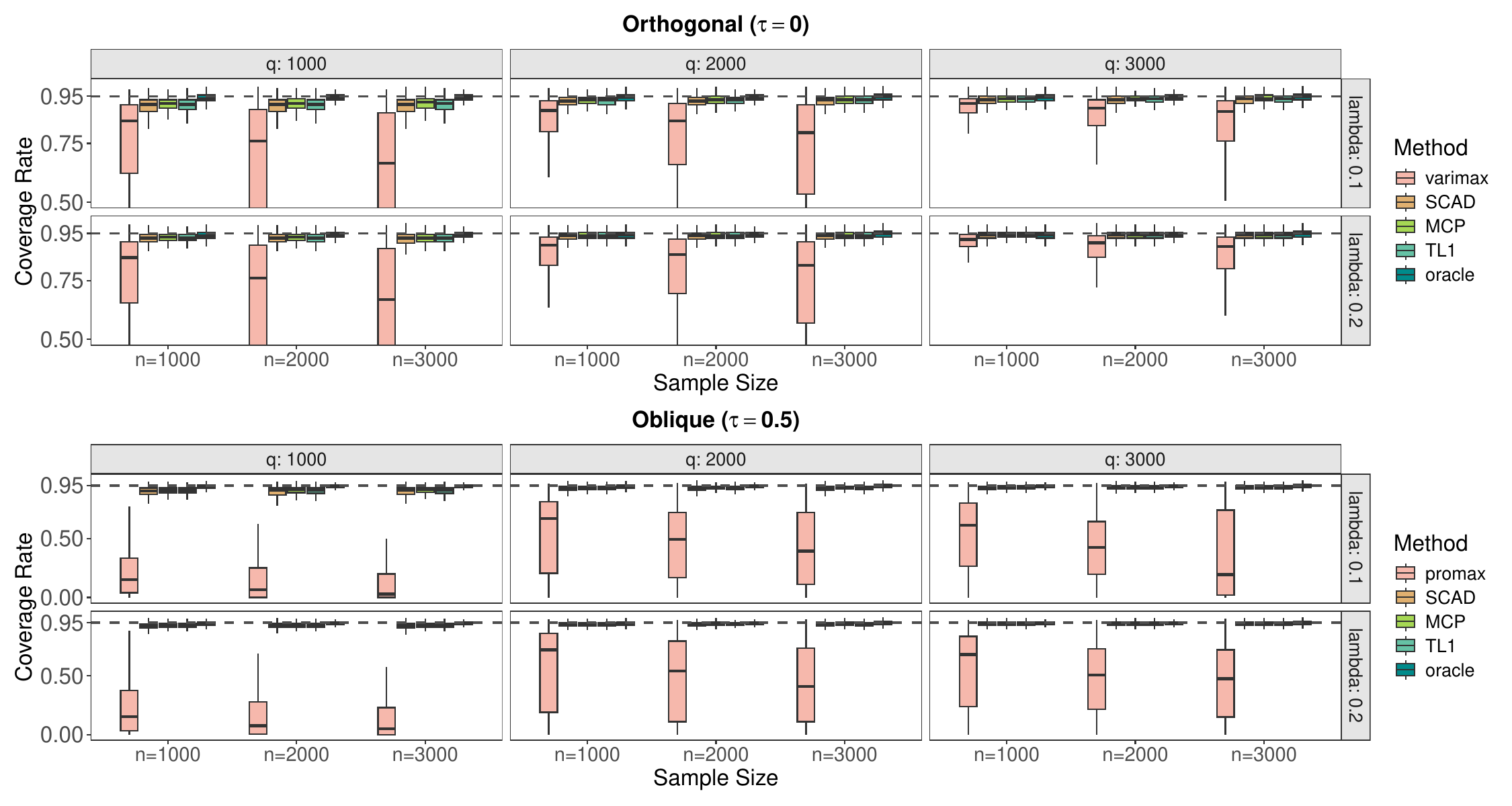}
    \caption{\small  Empirical coverage across 200 replications for $\bA^*$ under different settings of $n$, $q$, $\lambda$ and $\tau$. } 
    \label{fig_loadings_coverage}
\end{figure}
}
With the asymptotic distributions established in Theorem~\ref{thm_oracle} (Theorem~\ref{thm_lqa}), we can construct the 95\% Wald intervals for every entry of $\bZ$ and $\bA$, and the resulting empirical coverage is reported in Figure~\ref{fig_loadings_coverage}. Across scenarios, our method achieves empirical coverages close to the nominal 95\% level. When $n$ and $q$ exceed $2000$, the empirical coverages for both $\bA^*$ and $\bZ^*$ reach about 95\% and well approach the oracle benchmark. 
For comparison, we also use the plug-in covariance estimator in Theorem~\ref{thm_oracle}, evaluated at the vintage methods' estimator, as a naive approach to compute the variance for these methods. The results show that, with the plug-in covariance estimator, varimax and promax methods do not support valid inference regardless of the size of $n$ and $q$.

To further demonstrate that this issue is not due to underestimated variance, we conduct an ``infeasible de-biasing'' procedure for varimax. Using the true $\bA^*$, we compute its varimax optimum and a bias term $\Delta\bA^*$, which is independent of the data (see details in Section~\ref{supp_sec_simu_2}). We call it ``infeasible de-biasing'' as it requires the knowledge of the true $\bA^*$. With $\Delta\bA^*$, we construct a ``de-biased'' estimator, and construct 95\% Wald intervals for it with the plug-in covariance estimator in Theorem~\ref{thm_oracle}. As shown in Figure~\ref{fig_loadings_coverage_oracle} in the Supplementary Material, the resulting empirical coverage is close to 95\%, indicating that the under-coverage of the varimax method in Figure~\ref{fig_loadings_coverage} is inherent to the deterministic bias that is not estimable from the data. 

In summary, the simulation results demonstrate the effectiveness of the proposed method and provide empirical support for our theoretical guarantees. They also substantiate that the bias in vintage methods can be large and cannot be ignored for conducting valid statistical inference.

\subsection{Real Data Analysis}\label{subsec_real_data}
We apply our method to a personality dataset based on the International Personality Item Pool (IPIP) NEO inventory \citep{johnson2014measuring}\footnote{The dataset is publicly available at \href{https://osf.io/tbmh5/}{https://osf.io/tbmh5/}}, a publicly available version of the widely used NEO Personality Inventory \citep{costa2008revised}. The dataset contains 300 survey items designed to measure the Big Five personality traits: Neuroticism, Extraversion, Conscientiousness, Agreeableness, and Openness. The items are organised into five domains, with 60 items per domain. Items in the same domain are designed primarily to assess one single trait, while in reality, responses to them are potentially affected by other traits. We use a subset comprising 335 individuals from the UK who answered all 300 items. For individual $i\in[335]$, we represent the five traits by a $5$-dimensional latent variable $\bz_i$, and we model the 300 items through the representation matrix $\bA = (a_{jl})_{300\times 5}$, where $a_{jl}$ quantifies the association between the response to item $j$ and trait $l$. Responses are recorded on a five-category rating scale. We centre the responses to each item so that their sample mean across individuals is zero. 

We adopt the $\ell_2$ risk function $\ell_{ij}(\theta) = (\theta - Y_{ij})^2$ and we apply our framework in Section~\ref{sec_comp_guarantee} to obtain estimators for $\bA$ and $\bZ$. 
In Figure~\ref{fig_loadings_data}, we show a heatmap for the estimated $\bA^\T$, with the columns ordered into 5 contiguous blocks according to their intended domains. The approximate block diagonal pattern suggests that items in each domain are affected primarily by their target trait while also exhibiting cross-domain effects. 

{\spacingset{1}
    \begin{figure}[h]
\centering    
        \includegraphics[width=5.5in]{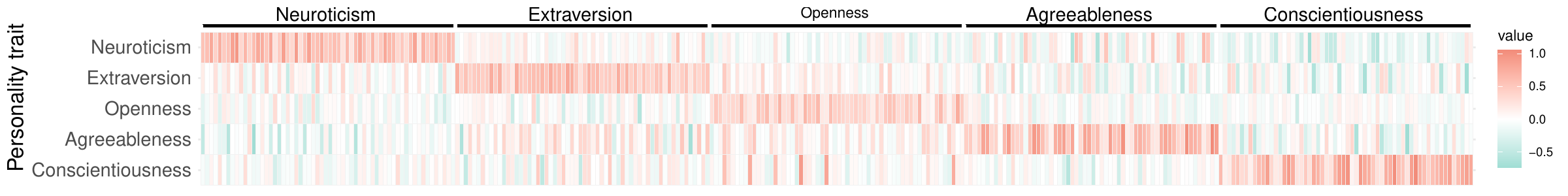}
  \caption{\small Heatmap for the (transposed) estimated representation matrix for the IPIP-NEO dataset. Each row reflects the dependence of 300 items on a certain personality trait. The columns are grouped and labelled by the item pools in each personality domain.}\label{fig_loadings_data}
\end{figure}
    \begin{figure}[h]
\centering    
        \includegraphics[width=5.5in]{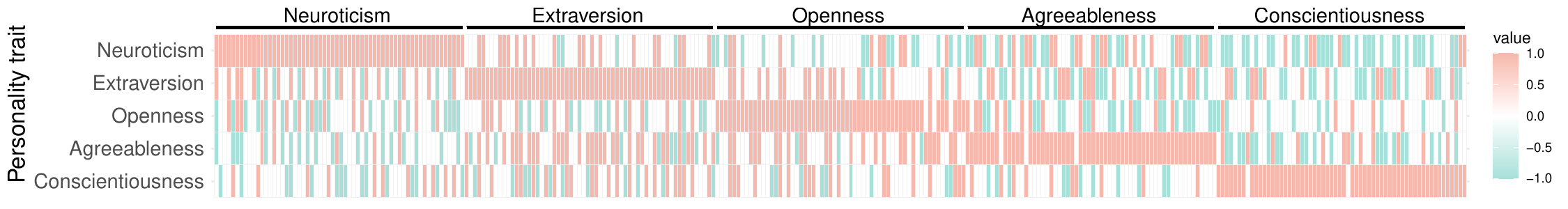}
  \caption{\small Heatmap for the testing results of the (transposed) representation matrix for the IPIP-NEO dataset, with Benjamini-Hochberg correction at significance level $0.05$. Red indicates a significant positive entry, blue indicates a significant negative entry, and white indicates a non-significant entry.}\label{fig_loadings_pvalue}
\end{figure}
 }
 
Using the asymptotic distribution in Theorem~\ref{thm_oracle} (Theorem~\ref{thm_lqa}), we compute $p$-values to test whether each entry of $\bA$ is zero. With the entry-wise asymptotic independence within each column, we treat the problem as five separate multiple testing analyses, one per trait, each involving 300 hypotheses. We then apply the Benjamini-Hochberg procedure to adjust the resulting $p$-values. 
Figure~\ref{fig_loadings_pvalue} reports the adjusted testing results. 
We also perform a more conservative analysis using Bonferroni correction, shown in Figure~\ref{fig_loadings_pvalue_bon} in Section~\ref{supp_sec_data} of the Supplementary Material. The $95$\% Wald confidence intervals for each entry of $\bA$ are also reported in the same section.
One item in the Openness domain shows a notably large dependence on the Conscientiousness trait, with the corresponding entry having a $z$-score of $19.59$. The item, described as ``Like to begin new things'', suggests that the Conscientiousness trait is associated with respondents’ tendency to initiate new activities. Additional examples of large cross-domain dependence are provided in Section~\ref{supp_sec_data} of the Supplementary Material.
Overall, the findings indicate a sparse representation matrix with many simple rows, although the entire structure may not be perfectly simple. This pattern aligns with the $(\lambda,\epsilon)$-sparsity notion, underscoring the applicability of our results.

\section{Discussion}\label{sec_conclude}

In this paper, we propose a novel rotation method, Folomin, and establish its statistical and computational guarantees. Our work is motivated by the observation that many vintage rotation methods yield estimates with a non-estimable and non-negligible bias, which fundamentally prevents valid statistical inference in practical sparsity regimes, including confidence intervals and hypothesis tests. To address this issue, we introduce the Folomin rotation framework based on a family of folded concave losses. We show that the Folomin estimator attains the same asymptotic distribution as the oracle estimator, enabling valid inference for both the representation matrix and the latent variables. For computation, we provide an efficient framework to approximate the Folomin estimator and prove that its output enjoys the same oracle inference property. 

Varimax and other vintage rotations are remarkably successful in practice, and the bias we study in this paper is not a critique of their practical value. Rather, our contribution is to clarify why these methods generally do not support valid statistical inference, including confidence intervals and hypothesis tests, and to make the source of this limitation explicit. We then address this issue with a novel rotation method that achieves the oracle inference property. At the same time, vintage rotations, particularly varimax, remain powerful tools for point estimation of the sparse representations. In our simulation studies, even in the presence of non-estimable bias, varimax can still yield accurate point estimates,  consistent with findings in the existing literature~\citep{Rohe2020VintageFA}. This also motivates a future direction to investigate whether other rotation methods, especially oblique rotations, can yield consistent estimation and thereby can further serve as initialisations for Folomin, enabling valid inference within our framework.

Besides the questions raised above, several further directions merit investigation.
First, our inference results support a range of downstream inference tasks, such as testing block structures for the representation matrix~\citep{brown2015confirmatory}. { Second, our results are established in the asymptotic regime $n,q\to\infty$. If the goal is only estimation and inference for $\bA$, $q$ need not diverge. Establishing inference results in the fixed-$q$ regime remains an interesting direction, where one can use an empirical Bayes formulation that treats the latent variables as random \citep{chen2025item}. Statistical inferences on $\bA$ may be drawn  
based on a likelihood function where the latent variables are marginalised out.}
{Third, the learned sparse representation may serve as more interpretable features in factor-augmented regression and related downstream inference settings~\citep{bing2022inference,bing2025kernel}.}
Moreover, our theory may serve as a foundation for interpretable pattern discovery and uncertainty quantification for more complex latent variable models, {such as latent space models~\citep{hoff2002latent,young2007random,athreya2018statistical}}, deep autoencoders~\citep{ghosh2025effect}, and deep generative models~\citep{ho2020denoising}.

\bibliography{reference}

\clearpage
\appendix
\renewcommand{\theequation}{S\arabic{equation}} 
\setcounter{equation}{0}  

\renewcommand{\thefigure}{S\arabic{figure}} 
\setcounter{figure}{0} 
\renewcommand{\theproposition}{\thesection\arabic{proposition}}
\renewcommand{\thecorollary}{\thesection\arabic{corollary}}

\begin{center}
\huge\bf Supplementary Material
\end{center}

\setcounter{tocdepth}{1}
\addtocontents{toc}{\protect\setcounter{tocdepth}{1}}
\tableofcontents

We introduce notation that will be used in this Supplementary Material. We use $A \lesssim B$ when there exists some constant $C$ such that $A\le CB$ and $A\asymp B$ if $A\lesssim B$ and $B\lesssim A$.  We define $\bm{e}_x^{(y)}$ as the $y$-dimensional vector with the $x$-th entry equal to 1 and all other entries equal to 0.  When $y=r$, we omit the superscript and denote the indicator vector as $\be_x$. Denote $\zero_{n\times m}$ for the $n\times m$ zero matrix. For any matrix $\bM\in\RR^{n\times m}$, let $\sigma_{\max}(\bM)$, $\sigma_{\min}(\bM)$, and $\sigma_{i}(\bM)$ be the largest, smallest, and the $i$th largest singular value of $\bM$, respectively. For any square matrix $\bM\in\RR^{n\times n}$, let $\lambda_{\max}(\bM)$ and $\lambda_{\min}(\bM)$ be the largest and smallest eigenvalues of $\bM$, respectively.
For matrix $\bM\in \mathbb{R}^{n \times m}$, $\mathrm{vec}(\bM)$ is the $nm$ vector obtained by stacking the columns of the matrix $\bM$ on top of one another. For a square matrix $\bM\in\RR^{n\times n}$, we define  $\mathrm{ndiag}(\bM)$ as the operator that preserves only the non-diagonal elements in $\bM$. 
For two vectors $\bx$ and $\by$, define the sine angle as $\sin\angle( \bx,\by) = \{1-\cos\angle( \bx,\by)^2\}^{1/2}$. We denote the vectorized parameters $(\bZ,\bA)$ as $\bphi=(\bphi_z^\T,\bphi_a^\T)^\T$ with $\bphi_z=(\bz_1^\T,\dots,\bz_n^\T)^\T$ and $\bphi_a = (\ba_1^\T,\dots,\ba_q^\T)^\T$, and the vectorized true parameter is denoted by $\bphi^*$ in a similar way.



\section{Remark~\ref{remark_distributional} Continued}\label{supp_sec_distributional}
\begin{proposition}[{ $(\lambda,\epsilon)$-sparsity with Distributional Assumption}]\label{prop_dist}\it 
 Suppose $\{a_{jk}\}_{j\in[q],k\in[r]}\overset{i.i.d}{\sim} F$, where $F$ satisfies $(i)$ $\mathrm{supp}(F) \subset [-M/\sqrt{r},M/\sqrt{r}]$ with $M$ given in Definition~\ref{def_sparsity}; $(ii)$ $F$ has a positive point mass at $0$ with $1/2<F(\{0\})<1$, and no other support in a neighborhood of $0$. Then for some $\lambda,\epsilon>0$, as $q\to\infty$, $\bA$ is $(\lambda,\epsilon)$-sparse with probability approaching 1.
\end{proposition}
\begin{proof}
   By assumption $(ii)$, there is a constant $\lambda>0$ such that every nonzero entry has magnitude at least $\lambda$. Set $p_0=F(\{0\})\in(1/2,1)$ and $\epsilon_0=\lambda/M$. Assumption $(i)$ ensures $\|\ba_j\|\le M$, so every nonzero coordinate of $\ba_j/\|\ba_j\|$ has magnitude at least $\epsilon_0$.
Choose $0<\epsilon<\epsilon_0^2/2$. If two nonzero rows have different supports, their normalised vectors $\bu,\bv$ satisfy $\min\{\|\bu-\bv\|,\|\bu+\bv\|\}\ge\epsilon_0$, and hence $|\bu^\T\bv|<1-\epsilon$. Thus every row in $\cC_\epsilon(\ba_j;\bA)$ has the same support as $\ba_j$.

For $S\subseteq[r]$, let $N_S^{(q)}=q^{-1}\sum_{j=1}^q1_{\{\mathrm{supp}(\ba_j)=S\}}$. The law of large numbers, applied to the finitely many support patterns, gives
\begin{equation*}
N_S^{(q)}\to p_0^{r-|S|}(1-p_0)^{|S|},
\end{equation*}almost surely, simultaneously for all $S\subseteq[r]$.
For $r\ge2$,
\begin{equation}\label{lemma_dis_ineq}
\max_{|S|\ge2}p_0^{r-|S|}(1-p_0)^{|S|}
=p_0^{r-2}(1-p_0)^2<p_0^{r-1}(1-p_0).
\end{equation}
Consequently, almost surely for all sufficiently large $q$,
\begin{equation*}
q^{-1}\max_{j\notin\cS(\bA)}|\cC_\epsilon(\ba_j;\bA)|
\le\max_{|S|\ge2}N_S^{(q)}
<\min_{k\in[r]}N_{\{k\}}^{(q)}
=q^{-1}\min_{k\in[r]}|\cS_k(\bA)|.
\end{equation*}
Zero rows have empty cone neighbourhoods. When $r=1$, every nonzero row is simple and the conclusion follows directly from $p_0<1$. Together with the entry and row norm bounds, this verifies Definition~\ref{def_sparsity}.
\end{proof}

{
\section{Independent Component Analysis}\label{supp_sec_ica}

Our framework is closely related to independent component analysis
(ICA)~\citep{comon1994independent,hyvarinen2001independent}. In ICA, one observes linear mixtures of latent components and seeks to recover both the mixing matrix and the latent components. 
Consider the linear noisy ICA model
\begin{equation}\label{eq_sparse_noisy_ica}
    \bY_i = \bA^*\bz_i^* + \be_i,\qquad i\in[n],
\end{equation}
where $\bA^*\in\mathbb R^{q\times r}$ is the mixing matrix, the coordinates of $\bz_i^*$ are mutually independent, and $\be_i = (e_{i1},\dots,e_{iq})^\T$ is additive noise. 
This fits the setup of Section~\ref{sec_setup} by taking the squared risk $\ell_{ij}(\theta;Y_{ij})=(\theta-Y_{ij})^2$.

ICA resolves the indeterminacy in a different way. Note that \eqref{eq_sparse_noisy_ica} is invariant under simultaneous invertible transformations of the mixing matrix $\bA^*$ and $\bz^*_i$. Rather than assuming sparsity in the representation matrix, as formulated in Definition~\ref{def_sparsity}, ICA imposes independence and non-Gaussianity on the latent components~\citep{hyvarinen2001independent}. The representative is typically identified through distributional features of the latent components, such as kurtosis, negentropy, or higher-order cumulants.

Despite the different perspectives in resolving indeterminacy, our method can also be applied to ICA settings where $\bZ = (\bz_1,\cdots,\bz_n)^\T$ admits a sparse structure. For instance, we may assume that the entries of $\bZ$ are independently generated from a distribution specified in Proposition~\ref{prop_dist}. This setting is closely related to the sparse ICA literature~\citep{zibulevsky2001blind,wang2024sparse}. By letting the mixing matrix $\bA$ play the role of latent variables and latent components $\bZ$ play the role of the representation matrix, we can apply the Folomin rotation to $\hat\bZ$, with the normalisation constraint imposed through $\hat\bA$ where $(\hat\bZ,\hat\bA)$ is obtained from \eqref{eq_theory_init}. Specifically, define
\begin{equation*}
    \widehat\bG_Z \in \argmin_{\bG\in \Xi_Z(\epsilon'_{nq},\hat\bA)}
    Q_{\rho_\gamma}\bigl(\hat\bZ\bG^{-1}\bigr),
\end{equation*}
where
\begin{equation*}
    \Xi_Z(c,\hat\bA) := \left\{ \bG\in\mathbb R^{r\times r}: \|\bG-\bI_r\|\le c,\  \operatorname{diag}\!\big\{\bG(q^{-1}\hat\bA^\T\hat\bA)\bG^\T
        \big\}=\bI_r\right\}.
\end{equation*}
The resulting ICA estimators are $\widehat\bZ_{\rm ICA}:=\hat\bZ\widehat\bG_Z^{-1}$ and $\widehat\bA_{\rm ICA}:=\hat\bA\widehat\bG_Z^\T$, for which we have the following result.

\begin{corollary}[Oracle inference for sparse ICA]
\label{cor_sparse_source_ica}
For model~\eqref{eq_sparse_noisy_ica}, we assume the following conditions.
\begin{enumerate}[label = $(\roman*)$]
    \item \label{assump_ica_1}$\bz_i^*\overset{\mathrm{i.i.d.}}{\sim}\mathcal P_{\bz}$
    have mutually independent coordinates such that $(1)$ $\EE_{\mathcal P_{\bz}}[\bz\bz^\T]\succ 0 $; and $(2)$ $\PP\big\{\bZ^*\text{ is }(\lambda_0,\epsilon_0)\text{-sparse with row index set }[n]\big\}\to 1$ as $n\to\infty$ for some $\lambda_0>0$ and $\epsilon_0>0$. Here, $\bA\succ 0$ means $\bA$ is positive definite.
    \item\label{assump_ica_2} $\|\bA^*\|_{2\to\infty}\le M$, ${\rm diag}(q^{-1}(\bA^*)^\T\bA^*)=\bI_r$ and $q^{-1}(\bA^*)^\T\bA^* \to \bSigma_a^*\succ 0 $.

    \item\label{assump_ica_3} $\{e_{ij}\}_{i\in[n],j\in[q]}$ are independent of
    $\bZ^*$, independent across $(i,j)$, mean zero, and uniformly sub-exponential with variance $\sigma$.

    \item\label{assump_ica_4} As $n,q\to\infty$, $\delta_{nq}:={\log(n\vee q)}{(n\wedge q)^{-1/2}}\to 0$.
    \item\label{assump_ica_5} Let $T_{\ell}^{Z,*}:=\{i\in[n]:z_{i\ell}^*=0\}$ and $
        (\sigma_n^Z)^{-1} := \min_{\ell\in[r]}\lambda_{r-1}\big\{n^{-1}(\bZ^*_{T_{\ell}^{Z,*},\,})^\T \bZ^*_{T_{\ell}^{Z,*},\,}\big\}$. Assume $\delta_{nq}=o(\epsilon_{nq})$, $\epsilon_{nq}=o(\epsilon'_{nq})$, and $\epsilon'_{nq}
        =o\left(\min\{\gamma,(n\wedge q)^{-c},(\sigma_n^Z)^{-1}\}\right)$ for  some constant $0<c<1/2$, with
    $\gamma\le \lambda_0/(a_3+1)$.
\end{enumerate}

Let $(\hat\bZ,\hat\bA)$ be obtained from \eqref{eq_theory_init} and $\widehat\bG_Z$ be defined as above. Then the asymptotic distribution result in Theorem~\ref{thm_oracle} holds for $(\widehat\bZ_{\rm ICA},\widehat\bA_{\rm ICA})$. Moreover, by applying the algorithms in Section~\ref{sec_comp_guarantee} with the roles of $\bA$ and $\bZ$ similarly interchanged, analogous initialisation consistency in Theorem~\ref{thm_init} and one-step update result in Theorem~\ref{thm_lqa} can be established.
\end{corollary}

\begin{proof}
Let $\mathcal E_n$ be the event on which $\bZ^*$ is $(\lambda_0,\epsilon_0)$-sparse. By \ref{assump_ica_1}, $\Pr(\mathcal E_n)\to1$, and $n^{-1}(\bZ^*)^\T\bZ^*\to \bSigma_z^* = \EE_{\mathcal P_{\bz}}[\bz\bz^\T]$ almost surely. Conditional on $\bZ^*$ and on $\mathcal E_n$, assumptions of Theorem~\ref{thm_oracle} hold with $\bA$ and $\bZ$ switched: the sparse representation matrix is $\bZ^*$; the corresponding latent matrix $\bA^*$ satisfies $\|\bA^*\|_{2\to\infty}\le M$ and $\operatorname{diag}\{q^{-1}(\bA^*)^\T\bA^*\}=\bI_r$; and the two limiting covariance matrices are $\bSigma_z^*$ and $\bSigma_a^*$, both positive definite. This verifies Assumption~\ref{assump_psd_covariance}. Assumption~\ref{assump_scaling} is assumed in \ref{assump_ica_3}.
For the squared risk, note that $\ell'_{ij}(\theta_{ij}^*)=-2e_{ij}$, $\ell''_{ij}(\theta_{ij}^*)=2$, and $\ell'''_{ij}(\theta)=0$. Together with \ref{assump_ica_3}, Assumption~\ref{assump_limit} is verified. Since that
\begin{equation*}
    \bPhi_{j}^*=2\bSigma_z^*,\qquad
    \bOmega_{j,a}^*=4\sigma^2\bSigma_z^*,\qquad
    \bPsi_{i}^*=2\bSigma_a^*,\qquad
    \bOmega_{i,z}^*=4\sigma^2\bSigma_a^*,
\end{equation*}
Assumption~\ref{assump_clt} naturally holds.
The other scaling conditions stated in Theorem~\ref{thm_oracle} is imposed in \ref{assump_ica_5}. Therefore we can apply Theorem~\ref{thm_oracle} to show the same asymptotic expansions and thus the same limiting distributions. The algorithms' results can be derived analogously.
\end{proof}

}

\section{Bias in Rotation with \texorpdfstring{$\ell_1$}{ell_1} loss}\label{supp_sec_l1_discussion}


We show that the rotation with $\ell_1$ loss is rotationally biased. 
Consider a representation matrix
\begin{equation}
    \bA = \begin{pmatrix}
        \one_{q/3} & \zero_{q/3} & \zero_{q/3}\\
        \zero_{q/3} & \one_{q/3} & \zero_{q/3}\\
        \zero_{q/3} & \zero_{q/3} & \one_{q/3}\\
        \one_{q^{\prime}} & \one_{q^{\prime}} & \bx
    \end{pmatrix},\label{eq_counter_example_A_supp}
\end{equation}
where $\bx\in\RR^{q^{\prime}}$ with its component $i.i.d.$ from $\cN(0,1)$. Clearly for any integers $q,q^{\prime}$, $\bA$ is $(\lambda,\epsilon)$-sparse with $\lambda<1/2$ and some sufficiently small $\epsilon$ with high probability. 
Define $\ell_1$-loss as $Q_{\ell_1}(\bA) = \sum_{j=1}^{q+q^{\prime}}\sum_{k=1}^r|a_{jk}|$. Let 
\begin{equation*}
    \bG = \begin{pmatrix}
        1-\epsilon & \sqrt{2\epsilon - \epsilon^2} & 0 \\ 0 & 1 & 0 \\ 0 & 0 & 1
    \end{pmatrix}\text{ with }\bG^{-1} = \begin{pmatrix}
        (1-\epsilon)^{-1} & -(1-\epsilon)^{-1}\sqrt{2\epsilon - \epsilon^2} & 0 \\ 0 & 1 & 0 \\ 0 & 0 & 1
    \end{pmatrix}.
\end{equation*}Note that $\bG$ is well-defined real matrix when $\epsilon<1$.
When $q^{\prime}>q/3$ and $\epsilon$ is small enough, we have
\begin{align*}
    Q_{\ell_1}(\bA\bG^{-1}) - Q_{\ell_1}(\bA) =&\, \frac{q}{3}\times \Big(\frac{1 }{1-\epsilon} + \frac{ \sqrt{2\epsilon - \epsilon^2}}{1-\epsilon} - 1\Big)\\&\, + q^{\prime}\times\Big(\frac{1 }{1-\epsilon} +1- \frac{ \sqrt{2\epsilon - \epsilon^2}}{1-\epsilon} - 2\Big)\\ = &\,q\times\frac{(\epsilon + \sqrt{2\epsilon - \epsilon^2})}{3(1-\epsilon)} + q^{\prime}\times\frac{\epsilon - \sqrt{2\epsilon - \epsilon^2}}{(1-\epsilon)}\\= &\,q\times\frac{\sqrt{2\epsilon}}{3(1-\epsilon)} - q^{\prime}\times\frac{\sqrt{2\epsilon}}{(1-\epsilon)} + O(\epsilon)\\<&\,0.
\end{align*}
Thus $\bI_r$ is not a stationary point for $\argmin_{\bG:\mathrm{diag}(\bG\bG^\T) = \bI_r}Q(\bA\bG^{-1})$ and thus $\ell_1$ loss based rotation is rotationally biased.

\section{Algorithm Details}\label{supp_sec_algorithm}

\subsection{Algorithm for initialization (\ref{eq_init_of_init_est}) and Theoretical Guarantees}\label{supp_sec_gda}

We provide a gradient descent algorithm for solving problem~\eqref{eq_init_of_init_est}, summarized in Algorithm~\ref{alg_gradient_descent}. To handle the constraints $n^{-1}\bZ^\T\bZ = \bI_r$ and $q^{-1}\bA^\T\bA = \mathrm{diag}(q^{-1}\bA^\T\bA )$, we employ a Lagrangian-type objective function with quadratic penalty, defined as 
\begin{equation}
    \cL_{G}(\bZ,\bA) = \cL(\bZ,\bA) + \frac{c nq}{2}\big\{\|n^{-1}\bZ^\T\bZ-\bI_r\|_{\Fn}^2 + \|\mathrm{ndiag}(q^{-1}\bA^\T\bA)\|_{\Fn}^2\big\}.\label{eq_langrangian_function}
\end{equation}
Here, $c>0$ is any sufficiently small constant specified in Proposition~\ref{prop_linear_convergence} below, and $\mathrm{ndiag}(\bM) = \bM - \mathrm{diag}(\bM)$. In our numerical implementation, we set $c = 0.1$. Quadratic penalty functions have been used to construct Lagrangian-type formulations in several studies~\citep[e.g.][]{chen2021nonlinear,wang2022maximum,li2023statistical}, which serve as a useful device for analyzing non-convex objectives such as $\cL(\bZ,\bA)$. Different from these studies, we employ the penalty primarily as an algorithmic mechanism to promote convergence, as is shown in Proposition~\ref{prop_linear_convergence} below.

Starting at an initialization $(\tilde\bZ^{(0)},\tilde\bA^{(0)})$, the algorithm iteratively updates all estimators along the gradient direction with step size
\begin{equation*}
    \alpha_{z} = \alpha/q,\quad\alpha_{a} = \alpha/n,
\end{equation*} for some small prespecified constant $\alpha>0$. It is easy to check that the gradient of $\cL_{G}(\bZ,\bA)$ can be computed as follows.
\begin{equation*}
    \partial_{\bZ}\cL_{G}(\bZ,\bA) = \{\ell_{ij}^{\prime}(\ba_j^\T\bz_i)\}_{n\times q}\bA + cnq\partial_{\bZ}\|n^{-1}\bZ^\T\bZ-\bI_r\|_{\Fn}^2/2,
\end{equation*}
and \begin{equation*}
    \partial_{\bA}\cL_{G}(\bZ,\bA) = \{\ell_{ij}^{\prime}(\ba_j^\T\bz_i)\}_{n\times q}^\T\bZ + cnq\partial_{\bA}\|\mathrm{ndiag}(q^{-1}\bA^\T\bA)\|_{\Fn}^2/2,
\end{equation*}
where $\partial_{\bZ}f(\bZ)$ denotes an $n\times r$ matrix $\{\partial_{z_{il}}f(\bZ)\}_{n\times r}$. To satisfy the inequality constraints in \eqref{eq_init_of_init_est}, a row-wise truncation operator $\cT_M(\cdot)$ is then applied using the same constant $M$  as in \eqref{eq_init_of_init_est}. 

The initialization for Algorithm~\ref{alg_gradient_descent} depends on the choice of risk function. For example, under the quadratic risk $\ell_{ij}(\theta)=(Y_{ij}-\theta)^2$, spectral methods are typically used to provide a convenient initialization with theoretical guarantees under moderate conditions~\citep{chen2021spectral,zhang2022heteroskedastic,lyu2025degree}. For binary response models, one may use the SVD-based procedure as in~\citet{zhang2020note}. For a general risk, a practical approach is to generate multiple random initializations, run Algorithm~\ref{alg_gradient_descent} from each starting point, and select the solution that yields the smallest objective value~\citep{collins2001generalization,wang2022maximum}.


\begin{algorithm}[h]
\caption{Projected gradient descent for solving \eqref{eq_init_of_init_est}}
\label{alg_gradient_descent}

\KwIn{Initialization $(\tilde\bZ^{(0)},\tilde\bA^{(0)})$; step size $\alpha$; iterations $T$}
\For{$t = 0,1,\dots, T-1$}{
  \tcp{\textbf{Gradient step}}
Set $\bL^{(t)} = \{\ell_{ij}^{\prime}(\tilde\ba_j^{(t)}{}^\T\tilde\bz_i^{(t)})\}_{n\times q}$
  
  Set $\bZ_{\mathrm{temp}} = \tilde\bZ^{(t)} - \alpha_z \big\{\bL^{(t)} \tilde\bA^{(t)} + cnq\partial_{\bZ}\|n^{-1}\bZ^\T\bZ-\bI_r\|_{\Fn}^2/2\big\}$ and $\bA_{\mathrm{temp}} = \tilde\bA^{(t)} - \alpha_a \big\{\bL^{(t)}{}^\T \tilde\bZ^{(t)} + cnq\partial_{\bA}\|\mathrm{ndiag}(q^{-1}\bA^\T\bA)\|_{\Fn}^2/2\big\}$;

  \tcp{\textbf{Projection}}
  Set $\tilde\bZ^{(t+1)} = \cT_M(\bZ_{\mathrm{temp}})$ and $\tilde\bA^{(t+1)} = \cT_M(\bA_{\mathrm{temp}})$ where $\cT_M(\bX)$ denotes the row-wise truncation operator defined by \begin{equation}(\cT_M(\bX))_{i,}=\left\{\begin{aligned}&\bX_{i,},\text{ if }\|\bX_{i,}\|\le M\\ & \bX_{i,} M/\|\bX_{i,}\|,\text{ if } \|\bX_{i,}\|> M\end{aligned}\right.;\label{eq_ct_row_wise_truncation}\end{equation}
}

\KwOut{$
(\tilde\bZ^{(T)},\tilde\bA^{(T)})$}
\end{algorithm}

To characterize the convergence of Algorithm~\ref{alg_gradient_descent}, we define the $\ell_2$- and $\ell_{\infty}$-error between the optimizer $(\hat\bZ^0,\hat\bA^0)$ and the estimates at the $t$-th iteration as 
\begin{align*}
    \tilde e_{\ell, 2}^{(t)} = n^{-1}\|\tilde\bZ^{(t)} - \hat\bZ^0\|_{\Fn}^2 + q^{-1}\|\tilde\bA^{(t)} - \hat\bA^0\|_{\Fn}^2;\\\tilde e_{\ell, \infty}^{(t)} = \max\big\{\|\tilde\bZ^{(t)} - \hat\bZ^0\|_{2\to\infty} , \|\tilde\bA^{(t)} - \hat\bA^0\|_{2\to\infty}\big\}.
\end{align*}
\begin{proposition}[{Linear Convergence}]\label{prop_linear_convergence}
    Suppose $\tilde e_{\ell,\infty}^{(0)} =o\big(\log(n\wedge q)^{-1}\big) $ and $c$ in \eqref{eq_langrangian_function} satisfies $c<b_L$. Under Assumptions~\ref{assump_psd_covariance}$(ii)$, $(iii)$, Assumptions~\ref{assump_scaling}--\ref{assump_clt}, there exists some absolute constant $c_0$ such that, for any $\alpha < [c_0b_L\min\{\lambda_{\min}(\bSigma_z^*),\lambda_{\min}(\bSigma_a^*)\}]^{-1}$ and any $\delta>0$, with probability approaching $1$, the iterates produced by Algorithm~\ref{alg_gradient_descent} converge linearly to $(\hat\bZ^0,\hat\bA^0)$: for all $t\ge 0$, 
    \begin{equation}
             \tilde e_{\ell, 2}^{(t)}~\le~ C_0\big(1-C_1\alpha \big)^{2t}\tilde e_{\ell,2}^{(0)} ,\;\text{ and }~\;\tilde e_{\ell, \infty}^{(t)}~\le~ C_0\big(1-C_1\alpha\big)^{t}\log(n\vee q)\tilde e_{\ell,\infty}^{(0)},\label{eq_l_2_cont_star}\end{equation}
    for some constants $C_0,C_1$ depending only on $\delta$ with $C_1 = c_0b_L\min\{\lambda_{\min}(\bSigma_z^*),\lambda_{\min}(\bSigma_a^*)\}/4$.
\end{proposition}
\begin{proof}
    See Section~\ref{supp_sec_prove_prop_linear_convergence}.
\end{proof}

\subsection{Hyper-parameters in Algorithm~\ref{alg_init}}\label{supp_subsec_tuning_para}
Theorem~\ref{thm_init} specifies the theoretical rates required for $\delta$ and $\delta^{\prime}$, but only up to constants that depend on the underlying data-generating mechanism. We therefore adopt a data-driven approach to choose $\delta$ and $\delta^{\prime}$. For $\delta$, Theorem~\ref{thm_init} requires that $ \delta_{nq}\ll\delta\le\lambda_0^2\lambda_{\min}(\bSigma_z^*)/16$. In practice, we set $\delta = \{\log (n\vee q)\}^2/\{10(n\wedge q)^{1/2}\}$. Since $\delta$ only plays the role of screening out very small entries, and we are primarily focusing on large values of $\cos\angle(\hat\bZ^0{}^\T\hat\ba_{j_1}^0,\hat\bZ^0{}^\T\hat\ba_{j_2}^0)$, it can be chosen to be moderately large within the admissible range.

For $\delta^{\prime}$, we use repeated random sample splitting. Specifically, for $m=1,\dots,K$, we randomly draw index set $I_1^{(m)},I_{2}^{(m)}\subseteq[n]$ of size $n_{\mathrm{val}}$, which we set as $4n/5$ to ensure that, after splitting, the dimensions $n,q$ are still balanced. Let $\bY_1^{(m)} = \bY_{I_{1}^{(m)},}$ be the training set and $\bY_2^{(m)} = \bY_{I_{2}^{(m)},}$ be the validation set. 
We then choose a grid of values $\delta^{\prime(t)} $ in $(0,1)$ for $1\le t\le M$. For each split $m$, we use $\bY_{I_{1}^{(m)},}$ to obtain estimate $(\hat\bZ^{(m)}_1,\hat\bA^{(m)}_1)$ via \eqref{eq_init_of_init_est}, and for each $\delta^{\prime(t)}$, we use these estimates to obtain $\{\hat\cS_k^{(m,t)}\}_{k\in[r]}$ from Algorithm~\ref{alg_init}. 

Then we use the validation set $\bY_2^{(m)}$ to obtain $(\hat\bZ^{(m)}_2,\hat\bA^{(m)}_2)$ via \eqref{eq_init_of_init_est}.
Define
\begin{align*}
    \mathrm{CV}^{(m)}_t = &\sum_{k=1}^r\sum_{\substack{j_1,j_2\in\hat\cS_k^{(m,t)}\\j_1<j_2}}\frac{1}{|\hat\cS_k^{(m,t)}|(|\hat\cS_k^{(m,t)}| - 1)}\big|1 - \cD_{(j_1,j_2)}^{\delta}\big(\hat\bZ^{(m)}_2,\hat\bA^{(m)}_2\big)\big|,
\end{align*}and for each $t\in[M]$,
\begin{equation*}
    \mathrm{CV}_t = \sum_{s=1}^K\mathrm{CV}^{(s)}_t.
\end{equation*}
Then we pick $\delta^{\prime(t)}$ with $t$ minimizing $\mathrm{CV}_t$.

\subsection{Hyper-parameters in Algorithm~\ref{alg_lqa}}\label{supp_subsec_tuning_para_2}

Theorem~\ref{thm_lqa} characterizes the theoretical rates required for the tuning parameters $\gamma$ and $\eta$. These conditions are relatively mild, and in our experiments, we find that the empirical performance is not sensitive to their choices. Throughout all numerical studies, we fix $R = 1/2$, $\gamma = 0.3$, $\eta = 10^{-6}$ and $T=200$. For $\gamma$, while the theory requires $\gamma(a_3+1)\le \lambda_0$, the algorithm is quite robust with respect to this tuning parameter and continues to produce accurate results even when this condition is only approximately satisfied. We attribute this robustness to the decay of $\rho_{\gamma}^{\prime}(x)$ for large $x$.

\section{Additional Empirical Studies}
{\subsection{Empirical Evaluation of the Bias}\label{supp_sec_bias}
\begin{figure}
    \centering
    \includegraphics[width=0.9\linewidth]{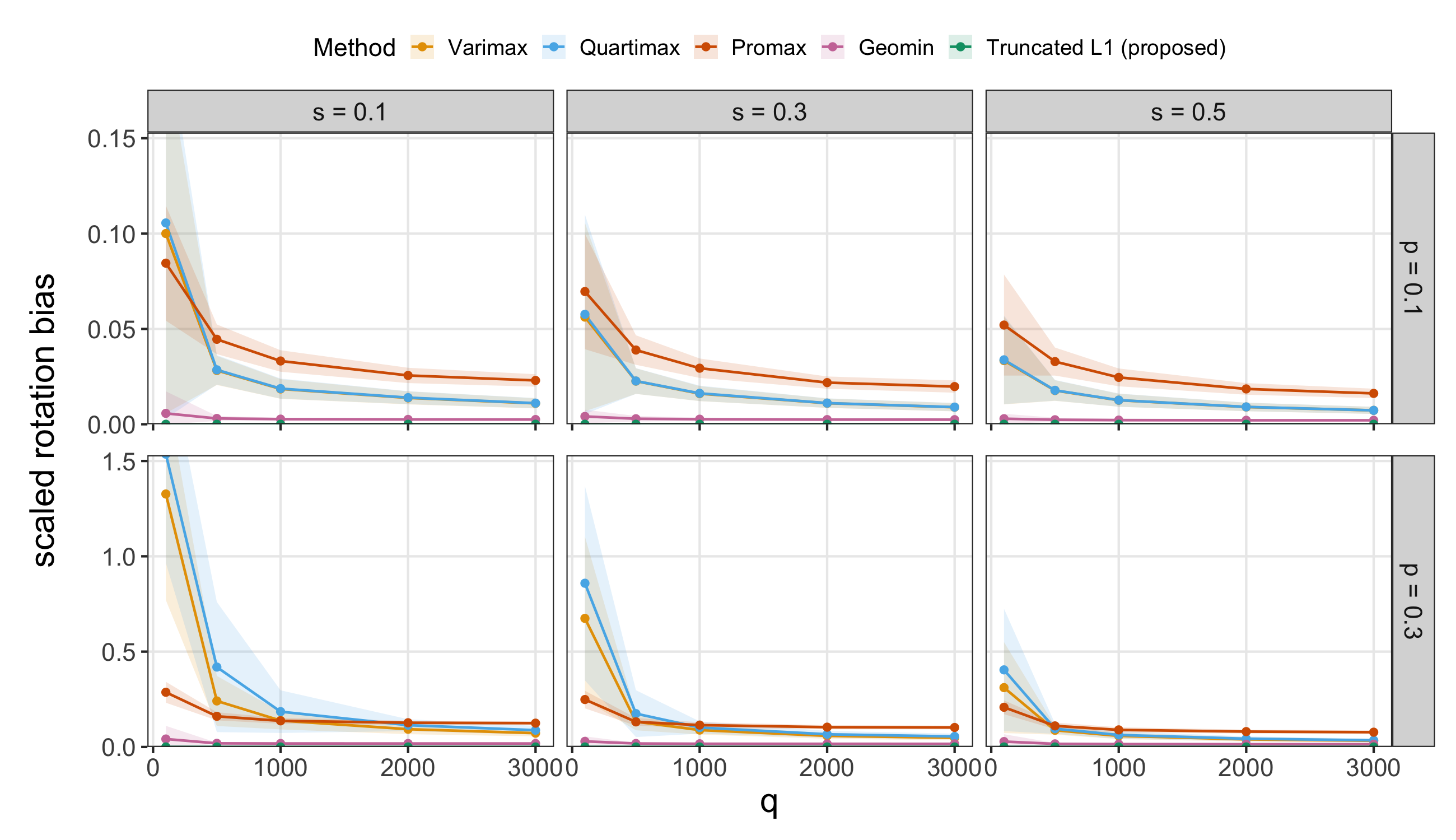}
    \caption{Scaled rotation bias $q^{-1}\|\bA^*-\bA^{\rm rotate}\|_{1,1}$ for vintage rotation methods and the Folomin rotation, under varying matrix size $q$, sparsity level $p$, and proportion of simple rows $s$.}
    \label{fig_bias}
\end{figure}
We conduct a simulation study to empirically examine the intrinsic bias of vintage rotation methods. We set the latent dimension to be $r=5$ and generate the target loading matrix $\bA^*\in\mathbb{R}^{q\times r}$ as follows. A proportion $s$ of rows are generated as simple rows and evenly assigned to the $r$ dimensions; each such row has exactly one nonzero entry sampled from ${\rm Unif}(1,2)$. For the remaining rows, the entries are independently generated from ${\rm Bernoulli}(p)\times{\rm Unif}(1,2)$, where $p$ controls the sparsity level. We consider $q\in\{100,500,1000,2000,3000\}$, $p\in\{0.1,0.3\}$, and $s\in\{0.1,0.3,0.5\}$.

For each generated $\bA^*$, we apply several vintage rotation methods directly to it, including two orthogonal rotations, varimax~\citep{kaiser1958varimax} and quartimax~\citep{ferguson1954concept}, and two commonly used oblique rotations, promax~\citep{hendrickson1964promax} and geomin~\citep{yates1987multivariate}. We compare these methods with the  Folomin rotation with truncated-$\ell_1$ loss as a benchmark. Since the input is the true $\bA^*$, any discrepancy after rotation reflects the intrinsic bias of the rotation criterion rather than statistical estimation error. The bias is evaluated by $q^{-1}\|\bA^*-\bA^{\rm rotate}\|_{1,1}$,
where $\|\bA\|_{1,1}:=\sum_{j=1}^q\sum_{l=1}^r |a_{jl}|$ and $\bA^{\rm rotate}$ denotes the rotated version of $\bA^*$ obtained under a given rotation criterion.

\begin{figure}[H]
    \centering
    \small
    \begin{minipage}{0.9\textwidth}
        \centering
        \includegraphics[width=\textwidth]{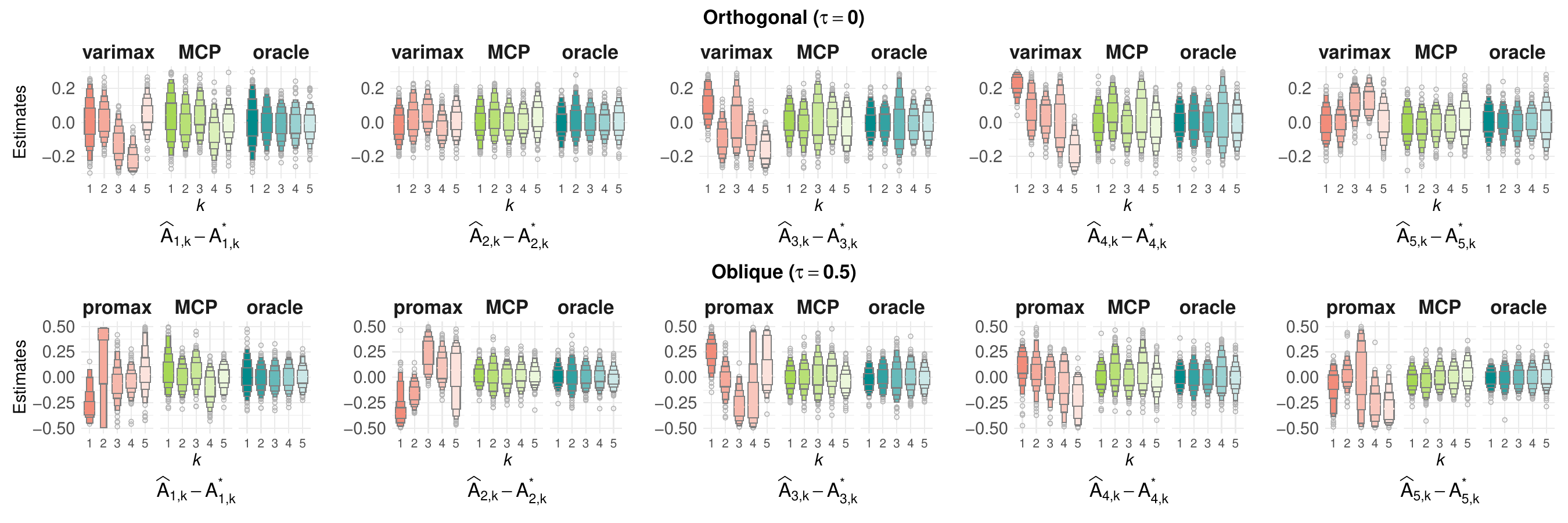}
        \captionof{figure}{Bias in estimating $\bA^*_{1:5,}$ under $n = 1000, q = 1000, \lambda = 0.1$.}
        \label{fig_loading_bias_n1_q1_l1}
        \vfill 

        \includegraphics[width=\textwidth]{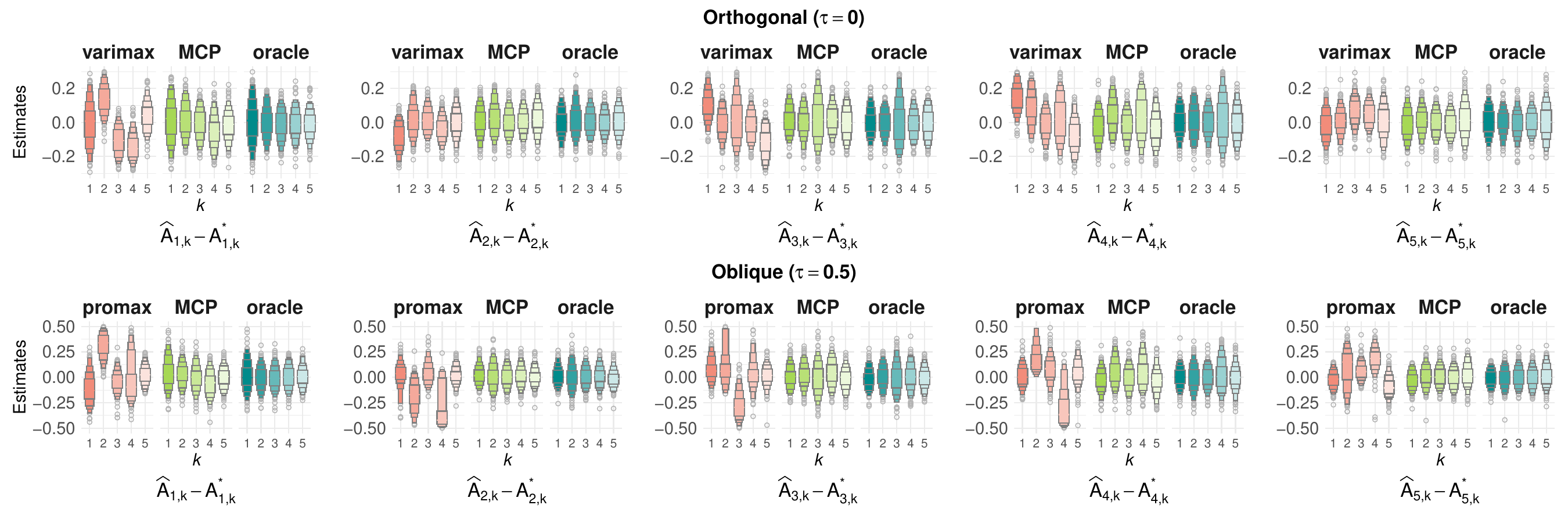}
        \captionof{figure}{Bias in estimating $\bA^*_{1:5,}$: $n=1000, q=2000, \lambda = 0.1$.}
        \label{fig_loading_bias_n1_q2_l1}
        \vfill

        \includegraphics[width=\textwidth]{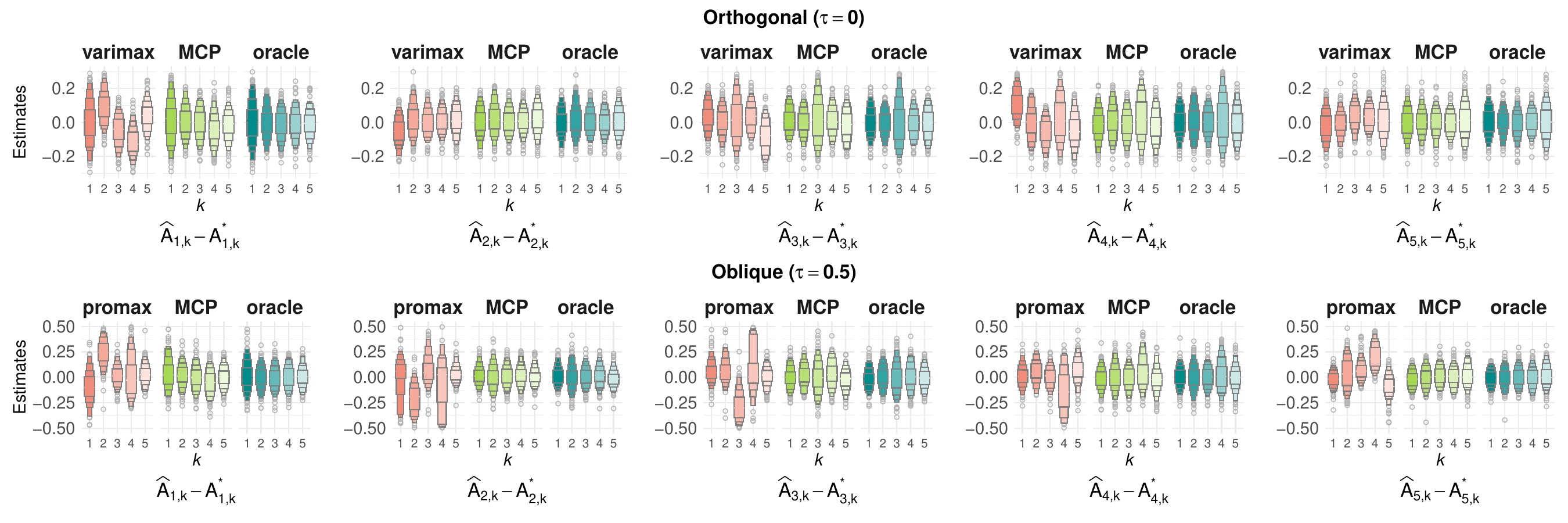}
        \captionof{figure}{Bias in estimating $\bA^*_{1:5,}$: $n=1000, q=3000, \lambda = 0.1$.}
        \label{fig_loading_bias_n1_q3_l1}
    \vfill
    
        \includegraphics[width=\textwidth]{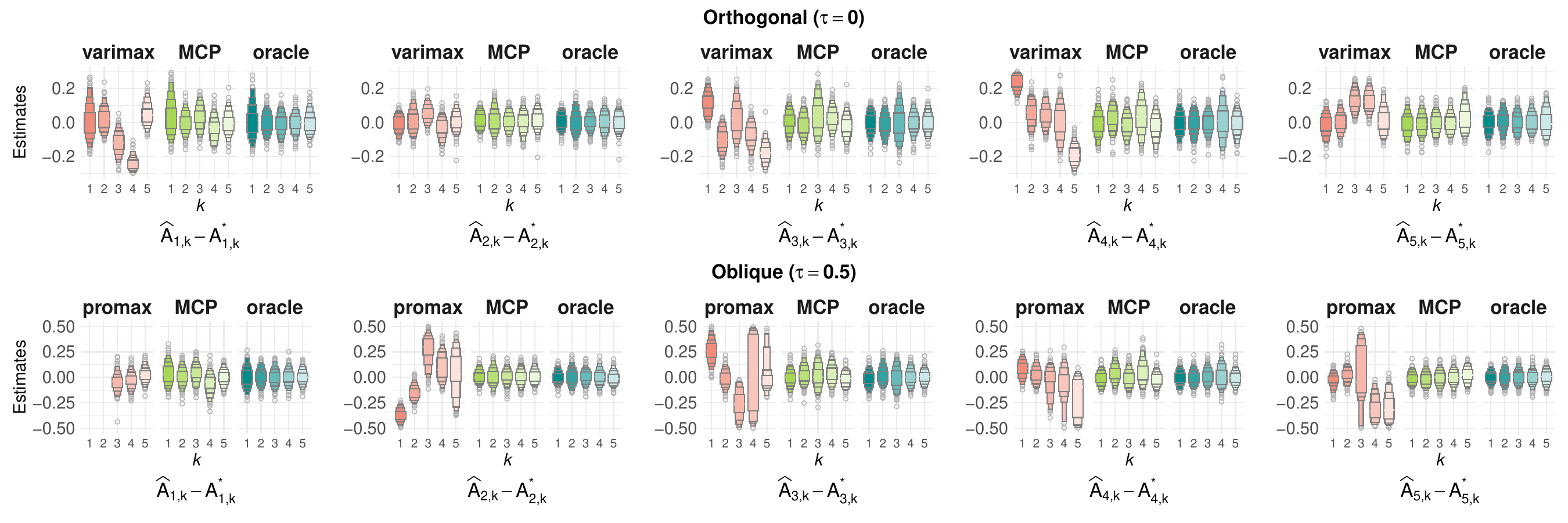}
        \captionof{figure}{Bias in estimating $\bA^*_{1:5,}$: $n=2000, q=1000, \lambda = 0.1$.}
        \label{fig_loading_bias_n2_q1_l1}

    \end{minipage}
\end{figure}

\begin{figure}[H]
    \centering
    \small
    \begin{minipage}{0.9\textwidth}
        \centering

        \includegraphics[width=\textwidth]{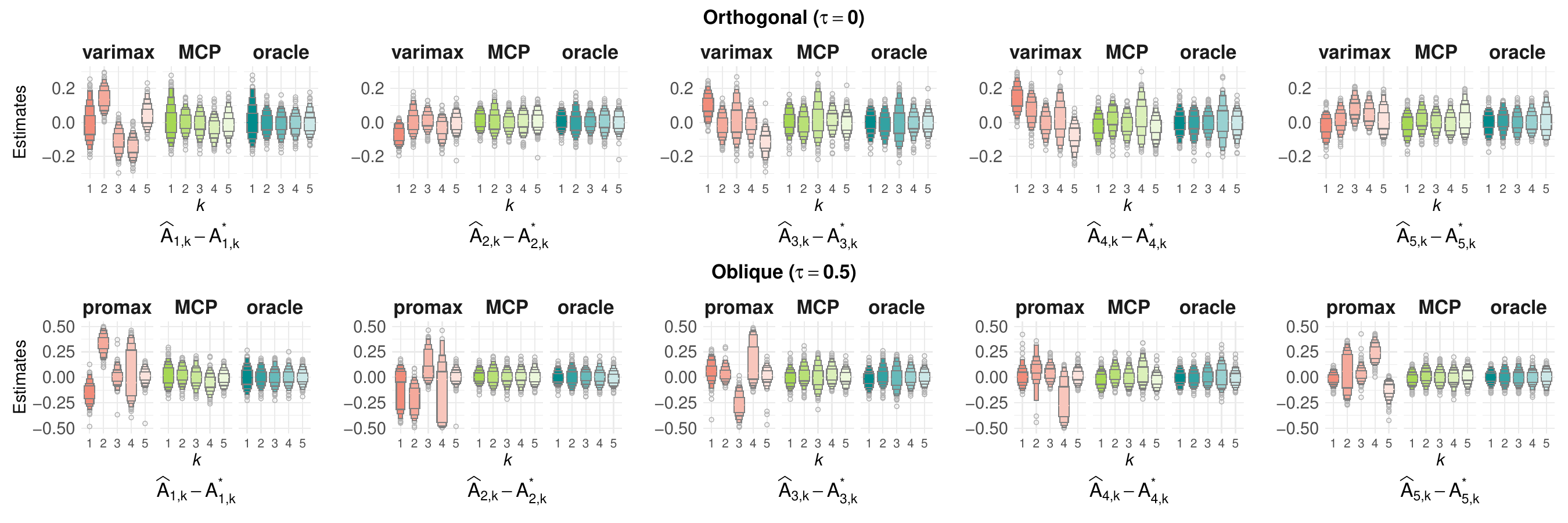}
        \captionof{figure}{Bias in estimating $\bA^*_{1:5,}$: $n=2000, q=2000, \lambda = 0.1$.}
        \label{fig_loading_bias_n2_q2_l1}
        \vfill
        
        \includegraphics[width=\textwidth]{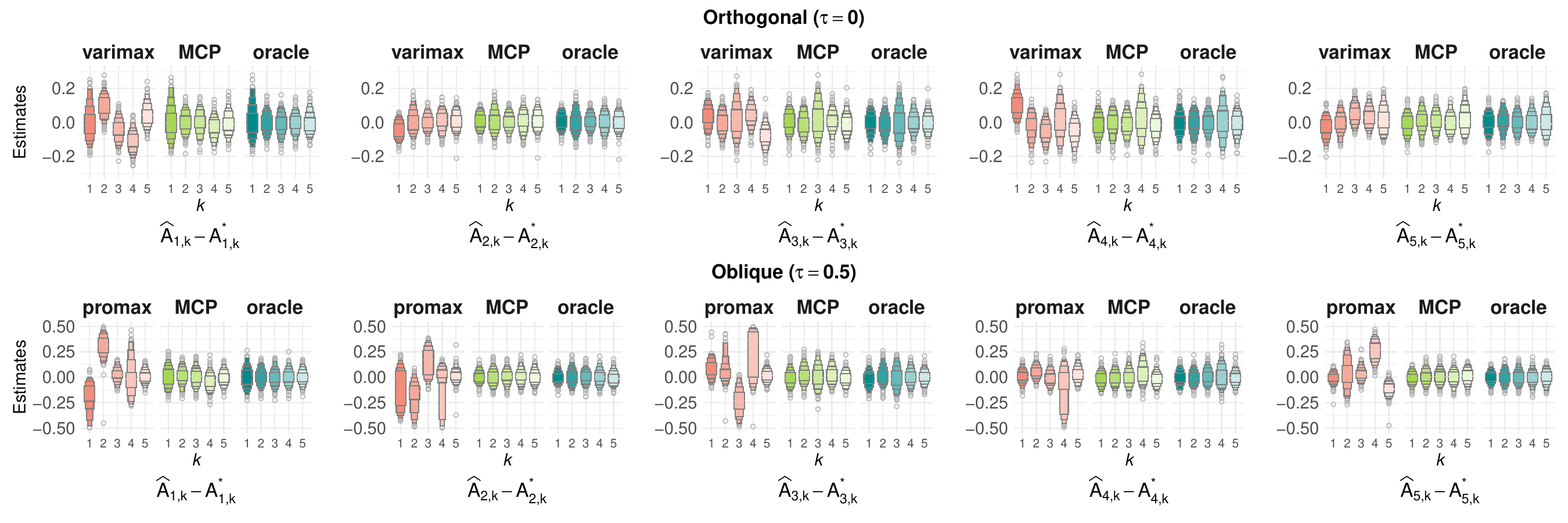}
        \captionof{figure}{Bias in estimating $\bA^*_{1:5,}$: $n=2000, q=3000, \lambda = 0.1$.}
        \label{fig_loading_bias_n2_q3_l1}
        \vfill
        
        \includegraphics[width=\textwidth]{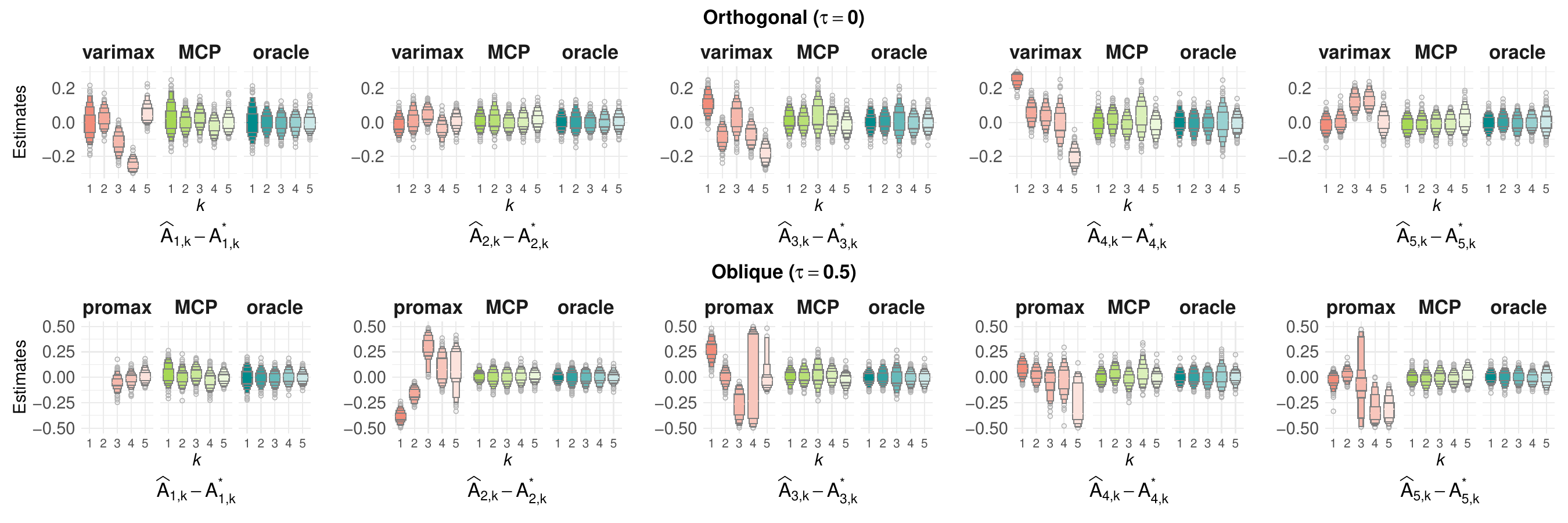}
        \captionof{figure}{Bias in estimating $\bA^*_{1:5,}$: $n=3000, q=1000, \lambda = 0.1$.}
        \label{fig_loading_bias_n3_q1_l1}
        \vfill

        \includegraphics[width=\textwidth]{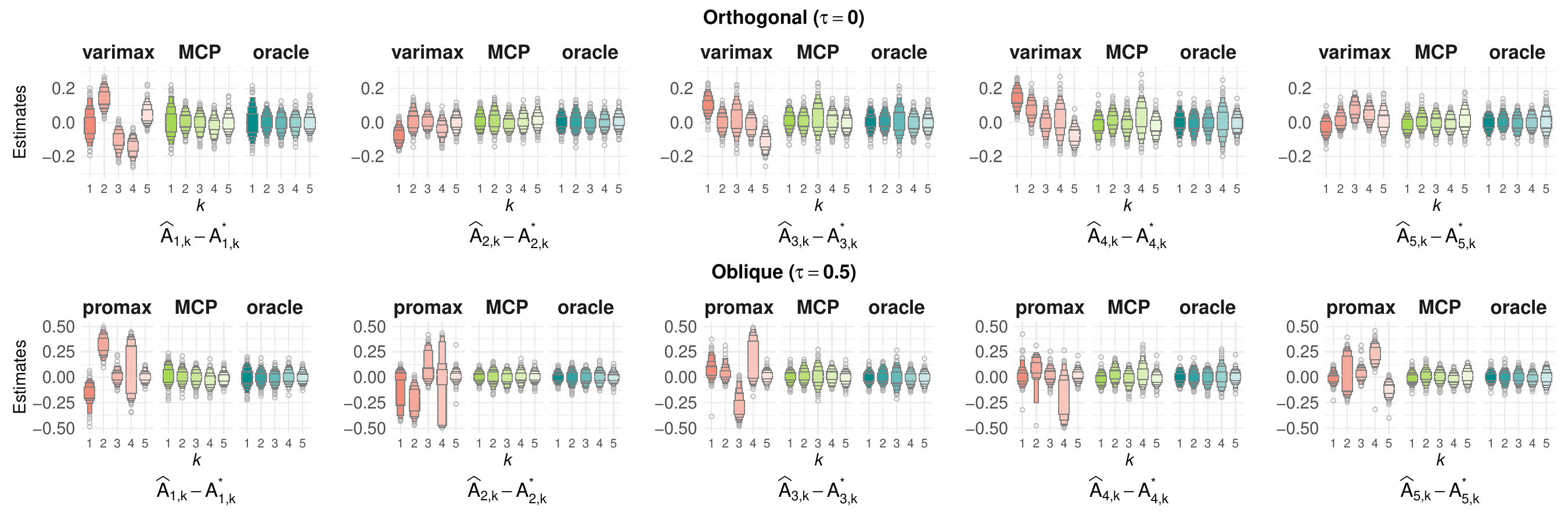}
        \captionof{figure}{Bias in estimating $\bA^*_{1:5,}$: $n=3000, q=2000, \lambda = 0.1$.}
        \label{fig_loading_bias_n3_q2_l1}
        \vfill

    \end{minipage}
\end{figure}

\begin{figure}[H]
    \centering
    \small
    \begin{minipage}{0.9\textwidth}
        \centering

        \includegraphics[width=\textwidth]{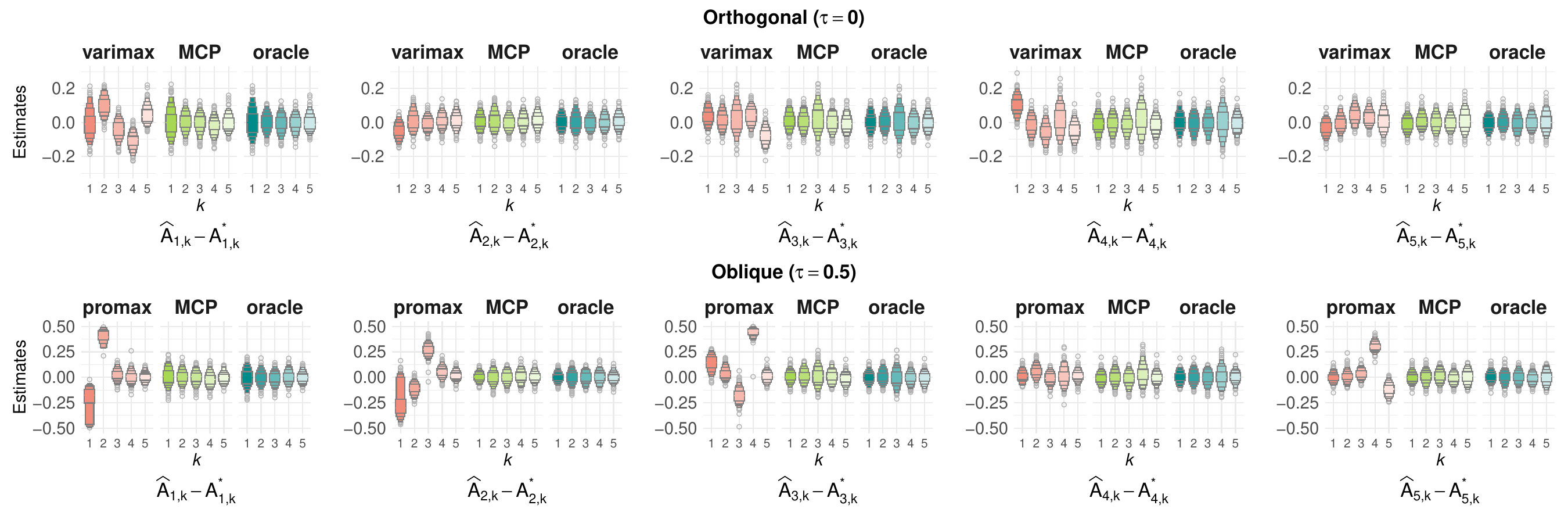}
        \captionof{figure}{Bias in estimating $\bA^*_{1:5,}$: $n=3000, q=3000, \lambda = 0.1$.}
        \label{fig_loading_bias_n3_q3_l1}
\vfill
        
        \includegraphics[width=\textwidth]{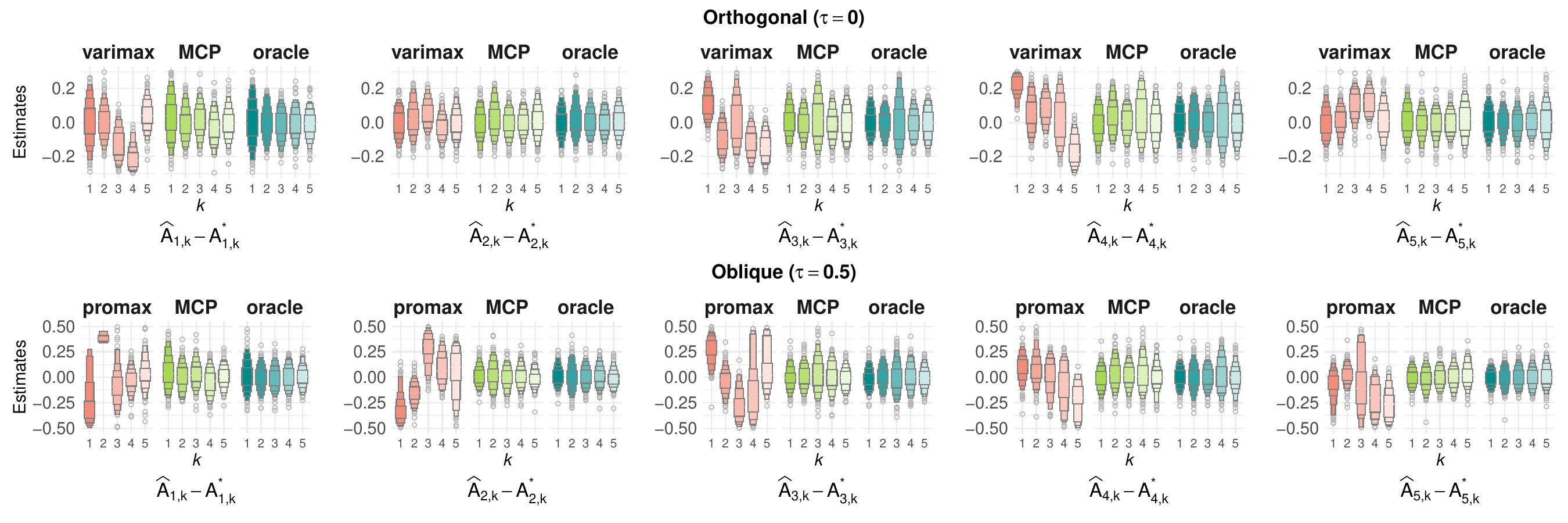}
        \captionof{figure}{Bias in estimating $\bA^*_{1:5,}$: $n=1000, q=1000, \lambda = 0.2$.}
        \label{fig_loading_bias_n1_q1_l2}
        \vfill 
        
        \includegraphics[width=\textwidth]{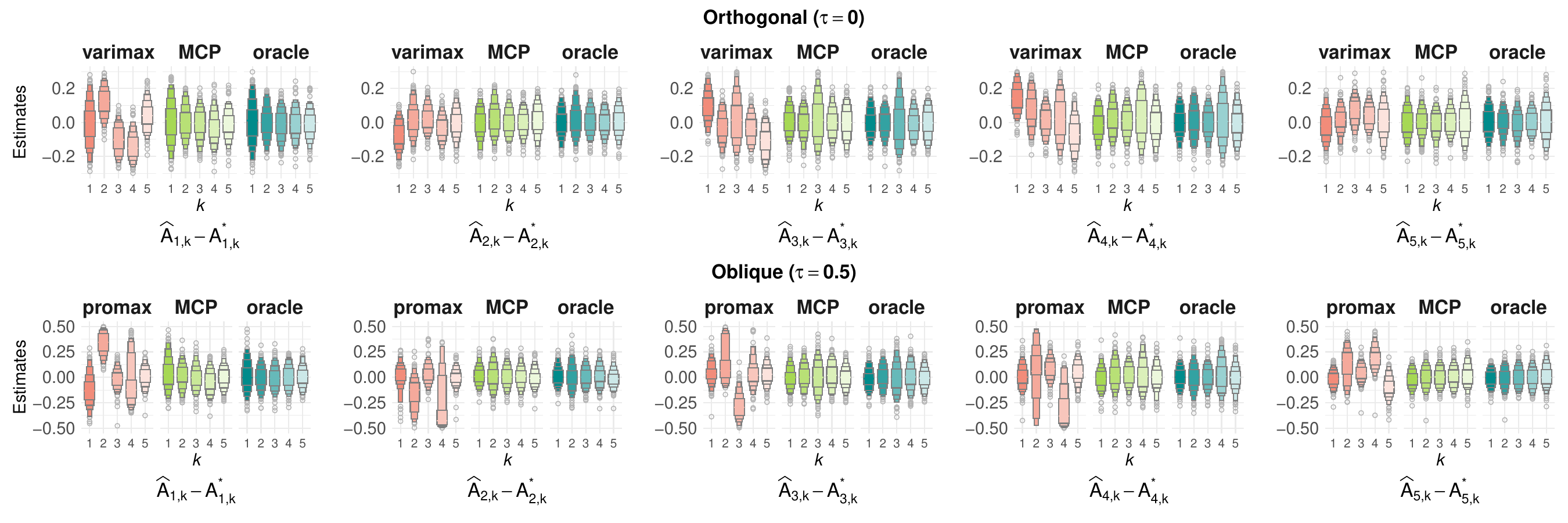}
        \captionof{figure}{Bias in estimating $\bA^*_{1:5,}$: $n=1000, q=2000, \lambda = 0.2$.}
        \label{fig_loading_bias_n1_q2_l2}
        \vfill

        \includegraphics[width=\textwidth]{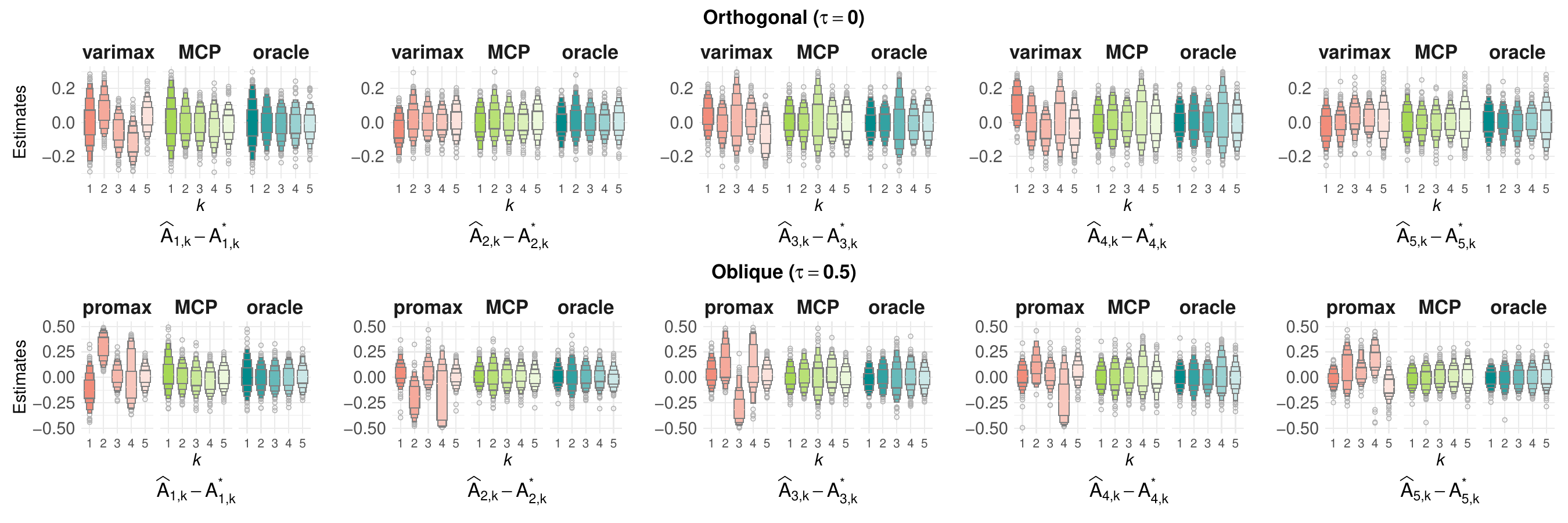}
        \captionof{figure}{Bias in estimating $\bA^*_{1:5,}$: $n=1000, q=3000, \lambda = 0.2$.}
        \label{fig_loading_bias_n1_q3_l2}
    \end{minipage}
    \end{figure}

\begin{figure}[H]
    \centering
    \small
    \begin{minipage}{0.9\textwidth}
        \centering
         \includegraphics[width=\textwidth]{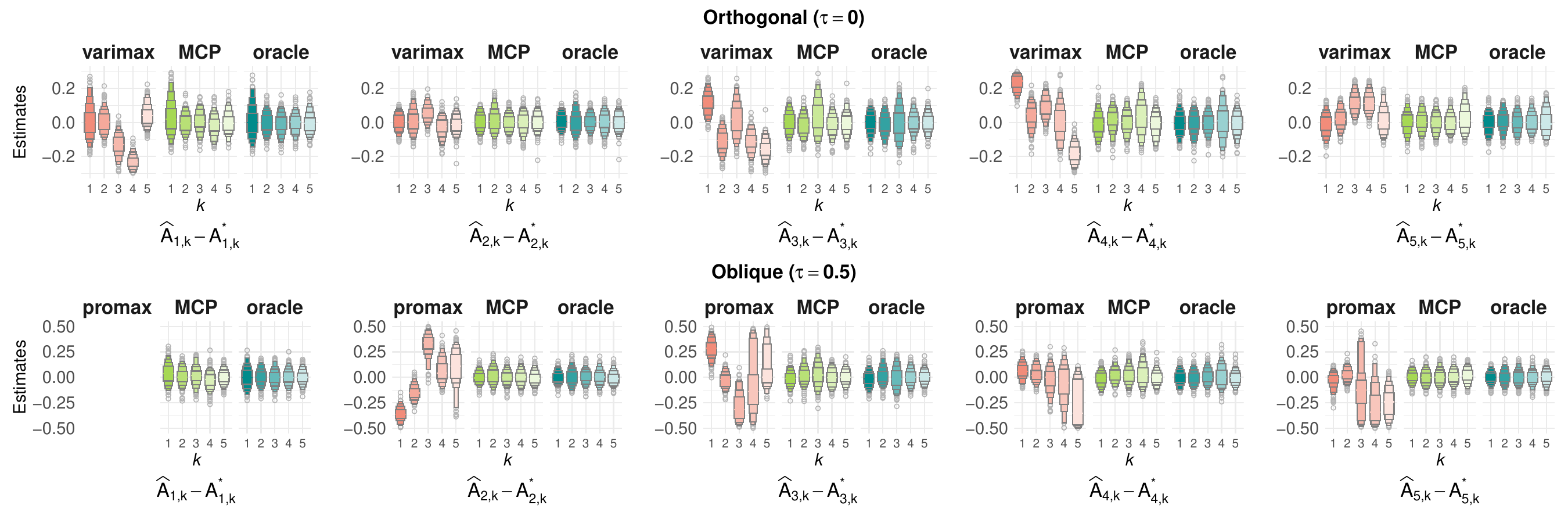}
        \captionof{figure}{Bias in estimating $\bA^*_{1:5,}$: $n=2000, q=1000, \lambda = 0.2$.}
        \label{fig_loading_bias_n2_q1_l2}
        \vfill
        
        \includegraphics[width=\textwidth]{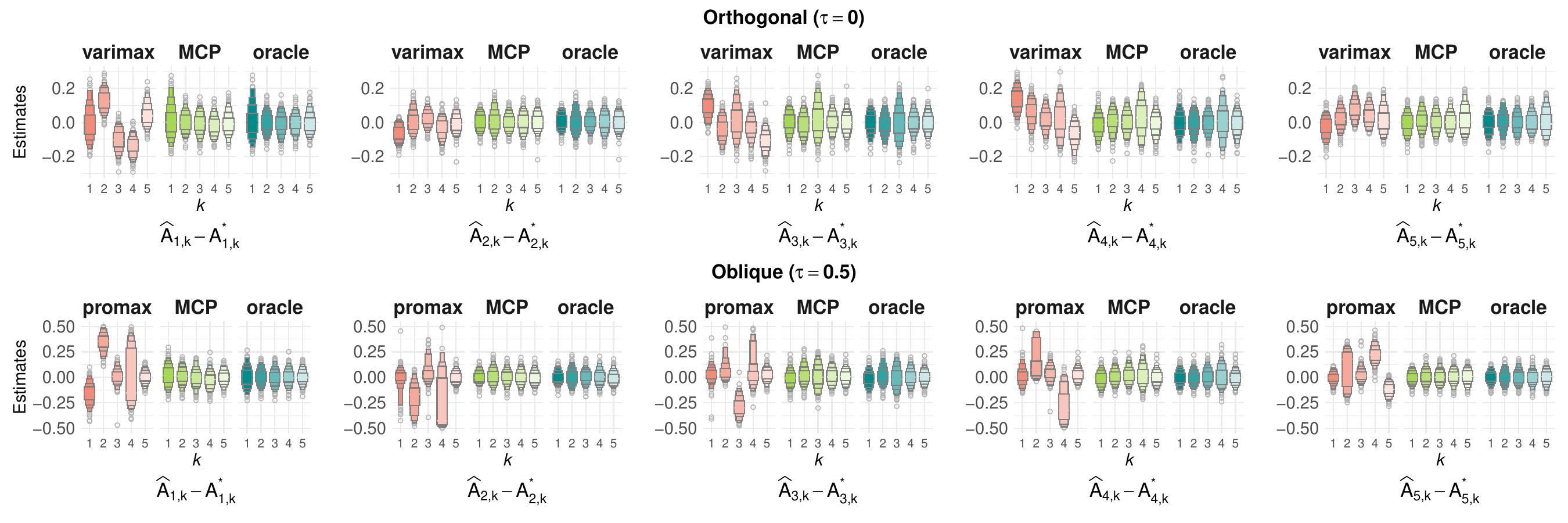}
        \captionof{figure}{Bias in estimating $\bA^*_{1:5,}$: $n=2000, q=2000, \lambda = 0.2$.}
        \label{fig_loading_bias_n2_q2_l2}
        \vfill

        \includegraphics[width=\textwidth]{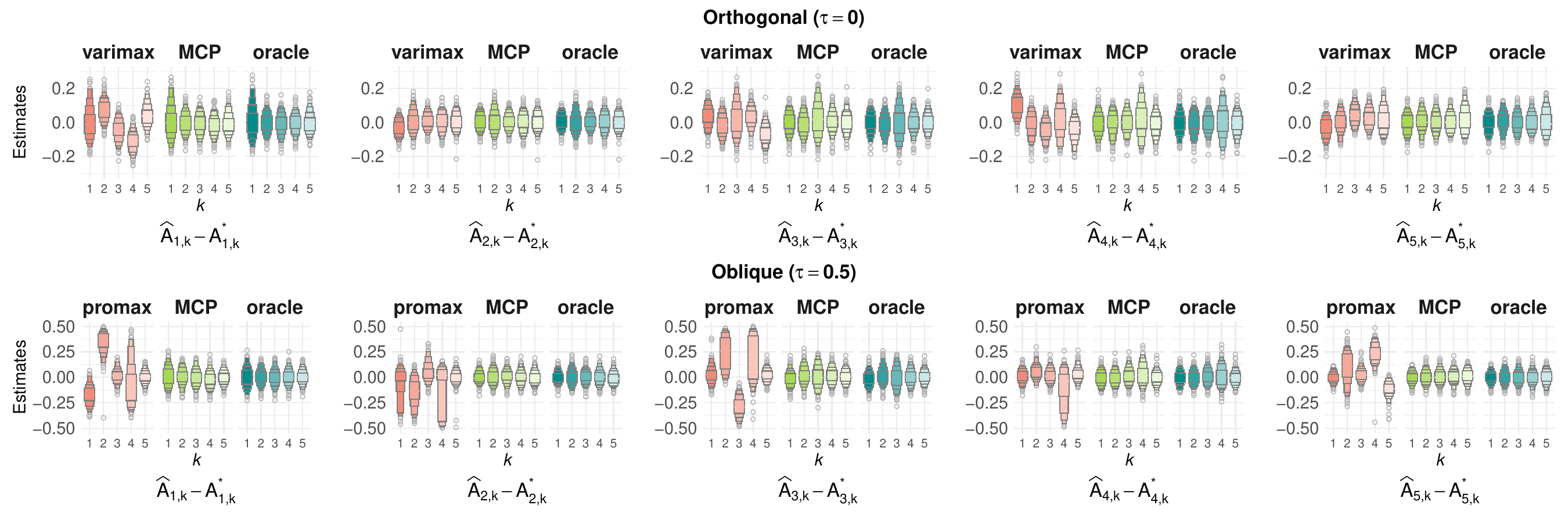}
        \captionof{figure}{Bias in estimating $\bA^*_{1:5,}$: $n=2000, q=3000, \lambda = 0.2$.}
        \label{fig_loading_bias_n2_q3_l2}
        \vfill 
        
        \includegraphics[width=\textwidth]{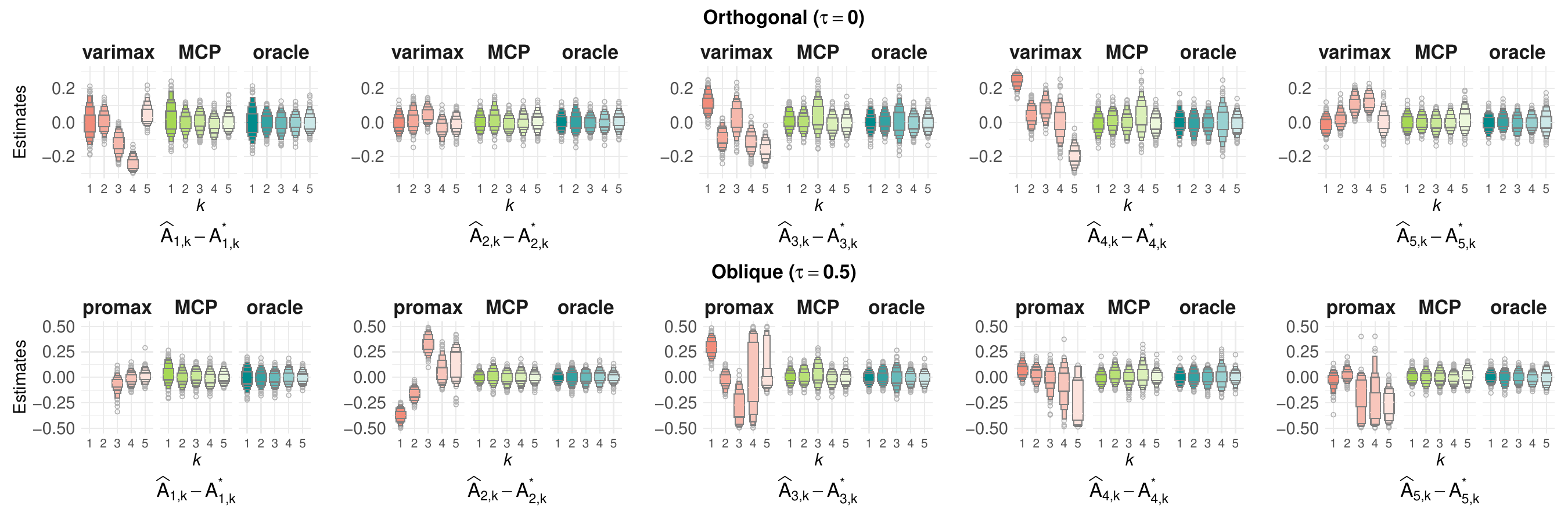}
        \captionof{figure}{Bias in estimating $\bA^*_{1:5,}$: $n=3000, q=1000, \lambda = 0.2$.}
        \label{fig_loading_bias_n3_q1_l2}
    \end{minipage}
\end{figure}

\begin{figure}[h]
    \centering
    \small
    \begin{minipage}{0.9\textwidth}
        \centering

            \includegraphics[width=\textwidth]{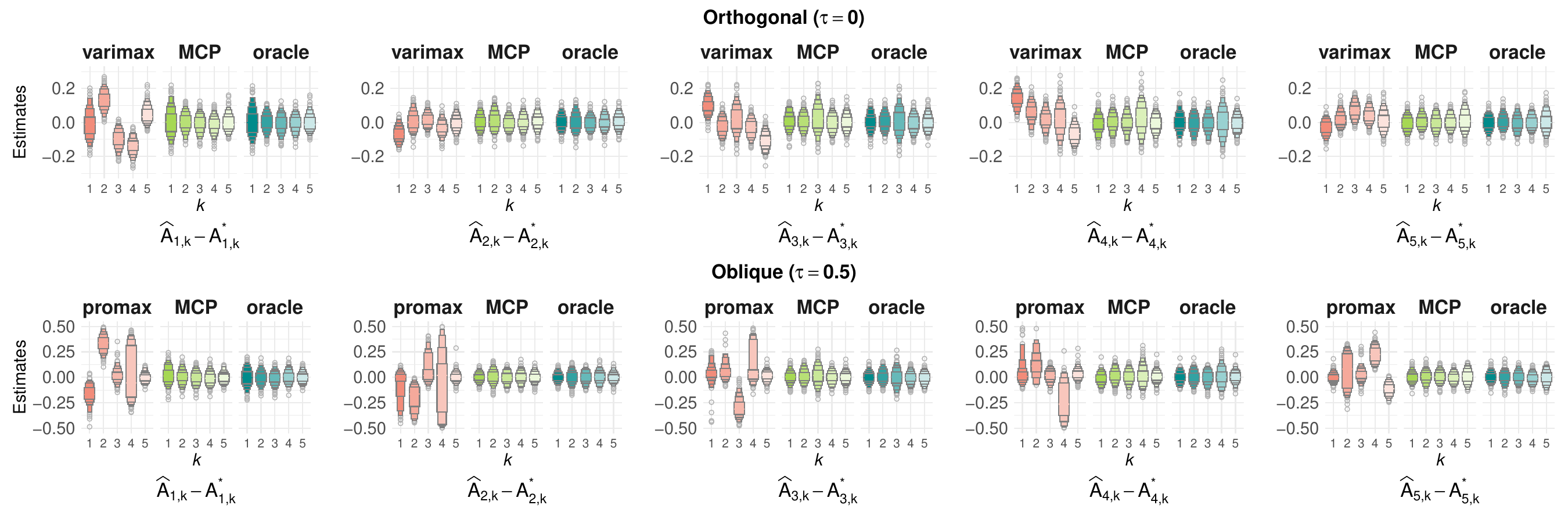}
        \captionof{figure}{Bias in estimating $\bA^*_{1:5,}$: $n=3000, q=2000, \lambda = 0.2$.}
        \label{fig_loading_bias_n3_q2_l2}
        \vfill 
        
        \includegraphics[width=\textwidth]{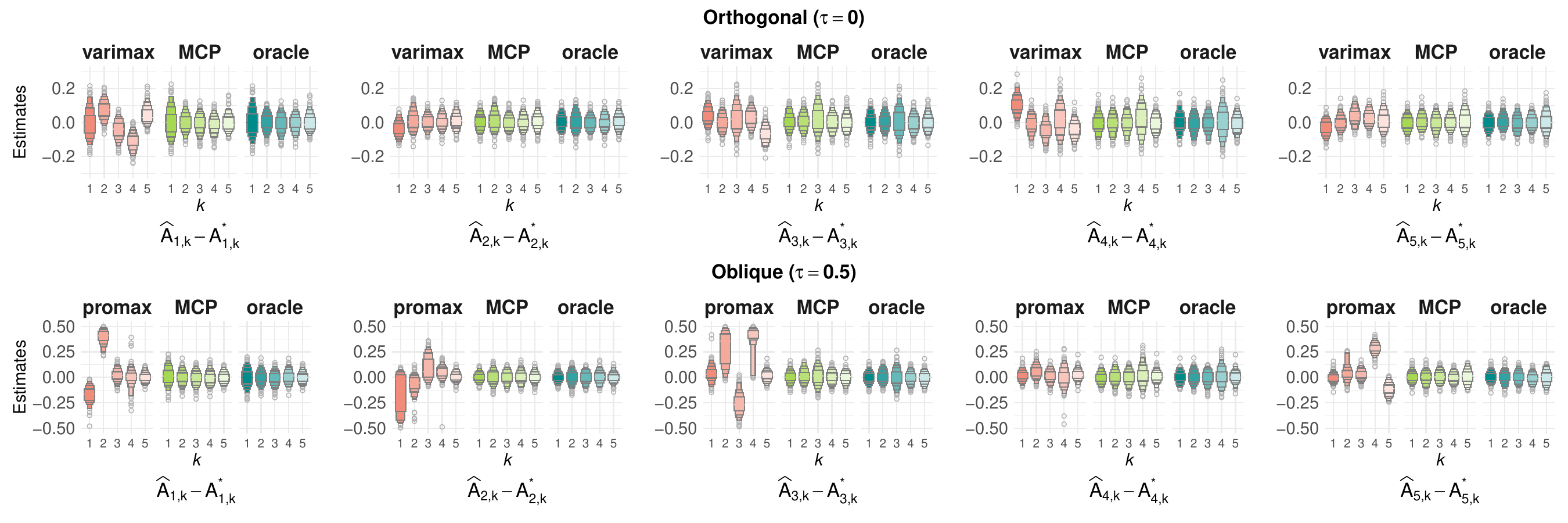}
        \captionof{figure}{Bias in estimating $\bA^*_{1:5,}$: $n=3000, q=3000, \lambda = 0.2$.}
        \label{fig_loading_bias_n3_q3_l2}

    \end{minipage}
\end{figure}

The results are reported in Figure~\ref{fig_bias}. Overall, the bias of vintage rotations generally decreases as $q$ increases, but remains nonzero in all finite-sample settings considered. The bias also tends to be smaller when the proportion of simple rows is larger and when the loading matrix is sparser. Among the compared vintage methods, geomin exhibits relatively small bias in our experiments. This is consistent with the fact that its idealized formulation contains a nonsmooth component at zero, although practical implementations often introduce a smoothing constant for computational stability. Our method yields zero bias up to numerical precision. We emphasize that this experiment is designed to demonstrate the presence of intrinsic rotation bias. Whether such bias is large enough to affect valid inference is further investigated in Section~\ref{supp_sec_simu_2}.
}

\subsection{Additional Results for Simulation Study I}\label{supp_sec_simu_1}
In this section, we report additional simulation results. Figures~\ref{fig_loading_bias_n1_q1_l1}--\ref{fig_loading_bias_n3_q3_l2} report the estimation bias for the first five rows of $\bA^*$, shown in the same layout as Figure~\ref{fig_loading_bias}, for the remaining settings. The patterns are consistent with those reported in Section~\ref{subsec_simu} of the main text.
Figure~\ref{fig_factors_coverage} summarizes the empirical coverage for $\bZ^*$ across methods and parameter settings.

{\small \spacingset{1}
\begin{figure}[h]
    \centering
        
    \includegraphics[width=6.2in]{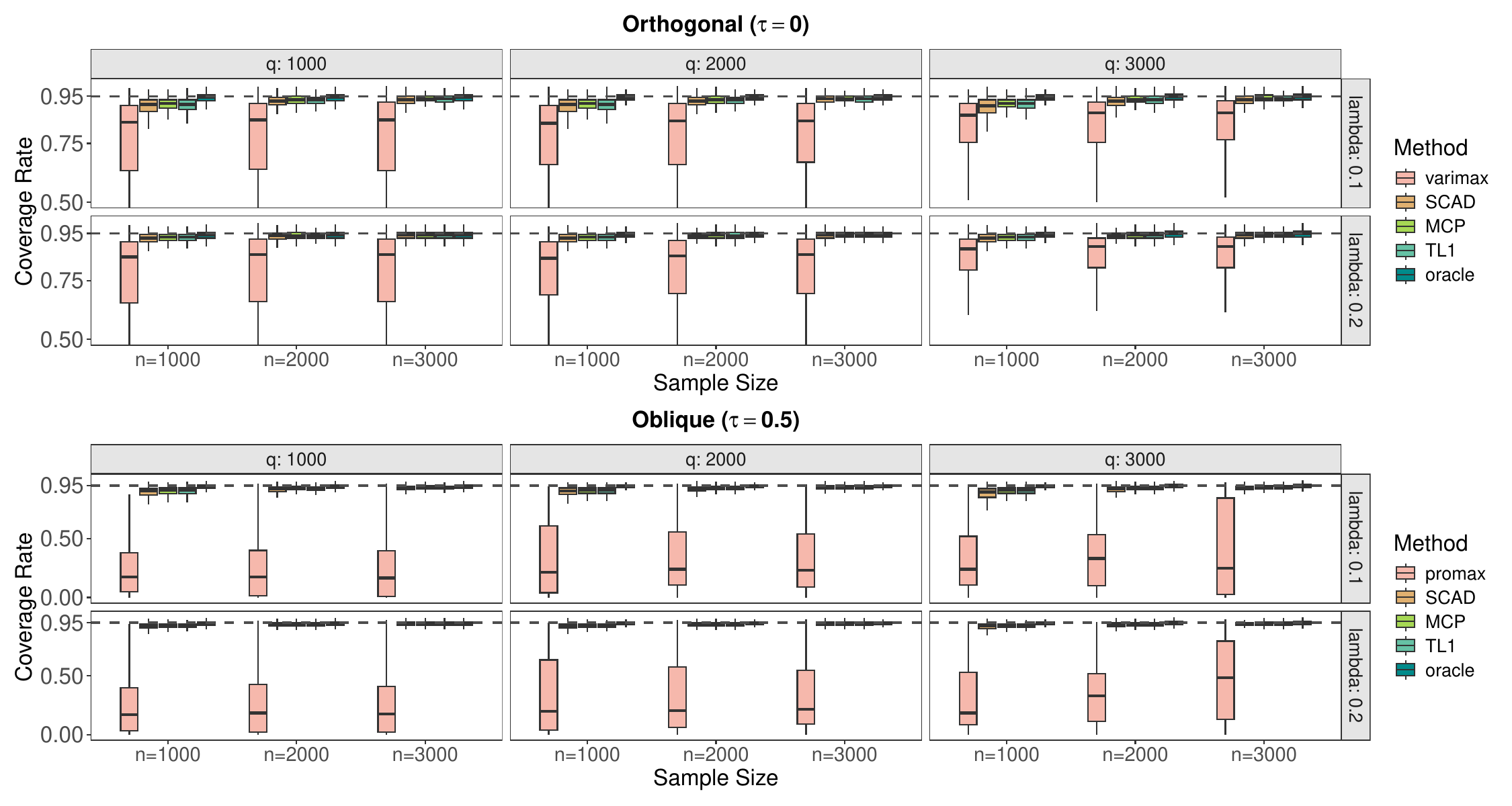}
    \caption{Empirical Coverage for $\bZ^*$ under different settings of $n$, $q$, $\lambda$ and $\tau$. } 
    \label{fig_factors_coverage}
\end{figure}
}

\subsection{Simulation Results of the ``infeasible de-biasing'' procedure}\label{supp_sec_simu_2}
Here, we describe the ``infeasible de-biasing'' procedure and present the resulting coverage. With the true representation matrix $\bA^*$, we compute {
\begin{equation*}
    \bG^*_{\rm varimax} = \argmin_{\bG\in\{\bG:\bG^\T\bG = \bI_r\}}Q_{\rm varimax}(\bA^*\bG),
\end{equation*}}
where {$Q_{\rm varimax}(\cdot)$} is defined in Example~\ref{prop_anti_varimax} in the main text. Then $\Delta\bA^*$ is constructed as {$\Delta\bA^* = \bA^*\bG^*_{\rm varimax} - \bA^*$}. The ``de-biased'' estimator is defined as {$\hat\bA_{\rm varimax} - \Delta\bA^*$} where {$\hat\bA_{\rm varimax}$} is the estimator provided by varimax method. Note that $\Delta\bA^*$ is constructed only when the true parameter $\bA^*$ is assumed to be known, and thus we call this de-biasing procedure infeasible. The empirical coverages of the ``de-biased estimator'' for $\bA^*$ and $\bZ^*$ with the plug-in covariance estimator in Theorem~\ref{thm_oracle} are presented in Figures~\ref{fig_loadings_coverage_oracle} and~\ref{fig_factors_coverage_oracle}, respectively. We can see that after accounting for the error $\Delta\bA^*$, the empirical coverages reach the nominal level across different scenarios. This indicates that the bias $\Delta\bA^*$ indicates the infeasibility of valid inference under varimax estimator.

{\small \spacingset{1}
\begin{figure}[h]
    \centering
    \includegraphics[width=6.2in]{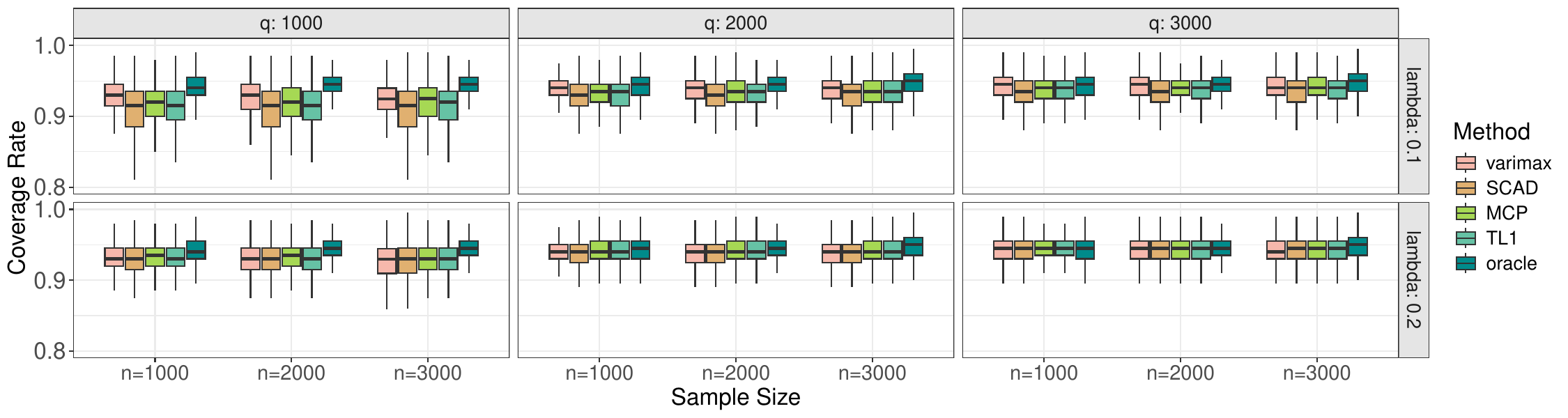}
    \caption{Empirical Coverage for $\bA^*$ under different settings of $n$, $q$, $\lambda$ and $\tau$, where varimax is obtained after the ``infeasible de-biasing'' procedure as described in Section~\ref{supp_sec_simu_2}.} 
    \label{fig_loadings_coverage_oracle}
\end{figure}
\begin{figure}[h]
    \centering
    \includegraphics[width=6.2in]{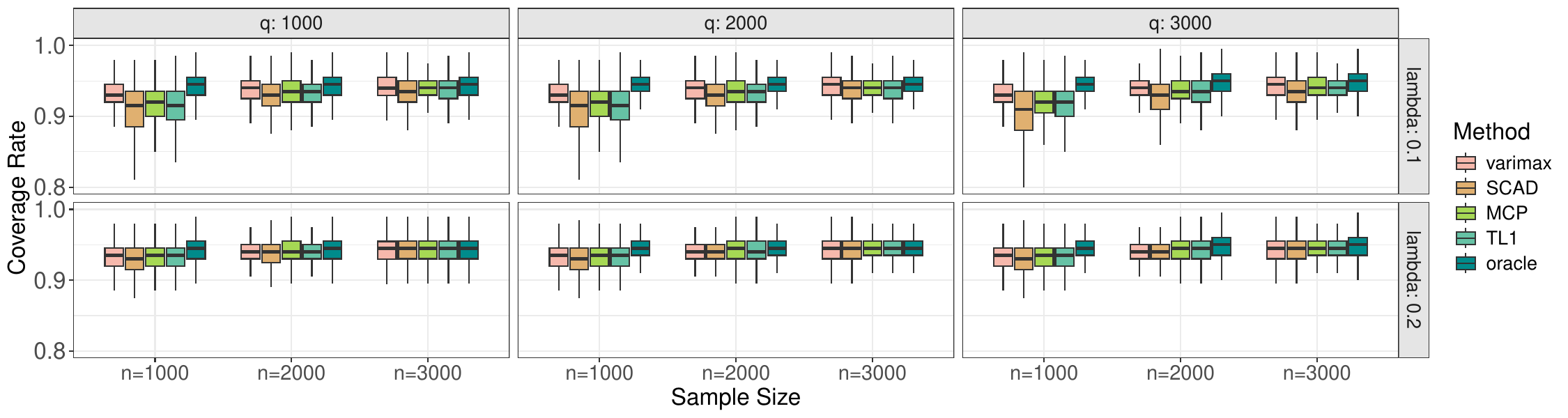}
    \caption{Empirical Coverage for $\bZ^*$ under different settings of $n$, $q$, $\lambda$ and $\tau$, where varimax is obtained after the ``infeasible debias'' procedure as described in Section~\ref{supp_sec_simu_2}.} 
    \label{fig_factors_coverage_oracle}
\end{figure}
}

\subsection{Additional Real Data Results}\label{supp_sec_data}
\begin{figure}[h]
\centering    
        \includegraphics[width=5.4in, height=2in]{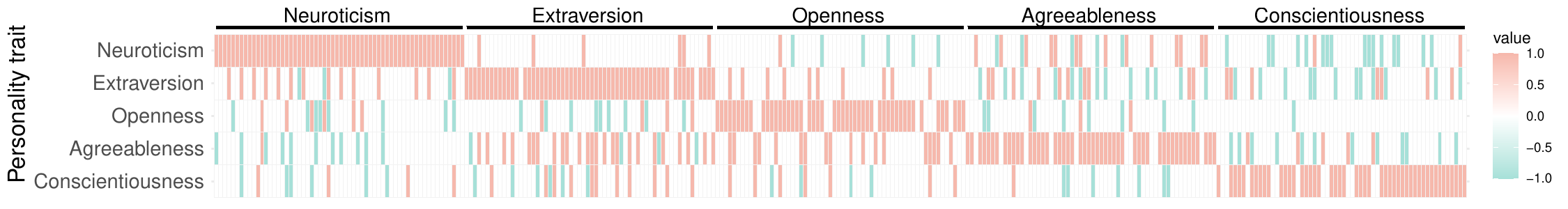}
  \caption{\small Heatmap for the (transposed) testing results of the representation matrix for the IPIP-NEO dataset with Bonferroni correction at significance level $0.05$. Red indicates a significant positive entry, blue indicates a significant negative entry, and white indicates a non-significant entry.}\label{fig_loadings_pvalue_bon}
\end{figure}

In this section, we provide additional results for the data analysis in Section~\ref{subsec_real_data} of the main text. Figure~\ref{fig_loadings_pvalue_bon} reports hypothesis testing results analogous to Figure~\ref{fig_loadings_pvalue}, but using the more conservative Bonferroni correction for the 300 simultaneous tests per trait at significance level $0.05$. Figures~\ref{fig:ci_loading1}--\ref{fig:ci_loading5} display the estimated values of five columns of $\bA^*$, respectively, along with the corresponding asymptotic confidence intervals, constructed using the asymptotic normality results in Theorem~\ref{thm_oracle} (and Theorem~\ref{thm_lqa}). Red intervals denote loadings that remain significant after multiple testing correction at the 0.05 level, using the Benjamini-Hochberg procedure. Table~\ref{tab:personality_items} presents four items in each domain whose responses exhibit significant dependence on personality traits from other domains. For instance, the item described as ``Would never go hang gliding or bungee jumping'' is designed to measure the Neuroticism trait, yet its response is shown to depend significantly on the Extraversion trait, with a z-score of 10.16.
Factor correlations

{\spacingset{1}
 \begin{figure}[htbp!]
\centering    
        \includegraphics[width=5.4in, height=1.6in]{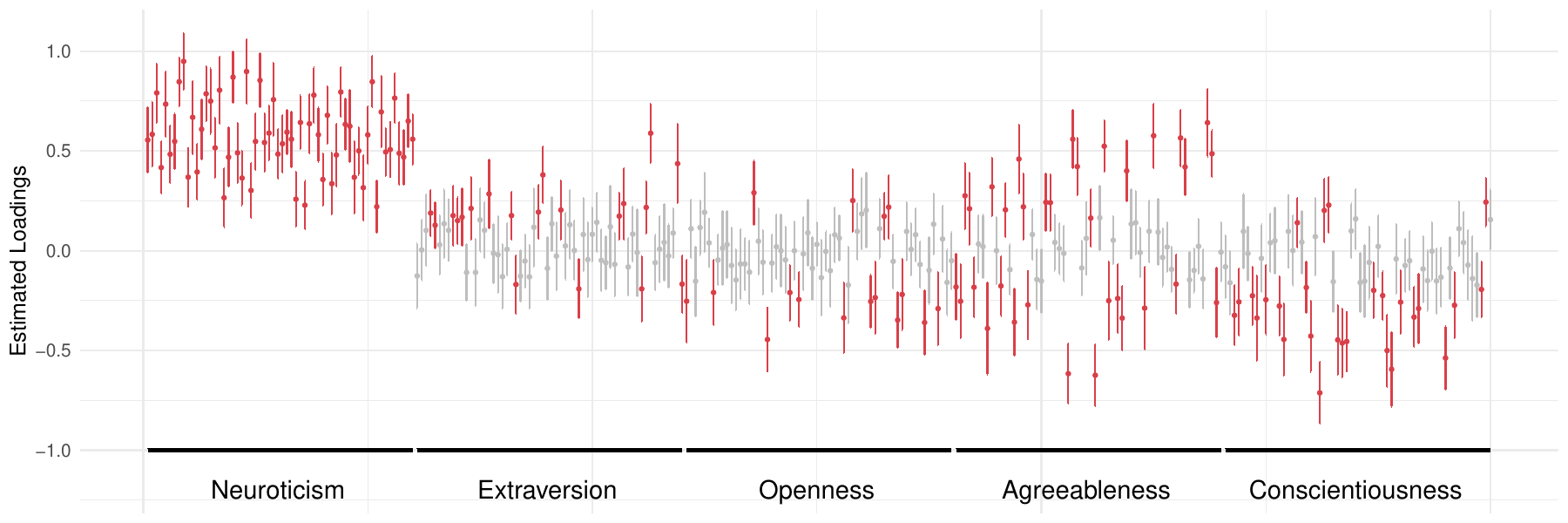}
   \caption{\small Confidence intervals for the first column of the loadings matrix corresponding to the ``Neuroticism'' facet. Red intervals represent significant loadings after Benjamini–Hochberg correction.} \label{fig:ci_loading1}
\end{figure}
 \begin{figure}[htbp!]
\centering    
        \includegraphics[width=5.4in, height=1.6in]{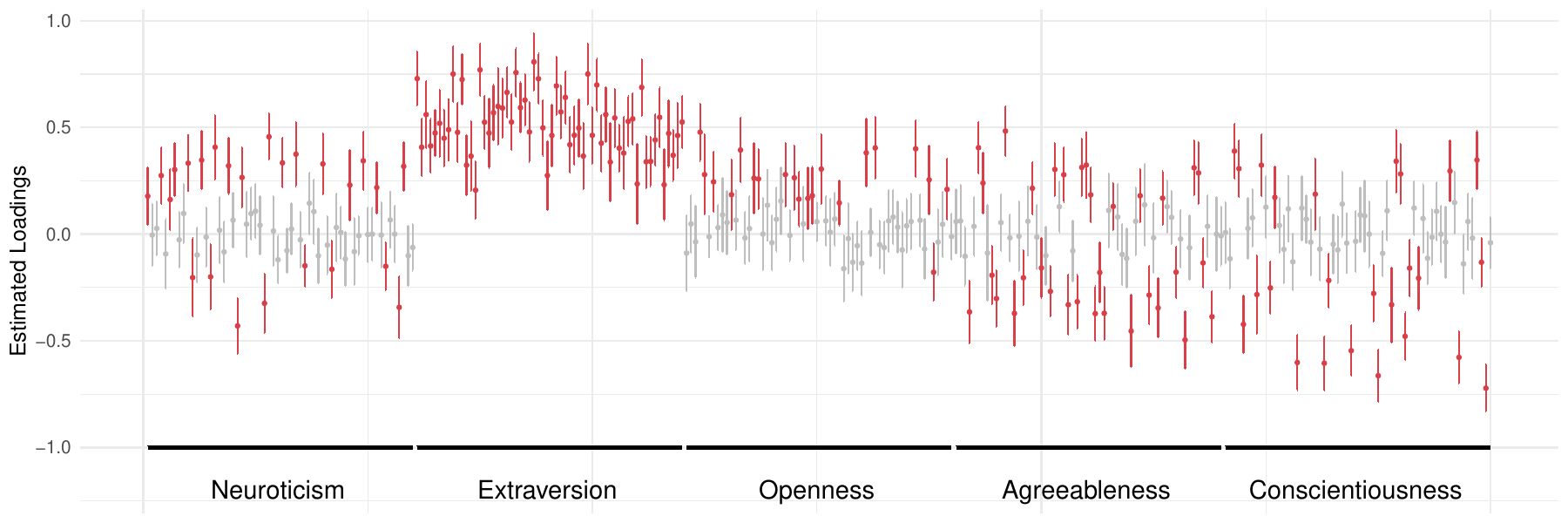}
   \caption{\small Confidence intervals for the second column of the loadings matrix corresponding to the ``Neuroticism'' facet. Red intervals represent significant loadings after Benjamini–Hochberg correction.} \label{fig:ci_loading2}
\end{figure}

 \begin{figure}[htbp!]
\centering    
        \includegraphics[width=5.4in, height=1.6in]{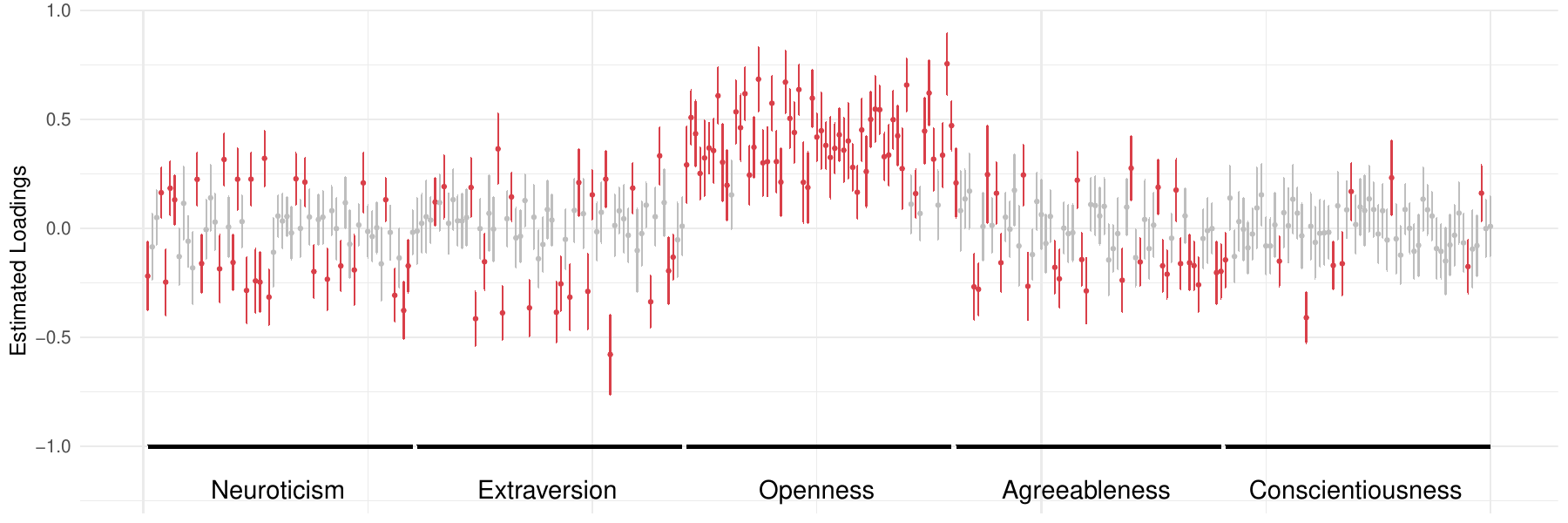}
   \caption{\small Confidence intervals for the third column of the loadings matrix corresponding to the ``Neuroticism'' facet. Red intervals represent significant loadings after Benjamini–Hochberg correction.} \label{fig:ci_loading3}
\end{figure}
 \begin{figure}[htbp!]
\centering    
        \includegraphics[width=5.4in, height=1.6in]{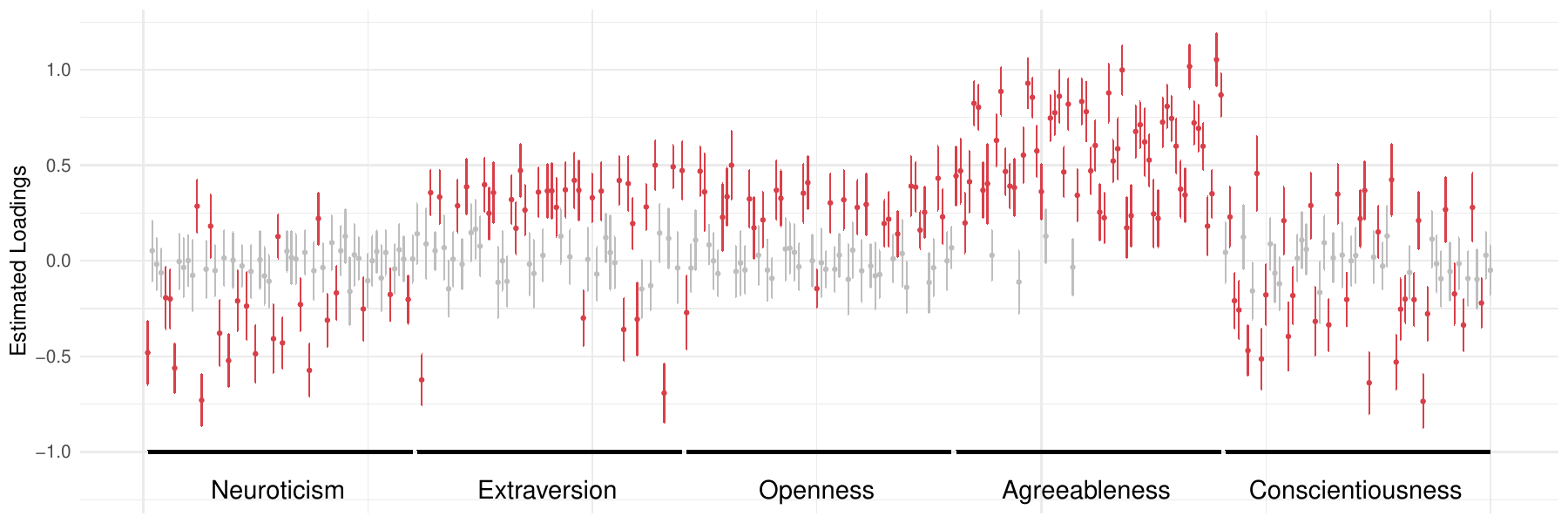}
   \caption{\small Confidence intervals for the fourth column of the loadings matrix corresponding to the ``Neuroticism'' facet. Red intervals represent significant loadings after Benjamini–Hochberg correction.} \label{fig:ci_loading4}
\end{figure}
 \begin{figure}[htbp!]
\centering    
        \includegraphics[width=5.4in, height=1.6in]{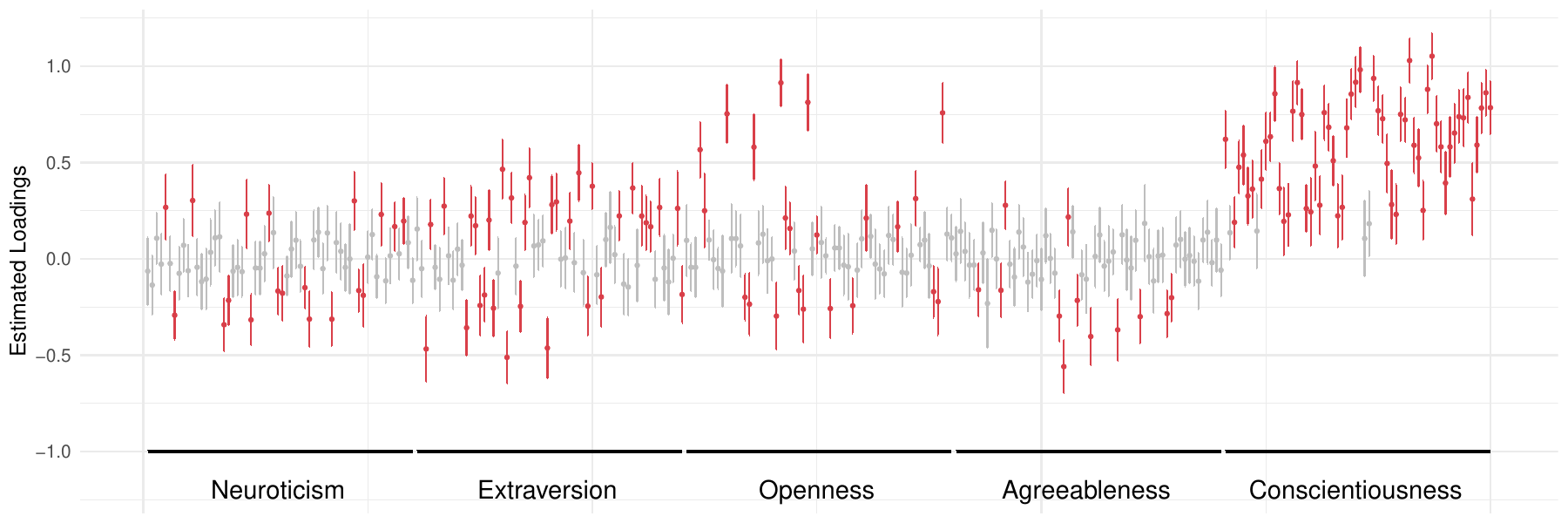}
   \caption{\small Confidence intervals for the fifth column of the loadings matrix corresponding to the ``Neuroticism'' facet. Red intervals represent significant loadings after Benjamini–Hochberg correction.} \label{fig:ci_loading5}
\end{figure}

\begin{table}[H]
\centering
\caption{Significant Personality Trait Items and Associated Values}
\label{tab:personality_items}
\begin{tabular}{lllc}
\toprule
\textbf{Domain} & \textbf{Trait} & \textbf{Item Description} & \textbf{z-score} \\ 
\hline\midrule
Neuroticism & E & Would never go hang gliding or bungee jumping & 10.16 \\
& O & Like to get lost in thought & 7.26 \\
& A & Hold a grudge & 10.67 \\
& C & Plunge into tasks with all my heart & 11.85 \\
\hline\addlinespace
Extraversion & N & Only feel comfortable with friends & 10.88 \\
& O & Prefer variety to routine & 9.20 \\
& A & Feel sympathy for those who are worse off & 10.67 \\
& & than myself\\
& C & Postpone decisions & 11.85 \\
\hline\addlinespace
Openness & N & Am able to stand up for myself & 7.37 \\
& E & Enjoy being part of a group & 8.57 \\
& A & Sympathize with the homeless & 6.04 \\
& C & Am sure of my ground & 9.05 \\
\hline\addlinespace
Agreeableness & N & Am afraid of many things & 13.80 \\
& E & Love large parties & 12.04 \\
& O & Prefer variety to routine & 9.36 \\
& C & Get others to do my duties & 13.42 \\
\hline\addlinespace
Conscientiousness & N & Feel that I'm unable to deal with things & 6.34 \\
& E & Seek to influence others & 9.86 \\
& O & Like to begin new things & 19.59 \\
& A & Believe in human goodness & 10.53 \\
\bottomrule
\end{tabular}
\end{table}






\section{Proof of Main Results}

\subsection{Proof of Proposition~\ref{prop_anti_smooth}}\label{sec_prove_prop_anti_smooth}
 The proof follows a constructive approach. Specifically, set $\bA^0 = (\bA_1^\T,\cdots,\bA_r^\T,\bA_{r+1}^\T)^\T$, where $\bA_{l} := \one_{q_l}\otimes \be_l^\T$ for $l\in[r]$ and $\bA_{r+1} := (\nu_{1},\dots,\nu_{r})\in\RR^{1\times r}$.  
    Apparently, $\bA^0$ is $(\lambda,\epsilon)$-sparse for any $\lambda\le 1\wedge \min\{\nu_l:l\in[r],\nu_l\neq 0\}$. We then show that for the following choices of $\bA_{r+1}$, $\bI_r$ is not a local optimum of $Q(\bA^0)$:
    \begin{enumerate}[label = $(\alph*)$]
    \item \textbf{Oblique Case:} For some $l,k
        \in[r]$, $\nu_k\nu_l\neq 0$ and $Q_{qk}(\bA^0)\nu_l\neq 0$
        \item \textbf{Orthogonal Case. } For some $l,k
        \in[r]$, $\nu_k\nu_l\neq 0$ and $Q_{ql}(\bA^0)/ Q_{qk}(\bA^0) \neq \nu_l/\nu_k$. 
    \end{enumerate}
Note that $\bA^0$ can always be constructed under both cases. For the orthogonal case, the only exception is when $Q_{ql}(\bA^0)\propto \nu_l$. In this case, it can be verified that any $\bG$ is a local optimum of $Q(\bA\bG^{-1})$ but not a strict optimum.
Without loss of generality, we assume $l=1$ and $k=2$.\vspace{0.5cm}\\ 
\noindent\textbf{Oblique Case. } We start with considering a more general case by assuming $\mathrm{diag}(\Mff) = \bI_r$ only.  Let $c_1 = [2 + 2\{\Mff\}_{1,2}]^{-1/2}$ and $c_2 = [2 - 2\{\Mff\}_{1,2}]^{-1/2}$. Consider $\bG$ of the following form
    \begin{equation}
        \bG = \begin{pmatrix}c_1\cos\alpha + c_2\sin\alpha & \begin{matrix}
            c_1\cos\alpha - c_2\sin\alpha & \zero_{r-2}^\T
        \end{matrix}\\\zero_{r-1} & \bI_{r-1}\end{pmatrix}.\label{eq_exm_G}
    \end{equation}
    Here, $\bG$ is designed to satisfy $\mathrm{diag}(\bG\Mff\bG^\T) = \bI_r$ for any $\alpha\in\RR$. We note that $\bG = \bI_r$ when $\alpha = \arctan(c_1/c_2)$.
    Next, we compute the derivatives of $\cP(\alpha) := Q(\bA^0\bG^{-1})$ at $\alpha = \alpha_0: = \arctan(c_1/c_2)$. First, we note
\begin{equation*}
    \cos\alpha_0 = \frac{1}{2c_1}\text{ and }\sin\alpha_0 = \frac{1}{2c_2}.
\end{equation*}
    The inverse of $\bG$ can be given as
    \begin{equation*}
        \bG^{-1} = \begin{pmatrix}(c_1\cos\alpha + c_2\sin\alpha)^{-1} & \begin{matrix}
           (c_1\cos\alpha + c_2\sin\alpha)^{-1}(c_2\sin\alpha - c_1\cos\alpha)  & \zero_{r-2}^\T
        \end{matrix}\\\zero_{r-1} & \bI_{r-1}\end{pmatrix}.
    \end{equation*}
    One can compute \begin{equation*}\big(1\;\zero_{r-1}^\T\big)\bG^{-1} = (c_1\cos\alpha + c_2\sin\alpha)^{-1}\big(1,\; c_2\sin\alpha - c_1\cos\alpha,\zero_{r-2}^\T\big),\end{equation*} 
    \begin{equation*}
        (\nu_{1},\nu_{2},\nu_{3},\dots,\nu_{r})\bG^{-1} = \left(\nu_{1}(c_1\cos\alpha + c_2\sin\alpha)^{-1},\;\nu_{2} + \nu_{1}\frac{c_2\sin\alpha - c_1\cos\alpha}{c_1\cos\alpha + c_2\sin\alpha},\nu_{3},\dots,\nu_{r}\right),
    \end{equation*} and other parts are unchanged. Then derivative $\partial_{\alpha}\cP(\alpha)$ at $\alpha_0$, ignoring the terms irrelevant to $\alpha$, can be expressed as
    \begin{align*}
        \partial_{\alpha}\cP(\alpha) \big|_{\alpha=\alpha_0}= &\,\sum_{j=1}^{q_1}Q_{j1}(1)\Big(\frac{c_1}{2c_2} - \frac{c_2}{2c_1}\Big) + Q_{j2}(0)\times 2c_1c_2\\ &\, + Q_{q1}(\nu_1)\Big(\frac{c_1}{2c_2} \nu_1- \frac{c_2}{2c_1}\Big) + Q_{q2}(\nu_2)\times 2c_1c_2\nu_1\\ = &\,-2c_1c_2\Big\{\{\Mff\}_{1,2}\sum_{j=1}^{q_1}Q_{j1}(1) + \{\Mff\}_{1,2}Q_{q1}(\nu_1)\nu_1 - Q_{q2}(\nu_2)\nu_1\Big\}.
    \end{align*}
    As $-1<\{\Mff\}_{1,2}<1$, we know $c_1c_2\neq 0$. Therefore, $\partial_{\alpha}\cP(\alpha) \big|_{\alpha=\alpha_0}\neq 0$ when
    \begin{equation*}
        \{\Mff\}_{1,2}\sum_{j=1}^{q_1}Q_{j1}(1) + \{\Mff\}_{1,2}Q_{q1}(\nu_1)\nu_1 \neq  Q_{q2}(\nu_2)\nu_1.
    \end{equation*} When $\Mff = \bI_r$, this immediately becomes $Q_{q2}(\nu_2)\nu_1\neq 0$. \vspace{0.5cm}\\ 
\noindent\textbf{Orthogonal Case.}
 Consider $\bG$ of the following form
    \begin{equation}
        \bG = \begin{pmatrix} \begin{matrix}
            \cos\alpha &\sin\alpha \\ - \sin\alpha & \cos\alpha
        \end{matrix}& \zero\\\zero& \bI_{r-2}\end{pmatrix}.\label{eq_exm_G_ortho}
    \end{equation} The obviously $\bG$ satisfies $\bG\bG^\T = \bI_r$ and $\bG = \bI_r$ when $\alpha = 0$. Next, we compute the derivative of $\cP_1(\alpha) = Q(\bA^0\bG)$ at $\alpha=0$, which, after ignoring the terms irrelevant to $\alpha$, can be expressed as
    \begin{align*}
        \partial_{\alpha}\cP_1(\alpha)|_{\alpha=0} = &\,\sum_{j=1}^{q_1}\big\{Q_{j1}(\cos\alpha)\sin\alpha + Q_{j2}(\sin\alpha)\cos\alpha\big\} \big|_{\alpha=0}\\&\,-\sum_{j = q_1+1}^{q_1+q_2}\big\{Q_{j1}(-\sin\alpha)\cos\alpha + Q_{j2}(\cos\alpha)\sin\alpha\big\}\big|_{\alpha=0}\\&\, + \frac{\partial Q(\bA^0\bG)}{\partial_{a_{q1}}}\frac{\partial(\nu_1\cos\alpha - \nu_2\sin\alpha)}{\partial\alpha}\Big|_{\alpha = 0} + \frac{\partial Q(\bA^0\bG)}{\partial_{a_{q2}}}\frac{\partial(\nu_1\sin\alpha + \nu_2\cos\alpha)}{\partial\alpha}\Big|_{\alpha=0}\\=&\,-Q_{q1}\nu_2 + Q_{q2}\nu_1.
    \end{align*}
    Here the second equality follows from $Q_{jk}(\bA) = 0$ when $ a_{jk} = 0$.
Thus, when $Q_{q1}/Q_{q2}\neq \nu_1/\nu_2$ at $\bA^0$, the derivative is nonzero and $\bI_r$ is not a stationary point.

Finally, for any $\lambda>0$ and $\epsilon>0$, since $\bA^0$ can always be $(\lambda,\epsilon)$-sparse while satisfying the condition that  $\nu_k\nu_l\neq 0$ and $Q_{qk}(\bA^0)\nu_l\neq 0$ in the oblique case, or $\nu_k\nu_l\neq 0$ and $Q_{ql}(\bA^0)/ Q_{qk}(\bA^0) \neq \nu_l/\nu_k$ in the orthogonal case, we conclude that any rotation method with the given criterion $Q(\cdot)$ is rotationally biased.

\subsubsection{Proof of Example~\ref{prop_anti_varimax}}

    In order for $\bI_r$ to be a stationary point of 
   {\begin{equation*}
        \argmin_{\bG\bG^\T = \bI_r}Q_{\rm varimax}(\bA^0\bG),
    \end{equation*}}
    it must hold that~\citep{chu1998orthomax}
    \begin{equation*}
        \bDelta_{\bA} : =(\bA^0)^\T\bA^0\mathrm{diag}(q^{-1}(\bA^0)^\T\bA^0) - (\bA^0)^\T\big(\bA^0\circ \bA^0\circ \bA^0\big)\text{ is symmetric,}
    \end{equation*} where $q = \sum_{l=1}^{r}q_l + 1$ denotes the number of rows of $\bA^0$.
    Note that by design of the loading matrix, one has
    \begin{align*}
        q^{-1}(\bA^0)^\T\bA^0 = &\,q^{-1}\sum_{l=1}^{r+1}\bA_l^\T\bA_l = q^{-1}\sum_{l=1}^r\bA_{l}^\T\bA_{l} + q^{-1}\bA_{r+1}^\T\bA_{r+1} \\ =&\, q^{-1}\mathrm{diag}(q_1,\dots,q_r) + q^{-1}\bA_{r+1}^\T\bA_{r+1}
    \end{align*}
    and 
    \begin{equation*}
        (\bA^0)^\T\big(\bA^0\circ \bA^0\circ \bA^0\big) = \mathrm{diag}(q_1,\dots,q_r) + \bA_{r+1}^\T(\bA_{r+1}\circ\bA_{r+1}\circ\bA_{r+1}).
    \end{equation*}
    Subsequently, we arrive at
    \begin{align*}
        \big(\bDelta_{\bA}\big)_{kl} = &\,(\bA_{,k}^0)^\T\bA_{,l}^0\big\{q^{-1}(\bA^0)^\T\bA^0\big\}_{ll} - (\bA_{,k}^0)^\T\big(\bA^0\circ \bA^0\circ \bA^0\big)_{,l}\\ = &\, \nu_{k}\nu_{l}\big(\frac{q_l}{q} + \frac{1}{q}\nu_{l}^2\big) - \nu_{k}\nu_{l}^3.
    \end{align*}
    Then $\bDelta_{\bA}$ is not symmetric as long as there is one pair $(k,l)$ such that 
    \begin{align*}
        0\neq &\,\nu_{k}\nu_{l}\big(\frac{q_l}{q} + \frac{1}{q}\nu_{l}^2\big) - \nu_{k}\nu_{l}^3 - \nu_{k}\nu_{l}\big(\frac{q_k}{q} + \frac{1}{q}\nu_{k}^2\big) + \nu_{l}\nu_{k}^3\\
        =&\,\nu_{l}\nu_{k}\big\{q_l-q_k + (q-1)(\nu_k^2 - \nu_l^2)\big\}/q,
    \end{align*}
    which is precisely when $\nu_l\nu_k\neq 0$ and $\nu_k^2 - \nu_l^2 \neq (q_k-q_l)/(q-1)$.

\subsection{Proof of Proposition~\ref{prop_idtfblty}}
\underline{\bf Oblique Setting.} We start with the oblique setting.
We prove \eqref{eq_population_solo} and then show it indicates that $Q_{\rho_{\gamma}}(\cdot)$ in \eqref{eq_solo} is rotationally bias-free.    Define $\Omega_0 = \{(j,k): a_{jk} = 0\}$ and $\Omega_1 = \{(j,k): a_{jk} \neq 0\}$. Then
    \begin{equation*}
        Q(\bA\bG^{-1}) - Q(\bA) = \underbrace{\sum_{(j,k)\in\Omega_0}\rho_{\gamma}\big((\bA\bG^{-1})_{j,k}\big) }_{\alpha_1}+ \underbrace{\sum_{(j,k)\in\Omega_1}\rho_{\gamma}\big((\bA\bG^{-1})_{j,k}\big) - \rho_{\gamma}( a_{jk})}_{\alpha_2}.
    \end{equation*}
    By \ref{item_concave} in Condition~\ref{assump_concave_function} and $\rho_{\gamma}(0) = 0$, we know $\alpha_1\ge 0$. In addition, $\alpha_1 = 0$ if and only if $(\bA\bG^{-1})_{j,k} = 0$ for all $(j,k)\in\Omega_0$. Next, we show $\alpha_1 = 0$ if and only if $\bG = \bI_r$. Recall that $\cS_k(\bA)$ is defined as $\{j: a_{jl} = 0\text{ for all }l\neq k\text{ and } a_{jk} \neq 0\}$. Then since $(\bA\bG^{-1})_{j,l} = (\ba_j)^\T(\bG^{-1})_{,l} = 0$ for any $j\in\cS_k$ and $l\neq k$, we know $(\bG^{-1})_{k,l} = 0$ for any $l\neq k$. Therefore, $\bG^{-1}$ must be diagonal. Since $\bG\in\Xi(c,\bZ)$, there is $\mathrm{diag}(\bG\Mff\bG^\T) = \bI_r$. With $\mathrm{diag}(\Mff) = \bI_r$, we know the diagonals of $\bG^{-1}$ must be $\pm 1$ when $\alpha_1 = 0$. But if the diagonal is $-1$, $\bG$ cannot belong to $\Xi(c,\bZ)$ for $c<1/2$. Therefore, $\alpha_1 = 0$ implies $\bG = \bI_r$.
    
    For $\alpha_2$, note that 
    \begin{align*}
        \min_{(j,k)\in\Omega_1}\Big|(\bA\bG^{-1})_{j,k}\Big| \ge  &\,\min_{(j,k)\in\Omega_1}\big|a_{jk}\big| -  \max_{(j,k)\in\Omega_1}\Big|\ba_j^\T(\bG^{-1} - \bI_r)_{,k}\Big|\\\ge&\,(a_3+1)\gamma - \max_{(j,k)\in\Omega_1}\|\bA_{j,}\|\times\|\bG^{-1} - \bI_r\|\\\ge&\, a_3\gamma.
    \end{align*}
    Here, the second inequality follows from $\min\{| a_{jk}|: a_{jk}\neq 0\}=\lambda_0\ge (a_3+1)\gamma$. For the last inequality, note that by $\|\bG- \bI_r\|\le c\le \min(\gamma/M,1)/2$, there is
    \begin{equation*}
        \|\bG^{-1} - \bI_r\|\le\|\bG - \bI_r\|\|\bG^{-1}\|\le \frac{1}{2}\|\bG^{-1} - \bI_r\| + \frac{\gamma}{2M},
    \end{equation*}
    which leads to $\|\bG^{-1} - \bI_r\|\le M^{-1}\gamma$. In addition, with Definition~\ref{def_sparsity}, we know for any $j\in[q]$, $\|\bA_j\|\le \|\bA\|_{2\to\infty}\le M$, which leads to the last inequality. Thus, we conclude that $\min_{(j,k)\in\Omega_1}\big|(\bA\bG^{-1})_{j,k}\big|\ge a_3\gamma$. 
    With $\min_{(j,k)\in\Omega_1}|a_{jk}|\ge \lambda_0\ge a_3\gamma$ and \ref{item_concave} of Condition~\ref{assump_concave_function}, we know $\alpha_2 = 0$ for every feasible $\bG$. Combine the analysis for $\alpha_1$ and $\alpha_2$, and we conclude that $\alpha_1 + \alpha_2 \ge 0$, with equality holding if and only if $\bG = \bI_r$.

Then, as an oblique criterion, we note that for any $(\bar\bZ,\bar\bA)$ that satisfy $\big(\bar\bZ\bar\bG^\T,\bar\bA\bar\bG^{-1}\big)=(\bZ^*, \bA^*)$ and $n^{-1}\bar\bZ^\T\bar\bZ = \bI_r$ as in \eqref{eq_solo}, within the local region $\{\bG\bar\bG:\|\bG - \bI_r\|\ll c\text{ and } \mathrm{diag}(\bG\bar\bG\bar\bG^\T\bG^\T)= \bI_r\}$, the optimum of 
\begin{equation}
    \argmin_{\bG\in\{\bG\bar\bG:\|\bG - \bI_r\|\ll c\text{ and } \mathrm{diag}(\bG\bar\bG\bar\bG^\T\bG^\T)= \bI_r\}}Q_{\rho_{\gamma}}(\bar\bA\bG^{-1})\label{eq_unique_local_solo_popu}
\end{equation}equals the optimum of 
\begin{equation*}
    \argmin_{\bG\in\{\bG:\|\bG - \bI_r\|\ll c\text{ and }\mathrm{diag}(\bG(n^{-1}\bZ^*{}^\T\bZ^*)\bG^\T)= \bI_r\}}Q_{\rho_{\gamma}}(\bA^*\bG^{-1}).
\end{equation*}
Here, the equivalence in the feasible set is due to $n^{-1}(\bZ^*)^\T\bZ^* = \bar\bG(n^{-1}\bar\bZ^\T\bar\bZ)\bar\bG^\T = \bar\bG\bar\bG^\T$. The second problem, by our previous analysis, has a unique solution $\bI_r$. Thus, we know \eqref{eq_unique_local_solo_popu} has a unique solution $\bar\bG$, which implies the criterion $Q_{\rho_{\gamma}}(\cdot)$ is rotationally bias-free as an oblique rotation criterion. 

{\underline{\bf Orthogonal Setting.} In the orthogonal setting, we similarly decompose $Q(\bA\bG^{-1}) - Q(\bA)$ as 
\begin{equation*}
        Q(\bA\bG^{-1}) - Q(\bA) = \underbrace{\sum_{(j,k)\in\Omega_0}\rho_{\gamma}\big((\bA\bG^{-1})_{j,k}\big) }_{\alpha_3}+ \underbrace{\sum_{(j,k)\in\Omega_1}\rho_{\gamma}\big((\bA\bG^{-1})_{j,k}\big) - \rho_{\gamma}( a_{jk})}_{\alpha_4}.
    \end{equation*}
   where $\Omega_0 = \{(j,k): a_{jk} = 0\}$ and $\Omega_1 = \{(j,k): a_{jk} \neq 0\}$. Again, we can show $\alpha_3 \ge 0$, with equality if and only if $\bG$ is diagonal. Because $\bG$ is orthogonal and $\|\bG - \bI_r\|\le c\le 1/2$, this indicates $\bG= \bI_r$. For $\alpha_4$, similarly we know with $c\le \lambda_0/M$ that $\alpha_4= 0$. This proves \eqref{eq_population_solo_ortho}. Then for any $(\bar\bZ,\bar\bA)$ that satisfy $(\bar\bZ\bar\bG^\T,\bar\bA\bar\bG^{-1}) = (\bZ^*,\bA^*)$ with $\bar\bG^\T\bar\bG = \bI_r$, within $\|\bG - \bI_r\| \le c$, we know 
   \begin{equation*}
   \argmin_{\bG\in\{\bar\bG^\T\bG:\bG\bG^\T = \bI_r,\;\|\bar\bG^\T\bG - \bI_r\|\le c\}}Q_{\rho_{\gamma}}(\bar\bA\bG) = \argmin_{\bG\in\{\|\bG - \bI_r\|\le c,\;\bG\bG^\T = \bI_r\}}Q_{\rho_{\gamma}}(\bA^*\bG) = \bI_r.
   \end{equation*}The first equality can be established by a similar argument as above and the second equality is due to \eqref{eq_population_solo_ortho}. Thus $\bar\bG$ is strict local optimum of $\argmin_{\bG\bG^\T = \bI_r}Q_{\rho_{\gamma}}(\bar\bA\bG)$.
This verifies that the criterion $Q_{\rho_{\gamma}}(\cdot)$ is rotationally bias-free as an orthogonal rotation criterion.}
\subsection{Proof of Theorem~\ref{thm_oracle}}\label{supp_sec_thm_oracle}

For $(\hat\bZ,\hat\bA)$ obtained from 
\begin{equation}
        (\hat\bZ,\hat\bA)\in \argmin_{(\bZ,\bA)\in\Theta^*(\epsilon_{nq})} \sum_{i=1}^n\sum_{j=1}^q\ell_{ij}(\ba_j^\T\bz_i),\tag{\ref{eq_theory_init}}
    \end{equation}
Let $\hat\bG^*$ be the optimal transformation for $(\hat\bZ,\hat\bA)$ as
\begin{equation}
    \hat\bG^* = \argmin_{\bG\text{ is invertible}}n^{-1}\big\|\hat\bZ \bG^\T - \bZ^*\big\|_{\Fn}^2 + q^{-1}\big\|\hat\bA\bG^{-1} - \bA^*\big\|_{\Fn}^2,\label{eq_opt_align}
\end{equation}with $\hat\bZ^* := \hat\bZ(\hat\bG^*)^\T$ and $\hat\bA^* := \hat\bA(\hat\bG^*)^{-1}$.

In what follows, we prove general results for $(\hat\bZ^*,\hat\bA^*)$, including consistency and asymptotic expansion, without imposing the $(\lambda,\epsilon)$-sparsity on $\bA^*$. Instead, we impose $\|\bA^*\|_{2\to\infty}\le M$ only.

\begin{lemma}\it\label{lemma_exist_g_star}
    Under Assumption~\ref{assump_psd_covariance}$(ii)$, $(iii)$ and $\|\bA^*\|_{2\to\infty}\le M$, with high probability, $\hat\bG^*$ defined above exists and 
    \begin{equation*}
       \|\hat\bG^* - \bI_r\| =O_p(\epsilon_{nq}).
    \end{equation*}
\end{lemma}
\begin{proof}
    See Section~\ref{sec_prove_lemma_exist_g_star}
\end{proof}
Next, we introduce the following lemmas on the consistency and asymptotic distributions of $\hat\bA(\hat\bG^*)^{-1}$ and $\hat\bZ(\hat\bG^*)^\T$. Recall that $\delta_{nq} = \log(n\vee q)/\sqrt{n\wedge q}$ as defined in Assumption~\ref{assump_scaling}. 
\begin{lemma}\it\label{lemma_consis_start}
    Under Assumption~\ref{assump_psd_covariance}$(ii)$, $(iii)$, $\|\bA^*\|_{2\to\infty}\le M$, Assumptions~\ref{assump_scaling} and~\ref{assump_limit} and $\delta_{nq}\ll\epsilon_{nq}$, for $(\hat\bZ,\hat\bA)$ obtained in \eqref{eq_theory_init} and $\hat\bG^*$ given in \eqref{eq_opt_align}, we have \begin{enumerate}[label = $(\roman*)$]
    \item $\sum_{j=1}^q\big\|(\hat\bG^*)^{-\T}\hat\ba_j - \ba_j^*\big\|^2 = O_p\big(q(n\wedge q)^{-1}\big)$ and $\sum_{i=1}^n\big\|\hat\bG^*\hat\bz_i - \bz_i^*\big\|^2 = O_p\big(n(n\wedge q)^{-1}\big)$;
    \item$\max_{j\in[q]}\big\|(\hat\bG^*)^{-\T}\hat\ba_j - \ba_j^*\big\| = O_p\big(\delta_{nq}\big)$ and $\max_{i\in[n]}\big\|\hat\bG^*\hat\bz_i - \bz_i^*\big\| = O_p\big(\delta_{nq}\big)$;
    \item $\big\|(\hat\bG^*)^{-\T}\hat\ba_j - \ba_j^*\big\| = O_p\big((n\wedge q)^{-1/2}\big)$ for each $j\in[q]$ and $\big\|\hat\bG^*\hat\bz_i - \bz_i^*\big\| = O_p\big((n\wedge q)^{-1/2}\big)$ for each $i\in[n]$.
    \end{enumerate}
\end{lemma}
\begin{proof}
    See Section~\ref{supp_sec_prove_ora_o}
\end{proof}
\begin{lemma}\label{lemma_oracle_asym}\it
    Under Assumptions~\ref{assump_psd_covariance}$(ii)$, $(iii)$, $\|\bA^*\|_{2\to\infty}\le M$, Assumptions~\ref{assump_scaling}--\ref{assump_clt} and $\delta_{nq}\ll\epsilon_{nq}$, for $(\hat\bZ,\hat\bA)$ obtained in \eqref{eq_theory_init} and $\hat\bG^*$ given in \eqref{eq_opt_align}, we have \begin{equation}
        \hat\bZ(\hat\bG^*)^\T - \bZ^* = \bN^{*}_z + \bR_z^*\label{eq_asympt_expand_z}
    \end{equation} where for each $i\in[n]$, $\bn^{z*}_{i} :=q^{1/2}(\bN^{*}_z)^\T\be_i^{(n)}$ obeys
    \begin{equation*}
         \bn^{z*}_{i}=-\sqrt{q}\big\{\sum_{j=1}^q\ell_{ij}^{\prime\prime}(\theta_{ij}^*)\ba_j^*\ba_j^*{}^\T\big\}^{-1}\sum_{j=1}^q\ell_{ij}^{\prime}(\theta_{ij}^*)\ba_j^*\overset{{ \rm d}}{\to}\cN(\zero_r,\bar\bPsi_{i}^{\star}),\text{ as }n,q\to\infty,
    \end{equation*} and the residual $\bR_z^*$ satisfies $\|\bR_z^*\|_{2\to\infty} = O_p\big(\log^{1/2}(n\vee q)(n\wedge q)^{-1}\big)$, $\|\bR_z^*\| = O_p(\sqrt{n}(n\wedge q)^{-1})$, and $\|\{\bR_z^*\}_{i,}\| = O_p((n\wedge q)^{-1})$ for $i\in[n]$.
    
    For $\hat\bA(\hat\bG^*)^{-1}$, we have\begin{equation}
        \hat\bA(\hat\bG^*)^{-1} - \bA^* = \bN^{*}_a + \bR^{*}_a,\label{eq_asympt_expand}
    \end{equation}
    where for each $j\in[q]$, $\bn^{a*}_{j} :=n^{1/2}(\bN^{*}_a)^\T\be_j^{(q)}$ obeys
    \begin{equation*}
         \bn^{a*}_{j} = -\sqrt{n}\big\{\sum_{i=1}^n\ell_{ij}^{\prime\prime}(\theta_{ij}^*)\bz_i^*{\bz_i^*}^\T\big\}^{-1}\sum_{i=1}^n\ell_{ij}^{\prime}(\theta_{ij}^*)\bz_i^* \overset{{ \rm d}}{\to}\cN(\zero_r,{\bar\bPhi}_{j}^{\star}),\text{ as }n,q\to\infty,
    \end{equation*}
     and the residual $\bR_a^*$ satisfies $\|\bR^{*}_a\|_{2\to\infty} = O_p\big(\log^{1/2}({n\vee q})(n\wedge q)^{-1}\big)$, $\|\bR_a^*\| = O_p(\sqrt{q}(n\wedge q)^{-1})$ and $\|\{\bR_a^*\}_{j,}\| = O_p((n\wedge q)^{-1})$ for $j\in[q]$. Here for convenience, define for each $i\in[n]$
     \begin{equation*}
         \bar\bPsi_{i}^{\star} = (\bPsi_{i}^*)^{-1}\bOmega_{i,z}^*(\bPsi_{i}^*)^{-1},
     \end{equation*}
     and for $j\in[q]$,
     \begin{equation*}
         \bar\bPhi_{j}^{\star} = (\bPhi_{j}^*)^{-1}\bOmega_{j,a}^*(\bPhi_{j}^*)^{-1}.
     \end{equation*}
    
    \label{lemma_asymp_start}
\end{lemma}
\begin{proof}
    See Section~\ref{supp_sec_prove_ora_o}.
\end{proof}
The following result is an immediate consequence of Lemma~\ref{lemma_asymp_start}.
\begin{corollary}\label{coro_asymp_start}\it
    Under the setup in Lemma~\ref{lemma_asymp_start}, one has
    \begin{equation*}
        \big\|n^{-1}\hat\bG^*\hat\bZ^\T\bZ^* - n^{-1}\bZ^*{}^\T\bZ^*\big\|_{\Fn} = O_p\big((n\wedge q)^{-1}\big);
    \end{equation*}
    and
    \begin{equation*}
        \big\|q^{-1}(\hat\bG^*)^{-\T}\hat\bA^\T\bA^* - q^{-1}\bA^*{}^\T\bA^*\big\|_{\Fn} = O_p\big((n\wedge q)^{-1}\big).
    \end{equation*}
\end{corollary}
\begin{proof}
   Apply Lemma~\ref{lemma_asymp_start}. Write $\bH_i=\sum_{j=1}^q\ell_{ij}^{\prime\prime}(\theta_{ij}^*)\ba_j^*(\ba_j^*)^\T$, $\bar\bH_i=\EE\bH_i$, and $\bs_i=\sum_{j=1}^q\ell_{ij}^{\prime}(\theta_{ij}^*)\ba_j^*$. Independence, bounded second derivatives, and bounded row norms give $\EE\|\bH_i-\bar\bH_i\|_{\Fn}^2=O(q)$ and $\EE\|\bs_i\|^2=O(q)$, uniformly in $i$. Since both inverse Hessians have norm $O(q^{-1})$, Cauchy--Schwarz yields $\EE\|\{\bH_i^{-1}-\bar\bH_i^{-1}\}\bs_i\|=O(q^{-1})$. The mean-zero vectors $\bar\bH_i^{-1}\bs_i$ are independent across $i$ and have second moments $O(q^{-1})$. Thus,
\begin{align*}
 \big\|n^{-1}(\hat\bZ(\hat\bG^*)^\T-\bZ^*)^\T\bZ^*\big\|_{\Fn}
 &\le \Big\|n^{-1}\sum_{i=1}^n\bar\bH_i^{-1}\bs_i(\bz_i^*)^\T\Big\|_{\Fn}
 +O_p(q^{-1})+n^{-1}\|\bR_z^*\|_{\Fn}\|\bZ^*\|\\
 &=O_p((nq)^{-1/2})+O_p(q^{-1})+O_p((n\wedge q)^{-1})
 =O_p((n\wedge q)^{-1}).
\end{align*} The bound for $\|\bR_z^*\|_{\Fn}$ follows from $\|\bR_z^*\|$ and that $r$ is fixed. Similarly, we can obtain the results for $\hat\bA(\hat\bG^*)^{-1}$.
\end{proof}

Define $(\hat\bZ^*,\hat\bA^*) = (\hat\bZ\hat\bG^*{}^\T,\hat\bA\hat\bG^*{}^{-1})$. Next, we consider the Folomin solution initialized at $(\hat\bZ^*,\hat\bA^*)$:
\begin{equation}
        \hat\bG_* = \argmin_{\bG\in \Xi(\epsilon_{nq}^{\prime}, \hat\bZ^*)}Q_{\rho_{\gamma}}\left(\hat\bA^*\bG^{-1}\right),\label{eq_theory_solo_align}\end{equation}where $\epsilon_{nq}^{\prime}$ is as specified in Theorem~\ref{thm_oracle}.
        
Our strategy is first showing that this solution $\hat\bG_*$ can be located in a region sufficiently close to $\bI_r$, by partitioning the feasible region $\Xi(\epsilon_{nq}^{\prime}, \hat\bZ^*)$ into many non-overlapping rings (in Lemma~\ref{lemma_oracle_folomin_solu}). This indicates that $\hat\bG^*$ can navigate $(\hat\bZ,\hat\bA)$ to $(\hat\bZ^*,\hat\bA^*)$, on which the Folomin would nearly identify $\bI_r$ as the local optimizer. Then we show that $\hat\bG^*$ can well approximate Folomin estimator $\hat\bG$, and combined with Lemma~\ref{lemma_asymp_start}, we prove Theorem~\ref{thm_oracle}.
        
We state a general lemma for the solution to \eqref{eq_theory_solo_align}. Specifically, we propose an alternative to Assumption~\ref{assump_psd_covariance} as follows.
\newenvironment{assumptionPrime}[1]{
    \renewcommand{\theassumption}{#1$^{\prime}$} 
    \begin{assumption}
}{
    \end{assumption}
    \renewcommand{\theassumption}{\arabic{assumption}} 
}
\begin{assumptionPrime}{1}\label{assump_psd_covariance_prime}
    \begin{enumerate}[label=$(\roman*)$] \item $\bA^*$ satisfies $\min\{|a_{jl}^*|:a_{jl}^*\neq 0\}\ge \lambda_0$ and $\|\bA^*\|_{2\to\infty}\le M$;
\item $\|\bZ^*\|_{2\to\infty}\le M$ for $M$ specified in Definition~\ref{def_sparsity}, and $\mathrm{diag}(n^{-1}\bZ^*{}^\T\bZ^*) = \bI_r$;
    \item $\bSigma_z^* = \lim_{n \rightarrow \infty}n^{-1}{\bZ^*}^\T\bZ^* $ and $\bSigma_{a}^* = \lim_{q \rightarrow \infty}q^{-1}{\bA^*}^\T\bA^* $ exist and are positive definite.
\end{enumerate}
\end{assumptionPrime}
Obviously, Assumption~\ref{assump_psd_covariance} implies Assumption~\ref{assump_psd_covariance_prime}.  Assumption~\ref{assump_psd_covariance_prime} does not require the identification of the radial streaks imposed by the $(\lambda,\epsilon)$-sparsity condition. In particular, the following lemma shows that, for $\hat\bG_*$ to be close to $\bI_r$, it is sufficient that the quantity $\sigma_q=\big[\min_{l\in[r]}\lambda_{r-1}\{q^{-1}(\bA^*_{\cT_l^*,})^\T\bA^*_{\cT_l^*,}\}\big]^{-1}$ is small with $\cT_k ^* = \{j\in[q]:a_{jl}^* = 0\}$. { This quantity measures the spectral strength supplied by the rows whose $l$th loading vanishes. In Lemma~\ref{lemma_oracle_folomin_solu} below, we show that, for $\hat\bG_*$ to be close to $\bI_r$, it is sufficient that $\sigma_q$ is sufficiently small, or equivalently that the zero pattern in $\bA^*$ carries enough spectral information in every latent direction. This condition can hold for general sparse configurations and is not tied to the number or magnitude of simple rows. In the main text, we present the theory under the $(\lambda,\epsilon)$-sparsity condition because it yields a computationally tractable and theoretically justified initialization scheme. This choice aligns with our goal of providing a complete theory--computation framework.} 

\begin{lemma}
    \label{lemma_oracle_folomin_solu}\it 
    Under the setting of Theorem~\ref{thm_oracle} with Assumption~\ref{assump_psd_covariance} replaced by Assumption~\ref{assump_psd_covariance_prime}, for $\hat\bG_*$ obtained in \eqref{eq_theory_solo_align}, we have
    \begin{equation*}
    \|\hat\bG_* - \bI_r\| = O_p(\sigma_q(n\wedge q)^{-1+\varepsilon}),
    \end{equation*}
    for any constant $\varepsilon>0$.
\end{lemma}
\begin{proof}\label{outline_proof_lemma_oracle_folomin_solu}
    The detailed proof of this lemma is in Section~\ref{sec_prove_lemma_oracle_folomin_solu}, and we outline the key part here. By the asymptotic result derived in Lemmas~\ref{lemma_consis_start} and~\ref{lemma_asymp_start}, we approximate the difference of $\rho_{\gamma}\big((\hat\bA^*\bG^{-1})_{j,l}\big) - \rho_{\gamma}(\hat\bA^*_{j,l})$ with $\big|\hat\bA^*_{j,l} + \bar\xi\big| - \big|\hat\bA^*_{j,l}\big|$ for each $(j,l)$ with $a_{jl}^* = 0$, where $\bar\xi$ is a normal term. Then, applying the property of the folded normal distribution (Lemma~\ref{lemma_fold_nor_dif}), we derive a sharp lower bound for the overall difference $Q_{\rho_{\gamma}}(\hat\bA^*\bG^{-1}) - Q_{\rho_{\gamma}}(\hat\bA^*)$ for any $\bG\in \Xi(\epsilon_{nq}^{\prime}, \hat\bZ^*)$: for any $\bG\in\{\bG\in \Xi(\epsilon_{nq}^{\prime},\hat\bZ^*), r_{nq}\le \|\bG-\bI_r\|\le R_{nq}\}$, with probability approaching $1$, there is
    \begin{equation}
        q^{-1}\gamma^{-1}\big\{Q_{\rho_{\gamma}}(\hat\bA^*\bG^{-1}) - Q_{\rho_{\gamma}}(\hat\bA^*)\big\} \ge C_1\frac{r_{nq}^2}{\sigma_qR_{nq}} - C_2\Big(R_{nq}^2 + \frac{1}{n\wedge q} + \frac{R_{nq}}{\sqrt{n\wedge q}}\Big),\label{eq_bound_diff_Q1_t1_t22_before}
    \end{equation}for some constant $C_1,C_2$. Then we can construct a finite series of $\big\{r_{nq}^{(t)},R_{nq}^{(t)}\big\}_{t=1}^{T}$ such that 
    \begin{enumerate}[label=$(\roman*)$]
        \item $R_{nq}^{(0)} = \epsilon_{nq}^{\prime}$;
        \item $r_{nq}^{(T)} = \sigma_q(n\wedge q)^{-1+\varepsilon}$;
        \item $r_{nq}^{(t)} = \sigma_q^{1/2}\Big\{(\Rdis^{(t)})^{3/2} + \frac{(\Rdis^{(t)})^{1/2}}{\sqrt{n\wedge q}}\Big\}\log (n\wedge q)\le R_{nq}^{(t)}$ for each $t\in[T]$.
    \end{enumerate}
    Because $\epsilon_{nq}^{\prime}\ll (n\wedge q)^{-c}$ for some constant $c>0$ and $\sigma_q\ll (\epsilon_{nq}^{\prime})^{-1}$, this construction is always valid. Then one can verify with the above lower bound \eqref{eq_bound_diff_Q1_t1_t22_before} that, for any $t\in[T]$, for any $\bG\in\{\bG\in \Xi(\epsilon_{nq}^{\prime},\hat\bZ^*), r_{nq}^{(t)}\le \|\bG-\bI_r\|\le R_{nq}^{(t)}\}$, with probability approaching $1$,
    \begin{equation}
        Q_{\rho_{\gamma}}(\hat\bA^*\bG^{-1}) > Q_{\rho_{\gamma}}(\hat\bA^*).
    \end{equation}Therefore, we know $\|\bG^* - \bI_r\|$ is of order $r_{nq}^{(T)} = \sigma_q(n\wedge q)^{-1+\varepsilon}$, with probability approaching $1$, which finishes the proof.
\end{proof}

Now we prove Theorem~\ref{thm_oracle}. 
 We aim to show that $\hat\bG$, as the optimum of \eqref{eq_theory_solo}, satisfies $\|\hat\bG - \hat\bG^*\| = O_p\big(\sigma_q(n\wedge q)^{-1 + \varepsilon}\big)$ for any small constant $\varepsilon$. Here, $\hat\bG^*$ is given in \eqref{eq_opt_align}. Note that $\hat\bG\in\{\bG:\|\bG - \bI_r\|\le \epsilon_{nq}^{\prime}\}$ by the definition of $\Xi(\epsilon_{nq}^{\prime},\hat\bZ)$, and $\|\hat\bG^* - \bI_r\| = O_p(\epsilon_{nq})$ by Lemma~\ref{lemma_exist_g_star}. As we have required $\epsilon_{nq}\ll\epsilon_{nq}^{\prime}$, it is possible for $\|\hat\bG - \hat\bG^*\| = O_p\big(\sigma_q(n\wedge q)^{-1 + \varepsilon}\big)$ to hold.

 Apply Lemma~\ref{lemma_oracle_folomin_solu} with radius $2\epsilon_{nq}^{\prime}$, which satisfies the same scaling requirements, and let $\bar\bG_*$ be a minimiser over $\Xi(2\epsilon_{nq}^{\prime},\hat\bZ^*)$. For sufficiently small $\varepsilon>0$, the lemma gives $\|\bar\bG_*-\bI_r\|=O_p(\sigma_q(n\wedge q)^{-1+\varepsilon})=o_p(\epsilon_{nq}^{\prime})$. Since $\|\hat\bG^*-\bI_r\|=O_p(\epsilon_{nq})=o_p(\epsilon_{nq}^{\prime})$, the matrix $\bar\bG_*\hat\bG^*$ belongs to $\Xi(\epsilon_{nq}^{\prime},\hat\bZ)$ with probability approaching one. Conversely, every $\bG\in\Xi(\epsilon_{nq}^{\prime},\hat\bZ)$ satisfies $\bG(\hat\bG^*)^{-1}\in\Xi(2\epsilon_{nq}^{\prime},\hat\bZ^*)$ with probability approaching one, because
\[
 \|\bG(\hat\bG^*)^{-1}-\bI_r\|\le\{\epsilon_{nq}^{\prime}+\|\hat\bG^*-\bI_r\|\}\|(\hat\bG^*)^{-1}\|<2\epsilon_{nq}^{\prime},
\]
and the normalisation constraint is preserved. We have the equivalence $Q_{\rho_\gamma}(\hat\bA\bG^{-1})=Q_{\rho_\gamma}\big(\hat\bA^*\{\bG(\hat\bG^*)^{-1}\}^{-1}\big)$. 
The two minimum values are therefore equal, and $\hat\bG(\hat\bG^*)^{-1}$ is a minimiser over the larger aligned ball. The lemma then gives $\|\hat\bG(\hat\bG^*)^{-1}-\bI_r\|=O_p(\sigma_q(n\wedge q)^{-1+\varepsilon})$.

Now we can conclude that $\|\hat\bG - \hat\bG^*\| = O_p(\sigma_q(n\wedge q)^{-1+\varepsilon})$. 
Next, note that 
\begin{equation*}
    \hat\bA\hat\bG^{-1} - \bA^* = \hat\bA\hat\bG^{-1} - \hat\bA^* + \hat\bA(\hat\bG^*)^{-1} - \bA^* = \hat\bA\big\{\hat\bG^{-1} - (\hat\bG^*)^{-1}\big\} + \hat\bA(\hat\bG^*)^{-1} - \bA^*.
\end{equation*}
The asymptotic expansion for $\hat\bA(\hat\bG^*)^{-1} - \bA^*$ is given in Lemma~\ref{lemma_asymp_start}. For the first term, note that
\begin{align*}
    \big\|\hat\bA\big\{\hat\bG^{-1} - (\hat\bG^*)^{-1}\big\}\big\|_{2\to\infty} \le &\,\|\hat\bA\|_{2\to\infty}\|\hat\bG^{-1}-(\hat\bG^*)^{-1}\|\\\le &\,\|\hat\bA\|_{2\to\infty}\|\hat\bG-\hat\bG^*\|\|\hat\bG^{-1}\|\|(\hat\bG^*)^{-1}\|\\=&\,O_p\big(\sigma_q(n\wedge q)^{-1+\varepsilon}\big).
\end{align*}
Here, the equality follows from $\|\hat\bG^* - \bI_r\| = o_p(1)$ from Lemma~\ref{lemma_exist_g_star} and $\|\hat\bG - \hat\bG^*\| = O_p(\sigma_q(n\wedge q)^{-1+\varepsilon})$. 
Then the asymptotic expansion for $\hat\bA\hat\bG^{-1} - \bA^*$ follows from Lemma~\ref{lemma_asymp_start} by  letting $\bR_a$ in Theorem~\ref{thm_oracle} be $\bR_a^{*} + \hat\bA\big\{\hat\bG^{-1} - (\hat\bG^*)^{-1}\big\}$ with $\bR_a^*$ defined in Lemma~\ref{lemma_asymp_start} and $\bN_a$ in Theorem~\ref{thm_oracle} be $\bN_a^{*}$ defined in Lemma~\ref{lemma_asymp_start}. Expansion for $\hat\bZ\hat\bG^\T$ can be obtained similarly.

Looking back to the proof, we note that $\epsilon_{nq}\ll\epsilon_{nq}^{\prime}$ is necessary, as otherwise inside the region $\Xi(\epsilon_{nq}^{\prime},\hat\bZ)$, the optimum would lie on the boundary and not be the one we aim for.


\subsection{Proof of Theorem~\ref{thm_init}}
Let 
\begin{equation}
\hat\bG^{\star} = \argmin_{\bG\text{ is invertible}}q^{-1}\big\|\hat\bA^0 \bG^{-1} - \bA^*\big\|_F^2 + n^{-1}\big\|\hat\bZ ^0\bG^\T - \bZ^*\big\|_F^2.\label{eq_opt_align_zero}
\end{equation}
To start, we show the well-posedness of $\hat\bG^{\star}$ defined as the solution of the following problem 
\begin{lemma}\it\label{lemma_exist_g_hat_star}
    Under the settings of Theorem~\ref{thm_init} and $\hat\bG^{\star}$ given in \eqref{eq_opt_align_zero}, as $n,q\to\infty$, with probability approaching $1$, $\hat\bG^{\star}$ exists and there is some absolute constant $C_g$ such that
    \begin{equation*}
        C_g^{-1}\le \sigma_{r}(\hat\bG^{\star}) \le \sigma_1(\hat\bG^{\star}) \le C_g.
    \end{equation*}
\end{lemma}
\begin{proof}
    See Section~\ref{sec_prove_lemma_exist_g_hat_star}
\end{proof}
For the true parameters $(\bZ^*,\bA^*)$, define $\bG^0 = (\bU^*_A)^\T(n^{-1}\bZ^*{}^\T\bZ^*)^{-1/2}$, where $\bU_A^*$ is the left singular matrix for $(n^{-1}\bZ^*{}^\T\bZ^*)^{1/2}q^{-1}(\bA^*)^\T\bA^*(n^{-1}\bZ^*{}^\T\bZ^*)^{1/2}$, and then $(\bZ^0,\bA^0) := \big(\bZ^*(\bG^0)^{\T},\bA^*(\bG^0)^{-1}\big)$ satisfy the constraints in \eqref{eq_init_of_init_est}, i.e.,
\begin{equation*}
    n^{-1}(\bZ^0)^\T\bZ^0 = \bI_r\text{ and }q^{-1}(\bA^0)^\T\bA^0 = \mathrm{diagonal}.
\end{equation*}
Here, $(n^{-1}\bZ^*{}^\T\bZ^*)^{-1/2}$ and $(n^{-1}\bZ^*{}^\T\bZ^*)^{-1/2}q^{-1}(\bA^*)^\T\bA^*(n^{-1}\bZ^*{}^\T\bZ^*)^{-1/2}$ is well-defined by Assumption~\ref{assump_psd_covariance} $(iii)$ for sufficiently large $n,q$, and thus $\bG^0$ is well-posed, i.e., $c\le \sigma_r(\bG^0)\le \sigma_1(\bG^0)\le C$ for some constant $c,C>0$. Without loss of generality, we assume that $M$ is selected large enough such that $\|\bZ^0\|_{2\to\infty}<M /2$ and $\|\bA^0\|_{2\to\infty}\le M/2$. Let $(\hat\bZ^{\star},\hat\bA^{\star}) = \big(\hat\bZ^0(\hat\bG^{\star})^\T,\hat\bA^0(\hat\bG^{\star})^{-1}\big)$. Then we have the following results.

\begin{lemma}\label{lemma_combine_zero}
    Under Assumption~\ref{assump_psd_covariance}(ii),(iii), $\|\bA^*\|_{2\to\infty}\le M$, Assumptions~\ref{assump_scaling}--\ref{assump_clt}, Lemmas~\ref{lemma_consis_start},~\ref{lemma_asymp_start}, and Corollary~\ref{coro_asymp_start} hold for $(\hat\bZ^{\star},\hat\bA^{\star})$.
\end{lemma}
\begin{proof}
See Section~\ref{supp_sec_prove_lemma_combine_zero}
\end{proof}

    In what follows, we assume the event holds for some absolute constant $C$:
    \begin{align}
        \Big\{\big\|\hat\bA^{\star} - \bA^*\big\|_{2\to\infty}\le C\delta_{nq}\text{ and }\big\|\hat\bZ^{\star} - \bZ^*\big\|_{2\to\infty}\le C\delta_{nq}\Big\},\label{eq_event_init_algorithm}
    \end{align}
    where $\delta_{nq} = \log(n\vee q)/\sqrt{n\wedge q}$. 
    The proof consists of three steps.\\
    In {\bf Step 1}, we show that the difference between $\cD_{(j_1,j_2)}^{\delta}(\hat\bZ^0,\hat\bA^0)$ and $\cD_{(j_1,j_2)}^{\delta}(\bZ^*,\bA^*)$ can be uniformly controlled over $j_1,j_2\in[q]$:
    \begin{equation}
        \max_{j_1,j_2\in[q]}\big|\cD_{(j_1,j_2)}^{\delta}(\hat\bZ^0,\hat\bA^0) - \cD_{(j_1,j_2)}^{\delta}(\bZ^*,\bA^*)\big| = O(\delta_{nq}).
    \end{equation}
    In {\bf Step 2}, we show that $\{\hat\cS_k\}_{k\in[r]}$ exactly recover $\{\cS_k(\bA^*)\}_{k\in[r]}$, up to a permutation. Lastly, in {\bf Step 3}, we derive $\|\hat\bG^0 - \hat\bG^{\star}\| = O_p(\tilde\sigma_q\delta_{nq})$, and Theorem~\ref{thm_init} is proved with Lemma~\ref{lemma_combine_zero}. 

    \medskip
    \underline{\textbf{Step 1.}} We first compute
    \begin{align*}
        \max_{j\in[q]}\big\|\hat\bZ^0\hat\ba_{j}^0 - \bZ^*\ba_j^*\big\| =&\, \max_{j\in[q]}\big\|\hat\bZ^{\star}\hat\ba_{j}^{\star} - \bZ^*\ba_j^*\big\| \\ 
        =&\, \max_{j\in[q]}\big\|\big(\hat\bZ^{\star} - \bZ^*\big)\ba_{j}^* \big\| + \max_{j\in[q]}\big\| \bZ^*\big(\hat\ba_{j}^{\star} - \ba_j^*\big)\big\|\\&\, + \max_{j\in[q]}\big\|\big(\hat\bZ^{\star} - \bZ^*\big)\big(\hat\ba_{j}^{\star} - \ba_j^*\big) \big\|\\\le &\, 3C\delta_{nq}\sqrt{n}M,
    \end{align*}
    given that $\|\bA^*\|_{2\to\infty}\le M$ and $\|\bZ^*\|_{2\to\infty}\le M$. Then with \eqref{eq_event_init_algorithm}, by telescoping we know \begin{equation*}\max_{j_1,j_2\in[q]}\big|(\hat\ba_{j_1}^0)^\T(\hat\bZ^{0})^\T\hat\bZ^{0}\hat\ba_{j_2}^{0} - (\ba_{j_1}^*)^\T(\bZ^*)^\T\bZ^*\ba_{j_2}^*\big| = O(n\delta_{nq}),\end{equation*}
    and $\max_{j\in[q]}\big|\|\hat\bZ^{0}\hat\ba_{j}^{0}\| - \|\bZ^*\ba_j^*\|\big|=O(\delta_{nq}\sqrt{n})$. 
    Next, when $\|\bZ^*\ba_j^*\| \neq 0$, with $\sigma_{\min}(\bZ^*)= n^{1/2}\big\{\lambda_{\min}(n^{-1}\bZ^*{}^\T\bZ^*)\big\}^{1/2} \ge n^{1/2}\{\lambda_{\min}(\bSigma_z^*)\}^{1/2}/2$ when $n$ is large enough, we know $\ba_j^*\neq \zero$. Since $\bA^*$ is $(\lambda_0,\epsilon_0)$-sparse, one has $\|\bZ^*\ba_j^*\| \ge \{n\lambda_{\min}(\bSigma_z^*)\}^{1/2}/2 \times\lambda_0$. Then
    \begin{equation*}
        \max_{j:\ba_j^*\neq\zero}\big\|\hat\bZ^0\hat\ba_j^0\big\|\ge \max_{j:\ba_j^*\neq\zero}\big\|\bZ^*\ba_j^*\big\| - \max_{j:\ba_j^*\neq\zero}\big\|\hat\bZ^0\hat\ba_j^0 - \bZ^*\ba_j^*\big\|\ge \big\{n\lambda_{\min}(\bSigma_z^*)\big\}^{1/2}\lambda_0/4 ,
    \end{equation*}
    with $\delta_{nq}\ll 1$.
    Thus, when $\|\bZ^*\ba_{j_1}^*\|\|\bZ^*\ba_{j_2}^*\|\neq 0$, we know
    \begin{equation*}
        n^{-1}\|\bZ^*\ba_{j_1}^*\|\|\bZ^*\ba_{j_2}^*\|\ge \delta\text{ and }n^{-1}\|\hat\bZ^0\hat\ba_{j_1}^0\|\|\hat\bZ^0\hat\ba_{j_2}^0\|\ge \delta,
    \end{equation*}uniformly over $j_1,j_2\in[q]$ due to $\delta\le \lambda_0^2\lambda_{\min}(\bSigma_z^*)/16$. 
    By telescoping one can obtain \begin{equation}\max_{j_1,j_2\in[q]}\big|\cD_{(j_1,j_2)}^{\delta}(\hat\bZ^0,\hat\bA^0) - \cD_{(j_1,j_2)}^{\delta}(\bZ^*,\bA^*)\big|=O(\delta_{nq})\label{eq_max_diff_d_ij}\end{equation}
    When $\|\bZ^*\ba_{j_1}^*\|\|\bZ^*\ba_{j_2}^*\|=0$, similarly, we can derive  $n^{-1}\|\hat\bZ^0\hat\ba_{j_1}^0\|\|\hat\bZ^0\hat\ba_{j_2}^0\|<\delta$ with $\delta\gg\delta_{nq}$ Therefore, \eqref{eq_max_diff_d_ij} still holds.

\medskip
\underline{\textbf{Step 2.}} 
We first characterize $\hat\cI_j$ computed from Algorithm~\ref{alg_init}. Let $\bM_z^* = (\bZ^*)^\T\bZ^*$. We begin with the following lemma
    \begin{lemma}\it \label{lemma_angular_conservation}
Let $\bM\in\mathbb R^{d\times d}$ be symmetric positive definite, with the largest and smallest eigenvalues given as
$0<\lambda_{\min}\le \lambda_{\max}$, and let $\kappa=\lambda_{\max}/\lambda_{\min}$.
Set $\delta\in(0,1]$. Given any $\bx,\by\in\RR^{d}\setminus\{\zero_d\}$, define
\[
\cos(\bx,\by)=\frac{|\bx^\T\by|}{\|\bx\|\|\by\|}\,,\qquad
\cos_{\bM}(\bx,\by)=\frac{|\bx^\T \bM \by|}{\sqrt{\bx^\T \bM \bx}\,\sqrt{\by^\T \bM \by}}.
\]
If $\cos(\bx,\by)\le 1-\delta$, then
\[
\cos_{\bM}(\bx,\by)\ \le\ 1-\delta',
\text{ with }
\delta'=\frac{\delta\,\lambda_{\min}}{(1-\delta/2)\lambda_{\max}+(\delta/2)\lambda_{\min}}
=\frac{\delta/\kappa}{\,1-\delta/2+\delta/(2\kappa)\,}\,>\,0.
\]
\end{lemma}
\begin{proof}
    See Section~\ref{sec_prove_lemma_angular_conservation}.
\end{proof}

We first consider the case where \underline{$j\in\cS_k(\bA^*)$}. One can easily check that $\cD_{(j,j^{\prime})}^{\delta}(\bZ^*,\bA^*) = 1$ for any $j^{\prime}$ that also belongs to $\cS_k(\bA^*)$ and $\delta>0$. Thus, $\cD_{(j,j^{\prime})}^{\delta}(\hat\bZ^0,\hat\bA^0) \ge 1 - O(\delta_{nq}) > 1-\delta^{\prime}$ according to \eqref{eq_max_diff_d_ij} and $\delta^{\prime}\gg \delta_{nq}$. Consequently, $\cS_k(\bA^*)\subseteq \hat\cI_j$. 

Next, we show $ \hat\cI_j\subseteq\cS_k(\bA^*)$. Given that $j^{\prime}\in\hat\cI_j$ with $\hat\cI_j$ returned by Algorithm~\ref{alg_init}, we know that 
$\cD_{(j,j^{\prime})}^{\delta}(\hat\bZ^0,\hat\bA^0) > 1-\delta^{\prime}$, and from \eqref{eq_max_diff_d_ij}, we conclude that $\cD_{(j,j^{\prime})}^{\delta}(\bZ^*,\bA^*) > 1-2\delta^{\prime}$. By Lemma~\ref{lemma_angular_conservation}, we know, for some constant $C$,
\begin{equation}
\frac{|(\ba_j^*)^\T\ba_{j^{\prime}}^*|}{\|\ba_j^*\|\|\ba_{j^{\prime}}^*\|}\ge 1 - C\delta^{\prime}.\label{eq_lambdajj_lower1}
\end{equation}
Suppose $j^{\prime}\notin\cS_k(\bA^*)$. Then there must be $\lambda_{j^{\prime}l}^*\neq 0$ for some $l\neq k$.
Since $\bA^*$ is $(\lambda_0,\epsilon_0)$-sparse, we know $\big|\lambda_{j^{\prime}l}^*\big|\ge \lambda_0$. Then
\begin{equation*}
    \frac{|(\ba_j^*)^\T\ba_{j^{\prime}}^*|}{\|\ba_j^*\|\|\ba_{j^{\prime}}^*\|} = \frac{ |a_{jk}^*\lambda_{j^{\prime}k}^*|}{ |a_{jk}^*|\times\|\ba_{j^{\prime}}^*\|}\le \frac{1}{\sqrt{1 + (\lambda_{j^{\prime}l}^*/\lambda_{j^{\prime}k}^*)^2}}\le \frac{1}{\sqrt{1 + \lambda_0^2/M^2}}<1.
\end{equation*}Here the last inequality follows from $\|\bA^*\|_{2\to\infty}\le M$. Since $\delta^{\prime}\ll 1$, this contradicts \eqref{eq_lambdajj_lower1}. So we conclude $j^{\prime}\in\cS_k(\bA^*)$, and therefore $\cS_k(\bA^*) = \hat\cI_j$.

Now we consider the case where \underline{$j\notin \cup_{k\in[r]}\cS_k(\bA^*)$}. In such setting, we show that $|\hat\cI_j|<\min_{k\in[r]}|\cS_k(\bA^*)|$. Since $\bA^*$ is $(\lambda_0,\epsilon_0)$-sparse, it suffices to show that 
\begin{equation*}
    \hat\cI_j\subseteq \left\{j^{\prime}:\frac{|(\ba_j^*)^\T\ba_{j^{\prime}}^*|}{\|\ba_j^*\|\|\ba_{j^{\prime}}^*\|}\ge 1-\epsilon_0\right\} = \cC_{\epsilon_0}(\ba_j^*;\bA^*).
\end{equation*}
With \eqref{eq_max_diff_d_ij} and $\delta_{nq}\ll\delta^{\prime}$, we know 
\begin{equation*}
    \left\{\cD_{(j,j^{\prime})}(\bZ^*,\bA^*)\le 1-2\delta^{\prime}\right\}\subseteq\left\{\cD_{(j,j^{\prime})}(\hat\bZ^0,\hat\bA^0)\le 1-\delta^{\prime}\right\} =[q]\setminus \hat\cI_j.
\end{equation*}
Then it suffices to show that
\begin{equation*}
    [q]\setminus\cC_{\epsilon_0}(\ba_j^*;\bA^*)=\left\{j^{\prime}:\frac{|(\ba_j^*)^\T\ba_{j^{\prime}}^*|}{\|\ba_j^*\|\|\ba_{j^{\prime}}^*\|}< 1-\epsilon_0\right\}\subseteq \left\{\cD_{(j,j^{\prime})}(\bZ^*,\bA^*)\le 1-2\delta^{\prime}\right\},
\end{equation*}
which is an immediate consequence of Lemma~\ref{lemma_angular_conservation} and $\delta^{\prime}\ll \epsilon_0$.

Finally, since $\cS_k(\bA^*)\cap \cS_l(\bA^*) = \varnothing$ for any $k\neq l$, we conclude that the top $r$ non-overlapping sets in $\{\hat\cI_j\}_{j\in[q]}$ equal $\{\cS_k(\bA^*)\}_{k\in[r]}$ up to a permutation, 

\medskip
\underline{\textbf{Step 3.}} Without loss of generality, we let $\hat\cS_k = \cS_k(\bA^*)$ for each $k\in[r]$, and write $\cS_k^*$ for short. Let $q_k = |\cup_{l\neq k}\cS_l^*|$. Define $\cS_{-k}^* = \cS^*\setminus\cS_k^* = \cup_{l\in[r],l\neq k}\cS_k^*$, $\bA_{-k}^* = \bA^*_{\cS_{-k}^*}$, $\hat\bA_{-k}^0 = \hat\bA^0_{\cS_{-k}^*}$ and $\hat\bA^{\star}_{-k} = \hat\bA^0_{\cS_{-k}^*}(\hat\bG^{\star})^{-1}$ for $\hat\bG^{\star}$ given in \eqref{eq_opt_align_zero}. By definition, 
\begin{equation}
    \hat\bA^{\star}_{-k} = \hat\bA_{-k}^0(\hat\bG^{\star})^{-1}.
\end{equation}
Since $\hat\bv_k$ is the $r$th right singular vector of $\hat\bA_{-k}^0$, we know
\begin{equation*}
    \hat\bv_k \in \argmin_{\bv:\|\bv\|=1}\big\|\hat\bA^{\star}_{-k}\hat\bG^{\star}\bv\big\|^2
\end{equation*}
We note that, by construction, 
\begin{equation*}
    q_k^{-1}(\bA^*_{-k})^\T\bA^*_{-k} = q_k^{-1}\sum_{l=1,l\neq k}^r(\bA^*_{\cS_l^*,})^\T\bA^*_{\cS_l^*,}
\end{equation*} is diagonal with the $l$-th diagonal entries being $q_k^{-1}\sum_{j\in\cS_l^*}( a_{jl}^*)^2\ge \tilde{\sigma}_q^{-1}$ except that the $k$-th diagonal entry is $0$.
Thus, the smallest singular value of $q_k^{-1/2}\bA^*_{-k}\hat\bG^{\star}$ is $0$, with corresponding right singular vector given as $(\hat\bG^{\star})^{-1}\be_k/\|(\hat\bG^{\star})^{-1}\be_k\|$. Here, $\|(\hat\bG^{\star})^{-1}\be_k\|$ is well-defined due to Lemma~\ref{lemma_exist_g_hat_star}. Furthermore, the $(r-1)$-th singular value of $q_k^{-1/2}\bA^*_{-k}\hat\bG^{\star}$ is lower bounded by $C\tilde{\sigma}_q^{-1}$ for some absolute constant $C$ according to Lemma~\ref{lemma_exist_g_hat_star}. With \eqref{eq_event_init_algorithm}, we can derive by simple telescoping that
\begin{equation*}
    \Big\|q_k^{-1}(\hat\bG^{\star})^\T(\bA^*_{-k})^\T\bA^*_{-k}\hat\bG^{\star} - q_k^{-1}(\hat\bG^{\star})^\T(\hat\bA^{\star}_{-k})^\T\hat\bA^{\star}_{-k}\hat\bG^{\star}\Big\| = O(\delta_{nq})
\end{equation*}
By the variant of Davis-Kahan theorem~\citep{yu2015useful}, we arrive at 
\begin{equation*}
    \left\|\hat\bv_k - \frac{(\hat\bG^{\star})^{-1}\be_k}{\|(\hat\bG^{\star})^{-1}\be_k\|}\right\|=O(\tilde\sigma_q\delta_{nq}),
\end{equation*}
Consequently, for $\tilde\bG = (\hat\bv_1,\dots,\hat\bv_r)$, there exists some diagonal matrix $\hat\bD$ such that 
\begin{equation*}
    \big\|\tilde\bG - (\hat\bG^{\star})^{-1}\hat\bD\big\| = O(\tilde\sigma_q\delta_{nq}).
\end{equation*}
By Lemma~\ref{lemma_exist_g_hat_star} and the definition of $\hat\bD$, we know the diagonal elements of $\hat\bD$ can be upper and lower bounded by some constant with probability approaching 1. Therefore, one has
\begin{align*}
    \big\|\tilde\bG^{-1} - \hat\bD^{-1}\hat\bG^{\star}\big\|\le &\,\big\|\tilde\bG^{-1}\big\|\big\|\tilde\bG - (\hat\bG^{\star})^{-1}\hat\bD\big\|\big\|(\hat\bG^{\star})^{-1}\hat\bD\big\|\\\le &\,\big\|\tilde\bG^{-1} - \hat\bD^{-1}\hat\bG^*\big\|\big\|\tilde\bG - (\hat\bG^{\star})^{-1}\hat\bD\big\|\big\|(\hat\bG^{\star})^{-1}\hat\bD\big\| \\&\,+ \big\|\tilde\bG - (\hat\bG^{\star})^{-1}\hat\bD\big\|\big\|(\hat\bG^{\star})^{-1}\hat\bD\big\| ^2.
\end{align*}
Since $\|\tilde\bG-(\hat\bG^{\star})^{-1}\hat\bD\|=o_p(1)$ and $(\hat\bG^{\star})^{-1}\hat\bD$ has singular values bounded above and away from zero, $\|\tilde\bG^{-1}\|=O_p(1)$. With Lemma~\ref{lemma_exist_g_hat_star}, this implies
\begin{equation}
    \big\|\tilde\bG^{-1} - \hat\bD^{-1}\hat\bG^{\star}\big\| = O_p(\tilde\sigma_q\delta_{nq}).\label{eq_init_G_inv}
\end{equation}
By telescoping, we know
\begin{equation*}
    \big\|\mathrm{diag}\big(\tilde\bG^{-1}\tilde\bG^{-\T} - \hat\bD^{-1}\hat\bG^{\star}(\hat\bG^{\star})^\T\hat\bD^{-\T}\big)\big\| = O_p(\tilde\sigma_q\delta_{nq}).
\end{equation*}
Finally, Corollary~\ref{coro_asymp_start} holds when $(\hat\bZ^*,\hat\bA^*)$ is replaced by $(\hat\bZ^{\star},\hat\bA^{\star})$ due to Lemma~\ref{lemma_combine_zero}, which yields that 
\begin{equation*}
    \big\|n^{-1}\hat\bG^{\star}(\hat\bZ^0)^\T\hat\bZ^0(\hat\bG^{\star})^\T - n^{-1}(\bZ^*)^\T\bZ^*\big\| = O_p\big((n\wedge q)^{-1}\big).
\end{equation*}
Together with $n^{-1}(\hat\bZ^0)^\T\hat\bZ^0 = \bI_r$ and $\mathrm{diag}(n^{-1}(\bZ^*)^\T\bZ^*) = \bI_r$ by $(ii)$ of Assumption~\ref{assump_psd_covariance}, we know 
\begin{equation*}
    \big\|\mathrm{diag}\big\{\hat\bG^{\star}(\hat\bG^{\star})^\T - \bI_r\big\}\big\| = O_p\big((n\wedge q)^{-1}\big).
\end{equation*}
Now we arrive at
\begin{align*}
    \big\|\big\{\mathrm{diag}(\tilde\bG^{-1}\tilde\bG^{-\T})\big\}^{-1/2} - \hat\bD\big\| \le &\, \big\|\big\{\mathrm{diag}(\hat\bD^{-1}\hat\bG^{\star}\hat\bG^{\star}{}^\T\hat\bD^{-\T})\big\}^{-1/2} - \hat\bD\big\| + O_p(\tilde\sigma_q\delta_{nq})\\ = &\, \big\|\big\{\hat\bD^{-1}\mathrm{diag}(\hat\bG^{\star}\hat\bG^{\star}{}^\T)\hat\bD^{-\T}\big\}^{-1/2} - \hat\bD\big\| + O_p(\tilde\sigma_q\delta_{nq})\\=&\,O_p\big(\tilde\sigma_q\delta_{nq}\big).
\end{align*}Here, the first equality is due to $\hat\bD$ being diagonal, and the second equality is due to \eqref{eq_init_G_inv} and $\delta_{nq}\gg (n\wedge q)^{-1}$ and $\tilde\sigma_q\gtrsim1$. Combine with \eqref{eq_init_G_inv} and we finish the proof.

\subsection{Proof of Theorem~\ref{thm_lqa}}
For any $(\hat\bZ,\hat\bA)$ obtained from \eqref{eq_theory_init}, by Lemma~\ref{lemma_exist_g_star} and $(ii)$ of Lemma~\ref{lemma_consis_start}, we know that 
\begin{align*}
    \big\|\hat\bA - \bA^*\big\|_{2\to\infty} \le &\, \max_{j\in[q]}\big\|(\hat\bG^*)^{-\T}\hat\ba_j - \ba_j^*\big\| + \big\|\hat\bA\{\bI_r - (\hat\bG^*)^{-1}\}\big\|_{2\to\infty}\\ \le &\, O_p(\delta_{nq}) + \|\hat\bA\|_{2\to\infty}\|\bI_r - (\hat\bG^*)^{-1}\|\\= &\,O_p(\epsilon_{nq}).
\end{align*}
Here, the last inequality is due to $\|\hat\bA\|_{2\to\infty}\le M$ by definition. Similarly, one can check that $\|\hat\bZ - \bZ^*\|_{2\to\infty} = O_p(\epsilon_{nq})$.

    In what follows, we assume that, for some absolute constant $C$, the following event holds for the initialization of Algorithm~\ref{alg_lqa}, given as $(\hat\bZ^{(0)},\hat\bA^{(0)}) = (\hat\bZ,\hat\bA)$ from \eqref{eq_theory_init}:
    \begin{align}
        \Big\{\big\|\hat\bA^{(0)} - \bA^*\big\|_{2\to\infty}\le C\epsilon_{nq}\text{ and }\big\|\hat\bZ^{(0)} - \bZ^*\big\|_{2\to\infty}\le C\epsilon_{nq}\Big\}.\label{eq_event_lqa_algorithm}
    \end{align}
    
 Under \eqref{eq_event_lqa_algorithm}, we note that for pair $(j,k)$ such that $ a_{jk}^*\neq 0$, by $\gamma(a_3+1) \le \lambda_0$,
    \begin{align*}
      \big|\hat\bA^{(0)}_{j,k}\big| \ge  \big| a_{jk}^*\big| - \Big|\big\{\hat\bA^{(0)} - \bA^*\big\}_{j,k}\Big| 
        \ge  \lambda_0 - C\epsilon_{nq} > a_3\gamma,
    \end{align*}
    uniformly under the event in \eqref{eq_event_lqa_algorithm} when $n$ and $q$ are large enough. Therefore, the weights can be updated as $w_{jk}^{(0)} = 0$ according to Condition~\ref{assump_concave_function}.
    
    Next, for pair $(j,k)$ such that $ a_{jk}^* = 0$, similarly, one has
    \begin{equation*}
         \big|\hat\bA^{(0)}_{j,k}\big|\le C\epsilon_{nq},
    \end{equation*}
    uniformly. Therefore, with Condition~\ref{assump_concave_function} and $\epsilon_{nq}\le\eta$, we know
    \begin{equation*}
        \frac{\rho_{\gamma}^{\prime}(0+)}{2\eta\sqrt{C^2+1}}\le w_{jk}^{(0)}\le \frac{\rho_{\gamma}^{\prime}(0+)}{\eta},\text{ when }a_{jk}^*=0
    \end{equation*}
    when $n$ and $q$ are large enough. Let $\omega_{jk} = w_{jk}^{(0)}\eta/\rho_{\gamma}^{\prime}(0+)$. Then we can rewrite the optimization problem in the first iteration as
   \begin{equation} \begin{aligned}
        \tilde\bH^{(1)} =&\, \mathop{\arg\min}_{\|\bH - \bI_r\|\le R\text{ and }\mathrm{diag}(\bH^\T\bH) = \bI_r}\sum_{j=1}^q\sum_{k=1}^r1_{( a_{jk}^*=0)}\omega_{jk}\left\{\hat\bA_{j,}^{(0)}\bH_{,k}\right\}^2
        \\=&\, \mathop{\arg\min}_{\|\bH - \bI_r\|\le R\text{ and }\mathrm{diag}(\bH^\T\bH) = \bI_r}\sum_{k=1}^r\big\{\hat\bA^{(0)}_{\cT_k^*,}\bh_k\big\}^\T\bOmega_k \big\{\hat\bA^{(0)}_{\cT_k^*,}\bh_k\big\},
    \end{aligned}\label{eq_lqa_opt_step}\end{equation}
    where recall $\cT_k^* = \{j: a_{jk}^* = 0\}$ and we denote $\bH =  (\bh_1,\dots,\bh_r) $ and $\bOmega_k = \mathrm{diag}(\omega_{jk})_{j\in\cT_k}$. Here, each diagonal entry of $\bOmega_k$ is bounded uniformly in $[2(C^2+1)^{-1/2},1]$ for any $k\in[r]$.

Recall that $(\hat\bZ^{(0)}, \hat\bA^{(0)}) = (\hat\bZ,\hat\bA) = (\hat\bZ^*(\hat\bG^*)^{-\T},\hat\bA^*\hat\bG^*)$ for $\hat\bG^*$ defined in \eqref{eq_opt_align}. We then consider the problem 
with $(\hat\bZ^{(0)},\hat\bA^{(0)})$ in \eqref{eq_lqa_opt_step} replaced by $\big(\bZ^*(\hat\bG^{*})^{-\T},\bA^{*}\hat\bG^{*}\big)$
\begin{equation}\begin{aligned}
    \bH^{*} = &\,\mathop{\arg\min}_{\|\bH - \bI_r\|\le R\text{ and }\mathrm{diag}\{\bH^\T\bH\} = \bI_r}\sum_{j=1}^q\sum_{k=1}^rw_{jk}^{(t)}1_{(a_{jk}^* = 0)}\left\{\bA_{j,}^*\hat\bG^{*}\bH_{,k}\right\}^2 \\=&\, \mathop{\arg\min}_{\|\bH - \bI_r\|\le R\text{ and }\mathrm{diag}\{\bH^\T\bH\} = \bI_r}\sum_{k=1}^r\big\{\bA^{*}_{\cT_k^*,}\hat\bG^{*}\bh_k\big\}^\T\bOmega_k \big\{\bA^{*}_{\cT_k^*,}\hat\bG^{*}\bh_k\big\}.
\end{aligned}\label{eq_lqa_opt_step_alt}\end{equation}
Write for short that
\begin{equation*}
    \hat \bM_{\lambda,k}^{(0)} = q^{-1}\big\{\hat\bA^{(0)}_{\cT_k^*,}\big\}^\T\bOmega_k \hat\bA^{(0)}_{\cT_k^*,}\text{ and }\bM_{\lambda,k}^{(0)} = q^{-1}\big\{\bA^{*}_{\cT_k^*,}\hat\bG^{*}\big\}^\T\bOmega_k \bA^{*}_{\cT_k^*,}\hat\bG^{*}.
\end{equation*}
We first note that \begin{align*}\big\|\hat \bM_{\lambda,k}^{(0)} - \bM_{\lambda,k}^{(0)}\big\| = \Big\|(\hat\bG^*)^{\T}\big\{q^{-1}(\hat\bA_{\cT_k^*,}^*)^\T\bOmega_k\hat\bA_{\cT_k^*,}^* - q^{-1}(\bA_{\cT_k^*,}^*)^\T\bOmega_k\bA_{\cT_k^*,}^*\big\}\hat\bG^*\Big\|\end{align*}
We first compute
    \begin{align*}
        \big\|q^{-1}(\hat\bA_{\cT_k^*,}^*)^\T\bOmega_k\hat\bA_{\cT_k^*,}^* - q^{-1}(\bA_{\cT_k^*,}^*)^\T\bOmega_k\bA_{\cT_k^*,}^*\big\| \le&\, 2\big\|q^{-1}(\hat\bA_{\cT_k^*,}^* - \bA^*_{\cT_k^*,})^\T\bOmega_k\hat\bA_{\cT_k^*,}^* \big\| \\&\, + \big\|q^{-1}(\hat\bA_{\cT_k^*,}^* - \bA^*_{\cT_k^*,})^\T\bOmega_k(\hat\bA_{\cT_k^*,}^* - \bA^*_{\cT_k^*,})\big\|.
    \end{align*}
    For $\big\|q^{-1}(\hat\bA_{\cT_k^*,}^* - \bA^*_{\cT_k^*,})^\T\bOmega_k\hat\bA_{\cT_k^*,}^* \big\| $, one has
    \begin{align}
        \big\|q^{-1}(\hat\bA_{\cT_k^*,}^* - \bA^*_{\cT_k^*,})^\T\bOmega_k\hat\bA_{\cT_k^*,}^* \big\|\le&\,  q^{-1}\big\|\sum_{j\in\cT_k^*}\big\{\sum_{i=1}^n\ell_{ij}^{\prime\prime}(\theta_{ij})\bz_i^*(\bz_i^*)^\T\big\}^{-1}\sum_{i=1}^n\ell_{ij}^{\prime}(\theta_{ij}^*)\bz_i^*\omega_{jk}(\ba_j^*)^\T\big\|\label{eq_weighted_sum_diff1} \\&\, + q^{-1}\big\|(\bR_a^*)_{\cT_k^*,}\big\|\|\bOmega_k\|\|\hat\bA^*_{\cT_k^*,}\|\label{eq_weighted_sum_diff2}.
    \end{align}
    For $j\in\cT_k^*$, define
\[
 h_\eta(x)=\frac{\eta\rho_\gamma'(|x|)}{\rho_\gamma'(0+)\sqrt{x^2+\eta^2}},
\]
By Condition~\ref{assump_concave_function}\ref{item_nonsmooth} and $\eta\ge\eta_0$, $h_\eta(\cdot)$ is uniformly bounded and Lipschitz near zero. All its arguments considered below lie in this neighbourhood with probability approaching one. Let $\omega_{jk}^{0}(\bG)=h_\eta\big((\ba_j^*)^\T\bG_{,k}\big)$. Since $\omega_{jk}=h_\eta((\hat\ba_j^*)^\T\hat\bG^*_{,k})$ and $\|\hat\bG^*\|=O_p(1)$, we have $|\omega_{jk}-\omega_{jk}^{0}(\hat\bG^*)|\lesssim\|\bE_{j,}\|$. Consequently, the contribution of this weight difference to \eqref{eq_weighted_sum_diff1} is $O_p(q^{-1}\|\bE\|_{\Fn}^2)=O_p((n\wedge q)^{-1})$.
Therefore, similar to the proof of Corollary~\ref{coro_asymp_start}, we know \eqref{eq_weighted_sum_diff1} can be bounded by $O_p(\sqrt{|\cT_k^*|/(nq^2)}) = O_p((nq)^{-1/2})$ with $|\cT_k^*| < q$. For \eqref{eq_weighted_sum_diff2}, with $\|\hat\bA^*_{\cT_k^*,}\| \le \sqrt{|\cT_k^*|}\|\hat\bA^*\|_{2\to\infty} = O\big(\sqrt{|\cT_k^*|}\big)$, invoke $\|\bR_a^*\| = O_p(\sqrt{q}(n\wedge q)^{-1})$ from Lemma~\ref{lemma_asymp_start} to get
    \begin{align*}
        q^{-1}\big\|(\bR_a^*)_{\cT_k^*,}\big\|\|\bOmega_k\|\|\hat\bA^*_{\cT_k^*,}\|\le &\,O\big(q^{-1} \sqrt{|\cT_k^*|}\big)\big\|(\bR_a^*)_{\cT_k^*,}\big\|\\ = &\,O_p\big((n\wedge q)^{-1}\big).
    \end{align*}
    For $\big\|q^{-1}(\hat\bA_{\cT_k^*,}^* - \bA^*_{\cT_k^*,})^\T\bOmega_k(\hat\bA_{\cT_k^*,}^* - \bA^*_{\cT_k^*,})\big\|$, by {Cauchy–Schwarz inequality} and $q^{-1/2}\|\hat\bA^*_{\cT_k^*,} - \bA_{\cT_k,}^*\|_{\Fn}\le q^{-1/2}\|\hat\bA^* - \bA^*\| = O_p((n\wedge q)^{-1/2})$, we know that it can be bounded by $O_p((n\wedge q)^{-1})$.
    
    Then we conclude that
\begin{equation}
    \big\|q^{-1}(\hat\bA_{\cT_k^*,}^*)^\T\bOmega_k\hat\bA_{\cT_k^*,}^* - q^{-1}(\bA_{\cT_k^*,}^*)^\T\bOmega_k\bA_{\cT_k^*,}^*\big\| = O_p\big((n\wedge q)^{-1} \big).\label{eq_local_update_hat_star_diff}
\end{equation}
With $\|\hat\bG^* - \bI_r\| =O_p(\epsilon_{nq})\ll R$ by Lemma~\ref{lemma_exist_g_star}, we know with \eqref{eq_local_update_hat_star_diff} that
\begin{equation*}
    \big\|\hat \bM_{\lambda,k}^{(0)} -  \bM_{\lambda,k}^{(0)}\big\| =  O_p\big((n\wedge q)^{-1} \big).
\end{equation*}
Let $d_k=\|(\hat\bG^*)^{-1}\be_k\|^{-1}$, $\bD=\mathrm{diag}(d_1,\dots,d_r)$, and $\bh^{(0)}_k = d_k^{-1}(\hat\bG^{*})^{-1}\be_k$. By definition, $\bh^{(0)}_k$ is an eigenvector of $\bM_{\lambda,k}^{(0)}$ for eigenvalue $0$. Due to $\big\|\hat\bG^{*} - \bI_r\big\| =O_p(\epsilon_{nq})=o_p(1)$ and $\sigma_q^{-1} = \min_{k\in[r]}\lambda_{r-1}\big\{q^{-1}(\bA^*_{\cT_k^*,})^\T\bA^*_{\cT_k^*,}\big\}$, we know that the first $r-1$ eigenvalues of $\bM_{\lambda,k}^{(0)}$ can be lowered bounded by $\sigma_q^{-1}$ uniformly over $k\in[K]$.
Now let $\hat\bh_k^{(0)}$ be the eigenvector of $\hat\bM^{(0)}_{\lambda,k}$ associated with the smallest eigenvalue. By the variant Davis-Kahan theorem in \citet{yu2015useful}, we know
\begin{equation*}
    \big\|\hat\bh_k^{(0)} - \bh_{k}^{(0)}\big\| = O_p\big(\sigma_q(n\wedge q)^{-1} \big).
\end{equation*}
Both matrices of eigenvectors lie within radius $R$ of $\bI_r$ with probability approaching one. Thus $\bH^*=(\hat\bG^*)^{-1}\bD$ and $\hat\bH^{(1)}=(\hat\bh_1^{(0)},\dots,\hat\bh_r^{(0)})$ solve \eqref{eq_lqa_opt_step_alt} and \eqref{eq_lqa_opt_step}, respectively, and
\begin{equation}
 \|(\hat\bH^{(1)})^{-1}(\hat\bG^*)^{-1}-\bD^{-1}\|=O_p(\sigma_q(n\wedge q)^{-1}).\label{eq_second_last_hat_H}
\end{equation}
For the normalisation step, set $\bV=(\hat\bH^{(1)})^{-1}\{n^{-1}(\hat\bZ^{(0)})^\T\hat\bZ^{(0)}\}(\hat\bH^{(1)})^{-\T}$. Corollary~\ref{coro_asymp_start}, together with Lemma~\ref{lemma_consis_start}$(i)$, implies $\mathrm{diag}\{n^{-1}(\hat\bZ^*)^\T\hat\bZ^*\}=\bI_r+O_p((n\wedge q)^{-1})$. Equation~\eqref{eq_second_last_hat_H} therefore gives $\mathrm{diag}(\bV)=\bD^{-2}+O_p(\sigma_q(n\wedge q)^{-1})$. Since $\bD=\bI_r+o_p(1)$, the algorithm's normalisation yields
\begin{equation}
 \hat\bG^{(1)}=\{\mathrm{diag}(\bV)\}^{-1/2}(\hat\bH^{(1)})^{-1}
 =\hat\bG^*+O_p(\sigma_q(n\wedge q)^{-1}).\label{eq_last_hat_H}
\end{equation}
In the proof of Theorem~\ref{thm_oracle}, we already know
        \begin{equation*}
            \big\|\hat\bG - \hat\bG^{*}\big\| = O_p\big(\sigma_q(n\wedge q)^{-1+\varepsilon}\big),
        \end{equation*}
        for any constant $\varepsilon>0$. Combined with \eqref{eq_last_hat_H}, we conclude that $\|\hat\bG^{(1)} - \hat\bG\| = O_p\big(\sigma_q(n\wedge q)^{-1+\varepsilon}\big)$, and with Lemma~\ref{lemma_combine_zero}, the asymptotic distributions follow by a similar argument in the proof of Theorem~\ref{thm_oracle}.

\section{Proof of Technical Lemmas}

\subsection{Proof of Lemmas~\ref{lemma_exist_g_star} and~\ref{lemma_exist_g_hat_star}}
To start, we introduce the following lemma
\begin{lemma}\it\label{lemma_exist_optimal_align}
    For any $\bZ,\bZ^*\in\RR^{n\times r}$ and $\bA,\bA^*\in\RR^{q\times r}$, suppose there exists some matrix $\bP\in\RR^{r\times r}$ and some constant $C$ such that $C^{-1}\le \sigma_{\min}(\bP) \le \sigma_{\max}(\bP) \le C$ and 
    \begin{equation*}
        n^{-1}\big\|\bZ\bP - \bZ^*\big\|_{\Fn}^2 + q^{-1}\big\|\bA\bP^{-\T} - \bA^*\big\|_{\Fn}^2\le \delta^2 \le C^{\prime}{\sigma_r(n^{-1/2}\bZ^*)^2},
    \end{equation*}
    where $C^{\prime}\le \min\{1/2,1/(64C^4)\}$. Then the solution
    \begin{equation}
        \argmin_{\bG\in\RR^{r\times r}\text{ is invertible}}n^{-1}\big\|\bZ\bG - \bZ^*\big\|_{\Fn}^2 + q^{-1}\big\|\bA\bG^{-\T} - \bA^*\big\|_{\Fn}^2\label{opt_align_AB}
    \end{equation}
    exists, denoted by $\bG$, and satisfies
    \begin{equation*}
        \big\|\bG - \bP\big\|_{\Fn}\le \frac{1}{2C\sqrt{C^{\prime}}}\,\frac{\delta}{\sigma_{r}(n^{-1/2}\bZ^*)}.
    \end{equation*}
\end{lemma}
\begin{proof}
    See Section~\ref{sec_prove_lemma_exist_optimal_align}
\end{proof}
Now we prove the two lemmas.
\subsubsection{Proof of  Lemma~\ref{lemma_exist_g_star}}\label{sec_prove_lemma_exist_g_star}

    Take $\bP = \bI_r$ in Lemma~\ref{lemma_exist_optimal_align} and we know $\delta^2\asymp \epsilon_{nq}^2$. Therefore with $r$ fixed and $\sigma_{r}(n^{-1/2}\bZ^*)\gtrsim 1$ by $(ii)$, $(iii)$ of Assumption~\ref{assump_psd_covariance}, we conclude
    \begin{equation*}
        \big\|\hat\bG^* - \bI_r\big\| = O_p(\epsilon_{nq})
    \end{equation*}

\subsubsection{Proof of Lemma~\ref{lemma_exist_g_hat_star}}\label{sec_prove_lemma_exist_g_hat_star}

    For $(\hat\bZ^0,\hat\bA^0)$ obtained in \eqref{eq_init_of_init_est}, by definition, one has $L(\hat\bZ^0,\hat\bA^0) = \sum_{i=1}^n\sum_{j=1}^q\ell_{ij}(\hat\ba_j^\T\hat\bz_i)\le \sum_{i=1}^n\sum_{j=1}^q\ell_{ij}\big((\ba_j^*)^\T\bz_i^*\big)$. Denote $\hat \theta_{ij}^0 = \hat\ba_j^0{}^\T\hat\bz_i^0$ for each $i\in[n]$ and $j\in[q]$.
    Expand the log-likelihood function, and we have 
    \begin{align*}
        \sum_{i=1}^n\sum_{j=1}^q\ell_{ij}\big( \theta_{ij}^*\big)\ge\sum_{i=1}^n\sum_{j=1}^q\ell_{ij}\big( \theta_{ij}^*\big)+\sum_{i=1}^n\sum_{j=1}^q\ell_{ij}^{\prime}( \theta_{ij}^*)(\hat \theta_{ij}^0 - \theta_{ij}^*)+\frac12\sum_{i=1}^n\sum_{j=1}^q\ell_{ij}^{\prime\prime}(\tilde \theta_{ij})(\hat \theta_{ij}^0 - \theta_{ij}^*)^2,
    \end{align*}
    which implies 
    \begin{equation*}
        \sum_{i=1}^n\sum_{j=1}^q\big\{\ell_{ij}^{\prime\prime}(\tilde \theta_{ij})\big\}(\hat \theta_{ij}^0- \theta_{ij}^*)^2\le -2\sum_{i=1}^n\sum_{j=1}^q\ell_{ij}^{\prime}( \theta_{ij}^*)(\hat \theta_{ij}^0 - \theta_{ij}^*).
    \end{equation*}
    Here $\tilde \theta_{ij}$ is between $ \theta_{ij}^*$ and $\hat \theta_{ij}^0$, and therefore $\ell_{ij}^{\prime\prime}(\tilde \theta_{ij}) = b^{\prime\prime}(\tilde\theta_{ij})\ge b_L$. The left side is lower bounded by $b_L\sum_{i=1}^n\sum_{j=1}^q(\hat \theta_{ij}^0 - \theta_{ij}^*)^2$. To bound the right side, we introduce the following lemma
\begin{lemma}\it
    We define $\bL_{\theta}^* = \big\{\ell_{ij}'(\theta_{ij}^*)\big\}_{n\times q}$ to be an $n\times q$ matrix. Under the Assumptions~\ref{assump_scaling} and~\ref{assump_limit}, for any constant $\delta>0$, with probability at least $1-(n\vee q)^{-\delta}$
    \begin{enumerate}[label=$(\roman*)$]
    \item for any $i\in[n]$, $\|(\bL_{\theta}^*)_{i,}\|\le C_{\delta} \sqrt{q}$ and $\|\bL_{\theta}^*\|_{2\to\infty} = \max_{1\le i\le n}\|(\bL_{\theta}^*)_{i,}\|\le C_{\delta}\sqrt{q\log n}$; 
    \item for any $j\in[q]$, $\|(\bL_{\theta}^*)_{,j}\| \le C_{\delta}\sqrt{n}$ and $\|(\bL_{\theta}^*)^\T\|_{2\to\infty} = \max_{1\le j\le q}\|(\bL_{\theta}^*)_{,j}\| \le C_{\delta}\sqrt{n\log q}$; 
        \item $\|\bL_{\theta}^*\| \le C_{\delta}\sqrt{n\vee q}$; 
    \end{enumerate}
     for some constant $C_{\delta}$ depending only on $\delta$.
    \label{lemma_concentration_l_mat}
\end{lemma}    
\begin{proof}See Section~\ref{supp_Sec_prove_lemma_concentration_l_mat}
\end{proof}

By Lemma~\ref{lemma_concentration_l_mat}, 
    \begin{align*}
        \Big|\sum_{i=1}^n\sum_{j=1}^q\ell_{ij}^{\prime}( \theta_{ij}^*)(\hat \theta_{ij}^0 - \theta_{ij}^*)\Big|=&\,\Big|\text{Trace}\Big[\big\{\ell_{ij}^{\prime}( \theta_{ij}^*)\big\}_{n\times q}\big\{\hat\bA^0\big(\hat\bZ^0\big)^\T -\bA^*\big(\bZ^*\big)^\T\big\}\Big]\Big|\\\le &\,\text{rank}\big\{\hat\bA^0\big(\hat\bZ^0\big)^\T-\bA^*\big(\bZ^*\big)^\T\big\}\big\|\hat\bA^0\big(\hat\bZ^0\big)^\T-\bA^*\big(\bZ^*\big)^\T\big\|_F\big\|\bL_{\theta}^*\big\|\\\le &\,O_p(r\sqrt{n\vee q}){\Big(\sum_{i=1}^n\sum_{j=1}^q(\hat \theta_{ij}^0 - \theta_{ij}^*)^2\Big)^{1/2}}.
    \end{align*}
    Here for the second inequality, we used $\text{Trace}(\Ab\Bb^\T) \le \text{rank}(\Bb)\|\Ab\|\|\Bb\|_F$.
    Therefore, 
    \begin{equation}
        \Big(\sum_{i=1}^n\sum_{j=1}^q(\hat \theta_{ij}^0 - \theta_{ij}^*)^2\Big)^{1/2}=O_p( \sqrt{n\vee q}),
    \end{equation}
    or
    \begin{equation}
        \Big\|\hat\bA^0\big(\hat\bZ^0\big)^\T -\bA^*\big(\bZ^*\big)^\T\Big\|_{\Fn}=O_p( \sqrt{n\vee q}).\label{eq_average_consis_w3}
    \end{equation}
Let $\hat\bG^{z} = (\bZ^*{}^\T\bZ^*)^{-1}\bZ^*{}^\T\hat\bZ^0$. By Assumption~\ref{assump_psd_covariance}$(ii)$, $(iii)$, $\hat\bG^{z}$ here is well-defined. Next, we find that
\begin{equation}\begin{aligned}
   \big\| \hat\bA^0(\hat\bG^z)^\T - \bA^* \big\| = &\, \big\|\hat\bA^0\hat\bZ^0{}^\T\bZ^*(\bZ^*{}^\T\bZ^*)^{-1} - \bA^*\big\|\\\le &\,\big\|\hat\bA^0\hat\bZ^0{}^\T - \bA^*(\bZ^*)^\T\big\| \big\|\bZ^*(\bZ^*{}^\T\bZ^*)^{-1}\big\|\\\lesssim &\,O_p\Big(\frac{\sqrt{n\vee q}}{\sigma_{r}(\bZ^*)}\Big) = O_p\Big(\frac{\sqrt{q}}{\sqrt{n\wedge q}}\Big),
\end{aligned}\label{eq_derive_lambda_gf}\end{equation}
by $(ii)$, $(iii)$ of Assumption~\ref{assump_psd_covariance} and $r$ being fixed. Next, for $\hat\bZ^0(\hat\bG^z)^{-1} - \bZ^*$, we first note that, similar to the derivation in \eqref{eq_derive_lambda_gf}, one has
\begin{align*}
    \big\|\hat\bG^z\hat\bA^0{}^\T\bA^*\big(\bA^*{}^\T\bA^*\big)^{-1} - \bI_r\big\| =&\, \Big\|(\bZ^*{}^\T\bZ^*)^{-1}\bZ^*{}^\T\big\{\hat\bZ^0\hat\bA^0{}^\T - \bZ^*\bA^*{}^\T\big\}\bA^*\big(\bA^*{}^\T\bA^*\big)^{-1}\Big\|\\ = &\,O_p\Big(\frac{\sqrt{n\vee q}}{\sqrt{nq}}\Big) = O_p\Big(\frac{1}{\sqrt{n\wedge q}}\Big).
\end{align*}
With $\sigma_{\max}\big\{\hat\bA^0{}^\T\bA^*\big(\bA^*{}^\T\bA^*\big)^{-1}\big\}\le \sigma_{\min}\big(\bA^*{}^\T\bA^*\big)^{-1}\sigma_{\max}\big(\hat\bA^0{}^\T\bA^*\big)\gtrsim 1$, we conclude with Weyl's inequality that $\sigma_{\min}(\hat\bG^z)\lesssim 1$.
Thus, we arrive at
\begin{equation*}
    \big\|\hat\bA^0{}^\T\bA^*\big(\bA^*{}^\T\bA^*\big)^{-1} - \big(\hat\bG^z\big)^{-1}\big\| =O_p\Big( \frac{1}{\sqrt{n\wedge q}}\Big).
\end{equation*}
Similar to the derivation of \eqref{eq_derive_lambda_gf}, one can derive
\begin{equation*}
    n^{-1/2}\big\|\hat\bZ^0\hat\bA^0{}^\T\bA^*\big(\bA^*{}^\T\bA^*\big)^{-1} - \bZ^*\big\| =O_p\Big( \frac{1}{\sqrt{n\wedge q}}\Big),
\end{equation*}
and therefore, 
\begin{equation*}
    n^{-1/2}\big\|\hat\bZ^0(\hat\bG^z)^{-1} - \bZ^*\big\|O_p\Big( \frac{1}{\sqrt{n\wedge q}}\Big).
\end{equation*}

Since $r$ is fixed, we know
\begin{equation*}
    q^{-1}\big\| \hat\bA^0(\hat\bG^z)^\T - \bA^* \big\|_{\Fn}^2 + n^{-1}\big\|\hat\bZ^0(\hat\bG^z)^{-1} - \bZ^*\big\|_{\Fn}^2O_p\Big( \frac{1}{{n\wedge q}}\Big),
\end{equation*}
Apply Lemma~\ref{lemma_exist_optimal_align}, and with
\begin{equation*}
    1\lesssim\sigma_r(\hat\bG^z)\le\sigma_{1}(\hat\bG^z)\lesssim 1
\end{equation*} with probability approaching $1$, we finish the proof.

\subsection{Proof of Lemmas~\ref{lemma_consis_start} and~\ref{lemma_oracle_asym}}\label{supp_sec_prove_ora_o}
To prove these two lemmas, we first consider the following optimization problem
\begin{equation}\begin{aligned}
   (\hat\bZ^{o},\hat\bA^{o}) =&\, \argmin_{(\bZ,\bA)\in\Theta^*(\epsilon)}\cL(\bZ,\bA) \\&\text{ subject to }\cP^*(\bZ,\bA)=0,\, \lambda_{\min}(n^{-1}\bZ^\T\bZ)\ge c\text{, and }\lambda_{\min}(q^{-1}\bA^\T\bA)\ge c,\end{aligned}\label{eq_problem_align}
\end{equation}
where $\epsilon>0$ is a fixed sufficiently small constant, $\Theta^*(\epsilon)$ is the local region defined in \eqref{sec_main} in the main text, and $c$ is some small constant satisfying $c < \min\{\lambda_{\min}(\bSigma_a^*),\lambda_{\min}(\bSigma_z^*)\}/2$, and
\begin{equation}
    \cP^*(\bZ,\bA) := \frac{1}{2}\|n^{-1}\bZ^\T\bZ - q^{-1}\bA^\T\bA - n^{-1}\bZ^\T\bZ^* + q^{-1}\bA^*{}^\T\bA\big\|_{\Fn}^2.\label{eq_define_key_penalty}
\end{equation}
We add superscript $^*$ to $\cP^*(\cdot,\cdot)$ to indicate its dependence on the true parameters $(\bZ^*,\bA^*)$. In what follows, we prove the consistency and asymptotic expansions in Lemmas~\ref{lemma_consis_start} and~\ref{lemma_oracle_asym} for $(\hat\bZ^o,\hat\bA^o)$ in Section~\ref{supp_sec_prove_ora_o_step1} and Section~\ref{supp_sec_prove_ora_o_step2}, respectively. Then in Section~\ref{supp_sec_prove_ora_o_step3}, we show that,
with probability approaching one, $(\hat\bZ^o,\hat\bA^o) = (\hat\bZ(\hat\bG^*)^{\T},\hat\bA(\hat\bG^*)^{-1})$ for any $(\hat\bZ,\hat\bA)$ obtained in \eqref{eq_theory_init} and $\hat\bG$ obtained in \eqref{eq_opt_align}. Combine these three steps, and we finish the proofs.

Before we start, we first introduce the following lemma to understand the motivation of constructing the penalty \eqref{eq_define_key_penalty}.
\begin{lemma}[Lemma 2 in~\citet{ma2021beyond}]\label{lemma_procruste_solution}
    For any $\bZ,\bZ^*\in\RR^{n\times r}$ and $\bA,\bA^*\in\RR^{q\times r}$ such that the solution
    \begin{equation}
        \argmin_{\bG\in\RR^{r\times r}\text{ is invertible}}n^{-1}\big\|\bZ\bG - \bZ^*\big\|_{\Fn}^2 + q^{-1}\big\|\bA\bG^{-\T} - \bA^*\big\|_{\Fn}^2
    \end{equation}
    exists and is denoted by $\tilde\bG$, it holds that $n^{-1}\tilde\bZ^\T(\tilde\bZ - \bZ^*) = q^{-1}(\tilde\bA - \bA^*)^\T\tilde\bA$ with $(\tilde\bZ,\tilde\bA) := (\bZ\tilde\bG,\bA\tilde\bG^{-\T})$.
\end{lemma}
\begin{proof}
    The proof of this lemma is similar to the proof of Lemma 2 in \citet{ma2021beyond}, and for completeness, we state the proof in Section~\ref{supp_sec_prove_lemma_procruste_solution}.
\end{proof}

Lemma~\ref{lemma_procruste_solution} implies that $\cP^*(\hat\bZ,\hat\bA) = 0$ if and only if $ \hat\bG^* = \bI_r$ with $\hat\bG^*$ given in  in \eqref{eq_opt_align}. 
Motivated by this observation, we study \eqref{eq_problem_align}, and show that it returns $(\hat\bZ^*,\hat\bA^*)$ with probability approaching $1$. Moreover, as established in Lemmas~\ref{lemma_hessian_property} and \ref{lemma_hessian_property_2}, the augmented (Lagrangian-type) objective $\cL(\bZ,\bA) + \nu\cP^*(\bZ,\bA)$ possesses favorable curvature and regularity properties, which enable a sharp asymptotic analysis of the resulting estimators.

\subsubsection{Step 1: Consistency}\label{supp_sec_prove_ora_o_step1}
Plug $(\bZ^*,\bA^*)$ into $\cP^*(\bZ,\bA)$ and one can find $\cP^*(\bZ^*,\bA^*) = 0$.
Then by definition of \eqref{eq_problem_align}, we know $\cL(\hat\bZ^o,\hat\bA^o)\le \cL(\bZ^*,\bA^*)$. Following a similar argument in deriving \eqref{eq_average_consis_w3}, we know
\begin{equation}
    \Big\|\hat\bA^o\big(\hat\bZ^o\big)^\T -\bA^*\big(\bZ^*\big)^\T\Big\|_{\Fn}=O_p( \sqrt{n\vee q}).\label{eq_average_consis_w4}
\end{equation}
With $\cP^*(\hat\bZ^o,\hat\bA^o) = 0$, we know
\begin{equation*}
    n^{-1}(\hat\bZ^o)^\T\hat\bZ^o - q^{-1}(\hat\bA^o)^\T\hat\bA^o = n^{-1}(\hat\bZ^o)^\T\bZ^* - q^{-1}(\bA^*)^\T\hat\bA^o 
\end{equation*}
Then, one has
\begin{align*}
n^{-1}\hat\bA^o{}^\T(\hat\bA^o - \bA^*)\hat\bZ^o{}^\T\hat\bZ^o =&\, n^{-1}\hat\bA^o{}^\T(\hat\bA^o\hat\bZ^o{}^\T-\bA^*\bZ^*{}^\T)\hat\bZ^o - n^{-1}\hat\bA^o{}^\T\bA^*(\hat\bZ^o - \bZ^*)^\T\hat\bZ^o\\=&\, n^{-1}\hat\bA^o{}^\T(\hat\bA^o\hat\bZ^o{}^\T-\bA^*\bZ^*{}^\T)\hat\bZ^o - q^{-1}\hat\bA^o{}^\T\bA^*\hat\bA^o{}^\T(\hat\bA^o - \bA^*).
\end{align*}
Add $q^{-1}\hat\bA^o{}^\T\hat\bA^o\hat\bA^o{}^\T(\hat\bA^o - \bA^*)$ to both sides to get
\begin{equation}\label{eq_aa_trans}\begin{aligned}
    &\,\Big\|\hat\bA^o{}^\T(\hat\bA^o - \bA^*)\big(n^{-1}\hat\bZ^o{}^\T\hat\bZ^o\big) +\big(q^{-1}\hat\bA^o{}^\T\hat\bA^o\big)\hat\bA^o{}^\T(\hat\bA^o - \bA^*) \Big\|_{\Fn}\\\le&\, n^{-1}\|\hat\bA^o\|_{\Fn}\|\hat\bZ^o\|_{\Fn}\big\|\hat\bA^o\hat\bZ^o{}^\T-\bA^*\bZ^*{}^\T\big\|_{\Fn} + q^{-1}\big\|\hat\bA^o{}^\T(\hat\bA^o - \bA^*)\big\|^2_{\Fn}.
\end{aligned}\end{equation}
For the first term on the right side, by $\|\hat\bA^o\|_{2\to\infty}\le M$, $\|\hat\bZ^o\|_{2\to\infty}\le M$, and \eqref{eq_average_consis_w4}, we know its Frobenius norm can be bounded by $O_p(\sqrt{(n\vee q)q/n})$. For the left side, we use the following lemma.

\begin{lemma}\label{lemma_symmetric_kronecker_prod}
Let $\bM_1\in\mathbb R^{r\times r}$ and $\bM_2\in\mathbb R^{r\times r}$ be symmetric positive definite. 
Define the linear operator $\mathcal L:\mathbb R^{r\times r}\to\mathbb R^{r\times r}$ by
\[
\mathcal L(\bA)\;:=\; \bA\bM_1 + \bM_2\bA.
\]
Then, for all $A\in\mathbb R^{r\times r}$,
\[
\|\bA\bM_1+\bM_2\bA\|_F \;\ge\; \big(\lambda_{\min}(\bM_1)+\lambda_{\min}(\bM_2)\big)\,\|\bA\|_F.
\]
\end{lemma}
\begin{proof}
See Section~\ref{supp_sec_lemma_symmetric_kronecker_prod}
\end{proof}
Therefore, with $\lambda_{\min}(n^{-1}\hat\bZ^o{}^\T\hat\bZ^o)\ge c\text{, and }\lambda_{\min}(q^{-1}\hat\bA^o{}^\T\hat\bA^o)\ge c$ from \eqref{eq_problem_align}, we conclude that
\begin{align*}
    2c\big\|\hat\bA^o{}^\T(\hat\bA^o - \bA^*)\big\|_{\Fn}\le q^{-1}\big\|\hat\bA^o{}^\T(\hat\bA^o - \bA^*)\big\|^2_{\Fn} + O_p\big(\sqrt{(n\vee q)q/n}\big),
\end{align*}
which is
\begin{equation*}
    q^{-1}\big\|\hat\bA^o{}^\T(\hat\bA^o - \bA^*)\big\|_{\Fn}\le O_p\big(c^{-1}(n\wedge q)^{-1/2}\big).
\end{equation*}
Next, we note that 
\begin{align*}
n^{-1}(\hat\bA^o - \bA^*)\hat\bZ^o{}^\T\hat\bZ^o =&\, n^{-1}(\hat\bA^o\hat\bZ^o{}^\T-\bA^*\bZ^*{}^\T)\hat\bZ^o - n^{-1}\bA^*(\hat\bZ^o - \bZ^*)^\T\hat\bZ^o\\=&\, n^{-1}(\hat\bA^o\hat\bZ^o{}^\T-\bA^*\bZ^*{}^\T)\hat\bZ^o - q^{-1}\bA^*\hat\bA^o{}^\T(\hat\bA^o - \bA^*).
\end{align*}
For the left side, with $\lambda_{\min}(n^{-1}\hat\bZ^o{}^\T\hat\bZ^o)\ge c$, we know
\begin{equation*}
    \big\|n^{-1}(\hat\bA^o - \bA^*)\hat\bZ^o{}^\T\hat\bZ^o\big\|_{\Fn}\ge c\|\hat\bA^o - \bA^*\|_{\Fn}.
\end{equation*}
For the right side, the Frobenius norm of the first term can be bounded by $O_p(\sqrt{(n\vee q)/n})$ similar to bounding the first term on the right side of \eqref{eq_aa_trans}. The Frobenius norm of the second term can be bounded by 
\begin{equation*}
    \big\|q^{-1}\bA^*\hat\bA^o{}^\T(\hat\bA^o - \bA^*)\big\|_{\Fn}\le \|\bA^*\|_{\Fn}q^{-1}\big\|\hat\bA^o{}^\T(\hat\bA^o - \bA^*)\big\|_{\Fn}\le O_p\big(c^{-1}\sqrt{q}(n\wedge q)^{-1/2}\big)
\end{equation*}
We then conclude that 
\begin{equation}
    q^{-1/2}\|\hat\bA^o - \bA^*\|_{\Fn}\le O_p\big(c^{-2}(n\wedge q)^{-1/2}\big).\label{eq_ave_consistency_Ao}
\end{equation}
Similarly, we can obtain
\begin{equation}
    n^{-1/2}\|\hat\bZ^o - \bZ^*\|_{\Fn}\le O_p\big(c^{-2}(n\wedge q)^{-1/2}\big).\label{eq_ave_consistency_Zo}
\end{equation}
By Weyl's inequality, one can derive from \eqref{eq_ave_consistency_Ao} and \eqref{eq_ave_consistency_Zo} that
\begin{align*}
    \lambda_{\min}(q^{-1}\hat\bA^o{}^\T\hat\bA^o)\ge&\, \lambda_{\min}(q^{-1}\bA^*{}^\T\bA^*) + \big\|q^{-1}\hat\bA^o{}^\T\hat\bA^o - q^{-1}\bA^*{}^\T\bA^*\big\|\\\ge &\,\lambda_{\min}(q^{-1}\bA^*{}^\T\bA^*) -O_p\big(c^{-2}(n\wedge q)^{-1/2}\big).
\end{align*}
Here, the second inequality follows by $\|\hat\bA^o\|\le\|\hat\bA^o\|_{\Fn}\le O(q^{1/2})$, \eqref{eq_ave_consistency_Ao} and telescoping. By $(ii)$, $(iii)$ of Assumption~\ref{assump_psd_covariance}, we know $\lambda_{\min}(q^{-1}\hat\bA^o{}^\T\hat\bA^o)\ge\lambda_{\min}(\bSigma_a^*)/2$ when $n$ and $q$ are large enough. Recall that $c < \min\{\lambda_{\min}(\bSigma_a^*),\lambda_{\min}(\bSigma_z^*)\}/2$.
With such a choice, the feasibility constraints $\lambda_{\min}(n^{-1}\bZ^\T\bZ)\ge c\text{, and }\lambda_{\min}(q^{-1}\bA^\T\bA)\ge c$ are strictly slack at the solution, so the optimizer does not lie on the constraint boundary.

Equations \eqref{eq_ave_consistency_Ao} and \eqref{eq_ave_consistency_Zo} indicate that $(i)$ in Lemma~\ref{lemma_consis_start} holds for $(\hat\bZ^o,\hat\bA^o)$. We then proceed to show $(ii)$ and $(iii)$. We first introduce the following results for the score function of $\cL(\bZ^*,\bA^*)$:
\begin{lemma}\it Suppose Assumption~\ref{assump_limit} holds. Let all weights below be deterministic and bounded in absolute value by $B$. For positive sequences $\epsilon_n,\epsilon_q,\epsilon_{nq}$ tending to infinity, the following bounds hold with probability at least $1-C\exp(-c\epsilon_n)$, $1-C\exp(-c\epsilon_q)$, and $1-C\exp(-c\epsilon_{nq})$, respectively, for constants $c,C>0$:
\begin{enumerate}[label=$(\roman*)$]
 \item For each fixed $j\in[q]$,
 $|\sum_{i=1}^n b_i\ell_{ij}^{\prime}(\theta_{ij}^*)|\lesssim B(\sqrt{n\epsilon_n}+\epsilon_n)$,
 and uniformly over $j\in[q]$,
 $\max_{j\in[q]}|\sum_{i=1}^n b_i\ell_{ij}^{\prime}(\theta_{ij}^*)|\lesssim B\{\sqrt{n(\epsilon_n+\log q)}+\epsilon_n+\log q\}$.
 \item For each fixed $i\in[n]$,
 $|\sum_{j=1}^q b_j\ell_{ij}^{\prime}(\theta_{ij}^*)|\lesssim B(\sqrt{q\epsilon_q}+\epsilon_q)$,
 and uniformly over $i\in[n]$,
 $\max_{i\in[n]}|\sum_{j=1}^q b_j\ell_{ij}^{\prime}(\theta_{ij}^*)|\lesssim B\{\sqrt{q(\epsilon_q+\log n)}+\epsilon_q+\log n\}$.
 \item For weights $\{b_{ij}\}_{i\in[n],j\in[q]}$, $|\sum_{i=1}^n\sum_{j=1}^q b_{ij}\ell_{ij}^{\prime}(\theta_{ij}^*)|\lesssim B(\sqrt{nq\epsilon_{nq}}+\epsilon_{nq})$ and $\max_{i\in[n],j\in[q]}|\ell_{ij}^{\prime}(\theta_{ij}^*)|\lesssim\epsilon_{nq}+\log(nq)$.
\end{enumerate}\label{lemma_concentration}
\end{lemma}
\begin{corollary}\label{coro_first_order_concern}\it
    Suppose Assumptions~\ref{assump_psd_covariance}$(ii)$, $(iii)$, $\|\bA^*\|_{2\to\infty}\le M$, and \ref{assump_limit} holds. On the true parameters $\bphi^{*}$, we have
    \begin{enumerate}[label = $(\roman*)$]
    \item $\big\|q^{-1/2}\partial_{\bphi_z}\cL(\bZ^*,\bA^*)\big\| = O_p(\sqrt{n})$;
        \item 
        $\big\|n^{-1/2}\partial_{\bphi_a}\cL(\bZ^*,\bA^*)\big\|= O_p(\sqrt{q})$;
    \item $\big\|\bS_{nq}\partial_{\bphi}\cL(\bZ^*,\bA^*)\big\|= O_p( \sqrt{n+q})$ with $\bS_{nq}$ defined below in \eqref{eq_define_snq}
    \end{enumerate}
\end{corollary}
\begin{proof}
    See Section~\ref{supp_sec_prove_lemma_concentration} for the proofs of Lemma~\ref{lemma_concentration} and Corollary~\ref{coro_first_order_concern}.\end{proof}

Let
\begin{equation*}
    \cL_{\nu}(\bZ,\bA) = \cL(\bZ,\bA) + \nu nq\cP^*(\bZ,\bA),
\end{equation*}with $\nu > 0  $ being some fixed constant. We now characterize the derivatives of $\cL_{\nu}(\bZ,\bA)$. Let 
\begin{equation}
    \bS_{nq} = \begin{pmatrix}
        q^{-1/2}\bI_{nr}& \zero\\\zero & n^{-1/2}\bI_{qr}
    \end{pmatrix}.\label{eq_define_snq}
\end{equation}

For the Hessian matrix of $\cL_{\nu}(\bZ,\bA)$, we have the following result:
\begin{lemma}\label{lemma_hessian_property}
    Let $\bphi_z = (\bz_1^\T,\cdots,\bz_n^\T)^\T$, $\bphi_a = (\ba_1^\T,\cdots,\ba_q^\T)^\T$, $\bphi = (\bphi_z^\T,\bphi_a^\T)^\T$, and 
    Under Assumption~\ref{assump_psd_covariance}$(ii)$, $(iii)$, $\|\bA^*\|_{2\to\infty}\le M$, Assumptions~\ref{assump_scaling} and~\ref{assump_limit}, there exist constants $\epsilon>0$, $\nu>0$, and $c>0$ such that
\begin{equation}
    \PP\big\{\min_{(\bZ,\bA)\in\Theta^*(\epsilon)}\lambda_{\min}(\bS_{nq}\partial_{\bphi\bphi}^2\cL_{\nu}(\bZ,\bA)\bS_{nq})\ge c\big\}\to 1,\label{eq_eigenvalue_lower_bound}
\end{equation}
as $n,q\to\infty$. Moreover, let $\bH_{Lz_iz_i}(\bZ,\bA) = \partial_{\bz_i\bz_i}^2\cL(\bZ,\bA)$ for $i\in[n]$, $\bH_{La_ja_j}(\bZ,\bA) = \partial_{\bz_j\bz_j}^2\cL(\bZ,\bA)$ for $j\in[q]$, and $\bH_{Lza}(\bZ,\bA) = \partial_{\bphi_z\bphi_a}^2\cL(\bZ,\bA)$. Then for some constant $c,C>0$ with $b_L\min\{\lambda_{\min}(\bSigma_z^*),\lambda_{\min}(\bSigma_{a}^*)\}/2\le c \le C$, 
\begin{align*}
    cq \!\le \! \min_{(\bZ,\bA)\in\Theta^*(\epsilon)}\min_{i\in[n]}\lambda_{\min}\{\bH_{Lz_iz_i}(\bZ,\bA)\}\!\le \!\max_{(\bZ,\bA)\in\Theta^*(\epsilon)}\max_{i\in[n]}\lambda_{\max}\{\bH_{Lz_iz_i}(\bZ,\bA)\}\!\le\!Cq;
\end{align*}
\begin{align*}
    cn\!\le \!  \min_{(\bZ,\bA)\in\Theta^*(\epsilon)}\min_{j\in[q]}\lambda_{\min}\{\bH_{La_ja_j}(\bZ,\bA)\}\!\le\!\max_{(\bZ,\bA)\in\Theta^*(\epsilon)}\max_{j\in[q]}\lambda_{\max}\{\bH_{La_ja_j}(\bZ,\bA)\}\!\le\!Cn.
\end{align*}
Moreover, for any $\delta>0$,  with probability at least $1-(n\vee q)^{-\delta}$,
\begin{equation*}
    \max_{(\bZ,\bA)\in\Theta^*(\epsilon)}\|\bH_{Lza}(\bZ,\bA)\|_{2\to\infty}\!\le\! C_{\delta}\sqrt{q\log n};\max_{(\bZ,\bA)\in\Theta^*(\epsilon)}\|\bH_{Laz}(\bZ,\bA)\|_{2\to\infty}\!\le\! C_{\delta}\sqrt{n\log q}.
\end{equation*}
Here, the constant $C_{\delta}$ depends only on $\delta$.
\end{lemma}

\begin{proof}
See Section~\ref{supp_sec_prove_lemma_hessian_property}.
\end{proof}

We consider the following problem
\begin{equation}
    (\tilde\bZ^o,\tilde\bA^o) = \argmin_{(\bZ,\bA)\in\Theta^*(\epsilon_{nq})}\big\|\bS_{nq}\partial_{\bphi}\cL_{\nu}(\bZ,\bA)\big\|_{\infty}^2,\label{eq_align_problem2}
\end{equation}
for some $\epsilon_{nq}$ satisfying $\log(n\vee q)/\sqrt{n\wedge q} = \delta_{nq}\ll\epsilon_{nq}\ll1$ where $\Theta^*(\epsilon_{nq})$ has been defined as the local region.$\big\{(\bZ,\bA):n^{-1/2}\|\bZ - \bZ^*\|_{\Fn} + q^{-1/2}\|\bA - \bA^*\|_{\Fn}\le \epsilon_{nq}, \|\bZ\|_{2\to\infty}\le M,\|\bA\|_{2\to\infty}\le M\big\}$. 
Next, we bound $n^{-1/2}\|\tilde\bZ^o - \bZ^*\|_{\Fn} + q^{-1/2}\|\tilde\bA^o - \bA^*\|_{\Fn}$ and $\max\{\|\tilde\bZ^o - \bZ^*\|_{2\to\infty} ,\|\tilde\bA^o - \bA^*\|_{2\to\infty}\}$ to show that $(\tilde\bZ^o,\tilde\bA^o)$ is an interior point of \eqref{eq_align_problem2}.

For $(\tilde\bZ^o,\tilde\bA^o) $ with the vector representation $\tilde\bphi^o$, by the integral mean value theorem,
\begin{equation}
    \bS_{nq}\big\{\partial_{\bphi}\cL_{\nu}(\tilde\bZ^o,\tilde\bA^o) - \partial_{\bphi}\cL_{\nu}(\bZ^*,\bA^*)\big\} = \bS_{nq}\big\{\int_{0}^{1}\bH(\bphi^* + s(\tilde\bphi^o-\bphi^*))ds\big\} (\tilde\bphi^o - \bphi^*),\label{eq_integral_expansion}
\end{equation}
where $\bH(\bphi)$ denotes the Hessian matrix $\partial_{\bphi\bphi}\cL_{\nu}(\bZ,\bA)$ with $\bphi$ being the vectorized parameter corresponding to $(\bZ,\bA)$. By \eqref{eq_eigenvalue_lower_bound} in Lemma~\ref{lemma_hessian_property} and $\bphi^*,\tilde\bphi^o\in\Theta^*(\epsilon)$ for any fixed constant $\epsilon>0$, we know
\begin{equation*}
    \lambda_{\min}\Big[\bS_{nq}\big\{\int_{0}^{1}\bH(\bphi^* + s(\tilde\bphi^o-\bphi^*))ds\big\}\bS_{nq}\Big]= \Theta_p(1).
\end{equation*}
Invert the integral in \eqref{eq_integral_expansion} and we obtain
\begin{equation}
    \begin{aligned}
        \big\|\bS_{nq}^{-1}(\tilde\bphi^o - \bphi^*)\big\|\le &\,\Big\|\Big[\bS_{nq}\big\{\int_{0}^{1}\bH(\bphi^* + s(\tilde\bphi^o-\bphi^*))ds\big\}\bS_{nq}\Big]^{-1}\Big\|\\&\quad\times\big\|\bS_{nq}\big\{\partial_{\bphi}\cL_{\nu}(\tilde\bZ^o,\tilde\bA^o) - \partial_{\bphi}\cL_{\nu}(\bZ^*,\bA^*)\big\}\big\|\\{\lesssim} &\,\big\|\bS_{nq}\big\{\partial_{\bphi}\cL_{\nu}(\tilde\bZ^o,\tilde\bA^o) - \partial_{\bphi}\cL_{\nu}(\bZ^*,\bA^*)\big\}\big\|\\\overset{(i)}{\lesssim} &\,2\sqrt{(n+q)r}\big\|\bS_{nq} \partial_{\bphi}\cL_{\nu}(\bZ^*,\bA^*)\big\|_{\infty}\\\overset{(ii)}{\le} &\,O_p(\sqrt{(n+q)\log(n\vee q)}).
    \end{aligned}
\end{equation}
Here, $(i)$ follows from $\big\|\bS_{nq}\big\{\partial_{\bphi}\cL_{\nu}(\tilde\bZ^o,\tilde\bA^o) - \partial_{\bphi}\cL_{\nu}(\bZ^*,\bA^*)\big\}\big\|\le \sqrt{(n+q)r}\big\|\bS_{nq}\partial_{\bphi}\cL_{\nu}(\tilde\bZ^o,\tilde\bA^o)\big\|_{\infty} +\sqrt{(n+q)r}\big\|\bS_{nq} \partial_{\bphi}\cL_{\nu}(\bZ^*,\bA^*)\big\|_{\infty}\le 2\sqrt{(n+q)r}\big\|\bS_{nq} \partial_{\bphi}\cL_{\nu}(\bZ^*,\bA^*)\big\|_{\infty}$ by the definition of \eqref{eq_align_problem2} and that $\bphi^*\in\Theta^*(\epsilon_{nq})$ for any $\epsilon_{nq}>0$; $(ii)$ follows from $(i)$ and $(ii)$ of  Lemma~\ref{lemma_concentration}, which gives $\max_{i\in[n]}q^{-1/2}\|\sum_{j=1}^q\ell_{ij}^{\prime}(\theta_{ij}^*)\ba_j^*\| = O_p(\log^{1/2} n) $ and $\max_{j\in[q]}n^{-1/2}\|\sum_{i=1}^n\ell_{ij}^{\prime}(\theta_{ij}^*)\bz_i^*\| = O_p(\log^{1/2} q) $. Therefore,
\begin{equation}
    n^{-1/2}\big\|\tilde\bZ^o - \bZ^*\big\|_{\Fn} + q^{-1/2}\big\|\tilde\bA^o - \bA^*\big\|_{\Fn}\le 2(nq)^{-1/2}\big\|\bS_{nq}^{-1}(\tilde\bphi^o - \bphi^*)\big\| = O_p\Big(\sqrt{\frac{{\log (n\vee  q)}}{n\wedge q}}\Big).\label{eq_tilde_average_consis}
\end{equation}
For $\max\{\|\tilde\bZ^o - \bZ^*\|_{2\to\infty} ,\|\tilde\bA^o - \bA^*\|_{2\to\infty}\}$, we note that 
\begin{equation*}
    \max\{\|\tilde\bZ^o - \bZ^*\|_{2\to\infty} ,\|\tilde\bA^o - \bA^*\|_{2\to\infty}\}\le \sqrt{r}\|\tilde\bphi^o - \bphi^*\|_{\infty}.
\end{equation*}
To upper bound $\|\tilde\bphi^o - \bphi^*\|_{\infty}$, we re-write the expansion \eqref{eq_integral_expansion} as 
\begin{equation*}
    \tilde\bH_{LD}(\tilde\bphi^o)(\tilde\bphi^o - \bphi^*) + \tilde\bH_{LOD}(\tilde\bphi^o)(\tilde\bphi^o - \bphi^*) + \tilde\bH_P(\tilde\bphi^o)(\tilde\bphi^o - \bphi^*) = \partial_{\bphi}\cL_{\nu}(\tilde\bZ^o,\tilde\bA^o) - \partial_{\bphi}\cL_{\nu}(\bZ^*,\bA^*),
\end{equation*}
or 
\begin{subequations}\label{eq_integral_expansion_inf}\begin{align}
    \bS_{nq}^2\tilde\bH_{LD}(\tilde\bphi^o)(\tilde\bphi^o - \bphi^*) =&\, -\bS_{nq}^2\tilde\bH_{LOD}(\tilde\bphi^o)(\tilde\bphi^o - \bphi^*)\label{eq_integral_expansion_w1}\\&\,- \bS_{nq}^2\tilde\bH_P(\tilde\bphi^o)(\tilde\bphi^o - \bphi^*) \label{eq_integral_expansion_w2}\\&\, + \bS_{nq}^2\big\{\partial_{\bphi}\cL_{\nu}(\tilde\bZ^o,\tilde\bA^o) - \partial_{\bphi}\cL_{\nu}(\bZ^*,\bA^*)\big\},\label{eq_integral_expansion_w3}
\end{align}\end{subequations}
where
\begin{equation*}
    \tilde\bH_{LD}(\bphi) := \int_0^1\begin{pmatrix}
        \bH_{Lzz}(\bphi^* + s(\bphi - \bphi^*)) & \zero \\ \zero & \bH_{Laa}(\bphi^* + s(\bphi - \bphi^*))
    \end{pmatrix}ds;
\end{equation*}
\begin{equation*}
    \tilde\bH_{LOD}(\bphi) := \int_0^1\begin{pmatrix}
        \zero&\bH_{Lza}(\bphi^* + s(\bphi - \bphi^*))  \\  \bH_{Laz}(\bphi^* + s(\bphi - \bphi^*))& \zero
    \end{pmatrix}ds;
\end{equation*}
\begin{equation*}
    \tilde\bH_P(\bphi) := \nu nq\int_0^1\partial_{\bphi\bphi}^2\cP^*(\bphi^* + s(\bphi - \bphi^*))ds.
\end{equation*}
Here, $\bH_{Lzz}(\bphi) = \partial_{\bphi_z\bphi_z}^2\cL(\bphi)$, $\bH_{Laa}(\bphi) = \partial_{\bphi_a\bphi_a}^2\cL(\bphi)$, and $\bH_{Lza}(\bphi) = \{\bH_{Laz}(\bphi)\}^\T = \partial_{\bphi_z\bphi_a}^2\cL(\bphi)$ with the vectorized parameter $\bphi$ representing $(\bZ,\bA)$. Next, we handle \eqref{eq_integral_expansion_w1}--\eqref{eq_integral_expansion_w3}.

\noindent\textbf{For Equation~\eqref{eq_integral_expansion_w1},} by Lemma~\ref{lemma_hessian_property}, we know $\|\bH_{Lza}(\bphi^* + s(\bphi - \bphi^*))\|_{2\to\infty} = O_p(\sqrt{q\log n})$ and $\|\bH_{Laz}(\bphi^* + s(\bphi - \bphi^*))\|_{2\to\infty} = O_p(\sqrt{n\log q})$ uniformly over $\bphi\in\Theta^*(\epsilon_{nq})$ and $s\in[0,1]$, and therefore with \eqref{eq_tilde_average_consis}, one has
\begin{align*}
    \|\eqref{eq_integral_expansion_w1}\|_{\infty}\le &\,O_p\big(\sqrt{\log n/q}\|\tilde\bphi^o_a - \bphi_a^*\| + \sqrt{\log q/n}\|\tilde\bphi^o_z - \bphi_z^*\| \big) 
    \\\le &\, O_p\Big(\sqrt{\log(n\vee q)}\big\{n^{-1/2}\|\tilde\bZ^o - \bZ^*\|_{\Fn} + q^{-1/2}\|\tilde\bA^o - \bA^*\|_{\Fn}\big\}\Big)\\\le &\,O_p\big(\log(n\vee q)/\sqrt{n\wedge q}\big).
\end{align*}

\vspace{0.2cm}

\noindent\textbf{For Equation~\eqref{eq_integral_expansion_w2}, } recall that $\cP^*(\bZ,\bA) = \frac{1}{2}\|n^{-1}\bZ^\T(\bZ - \bZ^*) - q^{-1}(\bA - \bA^*)^\T\bA \big\|_{\Fn}^2$. Then one has
\begin{align}\label{eq_first_order_expand_cp}
    \partial_{\bphi}\cP^*(\bZ,\bA) = \sum_{k=1}^r\sum_{l=1}^r(nq)^{-1/2}\bS_{nq}^{-1}\omega_{kl}(\bphi)\{\bv_{kl}(\bphi) + \bu_{kl}(\bphi)\},
\end{align}
where for each $k,l\in[r]$, $\omega_{kl}(\bphi) = n^{-1}\sum_{i=1}^nz_{ik}(z_{il} - z_{il}^*) - q^{-1}\sum_{j=1}^q(a_{jk} - a_{jk}^*)a_{jl}$, 
\begin{equation}
    \bv_{kl}(\bphi) := \big(n^{-1/2}z_{1k}\be_l^\T,\dots,n^{-1/2}z_{nk}\be_l^\T, -q^{-1/2}a_{1l}\be_k^\T,\dots,-q^{-1/2}a_{ql}\be_k^\T\big)^\T,\label{eq_define_vkl}
\end{equation}
and
\begin{equation}
    \bu_{kl}(\bphi) \!:=\!(n^{-1/2}(z_{1l} - z_{1l}^*)\be_k^\T,\dots,n^{-1/2}(z_{nl} - z_{nl}^*)\be_k^\T, q^{-1/2}( a_{1k}^* - a_{1k} )\be_l^\T,\dots,q^{-1/2}( a_{qk}^* - a_{qk} )\be_l^\T)^\T.\label{eq_define_ukl}
\end{equation}
Then the second-order derivative is computed as 
\begin{equation}\begin{aligned}
    \partial_{\bphi\bphi}^2\cP^*(\bZ,\bA) = &\,\sum_{k=1}^r\sum_{l=1}^r(nq)^{-1}\bS_{nq}^{-1}\{\bv_{kl}(\bphi) + \bu_{kl}(\bphi)\}\{\bv_{kl}(\bphi) + \bu_{kl}(\bphi)\}^\T\bS_{nq}^{-1} \\&\,+ (nq)^{-1}\bS_{nq}^{-1}\sum_{k=1}^r\sum_{l=1}^r\omega_{kl}(\bphi)\bE_{kl}\bS_{nq}^{-1},
    \end{aligned}\label{eq_define_second_cP}
\end{equation}
where 
\begin{equation}
    \bE_{kl} = (nq)^{1/2}\bS_{nq} \partial_{\bphi}\{\bv_{kl}(\bphi) + \bu_{kl}(\bphi)\} = \begin{pmatrix}\bI_n  & \zero \\ \zero&-\bI_q
    \end{pmatrix}\otimes (\be_l\be_k^\T + \be_k\be_l^\T).\label{eq_define_Ekl}
\end{equation}
When $\bphi\in\Theta^*(\epsilon_{nq})$, we know $n^{-1/2}\|\bphi_z - \bphi_z^*\| + q^{-1/2}\|\bphi_a - \bphi_a^*\| \le \epsilon_{nq}$. One can then verify with {Cauchy–Schwarz inequality} that $\|\bu_{kl}(\bphi)\| \lesssim \epsilon_{nq}$ and $|\omega_{kl}(\bphi)|\lesssim \epsilon_{nq}$ uniformly over $k,l\in[r]$. 
Since $\|\bphi\|_{\infty}$ is bounded by $M$ in $\Theta^*(\epsilon_{nq})$, we know $\|\bv_{kl}(\bphi)\| \lesssim 1$, $\|\bS_{nq}\bv_{kl}(\bphi)\|_{\infty} \lesssim (nq)^{-1/2}$ ,and $\|\bS_{nq}\bu_{kl}\|_{\infty}\lesssim (nq)^{-1/2}$. Consequently,
\begin{equation}\label{eq_approximation_p_hes_vv}\begin{aligned}
   &\, \big\|(nq)\bS_{nq}^2\partial_{\bphi\bphi}^2\cP^*(\bZ,\bA)\bS_{nq} - \bS_{nq}\sum_{k=1}^r\sum_{l=1}^r\{\bv_{kl}(\bphi) \}\{\bv_{kl}(\bphi)\}^\T\big\|_{2\to\infty}\\\le &\,\sum_{k=1}^r\sum_{l=1}^r\big\{\|\bS_{nq}\bv_{kl}(\bphi)\|_{\infty}\|\bu_{kl}(\bphi)\| + \|\bv_{kl}(\bphi)\|\|\bS_{nq}\bu_{kl}(\bphi)\|_{\infty} \\&\qquad\qquad+ \|\bS_{nq}\bu_{kl}(\bphi)\|_{\infty}\|\bu_{kl}(\bphi)\| + (nq)^{-1/2}(n\wedge q)^{-1/2}|\omega_{kl}(\bphi)|\big\}\\\lesssim &\,(nq)^{-1/2}\epsilon_{nq} + (nq)^{-1/2} + \epsilon_{nq}(nq)^{-1/2} + (nq)^{-1/2}(n\wedge q)^{-1/2}\epsilon_{nq}\\\lesssim &\,(nq)^{-1/2},
\end{aligned}\end{equation}
uniformly over $\bphi\in\Theta^*(\epsilon_{nq})$. Then this bound also holds when replacing $\partial_{\bphi\bphi}^2\cP^*(\bZ,bA)$ and $\{\bv_{kl}(\bphi) \}\{\bv_{kl}(\bphi)\}^\T$ with their integral from $\bphi^*$ to $\tilde\bphi^o$. Thus, denote $\tilde\bphi^s = \bphi^* + s(\tilde\bphi^o - \bphi^*)$ and we have
\begin{align*}
    \|\eqref{eq_integral_expansion_w2}\|_{\infty} = &\, (nq)^{1/2}\big\|\bS_{nq}^2\tilde\bH_P(\tilde\bphi^o)\bS_{nq}\bS_{nq}^{-1}(nq)^{-1/2}(\tilde\bphi^o - \bphi^*)\big\|_{\infty}\\\le &\,(nq)^{1/2}\big\|\bS_{nq}\int_0^1\{\bv_{kl}(\tilde\bphi^s) \}\{\bv_{kl}(\tilde\bphi^s)\}^\T ds\bS_{nq}^{-1}(nq)^{-1/2}(\tilde\bphi^o - \bphi^*)\big\|_{\infty}\\&\,+(nq)^{1/2}\big\|\int_0^1\big[(nq)\bS_{nq}^2\partial_{\bphi\bphi}^2\cP^*(\tilde\bphi^s)\bS_{nq} - \bS_{nq}\sum_{k=1}^r\sum_{l=1}^r\{\bv_{kl}(\tilde\bphi^s) \}\{\bv_{kl}(\tilde\bphi^s)\}^\T \big]ds\big\|_{2\to\infty}\\&\,\qquad\qquad\times\big\|\bS_{nq}^{-1}(nq)^{-1/2}(\tilde\bphi^o-\bphi^*)\big\|\\=&\,(nq)^{1/2}\big\|\bS_{nq}\int_0^1\sum_{k,l=1}^r\{\bv_{kl}(\tilde\bphi^s) \}\{\bv_{kl}(\tilde\bphi^s)\}^\T ds (nq)^{-1/2}\bS_{nq}^{-1}(\tilde\bphi^o - \bphi^*)\big\|_{\infty}\\&\, +O\big(\big\{n^{-1/2}\|\tilde\bZ^o - \bZ^*\|_{\Fn} + q^{-1/2}\|\tilde\bA^o - \bA^*\|_{\Fn}\big\}\big).
\end{align*}
Since $\bv_{kl}(\bphi)$ is linear with respect to $\bphi$, the integral $\int_0^1\{\bv_{kl}(\tilde\bphi^s) \}\{\bv_{kl}(\tilde\bphi^s)\}^\T ds$ can be calculated as
\begin{align*}
    \int_0^1\{\bv_{kl}(\tilde\bphi^s) \}\{\bv_{kl}(\tilde\bphi^s)\}^\T ds = &\, \{\bv_{kl}(\tilde\bphi^o) +\bv_{kl}(\bphi^*)  \}\{\bv_{kl}(\tilde\bphi^o) +\bv_{kl}(\bphi^*)  \}^\T/4 \\&\,+ \{\bv_{kl}(\tilde\bphi^o) - \bv_{kl}(\bphi^*)  \}\{\bv_{kl}(\tilde\bphi^o) -\bv_{kl}(\bphi^*)  \}^\T/12.
\end{align*}
When $\tilde\bphi^o,\bphi^*\in\Theta^*(\epsilon_{nq})$, one can check that $\|\bS_{nq}\{\bv_{kl}(\tilde\bphi^o) +\bv_{kl}(\bphi^*)  \}\|_{\infty}\lesssim (nq)^{-1/2}$, $\|\bS_{nq}\{\bv_{kl}(\tilde\bphi^o) - \bv_{kl}(\bphi^*)  \}\|_{\infty}\lesssim (nq)^{-1/2}$, $\|\{\bv_{kl}(\tilde\bphi^o) +\bv_{kl}(\bphi^*)  \}\|\lesssim 1$, $\|\{\bv_{kl}(\tilde\bphi^o) - \bv_{kl}(\bphi^*)  \}\|{}\lesssim \epsilon_{nq}$. Thus,
\begin{align*}
    &\,(nq)^{1/2}\big\|\bS_{nq}\int_0^1\sum_{k,l=1}^r\{\bv_{kl}(\tilde\bphi^s) \}\{\bv_{kl}(\tilde\bphi^s)\}^\T ds (nq)^{-1/2}\bS_{nq}^{-1}(\tilde\bphi^o - \bphi^*)\big\|_{\infty} \\\lesssim &\, (nq)^{1/2}\times (nq)^{-1/2}\|(nq)^{-1/2}\bS_{nq}^{-1}(\tilde\bphi^o - \bphi^*)\big\|\\&\,= \big\{n^{-1/2}\|\tilde\bZ^o - \bZ^*\|_{\Fn} + q^{-1/2}\|\tilde\bA^o - \bA^*\|_{\Fn}\big\}.
\end{align*}
Taking these bounds together and with \eqref{eq_tilde_average_consis}, we arrive at
\begin{equation*}
    \|\eqref{eq_integral_expansion_w2}\|_{\infty}= O_p\Big(\sqrt{\frac{{\log(n\vee q)}}{n\wedge q}}\Big).
\end{equation*}

\vspace{0.2cm}

\noindent\textbf{For Equation~\eqref{eq_integral_expansion_w3},} by definition of \eqref{eq_align_problem2} and Lemma~\ref{lemma_concentration_l_mat}, we have
\begin{align*}
    \|\eqref{eq_integral_expansion_w3}\|_{\infty}=&\,\big\|\bS_{nq}^2\big\{\partial_{\bphi}\cL_{\nu}(\tilde\bZ^o,\tilde\bA^o) - \partial_{\bphi}\cL_{\nu}(\bZ^*,\bA^*)\big\}\big\|_{\infty}\\\le&\, \frac{1}{\sqrt{n\wedge q}}\big\|\bS_{nq}\big\{\partial_{\bphi}\cL_{\nu}(\tilde\bZ^o,\tilde\bA^o) - \partial_{\bphi}\cL_{\nu}(\bZ^*,\bA^*)\big\}\big\|_{\infty}\\\le&\, \frac{2}{\sqrt{n\wedge q}}\big\|\bS_{nq}\partial_{\bphi}\cL_{\nu}(\bZ^*,\bA^*)\big\|_{\infty}\\\le &\,O_p\big(\sqrt{\log(n\vee q)/(n\wedge q)}\big)
\end{align*}

For $\bS_{nq}^2\tilde\bH_{LD}(\tilde\bphi^o)(\tilde\bphi^o - \bphi^*)$, by Lemma~\ref{lemma_hessian_property}, we know that 
\begin{equation*}
    \min_{i\in[n+q]}\lambda_{\min}\big[\{\bS_{nq}^2\tilde\bH_{LD}(\tilde\bphi^o)\}_{(i-1)r+1:ir,(i-1)r+1:ir}\big]= \Theta_p(1).
\end{equation*}
By the block-diagonal structure, we conclude that
\begin{equation*}
    \big\|\big\{\bS_{nq}^2\tilde\bH_{LD}(\tilde\bphi^o)\big\}^{-1}\big\|_{\infty} = O_p(1).
\end{equation*}
Invert $\bS_{nq}^2\tilde\bH_{LD}(\tilde\bphi^o)$ in \eqref{eq_integral_expansion_inf} and combine the above bounds to obtain
\begin{align}
    \big\|\tilde\bphi^o - \bphi^*\big\|_{\infty}\le &\,\big\|\big\{\bS_{nq}^2\tilde\bH_{LD}(\tilde\bphi^o)\big\}^{-1}\big\|_{\infty}\big\{\|\eqref{eq_integral_expansion_w1}\|_{\infty} + \|\eqref{eq_integral_expansion_w2}\|_{\infty} + \|\eqref{eq_integral_expansion_w3}\|_{\infty}\big\}\nonumber\\=&\,O_p\big(\log(n\vee q)/\sqrt{n\wedge q}\big).\label{eq_unif_consis_tilde_est}
\end{align}
Without loss of generality, we assume that $M$ is selected large enough such that $\|\bZ^*\|_{2\to\infty}\le M/2$ and $\|\bA^*\|_{2\to\infty}\le M/2$. Then, we know that under Assumption~\ref{assump_scaling}, $\|\tilde\bZ^o\|_{2\to\infty}\le\|\bZ^*\|_{2\to\infty} + \sqrt{r}\|\tilde\bphi^o - \bphi^*\|_{\infty}\le M/2 + o_p(1)$ and $\|\tilde\bA^o\|_{2\to\infty}\le \|\bA^*\|_{2\to\infty} + \sqrt{r}\|\tilde\bphi^o - \bphi^*\|_{\infty}\le M/2 + o_p(1)$. Combine with \eqref{eq_tilde_average_consis}, we know that $\log(n\vee q)/\sqrt{n\wedge  q}=\delta_{nq}\ll\epsilon_{nq}\ll1$ by Assumption~\ref{assump_scaling}, and thus $(\tilde\bZ^o,\tilde\bA^o)$ is an interior point of the feasible set in \eqref{eq_align_problem2} with probability approaching $1$. he Jacobian of $\bphi\mapsto\bS_{nq}\partial_{\bphi}\cL_\nu(\bphi)$ is invertible in this neighbourhood by Lemma~\ref{lemma_hessian_property}. If its value at the interior minimiser were nonzero, the inverse function theorem would provide nearby feasible points at which this vector is multiplied by $1-t$ for small $t>0$, strictly decreasing its squared infinity norm. Consequently,
\begin{equation*}
     \bS_{nq}\partial_{\bphi}\cL_\nu(\tilde\bZ^o,\tilde\bA^o)=\zero.
\end{equation*}
Clearly, we know that $(\tilde\bZ^o,\tilde\bA^o)$ minimises $\cL_{\nu}(\bZ,\bA)$ within $\Theta^*(\epsilon_{nq})$. We claim that $\cP^*(\tilde\bZ^o,\tilde\bA^o) = 0$. Suppose not ($\cP^*(\tilde\bZ^o,\tilde\bA^o)>0$). Note that by \eqref{eq_ave_consistency_Ao} and \eqref{eq_ave_consistency_Zo}, there is
\begin{equation*}
   n^{-1}\|\tilde\bZ^o - \bZ^*\|_{\Fn}^2 + q^{-1}\|\tilde\bA^o - \bA^*\|_{\Fn}^2 = O_p\big(\log(n\vee q)(n\wedge q)^{-1}\big).
\end{equation*}
By Lemma~\ref{lemma_exist_optimal_align}, we know that there exists some $\tilde\bG^o$ such that it is the minimizer of 
\begin{equation*}
    \argmin_{\bG\in\RR^{r\times r}\text{ is invertible}}n^{-1}\big\|\tilde\bZ^o\bG - \bZ^*\big\|_{\Fn}^2 + q^{-1}\big\|\tilde\bA^o\bG^{-\T} - \bA^*\big\|_{\Fn}^2,
\end{equation*}
and that $\|\tilde\bG^o-\bI_r\| = O_p(\log^{1/2}(n\vee q)(n\wedge q)^{-1/2})$. Then Lemma~\ref{lemma_procruste_solution} implies that $\cP^*(\tilde\bZ^o\tilde\bG^o,\tilde\bA^o(\tilde\bG^o)^{-\T}) = 0$. With $\|\tilde\bG^o-\bI_r\| = O_p(\log^{1/2}(n\vee q)(n\wedge q)^{-1/2})$, one can check that $(\tilde\bZ^o\tilde\bG^o,\tilde\bA^o(\tilde\bG^o)^{-\T})\in\Theta^*(\epsilon_{nq})$. But with $\cL(\tilde\bZ^o\tilde\bG^o,\tilde\bA^o(\tilde\bG^o)^{-\T}) = \cL(\tilde\bZ^o,\tilde\bA^o)$ and $\cP^*(\tilde\bZ^o,\tilde\bA^o)>0$, we know that for any $\nu>0$, $\cL_{\nu}(\tilde\bZ^o\tilde\bG^o,\tilde\bA^o(\tilde\bG^o)^{-\T})=\cL(\tilde\bZ^o\tilde\bG^o,\tilde\bA^o(\tilde\bG^o)^{-\T}) < \cL(\tilde\bZ^o,\tilde\bA^o) + \nu nq\cP^*(\tilde\bZ^o,\tilde\bA^o) = \cL_{\nu}(\tilde\bZ^o,\tilde\bA^o)$, which contradicts \eqref{eq_align_problem2}. Therefore, we conclude that, with probability approaching 1,
\begin{equation*}
    \cP^*(\tilde\bZ^o,\tilde\bA^o)=0.
\end{equation*}

Since $(\tilde\bZ^o,\tilde\bA^o)$ lies in the feasible set of \eqref{eq_problem_align}, we have
$\cL(\tilde\bZ^o,\tilde\bA^o)\ge \cL(\hat\bZ^o,\hat\bA^o)$, or equivalently,
\[
\cL_{\nu}(\tilde\bZ^o,\tilde\bA^o)\ge \cL_{\nu}(\hat\bZ^o,\hat\bA^o).
\]
Note that the optimizer $(\hat\bZ^o,\hat\bA^o)$ obtained from \eqref{eq_problem_align} belongs to
$\Theta^*(\epsilon_{nq})$ by \eqref{eq_ave_consistency_Ao} and \eqref{eq_ave_consistency_Zo}. Therefore, the inequality must in fact be an equality. Consequently, $(\hat\bZ^o,\hat\bA^o)$ also solves \eqref{eq_align_problem2} with probability approaching one, that is, $(\hat\bZ^o,\hat\bA^o)$ is asymptotically equivalent to $(\tilde\bZ^o,\tilde\bA^o)$.
We then arrive at the first order condition:
\begin{equation*}
    \PP\{\bS_{nq}\partial_{\bphi}\cL_{\nu}(\hat\bZ^o,\hat\bA^o) = \zero\} \to 1\text{ as }n,q\to\infty.
\end{equation*}
With $\cP^*(\hat\bZ^o,\hat\bA^o) = 0$, invoke \eqref{eq_first_order_expand_cp} to get $\partial_{\ba_j}\cP^*(\bZ,\bA)|_{(\bZ,\bA) = (\hat\bZ^o,\hat\bA^o)} = 0$. Then
\begin{equation}
    \PP\{\bS_{nq}\partial_{\bphi}\cL(\hat\bZ^o,\hat\bA^o) = \zero\} \to 1\text{ as }n,q\to\infty.\label{eq_first_order_condition}
\end{equation}

The equivalence $(\tilde\bZ^o,\tilde\bA^o) = (\hat\bZ^o,\hat\bA^o)$ and \eqref{eq_unif_consis_tilde_est} also imply that $(ii)$ in Lemma~\ref{lemma_consis_start} holds for $(\hat\bZ^o,\hat\bA^o)$. Finally, we show $(iii)$ for $(\hat\bZ^o,\hat\bA^o)$. By the first order condition, we have for each $j\in[q]$ that, with probability approaching $1$, \begin{equation*}\partial_{\ba_j}\cL(\bZ,\bA)|_{(\bZ,\bA) = (\hat\bZ^o,\hat\bA^o)} = \sum_{i=1}^n\ell_{ij}^{\prime}(\hat\theta_{ij}^o)\hat\bz_i^o = 0,\end{equation*}  where $\hat\theta_{ij}^o$ is defined as $(\hat\ba_j^o)^\T\hat\bz_i^o$ for each $i\in[n]$ and $j\in[q]$. With $\|\sum_{i=1}^n\ell_{ij}^{\prime}(\theta_{ij}^*)\bz_i^*\| = O_p(\sqrt{n})$ from $(i)$ of Lemma~\ref{lemma_concentration} and $\max_{i\in[n]}\|\bz_i\|\le \|\bZ\|_{2\to\infty}\le M$, we have
\begin{align}
    \big\|\sum_{i=1}^n\ell_{ij}^{\prime}(\hat\ba_j^o{}^\T\bz_i^*)\bz_i^* - \sum_{i=1}^n\ell_{ij}^{\prime}(\theta_{ij}^*)\bz_i^*\big\|\le &\,\big\|\sum_{i=1}^n\ell_{ij}^{\prime}(\hat\ba_j^o{}^\T\bz_i^*)\bz_i^* - \ell_{ij}^{\prime}(\hat\theta_{ij}^o) \hat\bz_i^o\big\| + \big\|\sum_{i=1}^n\ell_{ij}^{\prime}(\theta_{ij}^*)\bz_i^*\big\|\nonumber\\\le &\, \big\|\sum_{i=1}^n\ell_{ij}^{\prime}(\hat\ba_j^o{}^\T\bz_i^*)\bz_i^* - \ell_{ij}^{\prime}(\hat\theta_{ij}^o) \bz_i^*\big\| \label{eq_expand_for_ind_consis1} \\&\,+ \big\|\sum_{i=1}^n\ell_{ij}^{\prime}(\hat\theta_{ij}^o) \bz_i^* - \ell_{ij}^{\prime}(\hat\theta_{ij}^o) \hat\bz_i^o\big\|\label{eq_expand_for_ind_consis2} \\&\,+O_p(\sqrt{n}).\nonumber
\end{align}
For \eqref{eq_expand_for_ind_consis1}, we have
\begin{align*}
    \eqref{eq_expand_for_ind_consis1} \le \big\|\sum_{i=1}^n\{\max_{|\theta|\le M^2}\ell_{ij}^{\prime\prime}(\theta)\}\bz_i^*(\bz_i^*-\hat\bz_i^o)^\T\hat\ba_j^o\big\|\le b_U M\big\|\sum_{i=1}^n\bz_i^*(\hat\bz_i^o-\bz_i^*)\big\| = O_p\big(n/\sqrt{(n\wedge q)}\big),
\end{align*}
where the inequality follows from the boundedness of $\|\hat\ba_j^o\|\le \|\hat\bA^o\|_{2\to\infty}\le M$ and Assumptions~\ref{assump_limit}; and the last equality follows from {Cauchy–Schwarz inequality} and \eqref{eq_ave_consistency_Zo}. For \eqref{eq_expand_for_ind_consis2}, we have 
\begin{align*}
\eqref{eq_expand_for_ind_consis2} =&\, \big\|\sum_{i=1}^n\ell_{ij}^{\prime}(\hat\theta_{ij}^o) \bz_i^* - \ell_{ij}^{\prime}(\hat\theta_{ij}^o) \hat\bz_i^o\big\|\\\le &\,\Big[\sum_{i=1}^n\big\{\ell_{ij}^{\prime}(\hat\theta_{ij}^o)\big\}^2\Big]^{1/2}\big\{\sum_{i=1}^n\|\hat\bz_i^o - \bz_i^*\|^2\big\}^{1/2}\\\le &\,2\Big[ \sum_{i=1}^n\big\{\ell_{ij}^{\prime}(\hat\theta_{ij}^o) - \ell_{ij}^{\prime}(\theta_{ij}^*)\big\}^2 + \sum_{i=1}^n\big\{\ell_{ij}^{\prime}(\theta_{ij}^*)\big\}^2\Big]^{1/2}\times O_p\big(\sqrt{n/(n\wedge q)}\big)\\\le &\,2\Big[\sum_{i=1}^n\{\max_{|\theta|\le M^2}\ell_{ij}^{\prime\prime}(\theta)\}^2\big\{(\hat\ba_j^o{}^\T\hat\bz_i^o) -(\ba_j^*{}^\T\bz_i^*)\big\}^2+ O_p(n)\Big]^{1/2}\times O_p\big(\sqrt{n/(n\wedge q)}\big)\\= &\,O_p\big(n/\sqrt{n\wedge q}\big).
\end{align*}
The last equality follows from $\|\hat\ba_j^o\|\le M$, $\|\ba_j^*\|\le M$, $\max_{i\in[n]}\|\hat\bz_i^o\|\le M$, and $\max_{i\in[n]}\|\bz_i^*\|\le M$. Therefore, we have
\begin{align*}
    O_p\big((n\wedge q)^{-1/2}\big) \ge&\, \frac{1}{n}\big\|\sum_{i=1}^n\ell_{ij}^{\prime}(\hat\ba_j^o{}^\T\bz_i^*)\bz_i^* - \sum_{i=1}^n\ell_{ij}^{\prime}(\theta_{ij}^*)\bz_i^*\big\|\\=&\,\Big\|\int_0^1\frac{1}{n}\big[\sum_{i=1}^n\ell_{ij}^{\prime\prime}\big(\{\ba_j^* + s(\hat\ba_j^o-\ba_j^*)\}{}^\T\bz_i^*\big)\bz_i^*(\bz_i^*)^\T\big]ds(\hat\ba_j^o - \ba_j^*)\Big\|
\end{align*}
The second equality follows from the integral mean value theorem. Since
\begin{align*}
    \lambda_{\min}\Big(\frac{1}{n}\big[\sum_{i=1}^n\ell_{ij}^{\prime\prime}\big(\{\ba_j^* + s(\hat\ba_j^o-\ba_j^*)\}{}^\T\bz_i^*\big)\bz_i^*(\bz_i^*)^\T\big]\Big)\ge b_L\lambda_{\min}\Big(\frac{1}{n}\sum_{i=1}^n\bz_i^*(\bz_i^*)^\T\Big)\ge b_L\lambda_{\min}(\bSigma_z^*)/2,
\end{align*}
by $(ii)$, $(iii)$ of Assumption~\ref{assump_psd_covariance} and Assumption~\ref{assump_limit} when $n$ is large enough, we can further write
\begin{equation*}
    O_p\big((n\wedge q)^{-1/2}\big)\ge b_L\lambda_{\min}(\bSigma_z^*)/2\|\hat\ba_j^o - \ba_j^*\|,
\end{equation*}
which finally leads to 
\begin{equation}
    \|\hat\ba_j^o - \ba_j^*\| = O_p\big((n\wedge q)^{-1/2}\big).\label{eq_ind_consistency_Ao}
\end{equation}
By symmetry, we can show that for each $i\in[n]$,
\begin{equation}
    \|\hat\bz_i^o - \bz_i^*\| = O_p\big((n\wedge q)^{-1/2}\big).\label{eq_ind_consistency_Zo}
\end{equation}
This proves $(iii)$ in Lemma~\ref{lemma_consis_start} for $(\hat\bZ^o,\hat\bA^o)$

\subsubsection{Step 2: Asymptotic Distributions}\label{supp_sec_prove_ora_o_step2}

Now we prove the asymptotic expansions presented in Lemma~\ref{lemma_asymp_start} hold for $(\hat\bZ^o,\hat\bA^o)$. Expand $\partial_{\bphi}\cL_{\nu}(\cdot,\cdot)$ from $(\hat\bZ^o,\hat\bA^o)$ to $(\bZ^*,\bA^*)$ to get
\begin{equation}
    \partial_{\bphi}\cL_{\nu}(\hat\bZ^o,\hat\bA^o) - \partial_{\bphi}\cL_{\nu}(\bZ^*,\bA^*) = \partial_{\bphi\bphi}^2\cL_{\nu}(\bZ^*,\bA^*)(\hat\bphi^o - \bphi^*) + \bR,\label{eq_asymp_expand_start}
\end{equation} where $\bR$ is the residual term. For $i\le n$ and $l\in[r]$, the $\{(i-1)r+l\}$-th entry is given as $(\hat\bphi^o - \bphi^*)^\T\partial_{\bphi\bphi z_{il}}^3\cL_{\nu}(\tilde\bphi^{\flat})(\hat\bphi^o - \bphi^*)/2$ for some $\tilde\bphi^{\flat}$ on the line segment between $\bphi^*$ and $\hat\bphi^o$. For $n< i\le n+q$ and $l\in[r]$, the $\{(i-1)r+l\}$-th entry is given as $(\hat\bphi^o - \bphi^*)^\T\partial_{\bphi\bphi a_{i-n, l}}^3\cL_{\nu}(\tilde\bphi^{\flat})(\hat\bphi^o - \bphi^*)/2$. We use the vectorized parameter $\tilde\bphi^{\flat}$ to represent $(\tilde\bZ^{\flat},\tilde\bA^{\flat})$. The following two lemmas characterize the Hessian $\partial_{\bphi\bphi}^2\cL_{\nu}(\bZ^*,\bA^*)$ and the residual term $\bR$.

\begin{lemma}\label{lemma_hessian_property_2}
    Let 
    \begin{equation}
        \bH_{LD}(\bZ,\bA) = \begin{pmatrix}
            \partial_{\bphi_z\bphi_z}^2\cL(\bZ,\bA) & \zero \\ \zero & \partial_{\bphi_a\bphi_a}^2\cL(\bZ,\bA)
        \end{pmatrix}.\label{eq_define_h_ld}
    \end{equation}
    Under Assumptions~\ref{assump_psd_covariance}$(ii)$, $(iii)$, $\|\bA^*\|_{2\to\infty}\le M$, and~\ref{assump_limit}, for any $i\in[n+q]$,
    \begin{equation}
        \big\|\big[\{\partial_{\bphi\bphi}\cL_{\nu}(\bZ^*,\bA^*)\}^{-1} - \{\bH_{LD}(\bZ^*,\bA^*)\}^{-1}\big]_{(i-1)r+1:ir}\bS_{nq}^{-1}\big\| = O_p\big((nq)^{-1/2}\big),\label{lemma_hessian_property_2_eq1}
    \end{equation}
    \begin{equation}
        \big\|\big[\{\partial_{\bphi\bphi}\cL_{\nu}(\bZ^*,\bA^*)\}^{-1} - \{\bH_{LD}(\bZ^*,\bA^*)\}^{-1}\big]\bS_{nq}^{-1}\big\|_{2\to\infty} = O_p\big((nq)^{-1/2}{\log^{1/2}(n\vee q)}\big),\label{lemma_hessian_property_2_eq3}
    \end{equation}
\begin{equation}
        \text{ and }\big\|\bS_{nq}^{-1}\big[\{\partial_{\bphi\bphi}\cL_{\nu}(\bZ^*,\bA^*)\}^{-1} - \{\bH_{LD}(\bZ^*,\bA^*)\}^{-1}\big]\bS_{nq}^{-1}\big\| = O_p(1)\label{lemma_hessian_property_2_eq5}
    \end{equation}
    Furthermore, for any $i\in[n+q]$,
    \begin{equation}
        \big\|\big[\{\partial_{\bphi\bphi}\cL_{\nu}(\bZ^*,\bA^*)\}^{-1} - \{\bH_{LD}(\bZ^*,\bA^*)\}^{-1}\big]_{(i-1)r+1:ir}\partial_{\bphi}\cL(\bZ^*,\bA^*)\big\| = O_p\big((n\wedge q)^{-1}\big),\label{lemma_hessian_property_2_eq2}
    \end{equation}
    \begin{equation}
        \big\|\big[\{\partial_{\bphi\bphi}\cL_{\nu}(\bZ^*,\bA^*)\}^{-1} - \{\bH_{LD}(\bZ^*,\bA^*)\}^{-1}\big]\partial_{\bphi}\cL(\bZ^*,\bA^*)\big\|_{\infty} =O_p\big(\log^{1/2}(n\vee q)(n\wedge q)^{-1}\big).\label{lemma_hessian_property_2_eq4}
    \end{equation}
    \begin{equation}
        \text{ and }\big\|\bS_{nq}^{-1}\big[\{\partial_{\bphi\bphi}\cL_{\nu}(\bZ^*,\bA^*)\}^{-1} - \{\bH_{LD}(\bZ^*,\bA^*)\}^{-1}\big]\partial_{\bphi}\cL(\bZ^*,\bA^*)\big\|=O_p\big(\sqrt{nq}/(n\wedge q)\big).\label{lemma_hessian_property_2_eq6}
    \end{equation}
\end{lemma}
\begin{proof}See Section~\ref{supp_sec_prove_lemma_hessian_property_2}\end{proof}

\begin{lemma}
    Under Assumption~\ref{assump_psd_covariance}$(ii)$, $(iii)$, $\|\bA^*\|_{2\to\infty}\le M$, Assumptions~\ref{assump_scaling} and~\ref{assump_limit}, for $\bR$ in \eqref{eq_asymp_expand_start}, we have $\|(\bS_{nq}^2\bR)_{(i-1)r+1:ir}\| = O_p((n\wedge q)^{-1})$ for any $i\in[n+q]$ and $\|\bS_{nq}\bR\| = O_p\big(\sqrt{nq}/(n\wedge q)\big)$.\label{lemma_asymp_expand_res}
\end{lemma}
\begin{proof}
See Section~\ref{supp_sec_prove_lemma_asymp_expand_res}.
\end{proof}

Lemma~\ref{lemma_hessian_property_2} ensures that the second-order derivative matrix in \eqref{eq_asymp_expand_start} is invertible, and further implies that its inverse can be well approximated by a block-diagonal matrix. We now execute this approximation. Under the event in \eqref{eq_first_order_condition}, from Lemma~\ref{lemma_hessian_property_2}, one has
\begin{equation}
    (\hat\bphi^o - \bphi^*) = -\{\partial_{\bphi\bphi}^2\cL_{\nu}(\bZ^*,\bA^*)\}^{-1}\partial_{\bphi}\cL(\bZ^*,\bA^*) - \{\partial_{\bphi\bphi}^2\cL_{\nu}(\bZ^*,\bA^*)\}^{-1}\bR.\label{eq_asymp_expand_start_2}
\end{equation}
Here, we use $\partial_{\bphi}\cP^*(\bZ^*,\bA^*) = \zero$. To establish the expansion \eqref{eq_asympt_expand_z}, we need to bound $\|\bR_z^o\|_{2\to\infty}$ with $\bR_z^o$ defined as
\begin{equation*}
    \bR_z^o := \hat\bZ^o - \bZ^* - \bN_z^*,
\end{equation*}where $\bN_z^*$ is as defined in Lemma~\ref{lemma_asymp_start}. With $\bH_{LD}(\bZ^*,\bA^*)$ defined in Lemma~\ref{lemma_hessian_property_2}, we know $\mathrm{vec}((\bN_z^*)^\T)=-\big[\{\bH_{LD}(\bZ^*,\bA^*)\}^{-1}\cS^*\big]_{1:nr}.$, where $\cS^* = \partial_{\bphi}\cL(\bZ^*,\bA^*)\in\RR^{(n+q)\times r}$ denotes the vectorized score as
\begin{equation*}
    \cS^*_{i,} = \sum_{j=1}^q\ell_{ij}^{\prime}(\theta_{ij}^*)\ba_j^*{}^\T \text{ when }i\in[n],\text{ and }\cS^*_{j+n,} = \sum_{i=1}^n\ell_{ij}^{\prime}(\theta_{ij}^*)\bz_i^*{}^\T \text{ when }j\in[q].
\end{equation*}
With this notation, the difference of the latent variables in \eqref{eq_asymp_expand_start_2} can be rewritten as
\begin{equation*}
        \mathrm{vec}\big((\hat\bZ^o - \bZ^*)^\T\big) = -\big[\{\partial_{\bphi\bphi}^2\cL_{\nu}(\bZ^*,\bA^*)\}^{-1}\partial_{\bphi}\cL(\bZ^*,\bA^*)\big]_{1:nr,} - \big[\{\partial_{\bphi\bphi}^2\cL_{\nu}(\bZ^*,\bA^*)\}^{-1}\bR\big]_{1:nr,}.
\end{equation*}
Now we compute
\begin{align*}
    \|\bR_z^o\|_{2\to\infty} = &\,\max_{i\in[n]}\big\|\hat\bZ^o_{i,} - \bZ^*_{i,} - (\bN_z^*)_{i,}\big\|\\\le &\, \sqrt{r}\Big\|-\big[\{\partial_{\bphi\bphi}^2\cL_{\nu}(\bZ^*,\bA^*)\}^{-1}\partial_{\bphi}\cL(\bZ^*,\bA^*)\big]_{1:nr,} - \big[\{\partial_{\bphi\bphi}^2\cL_{\nu}(\bZ^*,\bA^*)\}^{-1}\bR\big]_{1:nr,}\\&\qquad +[\{\bH_{LD}(\bZ^*,\bA^*)\}^{-1}\partial_{\bphi}\cL(\bZ^*,\bA^*)]_{1:nr,}\Big\|_{\infty}\\\lesssim &\, \big\|\big[\{\partial_{\bphi\bphi}^2\cL_{\nu}(\bZ^*,\bA^*)\}^{-1} - \{\bH_{LD}(\bZ^*,\bA^*)\}^{-1}\big]\partial_{\bphi}\cL(\bZ^*,\bA^*)\big\|_{\infty}\\&\,+\big\|\big[\{\partial_{\bphi\bphi}^2\cL_{\nu}(\bZ^*,\bA^*)\}^{-1} - \{\bH_{LD}(\bZ^*,\bA^*)\}^{-1}\big]\bS_{nq}^{-1}\big\|_{2\to\infty}\big\|\bS_{nq}\bR\big\|\\&\,+\big\|\{\bH_{LD}(\bZ^*,\bA^*)\}^{-1}\bR\big\|_{2\to\infty}\\\overset{(i)}{\le}&\,O_p\Big({\log^{1/2}(n\vee q)}\big\{(n\wedge q)^{-1} + (nq)^{-1/2}\sqrt{nq}/(n\wedge q)\big\}\Big) \\&\,+ \max_{i\in[n]}\big\|\big[\{\bH_{LD}(\bZ^*,\bA^*)\}^{-1}\bR\big]_{(i-1)r+1:ir}\big\|\\\overset{(ii)}{\le} &\,\max_{i\in[n]}q\big\|\{\bH_{Lz_iz_i}(\bZ^*,\bA^*)\}^{-1}\big\|\big\|(\bS_{nq}^2\bR)_{(i-1)r+1:ir}\big\| +O_p\big(\log^{1/2}(n\vee q)/(n\wedge q)\big)\\
    \overset{(iii)}{\le}&\,O_p\big(\log^{1/2}(n\vee q)/(n\wedge q)\big).
\end{align*}
Here, $(i)$ follows from \eqref{lemma_hessian_property_2_eq3} and \eqref{lemma_hessian_property_2_eq4} in Lemma~\ref{lemma_hessian_property_2}; $(ii)$ follows from the block-diagonal structure of $\bH_{LD}(\cdot,\cdot)$; $(iii)$ follows from Lemmas~\ref{lemma_hessian_property} and~\ref{lemma_asymp_expand_res}. For $\|\{\bR_z^o\}_{o,}\|$ when $i\in[n]$, one can repeat the above calculation and invoke \eqref{lemma_hessian_property_2_eq1} and \eqref{lemma_hessian_property_2_eq2} in Lemma~\ref{lemma_hessian_property_2} to get
\begin{equation*}
    \|\{\bR_z^o\}_{i,}\| = O_p(1/(n\wedge q)),\text{ for each }i\in[n].
\end{equation*}
For $\|\bR_z^o\|$, repeat the derivation to obtain
\begin{align*}
    \|\bR_z^o\| \le &\, \Big\|-\big[\{\partial_{\bphi\bphi}^2\cL_{\nu}(\bZ^*,\bA^*)\}^{-1}\partial_{\bphi}\cL(\bZ^*,\bA^*)\big]_{1:nr,} - \big[\{\partial_{\bphi\bphi}^2\cL_{\nu}(\bZ^*,\bA^*)\}^{-1}\bR\big]_{1:nr,}\\&\qquad +[\{\bH_{LD}(\bZ^*,\bA^*)\}^{-1}\cS^*]_{1:nr,}\Big\|\\\le &\, \big\|\big[\{\partial_{\bphi\bphi}^2\cL_{\nu}(\bZ^*,\bA^*)\}^{-1} - \{\bH_{LD}(\bZ^*,\bA^*)\}^{-1}\big]_{1:nr,}\partial_{\bphi}\cL(\bZ^*,\bA^*)\big\|\\&\,+\big\|\big[\{\partial_{\bphi\bphi}^2\cL_{\nu}(\bZ^*,\bA^*)\}^{-1} - \{\bH_{LD}(\bZ^*,\bA^*)\}^{-1}\big]_{1:nr,}\bS_{nq}^{-1}\big\|\big\|\bS_{nq}\bR\big\|\\&\,+\big\|\big[\{\bH_{LD}(\bZ^*,\bA^*)\}^{-1}\big]_{1:nr,}\bR\big\|\\\overset{(i)}{\le}&\,O_p\big(q^{-1/2}\times \sqrt{nq}/(n\wedge q)\big) \\&\,+ \big\|\big[\{\bH_{LD}(\bZ^*,\bA^*)\}^{-1}\big]_{1:nr,}\bR\big\|\\\overset{(ii)}{\le} &\,\big\{q^{1/2}\big\|\{\bH_{Lzz}(\bZ^*,\bA^*)\}^{-1}\big\|\big\}\big\|\bS_{nq}\bR\big\| +O_p\big(\sqrt{n}/(n\wedge q)\big)\\
    \overset{(iii)}{\le}&\,O_p\big(\sqrt{n}/(n\wedge q)\big).
\end{align*}

This establishes \eqref{eq_asympt_expand_z} for $\hat\bZ^o$. Similarly, we can obtain \eqref{eq_asympt_expand}. 

Finally, we prove that $\bn_i^{z*} = (\bN_z^*)_{i,}\overset{{ \rm d}}{\to}\cN(\zero_r,\bar\bPsi_i^{\star})$ for each $i\in[n]$ and $\bn_j^{a*} = (\bN_a^*)_{j,}\overset{{ \rm d}}{\to}\cN(\zero_r,\bar\bPhi_i^{\star})$ for each $j\in[q]$. Since $\ell_{ij}^{\prime}(\theta_{ij}^*)$ are independent across $i,j$, mean zero, uniformly sub-exponential, and $\bz_i^*$ are deterministic and uniformly bounded, the triangular-array Lindeberg condition holds. With $q^{-1}\sum_{j=1}^q\big(\ell_{ij}^{\prime}(\theta_{ij}^*)\big)^2\ba_j^*{\ba_j^*}^\T\to \Omega_{i,z}$, we conclude with the multivariate Lindeberg–Feller CLT that
\begin{equation*}
    q^{-1/2}\sum_{j=1}^q\ell_{ij}^{\prime}(\theta_{ij}^*)\ba_j^*\overset{{ \rm d}}{\to}\cN(\zero_r,\Omega_{i,z})
\end{equation*}
By Slutsky's theorem and the convergence in probability that $n^{-1}\big\{\sum_{i=1}^n\ell_{ij}^{\prime\prime}(\theta_{ij}^*)\bz_i^*{\bz_i^*}^\T\big\}\overset{p}{\to}\bPhi_j^*$, we conclude that $\bn_i^{z*} = (\bN_z^*)_{i,}\overset{{ \rm d}}{\to}\cN(\zero_r,\bar\bPsi_i^{\star})$. Similarly, we can prove the convergence in distribution for $\bn_{j}^{a*}$. The independence between $\bn_i^{z*}$ and $\bn_{i^{\prime}}^{z*}$ for different $i,i^{\prime}\in[n]$ follows naturally from the independence of $\ell_{ij}^{\prime}(\theta_{ij}^*)$ across $i\in[n],j\in[q]$. The independence between $\bn_j^{a*}$ and $\bn_{j^{\prime}}^{a*}$ for different $j,j^{\prime}\in[q]$ follows similarly. For the dependence between $\bn_i^{z*}$ and $\bn_{j}^{a*}$, we note that 
\begin{align*}
    &\,\big\|\EE[\bn_i^{z*}(\bn_{j}^{a*})^\T]\big\| \\ =&\, \sqrt{nq} ~ \Big\|\EE\Big[\big\{\sum_{s=1}^q\ell_{is}^{\prime\prime}(\theta_{is}^*)\ba_s^*\ba_s^*{}^\T\big\}^{-1}\big\{\sum_{s=1}^q\ell_{is}^{\prime}(\theta_{is}^*)\ba_s^*\big\}\big\{\sum_{t=1}^n\ell_{tj}^{\prime}(\theta_{tj}^*)\bz_t^*{}^\T\big\}\big\{\sum_{t=1}^n\ell_{tj}^{\prime\prime}(\theta_{tj}^*)\bz_t^*{\bz_t^*}^\T\big\}^{-1}\Big]\Big\|\\ \overset{(i)}{\le} &\, (nq)^{-1/2} ~ \EE\big[\big\{\sum_{s=1}^q\ell_{is}^{\prime}(\theta_{is}^*)\ba_s^*\big\}\big\{\sum_{t=1}^n\ell_{tj}^{\prime}(\theta_{tj}^*)\bz_t^*{}^\T\big\}]\\ \overset{(ii)}{\le} &\, (nq)^{-1/2}M^2 ~ \EE\big[\{\ell_{ij}^{\prime}(\theta_{ij})\}^2]\\ = &\,o\big((nq)^{-1/2}\big).
\end{align*}
uniformly across $i\in[n]$ and $j\in[q]$. Here $(i)$ follows from Lemma~\ref{lemma_hessian_property}, $(ii)$ follows from the independence of $\ell_{ij}^{\prime}(\theta_{ij}^*)$ across $i\in[n],j\in[q]$, $\max_{i\in[n]}\|\bz_i^*\|\le M$, and $\max_{j\in[a]}\|\ba_j^*\|\le M$.

\subsubsection{Step 3: Equivalence}\label{supp_sec_prove_ora_o_step3}
Note that $(\hat\bZ,\hat\bA)\in\Theta^*(\epsilon_{nq})$ by definition with $\epsilon_{nq}\ll (n\wedge q)^{-\epsilon}$. Recall that Lemma~\ref{lemma_exist_g_star} guarantees the existence of $\hat\bG^*$ given in \eqref{eq_opt_align}.
With $(\hat\bZ^*, \hat\bA^*) = (\hat\bZ(\hat\bG^*)^\T,\hat\bZ\hat\bG^*{}^{-\T})$, invoke Lemma~\ref{lemma_procruste_solution} and we know
\begin{equation*}
    n^{-1}(\hat\bZ^*)^\T(\hat\bZ^* - \bZ^*) = q^{-1}\{\hat\bA^* - \bA^*\}^\T\hat\bA^*,
\end{equation*}
i.e., $\cP^*(\hat\bZ^*,\hat\bA^*) = 0$. Together with $(\hat\bZ^*,\hat\bA^*)\in\Theta^*(\epsilon)$ and $\|\hat\bG^* - \bI_r\| = O_p(\epsilon_{nq})$, one can check that $(\hat\bZ^*,\hat\bA^*)$ is in the feasible set of \eqref{eq_problem_align} with probability approaching $1$. Specifically, for the constraints in $\Theta=\big\{(\bZ,\bA):\|\bZ\|_{2\to\infty}\le M,\|\bA\|_{2\to\infty}\le M\big\}$, one can slightly enlarge the constant $M$ here to ensure the rotated solution $(\hat\bZ^*,\hat\bA^*)\in\Theta$. And for the constraints on the eigenvalues, by Weyl's inequality, one has
    \begin{equation*}\begin{aligned}
        \lambda_{\min}\big(n^{-1}\hat\bZ^*{}^\T\hat\bZ^* \big)\ge &\, \lambda_{\min}\big(n^{-1}\bZ^*{}^\T\bZ^* \big) - n^{-1}\big\|\hat\bZ^*{}^\T\hat\bZ^* - \bZ^*{}^\T\bZ^*\big\|\\\ge &\,\lambda_{\min}\big(n^{-1}\bZ^*{}^\T\bZ^* \big) - 2n^{-1}\big\|\hat\bZ^*{}^\T(\hat\bZ^*-\bZ^*)\big\| - n^{-1}\big\|(\hat\bZ^*-\bZ^*)^\T(\hat\bZ^*-\bZ^*)\big\|\\\ge &\,3\lambda_{\min}(\bSigma^*_z)/4 - O_p(\epsilon_{nq}),
    \end{aligned}\end{equation*}when $n$ is large enough. Here, the last inequality follows from $(ii)$, $(iii)$ of Assumption~\ref{assump_psd_covariance}, {Cauchy–Schwarz inequality}, and $n^{-1/2}\|\hat\bZ^* - \bZ^*\|_{\Fn}\le n^{-1/2}\|\hat\bZ - \bZ^*\|_{\Fn} + n^{-1/2}\|\hat\bZ^*\|\|\hat\bG^* - \bI_r\|_{\Fn} = O_p(\epsilon_{nq})$. This implies that $ \lambda_{\min}\big(n^{-1}\hat\bZ^*{}^\T\hat\bZ^* \big)\ge c$ with probability approaching $1$ when $c < \min\{\lambda_{\min}(\bSigma_a^*),\lambda_{\min}(\bSigma_z^*)\}/2$. Similarly, one can verify that $ \lambda_{\min}\big(q^{-1}\hat\bA^*{}^\T\hat\bA^* \big)\ge c$ with probability approaching $1$.

Since $(\hat\bZ^o,\hat\bA^o)$ is the solution of \eqref{eq_problem_align}, we know
\begin{equation*}
    \cL(\hat\bZ^o,\hat\bA^o)\le \cL(\hat\bZ^*,\hat\bA^*).
\end{equation*}
On the other side, since $(\hat\bZ^o,\hat\bA^o)\in\Theta^*(\epsilon_{nq})$ for $\epsilon_{nq}\gg \delta_{nq}$ by Step 1, we also have
\begin{equation*}
    \cL(\hat\bZ^*,\hat\bA^*)\le \cL(\hat\bZ^o,\hat\bA^o).
\end{equation*}
Together we know $\cL(\hat\bZ^*,\hat\bA^*)=\cL(\hat\bZ^o,\hat\bA^o)$, and the uniqueness of $(\hat\bZ^o,\hat\bA^o)$ implies that $(\hat\bZ^*,\hat\bA^*) = (\hat\bZ^o,\hat\bA^o)$ with probability approaching $1$. Taking the results established in Sections~\ref{supp_sec_prove_ora_o_step1} and~\ref{supp_sec_prove_ora_o_step2}, we finish the proof.

\subsection{Proof of Lemma~\ref{lemma_oracle_folomin_solu}}\label{sec_prove_lemma_oracle_folomin_solu}
First, for any $\bG$ with $\|\bG - \bI_r\|\le \epsilon_{nq}^{\prime}\ll \gamma \wedge  (n\wedge q)^{-c}\le O(1)$, we have
\begin{equation*}
    \|\bG^{-1} - \bI_r\|\le \|\bG - \bI_r\|\|\bG^{-1}\| \le \|\bG - \bI_r\|\|\bG^{-1} - \bI_r\| + \|\bG - \bI_r\|.
\end{equation*}
Therefore, when $n$ is large enough,
\begin{equation*}
    \|\bG^{-1} - \bI_r\|\le\frac{\|\bG - \bI_r\|}{1 - \|\bG - \bI_r\|}\le 2\epsilon_{nq}^{\prime},
\end{equation*}

In the following, we calculate a lower bound for $Q_{\rho_{\gamma}}(\hat\bA^*\bG^{-1}) - Q_{\rho_{\gamma}}(\hat\bA^*)$, and show that in order for this lower bound to be negative, we must have $\|\bG - \bI_r\| = O_p\big(\sigma_q(n\wedge q)^{-1+\varepsilon}\big)$. This immediately implies that, for $\hat\bG_*$ to be the optimum of \eqref{eq_theory_solo_align}, it must hold that $\|\hat\bG_* - \bI_r\| = O_p\big(\sigma_q(n\wedge q)^{-1+\varepsilon}\big)$. 

Let $\Omega_0 = \{(j,k): a_{jk}^* = 0\}$ and $\Omega_1 = \{(j,k): a_{jk}^* \neq 0\}$. Then, we can write $Q_{\rho_{\gamma}}(\hat\bA^*\bG^{-1}) - Q_{\rho_{\gamma}}(\hat\bA^*)$ as
\begin{align}
    &\,Q_{\rho_{\gamma}}(\hat\bA^*\bG^{-1}) - Q_{\rho_{\gamma}}(\hat\bA^*) \nonumber\\ = &\,\sum_{(j,k)\in\Omega_0}\Big[\rho_{\gamma}\left((\bA^*\bG^{-1})_{j,k} + (\hat\bA^* - \bA^*)_{j,k} + \big\{(\hat\bA^* - \bA^*)(\bG^{-1}-\bI_r)\big\}_{j,k} \right) \nonumber\\&\qquad- \rho_{\gamma}\big\{(\hat\bA^* - \bA^*)_{j,k}\big\}\Big]\label{eq_bound_diff_Q1_t1}\\&\, +\sum_{(j,k)\in\Omega_1}\Big[\rho_{\gamma}\left((\bA^*\bG^{-1})_{j,k} + (\hat\bA^* - \bA^*)_{j,k} + \big\{(\hat\bA^* - \bA^*)(\bG^{-1}-\bI_r)\big\}_{j,k} \right) \nonumber\\&\,\qquad\qquad- \rho_{\gamma}\big\{(\hat\bA^* - \bA^* + \bA^*)_{j,k}\big\}\Big].\label{eq_bound_diff_Q1_t2}
\end{align}
When $(j,k)\in\Omega_1$, with $\min\{| a_{jk}^*|: a_{jk}^*\neq0\}\ge\lambda_0 \ge \gamma(a_3+1)$ and $\|\bG - \bI_r\|\le \epsilon_{nq}^{\prime}$, one has
\begin{equation*}
    \big|\big(\bA^*\bG^{-1}\big)_{j,k}\big| \ge\gamma(a_3+1)-\big|(\ba^*_j)^\T(\bG^{-1} - \bI_r)_{,k}\big|\ge \gamma(a_3+1) - 2M{\epsilon_{nq}}^{\prime},
\end{equation*}
and 
\begin{equation*}
    \big|(\hat\bA^* - \bA^* + \bA^*)_{j,k}\big| \ge \gamma(a_3+1) - \big|\hat a_{jk}^* -  a_{jk}^*\big|,
\end{equation*}
uniformly over $(j,k)\in\Omega_1$. By Lemma~\ref{lemma_consis_start}, we know $\max_{j\in[q],k\in[r]}|\hat a_{jk} -  a_{jk}^*| = O_p\big(\log(n\vee q)(n\wedge q)^{-1/2}\big)  = O_p(\delta_{nq})\ll\gamma$. Together with $\epsilon_{nq}^{\prime}\ll \gamma$ and $\gamma \gg \delta_{nq}\gg n^{-1/2} $, we know, with probability approaching 1, $\big|(\bA^*\bG^{-1})_{j,k} + (\hat\bA^* - \bA^*)_{j,k} + \big\{(\hat\bA^* - \bA^*)(\bG^{-1}-\bI_r)\big\}_{j,k}\big|\ge \gamma a_3$ and $\big|(\hat\bA^* - \bA^* + \bA^*)_{j,k}\big|\ge \gamma a_3$ uniformly over $(j,k)\in\Omega_1$. We then conclude with \ref{item_concave} of Condition~\ref{assump_concave_function} that the second summation \eqref{eq_bound_diff_Q1_t2} is zero, with probability approaching 1 as $n,q\to\infty$.

Now we deal with \eqref{eq_bound_diff_Q1_t1}. Invoke Lemma~\ref{lemma_asymp_start} and we know \begin{equation}\hat a_{jk}^* -  a_{jk}^* = n^{-1/2}{\xi}_{jk} + r^{a*}_{jk}\label{eq_expand_lambda_jr}\end{equation}with $|r_{jk}^{a*}| := |(\bR_a^{*})_{j,k}| = O_p\big((n\wedge q)^{-1}\big)$ and 
\begin{equation}
   \xi_{jk}= (\bn_{j}^{a*})_{k} = -n^{1/2}\be_k^\T\big\{\sum_{i=1}^n\ell_{ij}^{\prime\prime}(\theta_{ij}^*)\bz_i^*{\bz_i^*}^\T\big\}^{-1}\sum_{i=1}^n\ell_{ij}^{\prime}(\theta_{ij}^*)\bz_i^*.\label{eq_z_jr_expand}
\end{equation}
For $\xi_{jk}$, we first note that 
\begin{equation}\begin{aligned}
    \max_{j\in[q]}\big\|\big\{\sum_{i=1}^n\ell_{ij}^{\prime\prime}(\theta_{ij}^*)\bz_i^*{\bz_i^*}^\T\big\}^{-1}\big\| = &\,\max_{j\in[q]}\Big[\lambda_{\min}\big\{\sum_{i=1}^n\ell_{ij}^{\prime\prime}(\theta_{ij}^*)\bz_i^*{\bz_i^*}^\T\big\}\Big]^{-1}\\\le &\,b_L^{-1}n^{-1}\max_{j\in[q]}\Big[\lambda_{\min}\big\{n^{-1}\sum_{i=1}^n\bz_i^*{\bz_i^*}^\T\big\}\Big]^{-1}\\\le &\,2b_L^{-1}n^{-1}\lambda_{\min}(\bSigma_z^*)^{-1},
\end{aligned}\label{eq_max_sub_e_sum}\end{equation}
where $n$ is large enough. Here, the first inequality follows from Assumption~\ref{assump_limit} and the second inequality follows from $(ii)$, $(iii)$ of Assumption~\ref{assump_psd_covariance}.
Since $\ell_{ij}^{\prime}(\theta_{ij}^*)$ is sub-exponential and independent across $i\in[n], j\in[q]$, by Bernstein inequality and $\max_{i\in[n]}\|\bz_i^*\|= \|\bZ^*\|_{2\to\infty}\le M$, we know $\max_{j\in[q]}\|\sum_{i=1}^n\ell_{ij}^{\prime}(\theta_{ij}^*)\bz_i^*\| = O_p(\sqrt{n\log q})$, and therefore,
\begin{equation}
    \max_{j\in[q],k\in[r]}|\xi_{jk}| = O_p(\sqrt{\log q})\le O_p( \delta_{nq}\sqrt{n}).\label{eq_max_sub_e}
\end{equation}
Plug the expansion \eqref{eq_expand_lambda_jr} into \eqref{eq_bound_diff_Q1_t1} and one has
\begin{align*}
    \eqref{eq_bound_diff_Q1_t1} = &\,\sum_{(j,k)\in\Omega_0}\rho_{\gamma}\left((\bA^*\bG^{-1})_{j,k} + (\hat\bA^* - \bA^*)_{j,k} + \big\{(\hat\bA^* - \bA^*)(\bG^{-1}-\bI_r)\big\}_{j,k} \right) \\&\,\qquad\quad- \rho_{\gamma}\big\{(\hat\bA^* - \bA^*)_{j,k}\big\}\\
    \overset{(i)}{\ge} &\,\sum_{(j,k)\in\Omega_0}{\rho_{\gamma}\left((\ba_{j}^*)^\T(\bG^{-1})_{,k} + \hat a_{jk}^* -  a_{jk}^* \right)- \rho_{\gamma}\big(\hat a_{jk}^* -  a_{jk}^*\big)}\\&\, - O_p\big(\gamma \epsilon_{nq}^{\prime} q/\sqrt{n\wedge q}\big)\\\overset{(ii)}{\ge} &\,\sum_{(j,k)\in\Omega_0}\underbrace{\Big\{\rho_{\gamma}\left((\bG^{-\T})_{k,}\ba_{j}^* + n^{-1/2}\xi_{jk} \right)- \rho_{\gamma}\big(n^{-1/2}\xi_{jk}\big)\Big\}}_{\delta_{jk}}\\&\, - O_p\Big(\frac{\gamma \epsilon_{nq}^{\prime} q}{\sqrt{n\wedge q}} + \frac{\gamma q}{n\wedge q}\Big).
\end{align*}
We explain inequalities $(i)$ and $(ii)$ as follows.\\
\noindent $(i):$ Note that $\|\hat\bA^* - \bA^*\|_{2\to\infty} = O_p(\delta_{nq})$ by Lemma~\ref{lemma_consis_start}, and $\|\bG^{-1} - \bI_r\| \le 2\epsilon_{nq}^{\prime} $. Then $\max_{j\in[q],k\in[r]}\big|\big\{(\hat\bA^* - \bA^*)(\bG^{-1}-\bI_r)\big\}_{j,k}\big| = O_p(\epsilon_{nq}^{\prime} \delta_{nq})$ with $\epsilon_{nq}^{\prime} \delta_{nq}\ll \gamma$. Existence of $\rho_{\gamma}(0+)$ and $\rho_{\gamma}^{\prime}(x)\ge a_0\gamma$ when $x\in(0,a_1\gamma]$ from Condition~\ref{assump_concave_function} implies $|\rho_{\gamma}(x) - \rho_{\gamma}(y)|\le  a_0\gamma|x-y|$. Therefore, with $\sum_{j=1}^q\|\hat\ba_j^* - \ba_j^*\|^2 = O_p\big(q(n\wedge q)^{-1}\big)$ by Lemma~\ref{lemma_consis_start}, we obtain \begin{align*}
    &\,\sum_{(j,k)\in\Omega_0}\rho_{\gamma}\left((\bA^*\bG^{-1})_{j,k} + (\hat\bA^* - \bA^*)_{j,k} + \big\{(\hat\bA^* - \bA^*)(\bG^{-1}-\bI_r)\big\}_{j,k} \right)\\&\, - \sum_{(j,k)\in\Omega_0}\rho_{\gamma}\left((\bA^*\bG^{-1})_{j,k} + (\hat\bA^* - \bA^*)_{j,k}\right)\\\ge &\,-\sum_{(j,k)\in\Omega_0}a_0\gamma \big|\big\{(\hat\bA^* - \bA^*)(\bG^{-1}-\bI_r)\big\}_{j,k}\big|\\\ge &\,-a_0\gamma \|\hat\bA^* - \bA^*\|_{\Fn}\|\bG^{-1}-\bI_r\| = -O_p(\gamma (n\wedge q)^{-1/2}\epsilon_{nq}^{\prime}) 
\end{align*}\\
\noindent $(ii):$ This part can be derived similarly to $(i)$ with expansion \eqref{eq_expand_lambda_jr}, $\max_{j\in[q],k\in[r]}|r_{jk}^{a}| = O_p(\log^{1/2}(n\vee q)/(n\wedge q))$ and $\sum_{j=1}^q\sum_{k=1}^r|r_{jk}^{a}|^2 =  O_p\big(q(n\wedge q)^{-2}\big)$ from Lemma~\ref{lemma_asymp_start}.


Define for $(j,k)\in\Omega_0$ that\begin{equation}\mu_{jk} := -\sqrt{n}(\bG^{-\T})_{k,}\ba_{j}^* = -\sqrt{n}(\bG^{-\T} - \bI_r)_{k,}\ba_{j}^*.\label{eq_mu_jr_expand}\end{equation} 
Since $\|\bG^{-\T} - \bI_r\|\le 2\epsilon_{nq} ^{\prime}$, we conclude that  with $\max_{j\in[q]}\|\ba_j^*\|\le M$ that
\begin{equation}
    \max_{j\in[q],k\in[r]}|\mu_{jk}|\le 2M\sqrt{n}\epsilon_{nq}^{\prime} .\label{eq_max_sub_mu}
\end{equation}
By Condition~\ref{assump_concave_function}, we know that, when $0 < x\le a_1\gamma$, there is
\begin{equation*}
    (a_0\gamma-a_2\gamma|x|)\le \rho^{\prime}(x)\le a_0\gamma
\end{equation*}
For $\delta_{jk} = \rho_{\gamma}(|\xi_{jk} - \mu_{jk}|/\sqrt{n}) - \rho_{\gamma}(|\xi_{jk}|/\sqrt{n})$, note that $\max_{j\in[q],k\in[r]}|\xi_{jk} - \mu_{jk}|/\sqrt{n}\le O_p(\delta_{nq} + 2M\epsilon^{\prime}_{nq})$ by \eqref{eq_max_sub_e} and \eqref{eq_max_sub_mu} with $\delta_{nq}\vee \epsilon^{\prime}_{nq}\ll a_1\gamma$ , and $\max_{j\in[q],k\in[r]}|\xi_{jk} |/\sqrt{n}\le O_p(\delta_{nq})$ by \eqref{eq_max_sub_e} with $\delta_{nq}\ll a_1\gamma$. Therefore,
with probability approaching $1$, $\delta_{jk}$ can be expressed as
\begin{align*}
    \delta_{jk} = &\, \rho_{\gamma}\big(|\xi_{jk} - \mu_{jk}|/\sqrt{n}\big) - \rho_{\gamma}\big(|\xi_{jk}|/\sqrt{n}\big) \\\ge&\,(a_0\gamma - a_2\gamma|\xi_{jk} - \mu_{jk}|/\sqrt{n}) |\xi_{jk} - \mu_{jk}|/\sqrt{n} - a_0\gamma |\xi_{jk}|/\sqrt{n}\\ \ge &\,\frac{a_0\gamma}{\sqrt{n}}\underbrace{\big(|\xi_{jk} - \mu_{jk}| - |\xi_{jk}|\big)}_{\delta_{jk}^L} - 2a_2\gamma(|\xi_{jk}|^2/{n} + 4M^2(\epsilon_{nq}^{\prime}) ^2),
\end{align*}uniformly over $j\in[q]$ and $k\in[r]$.
Here, the second inequality is due to $\max_{j\in[q],k\in[r]}|\mu_{jk}|\le 2M\sqrt{n}\epsilon_{nq}^{\prime} $ and {Cauchy–Schwarz inequality}. This shows that $\delta_{jk}$ can be lower bounded by the linearized difference $\delta_{jk}^L$ up to a small order term.

Note that with \eqref{eq_z_jr_expand},
\begin{align*}
    \sum_{(j,k)\in\Omega_0}|\xi_{jk}|^2 \le &\,\sum_{j=1}^q\sum_{k=1}^r|\xi_{jk}|^2\\=&\,n\sum_{j=1}^q\Big\|\big\{\sum_{i=1}^n\ell_{ij}^{\prime\prime}(\theta_{ij}^*)\bz_i^*{\bz_i^*}^\T\big\}^{-1}\sum_{i=1}^n\ell_{ij}^{\prime}(\theta_{ij}^*)\bz_i^*\Big\|^2\\\le&\,n\max_{j\in[q]}\Big[\lambda_{\min}\big\{\sum_{i=1}^n\ell_{ij}^{\prime\prime}(\theta_{ij}^*)\bz_i^*{\bz_i^*}^\T\big\}\Big]^{-2}\sum_{j=1}^q\Big\|\sum_{i=1}^n\ell_{ij}^{\prime}(\btheta^*)\bz_i^*\Big\|^2\\= &\,O_p(q).
\end{align*}
Here, the last inequality follows from \eqref{eq_max_sub_e_sum}. 
Then we arrive at
\begin{equation}
    \eqref{eq_bound_diff_Q1_t1}\ge \frac{a_0\gamma}{\sqrt{n}}\sum_{k\in[r]}\sum_{j\in\cT_k^*}\delta_{jk}^L - O_p\Big(q(\epsilon_{nq}^{\prime}) ^2 \gamma + \frac{q\gamma }{{n}}+\frac{\gamma \epsilon_{nq}^{\prime} q}{\sqrt{n\wedge q}} + \frac{q\gamma }{n\wedge q}\Big).\label{eq_bound_diff_Q1_t1_t1}
\end{equation} Here, $\cT_k^*$ has been defined as $\{j:a_{jk}^* = 0\}$.

Next, we focus on deriving the lower bound for the first term. We first introduce the following two technical lemmas.

\begin{lemma}\it\label{lemma_unif_sub_exp}
    Under Assumptions~\ref{assump_psd_covariance_prime} and~\ref{assump_limit}, for $\xi_{jk}$ given in \eqref{eq_z_jr_expand} and $\mu_{jk}$ given in \eqref{eq_mu_jr_expand}, $\bar\delta_{jk}^L := |\xi_{jk} - \mu_{jk}| - \EE\big[|\xi_{jk} - \mu_{jk}|\big]$ is sub-exponential, with sub-exponential norm bounded as {$\big\||\xi_{jk} - \mu_{jk}| - \EE\big[|\xi_{jk} - \mu_{jk}|\big]\big\|_{\psi_1}\le C$} for some constant $C$ uniformly over $j,k,\mu_{jk}$.
\end{lemma}

\begin{proof}
See Section~\ref{sec_prove_lemma_unif_sub_exp}.
\end{proof}

\begin{lemma}\it
    Suppose $\xi \sim \cN(0,\phi^2)$. For any $a\in\RR$, one has
    \begin{equation*}
        \EE[|\xi-a| - |\xi|]\ge \min\big\{c_1|a|^2 , c_2|a|\big\}
    \end{equation*}
    for some constants $c_1,c_2>0$ that depend on $\phi$.\label{lemma_fold_nor_dif}
\end{lemma}
\begin{proof}
See Section~\ref{sec_prove_lemma_fold_nor_dif}.
\end{proof}
By the definition of $\xi_{jk}$ and independence of $\ell_{ij}^{\prime}(\theta_{ij}^*)$ across $i\in[n],j\in[q]$, we know that, for any fixed $k\in[r]$, $\delta_{jk}^L$s are independent across $j\in[q]$. Then
Lemma~\ref{lemma_unif_sub_exp} allows us to apply the Bernstein bound to obtain that, for any fixed $\bG$ and $k\in[r]$,
\begin{align*}
    \PP\Big(\Big|\sum_{j\in\cT_k}\delta_{jk}^L - \EE\big[\sum_{j\in\cT_k}\delta_{jk}^L\big]\Big|\ge t\Big)\le 2\exp\left(-c\min\big\{\frac{t^2}{|\cT_k^*|C^2},\frac{t}{C}\big\}\right)
\end{align*}
for some absolute constant $c$ with $C$ given in Lemma~\ref{lemma_unif_sub_exp}. Take $t = \sqrt{Cq/c}$ and we get $\big|\sum_{j\in\cT_k}\delta_{jk}^L - \EE\big[\sum_{j\in\cT_k}\delta_{jk}^L\big]\big| =O_p(\sqrt{q})$ Since the rank $r$ is fixed, we conclude that
\begin{equation*}
    \sum_{(j,k)\in\Omega_0}\delta_{jk}^L \ge  \EE\big[\sum_{(j,k)\in\Omega_0}\delta_{jk}^L\big]-O_p(\sqrt{q}).
\end{equation*}
To further show uniformity over feasible $\bG$, one can first subtract the deterministic term $|\mu_{jk}|$ to get the class
$|x-\ba_j^*{}^\T\bbeta|-|\ba_j^*{}^\T\bbeta|-|x|$, with envelope $2|x|$ and polynomial covering numbers. By symmetrising this class and integrating the covering bound, one can show that the above holds uniformly.
Now we seek to lower bound $\EE[\delta_{jk}^L] = \EE[|\xi_{jk} - \mu_{jk}| - |\xi_{jk}|]$. First, since \begin{equation*}\big\|\{\sum_{i=1}^n\ell_{ij}^{\prime\prime}(\theta_{ij}^*)\bz_i^*(\bz_i^*)^\T\}^{-1}\big\| \le \min_{i\in[n],j\in[q]}b_L^{-1}\big[\lambda_{\min}\big\{\sum_{i=1}^n\bz_i^*(\bz_i^*)^\T\big\}\big]^{-1} = O(n^{-1})\end{equation*} uniformly over $j\in[q]$ and $\ell_{ij}^{\prime}(\theta_{ij}^*)\bz_i $ is sub-exponential with zero mean, we know $|\xi_{jk}|$ has uniformly bounded $L^p$ norm for $j\in[q]$ and $k\in[r]$. Let $\tilde\xi_{jk}=-\be_k^\T\big\{\EE[n^{-1}\sum_{i=1}^n\ell_{ij}^{\prime\prime}(\theta_{ij}^*)\bz_i^*(\bz_i^*)^\T]\big\}^{-1}n^{-1/2}\sum_{i=1}^n\ell_{ij}^{\prime}(\theta_{ij}^*)\bz_i^*$. Then $\EE|\tilde\xi_{jk} - \xi_{jk}| = O(n^{-1/2})$ uniformly over $j\in[q]$ and $k\in[r]$. Let $N_{jk}\sim\cN(0, \phi_{jk}^*)$ with $\phi_{jk}^* :={\rm Var}(\tilde\xi_{jk})$. By the proof of Lemma~\ref{lemma_asymp_start} in Section~\ref{supp_sec_prove_ora_o_step2}, we know the limiting distribution of $\tilde\xi_{jk}$ is $N_{jk}$ for all $j\in[q]$ and $k\in[r]$. Invoke the Berry-Esseen theorem~\citep{stein1986approximate}, we know 
\begin{equation*}
    \big|\EE[h(\xi_{jk})] - \EE[h(N_{jk})]\big|\le \frac{CL(h)}{\sqrt{n}},
\end{equation*}uniformly over $j,k$ for some absolute constant $C$ and Lipschitz function $h$ with Lipschitz constant $L(h)$. Therefore, 
\begin{equation*}
    \EE\big[\big|\xi_{jk} - \mu_{jk}\big| - \big|\xi_{jk}\big|\big]\ge \EE\big[\big|N_{jk} - \mu_{jk}\big| - \big|N_{jk}\big|\big] - 2C/\sqrt{n}.
\end{equation*}
Now apply Lemma~\ref{lemma_fold_nor_dif} and we have 
\begin{equation*}
    \EE\big[\sum_{(j,k)\in\Omega_0}\delta_{jk}^L\big]\ge \sum_{(j,k)\in\Omega_0}\Big[\min\big\{c_1\mu_{jk}^2, c_2|\mu_{jk}|\big\} - 2C/\sqrt{n}\Big].
\end{equation*} Here, we take universal constants $c_1$ and $c_2$ because by definition, there exist some uniform constants $c_{\phi,\min}$ and $C_{\phi,\max}$ such that  $c_{\phi,\min}\le |\phi_{jk}^*|\le C_{\phi,\max}$ holds uniformly. 
Therefore,
\begin{equation*}
    \sum_{(j,k)\in\Omega_0}\delta_{jk}^L\ge \sum_{(j,k)\in\Omega_0}\min\big\{c_1\mu_{jk}^2, c_2|\mu_{jk}|\big\} -O_p(\sqrt{q} + q/\sqrt{n}).
\end{equation*}
Combined with \eqref{eq_bound_diff_Q1_t1_t1} and we arrive at 
\begin{equation}
    \eqref{eq_bound_diff_Q1_t1}\ge \frac{a_0\gamma}{\sqrt{n}}\sum_{(j,k)\in\Omega_0}\min\big\{c_1\mu_{jk}^2, c_2|\mu_{jk}|\big\} - O_p\Big(\gamma\sqrt{\frac{q}{n}} + \gamma q(\epsilon_{nq}^{\prime}) ^2 + \frac{\gamma q\epsilon_{nq}^{\prime}}{\sqrt{n\wedge q}} + \frac{\gamma q}{n\wedge q}\Big).\label{eq_bound_diff_Q1_t1_t2}
\end{equation}
Finally, we introduce the following lemma.
\begin{lemma}\it\label{lemma_regularize_loadings}
    Let $\mu_{jk} = -\sqrt{n}(\bG^{-\T})_{k,}\ba_j^*$ and $\Omega_0$ contain all indices $(j,k)$ with $ a_{jk}^* = 0$. Under Assumptions~\ref{assump_psd_covariance_prime} and~\ref{assump_limit}, for any sequences $r_0,R_0\ll 1$ and constants $c_1,c_2>0$, one has 
    \begin{align*}
        \sum_{(j,k)\in\Omega_0}\min\big(c_1\mu_{jk}^2,c_2|\mu_{jk}|\big)\ge C_1\frac{\sqrt{n}q{r_0^2}}{R_0\sigma_q} - C_2\sqrt{n}q\sigma_q^{-1}\big\{R_0^2 + (n\wedge q)^{-1}\big\},
    \end{align*}
    for some constants $C_1,C_2>0$ and any $\bG\in\Xi(R_0,\hat\bZ^*)$ satisfying $
        r_0\le \big\|\bG - \bI_r\big\|\le R_0$. Here, $\sigma_q^{-1} = \min_{k\in[r]}\lambda_{r-1}\big\{q^{-1}(\bA^*_{\cT_k^*,})^\T\bA_{\cT_k^*,}^*\big\}$.
\end{lemma}
\begin{proof}
See Section~\ref{sec_prove_lemma_regularize_loadings}.
\end{proof}
Lemma~\ref{lemma_regularize_loadings} implies that, when $\rdis\le \|\bG - \bI_r\|\le \Rdis = \epsilon_{nq}^{\prime}$, \eqref{eq_bound_diff_Q1_t1_t2} can be further lower bounded as 
\begin{equation}
    \gamma^{-1}\eqref{eq_bound_diff_Q1_t1}\ge q\Bigg\{a_0C_1\frac{\rdis^2}{\sigma_q\Rdis} - O_p\Big( \Rdis ^2 + \frac{1}{n\wedge q}+\frac{ \Rdis }{\sqrt{n\wedge q}} \Big)\Bigg\},\label{eq_bound_diff_Q1_t1_t22}
\end{equation}as $\sigma_q^{-1}\lesssim1$.
This is \eqref{eq_bound_diff_Q1_t1_t22_before}. The remaining steps are detailed in the proof of Lemma~\ref{lemma_oracle_folomin_solu} therein.

\subsection{Proof of Lemma~\ref{lemma_combine_zero}}\label{supp_sec_prove_lemma_combine_zero}
Suppose $(\hat\bZ^o,\hat\bA^o)$ is obtained from \eqref{eq_problem_align}. Let $\hat\bG^0 = (\hat\bU^o_A)^\T(n^{-1}\hat\bZ^o{}^\T\hat\bZ^o)^{-1/2}$, where $\hat\bU_A^o$ is the left singular matrix for $(n^{-1}\hat\bZ^o{}^\T\hat\bZ^o)^{1/2}q^{-1}(\hat\bA^o)^\T\hat\bA^o(n^{-1}\hat\bZ^o{}^\T\hat\bZ^o)^{1/2}$. Then $(\hat\bZ^o(\hat\bG^0)^\T,\hat\bA^o(\hat\bG^0)^{-1})$ satisfy the constraints in \eqref{eq_init_of_init_est} as well. Moreover, since Lemma~\ref{lemma_consis_start} holds for $(\hat\bZ^o,\hat\bA^o)$, it can be checked that $\|\hat\bG^0\|$ and $\|(\hat\bG^0)^{-1}\|$ are bounded with probability approaching $1$, and $\|\hat\bG^0 - \bG^0\|=O_p((n\wedge q)^{-1/2})$. Without loss of generality, we assume that $M$ is selected large enough such that $\|\bZ^0\|_{2\to\infty}<M /2$ and $\|\bA^0\|_{2\to\infty}\le M/2$. Then, we know $(\hat\bZ^o(\hat\bG^0)^\T,\hat\bA^o(\hat\bG^0)^{-1})$ satisfy the constraints in \eqref{eq_init_of_init_est}, and thus, $\cL(\hat\bZ^0,\hat\bA^0)\le \cL(\hat\bZ^o(\hat\bG^0)^\T,\hat\bA^o(\hat\bG^0)^{-1}) = \cL(\hat\bZ^o,\hat\bA^o)$. 

By Lemma~\ref{lemma_exist_g_hat_star}, we know $\hat\bG^{\star}$ exists with probability approaching $1$, and by the proof of Lemma~\ref{lemma_exist_g_hat_star}, we know that
\begin{equation*}
    q^{-1}\big\| \hat\bA^0(\hat\bG^z)^\T - \bA^* \big\|_{\Fn}^2 + n^{-1}\big\|\hat\bZ^0(\hat\bG^z)^{-1} - \bZ^*\big\|_{\Fn}^2=O_p\Big( \frac{1}{{n\wedge q}}\Big).
\end{equation*}
The existence of $\hat\bG^{\star}$ together with Lemma~\ref{lemma_procruste_solution} implies that $\cP^*(\hat\bZ^0\hat\bG^{\star}{}^\T,\hat\bA^0\hat\bG^{\star}{}^{-1}) = 0$ for $\cP^*(\cdot,\cdot)$ defined in \eqref{eq_define_key_penalty}. The optimal alignment has no larger average squared error than the candidate above, and so its average error tends to zero.
Then we can enlarge $M$ in \eqref{eq_problem_align} such that $(\hat\bZ^0\hat\bG^{\star}{}^\T,\hat\bA^0\hat\bG^{\star}{}^{-1})$ is within the feasible set of \eqref{eq_problem_align}. Consequently, with \begin{equation*}\cL(\hat\bZ^0\hat\bG^{\star}{}^\T,\hat\bA^0\hat\bG^{\star}{}^{-1}) = \cL(\hat\bZ^0,\hat\bA^0)\le \cL(\hat\bZ^o,\hat\bA^o)\end{equation*} and uniqueness of the solution to \eqref{eq_problem_align}, we know $(\hat\bZ^0\hat\bG^{\star}{}^\T,\hat\bA^0\hat\bG^{\star}{}^{-1}) = (\hat\bZ^o,\hat\bA^o)$ with probability approaching $1$. Together with the results established in Section~\ref{supp_sec_prove_ora_o}, we finish the proof.

\subsection{Proof of Lemma~\ref{lemma_angular_conservation}}\label{sec_prove_lemma_angular_conservation}

Write $\bM=\bQ^\T \bUpsilon \bQ$ with
$\bUpsilon=\mathrm{diag}(\lambda_1,\dots,\lambda_d)$ and
$\lambda_{\min}\le \lambda_i\le \lambda_{\max}$. The quantities
$\cos(\cdot,\cdot)$ and $\cos_{\bM}(\cdot,\cdot)$ are invariant under the common orthogonal change of variables $\bx\mapsto \bQ\bx$, $\by\mapsto \bQ\by$, so it suffices to work with $\bUpsilon$.

Set $c=\cos(\bx,\by)\le 1-\delta$.
Consider the two dimensional subspace spanned by any eigenvectors $\bu_{\max},\bu_{\min}$ corresponding to $\lambda_{\max}$ and $\lambda_{\min}$.

For the following constrained optimization problem
\begin{equation*}
    \max_{\ba,\bb}\ \frac{\ba^\T \bUpsilon \bb}{\sqrt{\ba^\T \bUpsilon \ba}\sqrt{\bb^\T \bUpsilon \bb}}
\quad\text{s.t.}\quad \|\ba\|_2=\|\bb\|_2=1,\ \ba^\T \bb=c
\end{equation*}
One can observe that the maximum of $\cos_{\bUpsilon}(a,b)$ over the admissible set is attained within the span of two eigenvectors of ${\bUpsilon}$, and the symmetric choice below is optimal within that span. Simple calculations yield the solution 
\[
\tilde \bx=\alpha\bu_{\max}+\beta\bu_{\min},\qquad
\tilde \by=\alpha\bu_{\max}-\beta\bu_{\min}.
\]
with
\[
\alpha^2=\frac{1+c}{2}\le 1-\frac{\delta}{2},\qquad
\beta^2=\frac{1-c}{2}\ge \frac{\delta}{2},
\]
Hence, the optimal value is
\begin{equation}\label{eq:ratio-two-d}
\cos_{\bUpsilon}(\tilde \bx,\tilde \by)
=\frac{\alpha^2\lambda_{\max}-\beta^2\lambda_{\min}}{\alpha^2\lambda_{\max}+\beta^2\lambda_{\min}}.
\end{equation}

Using $\alpha^2\le 1-\delta/2$ and $\beta^2\ge \delta/2$ in \eqref{eq:ratio-two-d} yields
\[
\cos_{\bUpsilon}(a,b)\ \le\ \frac{(1-\delta/2)\lambda_{\max}-(\delta/2)\lambda_{\min}}
{(1-\delta/2)\lambda_{\max}+(\delta/2)\lambda_{\min}}
=1-\frac{\delta\,\lambda_{\min}}{(1-\delta/2)\lambda_{\max}+(\delta/2)\lambda_{\min}}.
\]
The claimed expression for $\delta'$ follows, and $\delta'>0$ since $\delta>0$ and $\lambda_{\min}>0$.

\subsection{Proof of Lemma~\ref{lemma_exist_optimal_align}}\label{sec_prove_lemma_exist_optimal_align}
Consider the following constrained problem
\begin{equation}\begin{aligned}
    \min_{\bQ\in\RR^{r\times r}\text{ is invertible}}&\,n^{-1}\big\|\bZ\bQ - \bZ^*\big\|_{\Fn}^2 + q^{-1}\big\|\bA\bQ^{-\T} - \bA^*\big\|_{\Fn}^2\\&\,\text{ subject to }\|\bQ - \bP\|\le \frac{1}{2C\sqrt{C^{\prime}}}\,\frac{\delta}{\sigma_{r}(n^{-1/2}\bZ^*)}.
\end{aligned}\label{opt_align_AB_constr}\end{equation}
Denote the objective function $f(\bQ) = n^{-1}\big\|\bZ\bQ - \bZ^*\big\|_{\Fn}^2 + q^{-1}\big\|\bA\bQ^{-\T} - \bA^*\big\|_{\Fn}^2$. Then by Weyl's inequality, one has for any $\bQ$, if exists, that solves the above problem,
\begin{equation*}
    \sigma_{\min}(\bQ) \ge \sigma_{\min}(\bP) - \|\bQ - \bP\|>C^{-1} - \frac{1}{2C\sqrt{C^{\prime}}}\,\frac{\delta}{\sigma_{r}(n^{-1/2}\bZ^*)}\ge \frac{1}{2C},
\end{equation*}
given that $\delta\le \sqrt{C^{\prime}}\sigma_r(n^{-1/2}\bZ^*)$. Therefore, the feasible set in \eqref{opt_align_AB_constr} is a compact set over the invertible matrix space, and by the Weierstrass extreme value theorem, the solution exists. Subsequently, it suffices to show that any $\bQ$ such that $f(\bQ)\le 2\delta^2$ lies in the feasible set of \eqref{opt_align_AB_constr}. As we already know the minimizer $\hat\bQ$ satisfies $f(\hat\bQ)\le 2\delta^2$, this immediately implies that the solution of \eqref{opt_align_AB_constr}, denoted by $\hat\bQ$, is also the solution to \eqref{opt_align_AB}.

When $f(\bQ)\le 2\delta^2$, one has
\begin{equation*}
    \big\|\bZ(\bQ-\bP)\big\|_{\Fn}\le \big\|\bZ\bQ - \bZ^*\big\|_{\Fn} + \big\|\bZ\bP - \bZ^*\big\|_{\Fn}\le (1+\sqrt{2})n^{1/2}\delta.
\end{equation*}
Next, note that
\begin{align*}
    \sigma_{r}(\bZ)\ge\sigma_{r}(\bZ\bP)/\sigma_{\max}(\bP)\ge &\,C^{-1}\big\{\sigma_{r}(\bZ^*) - \big\|\bZ\bP - \bZ^*\big\|\big\}\\\ge &\, 
    C^{-1}\big\{\sigma_{r}(\bZ^*) - \sqrt{n}\delta\big\}\\\ge &\, C^{-1}\big(1 - \sqrt{C^{\prime}}\big)\sigma_{r}(\bZ^*).
\end{align*}with $\delta\le \sqrt{C^{\prime}}\sigma_r(n^{-1/2}\bZ^*)$. Subsequently, 
\begin{equation*}
    \big\|\bQ - \bP\big\|_{\Fn}\le \frac{(1+\sqrt{2})\delta}{n^{-1/2}\sigma_{\min}(\bZ)}\le \frac{(1+\sqrt{2})C\delta}{(1 - \sqrt{C^{\prime}})\sigma_{r}(n^{-1/2}\bZ^*)}\le \frac{1}{2C\sqrt{C^{\prime}}}\frac{\delta}{\sigma_r(n^{-1/2}\bZ^*)}.
\end{equation*}
Here, the last inequality follows from $C^{\prime}\le 1/(64C^4)$. Therefore, $\bQ$ belongs to the feasible set of \eqref{opt_align_AB_constr}.

\subsection{Proof of Lemma~\ref{lemma_concentration_l_mat}}\label{supp_Sec_prove_lemma_concentration_l_mat}
The proof of this lemma relies on the following result:

\noindent{\textbf{Corollary 3.12 in \citet{bandeira2016sharp}}.}{\it Let $\bX$ be an $n \times n$ symmetric matrix whose entries $X_{ij}$ with $i\le j$ are independent symmetric random variables. Then there exists for any $0 < \varepsilon \leq 1/2$ a universal constant $\tilde{c}_\varepsilon$ such that for every $t \geq 0$
\begin{equation}
{\PP}\left\{\|\bX\| \geq (1 + \varepsilon) 2\tilde{\sigma} + t \right\} \leq n e^{-t^2 / \tilde{c}_\varepsilon \tilde{\sigma}_*^2},\label{eq_random_mat_nonasymp_bound}
\end{equation}
where we define
\[
\tilde{\sigma} := \max_{1\le i\le n} \sqrt{\sum_{j\in[n]} \mathbb{E}[X_{ij}^2]}, \quad
\tilde{\sigma}_* := \max_{1\le i,j\le n} \|X_{ij}\|_\infty.
\]}

We start with constructing a symmetric matrix with each entry bounded in the $\ell_\infty$ norm:
\begin{equation*}
    \bX^i := \begin{pmatrix}
    \zero &\text{trun}_\nu\Big(\bL_{\theta,i}^\T - \tilde\bL_{\theta,i}^\T\Big)\\\text{trun}_\nu\Big(\bL_{\theta,i} - \tilde\bL_{\theta,i}\Big)  &\zero
\end{pmatrix}
\end{equation*}
Here $\bL_{\theta,i}:=(\bL_{\theta}^*)_{i,}$ and $\tilde\bL_{\theta,i}$ is an independent copy of $\bL_{\theta,i}$, and $\text{trun}_\nu(\bL_{\theta,i})$ is the entry-wise truncation at threshold $\nu$. Since {$\|\ell_{ij}^{\prime}(\theta_{ij}^*)\|_{\psi_1}\le C$} uniformly over $i\in[n],j\in[q]$, it holds that $\EE|\ell_{ij}^{\prime}(\theta_{ij}^*)|^2\le cC^2$ for some constant $c$. Then
\begin{align*}
    \max_{1\le t\le q+1} \sqrt{\sum_{j=1}^{q+1} \mathbb{E}\big[(\bX^i_{tj})^2\big]} \le &\,2\max\left\{\Big(\sum_{j=1}^{q}\EE\big[\ell_{ij}^{\prime}(\theta_{ij}^*)^2\big]\Big)^{1/2}, \max_{1\le j\le q}\Big(\EE\big[\ell_{ij}^{\prime}(\theta_{ij}^*)^2\big]\Big)^{1/2}\right\}\\\le &\,2cC^{2} \sqrt{q}.
\end{align*}
Next for $\max_{1\le t,j\le q+1}\|\bX^i_{tj}\|$, by definition it can be bounded by $\nu$. And apply \eqref{eq_random_mat_nonasymp_bound} to get
 \begin{equation*}
     {\PP}\left\{\|\bX^i\| \geq 4c(1 + \varepsilon) C^{2} \sqrt{q} + t \right\} \leq (q+1) \exp\{-(\tilde{c}_\varepsilon \nu^2)^{-1}{t^2}\}.
 \end{equation*}
 Given a constant $c$, since $\ell_{ij}^{\prime}(\theta_{ij}^*)$s are sub-exponential with {$\|\ell_{ij}^{\prime}(\theta_{ij}^*)\|_{\psi_1}\le C$}, we have for any $0<\psi<(cC)^{-1}$ that
 \begin{align*}
     &\,\PP\left\{\big\|\bL_{\theta,i} - \tilde\bL_{\theta,i}\big\|\ge 4 (1 + \varepsilon) cC^2 \sqrt{q} + t \right\} \\\le  &\,{\PP}\left\{\|\bX^i\| \geq4 (1 + \varepsilon) cC^2 \sqrt{q} + t \right\} + 2\sum_{j=1}^q\PP\left\{|\ell_{ij}^{\prime}(\theta_{ij}^*)|\ge \nu/2\right\}\\\le &\, (q+1) \exp\{-(\tilde{c}_\varepsilon \nu^2)^{-1}{t^2}\} + 2q\PP\left\{\exp(\psi |\ell_{ij}^{\prime}(\theta_{ij}^*)|)\ge \exp(\psi \nu/2)\right\}\\\le &\, (q+1) \exp\{-(\tilde{c}_\varepsilon \nu^2)^{-1}{t^2}\} + 4q\exp\left\{c^2C^2\psi^2 - \nu\psi/2\right\},
 \end{align*}
  by union bound. For the second term, take $\psi = \min\{(cC)^{-1},(2cC)^{-2}\nu\}$ to bound it as
 \begin{equation*}
     4q\exp\left\{-\min\big(\frac{\nu}{4cC},\frac{\nu^2}{8c^2C^2}\big)\right\}
 \end{equation*}
 For any $\delta > 0$, take $\nu = 4cC\big\{(\delta + 1)\log (q)  + \log 4\big\}$ and take $t =2\tilde c_\epsilon^{1/2} \nu \big((\delta + 1)\log(1+q)\big)^{1/2}$, and we conclude that
 \begin{equation*}
     \PP\left\{\big\|\bL_{\theta,i} - \tilde\bL_{\theta,i}\big\|\ge  4c(1 + \varepsilon) C^2 \sqrt{ q} + 2\tilde c_\epsilon^{1/2}\nu\big((\delta + 1)\log(1+q)\big)^{1/2} \right\} \le q^{-\delta}.
 \end{equation*}
 By Jensen's inequality, \begin{align*}
     \PP\left\{\|\bL_{\theta,i}\|\ge t\right\}  =&\,\EE_{\bL_{\theta,i}}\left[ 1_{\{\|\bL_{\theta,i} - \EE[ \bL_{\theta,i}]\|\ge t\}}\right]\\= &\,\EE_{\bL_{\theta,i}}\left[ 1_{\{\|\bL_{\theta,i} - \EE[ \tilde\bL_{\theta,i}]\|\ge t\}}\right]\\\le &\,\EE_{\bL_{\theta,i},\tilde\bL_{\theta,i}}\left[ 1_{\{\|\bL_{\theta,i} -  \tilde\bL_{\theta,i}]\|\ge t\}}\right]\\\le &\,P\left\{\big\|\bL_{\theta,i} -  \tilde\bL_{\theta,i}]\big\|\ge t\right\},
 \end{align*}
 which finally contributes to
  \begin{equation*}
     \PP\left\{\big\|\bL_{\theta,i}\big\|\ge  4c(1 + \varepsilon) C^2 \sqrt{ q} + 2\tilde c_\epsilon^{1/2}\nu\big((\delta + 1)\log(1+q)\big)^{1/2}\right\} \le q^{-\delta}.
 \end{equation*}
 Since $\log^3 (q)\ll q$, with probability at least $1-q^{-\delta} $, $\|\bL_{\theta,i}\|\le C_{\delta}\sqrt{q}$ with constant $C_\delta$ that is only related to $\delta$, or $\|\bL_{\theta,i}\| = \|(\bL_{\theta}^*)_{i,}\| \le C_{\delta}\sqrt{q}$.
 If we take $\nu = 4cC\big\{(\delta + 1)\log (nq)  + \log 4\big\}$ and take $t =2\tilde c_\epsilon^{1/2} \nu \big((\delta + 1)\log(n+nq)\big)^{1/2}$, we can obtain
 \begin{equation*}
     \PP\left\{\big\|\bL_{\theta,i}\big\|\ge  4c(1 + \varepsilon) C^2 \sqrt{q} + 2\tilde c_\epsilon^{1/2}\nu\big((\delta + 1)\log(n+nq)\big)^{1/2} \right\} \le n^{-1-\delta}q^{-\delta}.
 \end{equation*}
 By union bound,
  \begin{equation*}
     \PP\left\{\max_{1\le i\le n}\big\|\bL_{\theta,i}\big\|\ge  4c(1 + \varepsilon) C^2 \sqrt{q} + 2\tilde c_\epsilon^{1/2}\nu\big((\delta + 1)\log(n+nq)\big)^{1/2} \right\} \le n^{-\delta}q^{-\delta}.
 \end{equation*}
 With $q\gg \log n$ by Assumption~\ref{assump_scaling}, we know $\sqrt{q\log n}\gg \log^{3/2}(nq)$. Therefore, we arrive at $\max_{i\in[n]}\|\bL_{\theta,i}\| = \max_{i\in[n]}\|(\bL_{\theta}^*)_{i,}\| \le C_{\delta}\sqrt{q\log n}$ with probability at least $1-(n q)^{-\delta}$.
 This proves $(i)$. Similarly, we can derive $(ii)$. Next we prove $(iii)$ following a similar procedure. We start by constructing the symmetric matrix with each entry bounded in the $\ell_\infty$ norm,
\begin{equation*}
\bX = \begin{pmatrix}
    \zero &\text{trun}_\nu\Big(\bL_{\theta}^\T - \tilde\bL_{\theta}^\T\Big)\\\text{trun}_\nu\Big(\bL_{\theta} - \tilde\bL_{\theta}\Big)  &\zero
\end{pmatrix}.
\end{equation*}
Here $\bL_{\theta}:=\bL_{\theta}^*$ and $\tilde\bL_{\theta}$ is an independent copy of $\bL_{\theta}$, and $\text{trun}_\nu(\bL_{\theta})$ is the entry-wise truncation at threshold $\nu$. Since $\EE|\ell_{ij}^{\prime}(\theta_{ij}^*)|^2\le cC^2$,
\begin{align*}
    \max_{1\le i\le n+q} \sqrt{\sum_{j=1}^{n+q} \mathbb{E}[X_{ij}^2]} \le &\,2\max\left\{\max_{1\le i\le n}\Big(\sum_{j=1}^q\EE\big[\ell_{ij}^{\prime}(\theta_{ij}^*)^2\big]\Big)^{1/2}, \max_{1\le j\le q}\Big(\sum_{i=1}^n\EE\big[\ell_{ij}^{\prime}(\theta_{ij}^*)^2\big]\Big)^{1/2}\right\}\\\le &\,2cC^2 \sqrt{n\vee q}.
\end{align*}
Similarly, apply \eqref{eq_random_mat_nonasymp_bound} to get
 \begin{equation*}
     {\PP}\left\{\|\bX\| \geq 4(1 + \varepsilon) cC^2 \sqrt{n\vee q} + t \right\} \leq (n+q) \exp\{-(\tilde{c}_\varepsilon \nu^2)^{-1}{t^2}\}.
 \end{equation*}
Given a constant $c$, since $\ell_{ij}^{\prime}(\theta_{ij}^*)$s are sub-exponential with {$\|\ell_{ij}^{\prime}(\theta_{ij}^*)\|_{\psi_1}\le C$}, we have for any $0<\psi<(cC)^{-1}$ that
 \begin{align*}
     &\,\PP\left\{\big\|\bL_{\theta} - \tilde\bL_{\theta}\big\|\ge  (1 + \varepsilon) 4cC^2 \sqrt{n\vee q} + t \right\} \\\le  &\,{\PP}\left\{\|\bX\| \geq (1 + \varepsilon) 4cC^2 \sqrt{n\vee q} + t \right\} + 2\sum_{i=1}^n\sum_{j=1}^q\PP\left\{|\ell_{ij}^{\prime}(\theta_{ij}^*)|\ge \nu/2\right\}\\\le &\, (n+q) \exp\{-(\tilde{c}_\varepsilon \nu^2)^{-1}{t^2}\} + 2nq\PP\left\{\exp(\psi |\ell_{ij}^{\prime}(\theta_{ij}^*)|)\ge \exp(\psi \nu/2)\right\}\\\le &\, (n+q) \exp\{-(\tilde{c}_\varepsilon \nu^2)^{-1}{t^2}\} + 4nq\exp\left\{c^2C^2\psi^2 - \nu\psi/2\right\},
 \end{align*}
 by union bound. For the second term, take $\psi = \min\{(cC)^{-1},(2cC)^{-2}\nu\}$ to bound it as
 \begin{equation*}
     4nq\exp\left\{-\min\big(\frac{\nu}{4cC},\frac{\nu^2}{8c^2C^2}\big)\right\}
 \end{equation*}
 For any $\delta > 0$, take $\nu = 4cC\big\{(\delta + 1)\log (nq)  + \log 4\big\}$ and take $t =2\tilde c_\epsilon^{1/2} \nu \big((\delta + 1)\log(n+nq)\big)^{1/2}$, and we conclude that
 \begin{align*}
     \PP\left\{\big\|\bL_{\theta} - \tilde\bL_{\theta}\big\|\ge  (1 + \varepsilon) 4cC^2 \sqrt{n\vee q} + 2\tilde c_\epsilon^{1/2}\nu\big((\delta + 1)\log(n+nq)\big)^{1/2} \right\} \le (n+q)^{-\delta} + (nq)^{-\delta}.
 \end{align*}
Therefore, we arrive at
 \begin{equation}\begin{aligned}
     \PP\left\{\big\|\bL_{\theta}^*\big\|\ge  (1 + \varepsilon) 4cC^2 \sqrt{n\vee q} + 2\tilde c_\epsilon^{1/2} \nu\big((\delta + 1)\log(n+q)\big)^{1/2} \right\} \le (n+q)^{-\delta} + (nq)^{-\delta}.\end{aligned}\label{eq_non_asymptotic_bound_1}
 \end{equation}
With $(n\vee q)\gg \log^3(n+q)$, we conclude that $\|\bL_{\theta}^*\| \le C_{\delta}\sqrt{n\vee q}$ with probability at least $1-(n+q)^{-\delta} - (nq)^{-\delta}$. Taking the above tail bounds together and applying a union bound yields probability at least $1-n^{-\delta}-q^{-\delta}-3(nq)^{-\delta}-(n+q)^{-\delta}$.

\subsection{Proof of Lemma~\ref{lemma_procruste_solution}}\label{supp_sec_prove_lemma_procruste_solution}
Let $f(\bQ) = n^{-1}\big\|\bZ\bQ - \bZ^*\big\|_{\Fn}^2 + q^{-1}\big\|\bA\bQ^{-\T} - \bA^*\big\|_{\Fn}^2$. Then one can compute the gradient is
\begin{equation*}
    \partial_{\bQ}f(\bQ) = \frac{2}{n}\bZ^\T(\bZ\bQ-\bZ^*) - \frac{2}{q}\bQ^{-\T}(\bA\bQ^{-\T}-\bA^*)^\T\bA\bQ^{-\T}.
\end{equation*}
When the minimizer $\tilde\bG$ exists, we know it must satisfy the first-order optimality condition $\partial_{\bQ}f(\tilde\bG) = \zero$, which yields
\begin{equation*}
    n^{-1}\tilde\bZ^\T(\tilde\bZ - \bZ^*) = q^{-1}(\tilde\bA - \bA^*)^\T\tilde\bA\text{ where }(\tilde\bZ,\tilde\bA) = (\bZ\tilde\bG,\bA\tilde\bG^{-\T})
\end{equation*}
\subsection{Proof of Lemma~\ref{lemma_symmetric_kronecker_prod}}\label{supp_sec_lemma_symmetric_kronecker_prod}
Note that by $\mathrm{vec}(\bA\bX\bB)=(\bB^\T\otimes \bA)\,\mathrm{vec}(\bX)$, we have
\begin{align*}
\mathrm{vec}(\bA\bM_1+\bM_2\bA)
 = &\, (\bM_1^\T\otimes \bI_r)\,\mathrm{vec}(\bA) + (\bI_r\otimes \bM_2)\,\mathrm{vec}(\bA)\\=&\,\big\{(\bM_1\otimes \bI_r) + (\bI_r\otimes \bM_2)\big\}\mathrm{vec}(\bA).
\end{align*}
The second equality follows from that $\bM_1$ is symmetric. 
Next, note that for symmetric matrices, the eigenvalues of the Kronecker sum satisfy
\[
\lambda\big(\bM_1\otimes \bI_r + \bI_r\otimes \bM_2\big)
=\{\lambda_i(\bM_1)+\lambda_j(\bM_2): 1\le i\le r,\ 1\le j\le r\}.
\]So $\bM_1\otimes \bI_r + \bI_r\otimes \bM_2$ is symmetric positive definite.
Therefore, let $\bK = \bM_1\otimes \bI_r + \bI_r\otimes \bM_2$ and we have
\begin{align*}
\|\mathrm{vec}(\bA\bM_1+\bM_2\bA)\|_2
=&\, \|\bK\,\mathrm{vec}(\bA)\|_2
\ge \lambda_{\min}(\bK)\,\|\mathrm{vec}(\bA)\|_2
\\= &\, \big(\lambda_{\min}(\bM_1)+\lambda_{\min}(\bM_2)\big)\,\|\mathrm{vec}(\bA)\|_2.
\end{align*}
With $\|\mathrm{vec}(\bX)\|_2=\|\bX\|_{\Fn}$, we finish the proof.

\subsection{Proofs of Lemma~\ref{lemma_concentration} and Corollary~\ref{coro_first_order_concern}}\label{supp_sec_prove_lemma_concentration}
\subsubsection{Proof of Lemma~\ref{lemma_concentration}}
    Follow the Bernstein-type probability bound as in Proposition 5.16 in \citet{vershynin2010introduction}:
\[
\mathbb{P} \left\{ \left| \sum_{i=1}^n b_i \ell_{ij}^{\prime}(\theta_{ij}^*) \right| \geq t \right\} \leq 2 \exp \left[ -c \min \left( \frac{t^2}{(CB)^2n}, \frac{t}{CB} \right) \right],\;\forall j\in[q],
\]where $c > 0$ is an absolute constant and $C$ is defined in Assumption~\ref{assump_limit} with {$\|\ell_{ij}^{\prime}(\theta_{ij}^*)\|_{\psi_1}\le C$}. 
Take $t = (\sqrt{n\epsilon_n} + \epsilon_n)^{1/2}KCB$ for a sufficiently large constant $K$, and we obtain the bound for $\big| \sum_{i=1}^N b_i \ell_{ij}^{\prime}(\theta_{ij}^*) \big| $ in $(i)$. Next by union bound, 
\[
\mathbb{P} \left\{\max_{1\le j\le q} \left| \sum_{i=1}^n b_i\ell_{ij}^{\prime}(\theta_{ij}^*) \right| \geq t \right\} \leq 2 q\exp \left[ -c \min \left( \frac{t^2}{(CB)^2n}, \frac{t}{CB} \right) \right].
\]Take $t = KCB\{\sqrt{n(\epsilon_n+\log q)}+\epsilon_n+\log q\}$ and the second part in $(i)$ is proved.
Similarly, we can derive $(ii)$ and $(iii)$.

\subsubsection{Proof of Corollary~\ref{coro_first_order_concern}}\label{sup_sec_prove_coro_first_order_concern}

    This corollary can be proven following a similar approach to the proof of Lemma~\ref{lemma_concentration_l_mat}, using the fact that each entry in $\bS_{nq}\partial_{\bphi}\cL(\bZ^*,\bA^*)$ is sub-exponential, as shown in Lemma~\ref{lemma_concentration}.

\subsection{Proof of Lemma~\ref{lemma_hessian_property}}\label{supp_sec_prove_lemma_hessian_property}
In this proof, we use the vectorized parameter $\bphi$ to represent $(\bZ,\bA)$, and accordingly write any function originally taking $(\bZ,\bA)$ as its argument as a function of $\bphi$.
    Recall that in \eqref{eq_approximation_p_hes_vv}, we approximate $\partial_{\bphi\bphi}\cP^*(\bphi)$ with $\sum_{k=1}^r\sum_{l=1}^r(nq)^{-1}\bS_{nq}^{-1}\bv_{kl}(\bphi)\{\bv_{kl}(\bphi)\}^\T\bS_{nq}^{-1}$ where $\bv_{kl}(\bphi)$ is defined in \eqref{eq_define_vkl} as 
    \begin{equation*}
    \bv_{kl}(\bphi) := \big(n^{-1/2}z_{1k}\be_l^\T,\dots,n^{-1/2}z_{nk}\be_l^\T, -q^{-1/2}a_{1l}\be_k^\T,\dots,-q^{-1/2}a_{ql}\be_k^\T\big)^\T,
\end{equation*}
for each $k,l\in[r]$.
    Inspired by this, we start by decomposing $\sum_{k=1}^r\sum_{l=1}^r\bv_{kl}(\bphi)\{\bv_{kl}(\bphi)\}^\T$ as 
    \begin{align*}
        \sum_{k=1}^r\sum_{l=1}^r\bv_{kl}(\bphi)\{\bv_{kl}(\bphi)\}^\T = &\,\begin{pmatrix}
            \bV_z(\bphi)\{\bV_z(\bphi)\}^\T & \zero \\\zero & \bV_a(\bphi)\{\bV_a(\bphi)\}^\T
        \end{pmatrix}\\&\,+\begin{pmatrix}
            \bI_n\otimes q^{-1}\sum_{j=1}^q\ba_j\ba_j^\T & \zero
            \\\zero & \bI_q\otimes n^{-1}\sum_{i=1}^n\bz_i\bz_i^\T
        \end{pmatrix}\\&\,-\sum_{i=1}^n\sum_{j=1}^q\bS_{nq}\begin{pmatrix}
            \be_i^{(n)}\otimes \ba_j\\\be_j^{(q)}\otimes \bz_i
        \end{pmatrix}\begin{pmatrix}
            \be_i^{(n)}\otimes \ba_j\\\be_j^{(q)}\otimes \bz_i
        \end{pmatrix}^\T\bS_{nq},
    \end{align*}
    where $\bV_z(\bphi) = \{\bV(\bphi)\}_{1:nr,}$ and $\bV_a(\bphi) = \{\bV(\bphi)\}_{nr+1:nr+qr,}$ with \[\bV(\bphi) := (\bv_{11}(\bphi), \bv_{12}(\bphi),\cdots,\bv_{1r}(\bphi),\bv_{21}(\bphi),\cdots,\bv_{rr}(\bphi)).\]
    Thus, one has
    \begin{align}
        &\,\bS_{nq}\big\{\partial_{\bphi\bphi}^2\cL(\bphi) + \nu\sum_{k=1}^r\sum_{l=1}^r\bS_{nq}^{-1}\bv_{kl}(\bphi)\{\bv_{kl}(\bphi)\}^\T\bS_{nq}^{-1}\big\}\bS_{nq}\nonumber\\=&\,\sum_{i=1}^n\sum_{j=1}^q\big\{\ell_{ij}^{\prime\prime}(\theta_{ij}) -\nu\big\}\bS_{nq}\begin{pmatrix}
            \be_i^{(n)}\otimes \ba_j\\\be_j^{(q)}\otimes \bz_i
        \end{pmatrix}\begin{pmatrix}
            \be_i^{(n)}\otimes \ba_j\\\be_j^{(q)}\otimes \bz_i
        \end{pmatrix}^\T\bS_{nq}\label{eq_hessian_positive_expand1}\\&\,+\nu\begin{pmatrix}
            \bV_z(\bphi)\{\bV_z(\bphi)\}^\T & \zero \\\zero & \bV_a(\bphi)\{\bV_a(\bphi)\}^\T
        \end{pmatrix}\label{eq_hessian_positive_expand2}\\&\,+\nu\begin{pmatrix}
            \bI_n\otimes q^{-1}\sum_{j=1}^q\ba_j\ba_j^\T & \zero
            \\\zero & \bI_q\otimes n^{-1}\sum_{i=1}^n\bz_i\bz_i^\T
        \end{pmatrix}+ \bS_{nq}\bR_L(\bphi)\bS_{nq}.\label{eq_hessian_positive_expand3}
    \end{align}Here, $\theta_{ij}$ is defined as $\ba_j^\T\bz_i$ for each $i\in[n]$ and $j\in[q]$; and we let $\bR_L(\bphi)=\begin{pmatrix}\zero&\bL_\theta(\bphi)\otimes\bI_r\\\bL_\theta(\bphi)^\T\otimes\bI_r&\zero\end{pmatrix}$, where $(\bL_\theta(\bphi))_{ij}=\ell_{ij}^{\prime}(\theta_{ij})$. By Lemma~\ref{lemma_concentration_l_mat}, the mean value theorem, and $\|\bA\bZ^\T-\bA^*\bZ^*{}^\T\|_{\Fn}=O(\epsilon\sqrt{nq})$, we have $\|\bS_{nq}\bR_L(\bphi)\bS_{nq}\|\le \|\bL_\theta(\bphi^*)\|/\sqrt{nq}+O(\epsilon)=o_p(1)+O(\epsilon)$ uniformly over $\bphi\in\Theta^*(\epsilon)$.
    By construction, \eqref{eq_hessian_positive_expand1} is positive semi-definite. By Assumption~\ref{assump_limit} and $0<\nu<b_L$, \eqref{eq_hessian_positive_expand2} is positive semi-definite. Consequently,
    \begin{align*}
        &\,\lambda_{\min}\Big[\bS_{nq}\big\{\partial_{\bphi\bphi}^2\cL(\bphi) + \nu\sum_{k=1}^r\sum_{l=1}^r\bS_{nq}^{-1}\bv_{kl}(\bphi)\{\bv_{kl}(\bphi)\}^\T\bS_{nq}^{-1}\big\}\bS_{nq}\Big]\\\ge&\, \nu\min\Big\{\lambda_{\min}\big(q^{-1}\sum_{j=1}^q\ba_j\ba_j^\T \big),\lambda_{\min}\big(n^{-1}\sum_{i=1}^n\bz_i\bz_i^\T \big)\Big\} - \|\bS_{nq}\bR_l(\bphi)\bS_{nq}\|
    \end{align*}
    For this lower bound, by Weyl's inequality, one has
    \begin{equation*}\begin{aligned}
        \lambda_{\min}\big(q^{-1}\sum_{j=1}^q\ba_j\ba_j^\T \big)\ge &\, \lambda_{\min}\big(q^{-1}\sum_{j=1}^q\ba_j^*\ba_j^*{}^\T \big) - q^{-1}\big\|\sum_{j=1}^q\ba_j\ba_j^\T - \sum_{j=1}^q\ba_j^*\ba_j^*{}^\T\big\|\\\ge &\,\lambda_{\min}\big(q^{-1}\bA^*{}^\T\bA^* \big) - 2q^{-1}\big\|\bA^\T(\bA-\bA^*)\big\| - q^{-1}\big\|(\bA-\bA^*)^\T(\bA-\bA^*)\big\|\\\ge &\,\lambda_{\min}\big(q^{-1}\bA^*{}^\T\bA^* \big)-2q^{-1}\|\bA\|\|\bA-\bA^*\|-q^{-1}\|\bA-\bA^*\|^2.
    \end{aligned}\end{equation*}
    When $\bphi^*,\bphi\in\Theta^*(\epsilon)$, we know $\|\bA\|\le\|\bA\|_{\Fn}\le M\sqrt{qr}$ and $\|\bA-\bA^*\|\le \|\bA-\bA^*\|_{\Fn}\le \epsilon\sqrt{q}$. By Assumption~\ref{assump_psd_covariance}$(iii)$, we know $\lambda_{\min}(q^{-1}\bA^*{}^\T\bA^* )\ge 3\lambda_{\min}(\bSigma_a^*)/4$ when $q$ is large enough. Therefore, with $\epsilon\ll 1$, one arrives at
    \begin{equation}
        \lambda_{\min}\big(q^{-1}\sum_{j=1}^q\ba_j\ba_j^\T \big)> 3\lambda_{\min}(\bSigma_a^*)/4 -O( \epsilon)\ge \lambda_{\min}(\bSigma_a^*)/2,\label{eq_lower_bound_aaT}
    \end{equation}uniformly over $\bphi\in\Theta^*(\epsilon)$, given that $\epsilon$ is sufficiently small. Similarly we can derive a lower bound for $\lambda_{\min}(n^{-1}\sum_{i=1}^n\bz_i\bz_i^\T)$. We then conclude that 
    \begin{equation}
        \lambda_{\min}\Big[\bS_{nq}\big\{\partial_{\bphi\bphi}^2\cL(\bphi) + \nu\sum_{k=1}^r\sum_{l=1}^r\bS_{nq}^{-1}\bv_{kl}(\bphi)\{\bv_{kl}(\bphi)\}^\T\bS_{nq}^{-1}\big\}\bS_{nq}\Big]\ge\frac{\nu}4\min\{\lambda_{\min}(\bSigma_a^*),\lambda_{\min}(\bSigma_z^*)\}.\label{eq_eigenvalue_lower_bound_replace}
    \end{equation}
    Now we compute $\bS_{nq}\partial_{\bphi\bphi}^2\{nq\cP^*(\bphi)\}\bS_{nq} - \sum_{k=1}^r\sum_{l=1}^r\bv_{kl}(\bphi)\{\bv_{kl}(\bphi)\}^\T$. Specifically, in \eqref{eq_define_ukl} we have computed
    \begin{align*}
        \partial_{\bphi\bphi}^2\cP^*(\bZ,\bA) = &\,\sum_{k=1}^r\sum_{l=1}^r(nq)^{-1}\bS_{nq}^{-1}\{\bv_{kl}(\bphi) + \bu_{kl}(\bphi)\}\{\bv_{kl}(\bphi) + \bu_{kl}(\bphi)\}^\T\bS_{nq}^{-1} \\&\,+ (nq)^{-1}\bS_{nq}^{-1}\sum_{k=1}^r\sum_{l=1}^r\omega_{kl}(\bphi)\bE_{kl}\bS_{nq}^{-1},
    \end{align*}
    where for each $k,l\in[r]$, $\omega_{kl}(\bphi) = n^{-1}\sum_{i=1}^nz_{ik}(z_{il} - z_{il}^*) - q^{-1}\sum_{j=1}^q(a_{jk} - a_{jk}^*)a_{jl}$, $\bu_{kl}$ is given in \eqref{eq_define_ukl}, $\bE_{kl}$ is given in \eqref{eq_define_Ekl}. When $\bphi\in\Theta^*(\epsilon)$, we know $n^{-1/2}\|\bphi_z - \bphi_z^*\| + q^{-1/2}\|\bphi_a - \bphi_a^*\| \le \epsilon$. One can then verify with {Cauchy–Schwarz inequality} that $\|\bu_{kl}(\bphi)\| \lesssim \epsilon$ and $|\omega_{kl}(\bphi)|\lesssim \epsilon$ uniformly over $k,l\in[r]$. 
Since $\|\bphi\|_{\infty}$ is bounded by $M$ in $\Theta^*(\epsilon)$, we know $\|\bv_{kl}(\bphi)\| \lesssim 1$. Therefore, by the triangle inequality, we conclude that
\begin{equation*}
    \big\|\bS_{nq}\partial_{\bphi\bphi}^2\{nq\cP^*(\bphi)\}\bS_{nq} - \sum_{k=1}^r\sum_{l=1}^r\bv_{kl}(\bphi)\{\bv_{kl}(\bphi)\}^\T\big\|\lesssim \epsilon.
\end{equation*}
Since $\epsilon$ can be selected to be sufficiently small, we prove \eqref{eq_eigenvalue_lower_bound} with \eqref{eq_eigenvalue_lower_bound_replace} and that $\bSigma_{a}^*,\bSigma_{z}^*$ are positive definite with fixed dimension from Assumption~\ref{assump_psd_covariance}$(iii)$.

For the smallest eigenvalues of $\bH_{Lz_iz_i}(\bphi)$, we note that
\begin{align*}
    \min_{(\bZ,\bA)\in\Theta^*(\epsilon)}\lambda_{\min}\big\{\bH_{Lz_iz_i}(\bphi)\big\} = &\,\min_{(\bZ,\bA)\in\Theta^*(\epsilon)}\lambda_{\min}\big\{\sum_{j=1}^q\ell_{ij}^{\prime\prime}(\theta_{ij})\ba_j\ba_j^\T\big\}\\\ge&\,b_L\min_{(\bZ,\bA)\in\Theta^*(\epsilon)}\lambda_{\min}\big\{\sum_{j=1}^q\ba_j\ba_j^\T\big\}.
\end{align*}
Here, the inequality follows from $\ell_{ij}^{\prime\prime}(\theta)\ge b_L$ when $|\theta|\le M^2$ by Assumption~\ref{assump_limit} and that $(\bZ,\bA)\in\Theta^*(\epsilon)$, where $\max_{i\in[n],j\in[q]}|\theta_{ij}|\le \|\bZ\|_{2\to\infty}\|\bA\|_{2\to\infty}\le M^2$. Then with \eqref{eq_lower_bound_aaT}, one can verify that $\min_{(\bZ,\bA)\in\Theta^*(\epsilon)}\lambda_{\min}\big\{\bH_{Lz_iz_i}(\bphi)\big\}\ge cq$ for some $c\ge b_L\min\{\lambda_{\min}(\bSigma_z^*),\lambda_{\min}(\bSigma_{a}^*)\}/2$. For the upper bound, similar to \eqref{eq_lower_bound_aaT}, one can show that $\max_{(\bZ,\bA)\in\Theta^*(\epsilon)}\lambda_{\max}\big\{\sum_{j=1}^q\ba_j\ba_j^\T\big\}\le 2\lambda_{\max}(\bSigma_a^*)$, and following the same procedure, one can check $\max_{(\bZ,\bA)\in\Theta^*(\epsilon)}\lambda_{\max}\big\{\bH_{Lz_iz_i}(\bphi)\big\}\le Cq$. The upper and lower bounds for the eigenvalues of $\bH_{La_ja_j}(\bphi)$ follow by symmetry.

Finally, we bound $\max_{(\bZ,\bA)\in\Theta^*(\epsilon)}\|\bH_{Lza}(\bphi)\|_{2\to\infty}$. First, we write $\bH_{Lza}(\bphi)$ as $\bH_{Lza}^{(1)}(\bphi) + \bH_{Lza}^{(2)}(\bphi)$ where for each $i\in[n],j\in[q]$,
\begin{equation*}
    \{\bH_{Lza}^{(1)}(\bphi)\}_{(i-1)r+1:ir,(j-1)r+1:jr} = \ell_{ij}^{\prime\prime}(\theta_{ij})\ba_j\bz_i^\T\!\text{ and }\! \{\bH_{Lza}^{(2)}(\bphi)\}_{(i-1)r+1:ir,(j-1)r+1:jr} = \ell_{ij}^{\prime}(\theta_{ij})\bI_r.
\end{equation*}
Since $\ell_{ij}^{\prime\prime}(\theta_{ij})\le b_U$, $\max_{j\in[q]}\|\ba_j\|\le M$, and $\max_{i\in[n]}\|\bz_i\|\le M$, we know
\begin{equation*}
   \max_{(\bZ,\bA)\in\Theta^*(\epsilon)}\max_{i\in[n],j\in[q]} \big\|\{\bH_{Lza}^{(1)}(\bphi)\}_{(i-1)r+1:ir,(j-1)r+1:jr}\big\|_{\max}\lesssim 1,
\end{equation*}
and we then conclude that for the first term, $\max_{(\bZ,\bA)\in\Theta^*(\epsilon)}\|\bH_{Lza}^{(1)}(\bphi)\|_{2\to\infty} = O(\sqrt{q})$. For $\bH_{Lza}^{(2)}(\bphi)$, invoke $(i)$ of Lemma~\ref{lemma_concentration_l_mat} to get
\begin{equation*}
    \big\|\bH_{Lza}^{(2)}(\bphi^*)\big\|_{2\to\infty} \le C_{\delta}\sqrt{q\log n}.
\end{equation*}Next, for any $\bphi\in\Theta^*(\epsilon)$, one has
\begin{align*}
    \big\|\bH_{Lza}^{(2)}(\bphi)\big\|_{2\to\infty}\le &\, \big\|\bH_{Lza}^{(2)}(\bphi) - \bH_{Lza}^{(2)}(\bphi^*)\big\|_{2\to\infty} + \big\|\bH_{Lza}^{(2)}(\bphi^*)\big\|_{2\to\infty} \\= &\,\max_{i\in[n]}\big\|\{\bH_{Lza}^{(2)}(\bphi) - \bH_{Lza}^{(2)}(\bphi^*)\}_{(i-1)r+1:ir,}\big\| + C_{\delta}\sqrt{q\log n}\\\le &\,\max_{i\in[n]}\Big[\sum_{j=1}^q\{\ell_{ij}^{\prime}(\theta_{ij}) - \ell_{ij}^{\prime}(\theta_{ij}^*)\}^2\Big]^{1/2} + C_{\delta}\sqrt{q\log n}\\\le &\,\max_{i\in[n]}\|\bz_i\|\Big[\sum_{j=1}^q\{\ell_{ij}^{\prime\prime}(\tilde\theta_{ij})\}^2|\theta_{ij} - \theta_{ij}^*|^2\Big]^{1/2}+ C_{\delta}\sqrt{q\log n}\\\le &\,C_{\delta}\sqrt{q\log n}.
\end{align*}
Here, $\tilde\theta_{ij}$ is between $\theta_{ij}$ and $\theta_{ij}^*$, and thus $\ell_{ij}^{\prime\prime}(\tilde\btheta_{ij})\le b_U$ is bounded by $b_U$ by Assumption~\ref{assump_limit}; and the last inequality follows from the boundedness of $\bphi$ in $\Theta^*(\epsilon)$. Combine the above results and we know $\max_{(\bZ,\bA)\in\Theta^*(\epsilon)}\|\bH_{Lza}(\bZ,\bA)\|_{2\to\infty}\!\le\! C_{\delta}\sqrt{q\log n}$. By symmetry, we can derive $\max_{(\bZ,\bA)\in\Theta^*(\epsilon)}\|\bH_{Laz}(\bZ,\bA)\|_{2\to\infty}\!\le\! C_{\delta}\sqrt{n\log q}$.

\subsection{Proof of Lemma~\ref{lemma_hessian_property_2}}\label{supp_sec_prove_lemma_hessian_property_2}

    In this proof, we use the vectorized parameter $\bphi$ to represent $(\bZ,\bA)$, and accordingly write any function originally taking $(\bZ,\bA)$ as its argument as a function of $\bphi$. Let $\cR_i = \{(i-1)r+1,(i-1)r+2,\dots,ir\}$
    
Recall that we have the block-wise form of the Hessian matrix $\bH(\bphi) := \partial_{\bphi\bphi}^2\cL_{\nu}(\bphi)$ given as
\begin{equation*}
    \bH(\bphi) = \begin{pmatrix}
        \bH_{Lzz}(\bphi) & \bH_{Lza}(\bphi)\\
        \bH_{Laz}(\bphi) & \bH_{Laa}(\bphi)
    \end{pmatrix} + \nu nq\partial^2_{\bphi\bphi}\cP^*(\bphi),
\end{equation*}
where $\bH_{Lzz}(\bphi) = \partial_{\bphi_z\bphi_z}^2\cL(\bphi)$, $\bH_{Laa}(\bphi) = \partial_{\bphi_z\bphi_a}^2\cL(\bphi)$, and $\bH_{Lza}(\bphi) =\{\bH_{Laz}(\bphi)\}^\T = \partial_{\bphi_z\bphi_z}^2\cL(\bphi)$. At the true parameter $\bphi^* = (\bZ^*,\bA^*)$, invoke the expression for $\partial^2_{\bphi\bphi}\cP^*(\cdot)$ in \eqref{eq_define_second_cP} and we know
\begin{equation*}
    nq\partial^2_{\bphi\bphi}\cP^*(\bphi)|_{\bphi^*} = \sum_{k=1}^r\sum_{l=1}^r\bS_{nq}^{-1}\bv_{kl}(\bphi^*)\{\bv_{kl}(\bphi^*)\}^\T\bS_{nq}^{-1}.
\end{equation*}
We define $\bv_{kl}^* = \bv_{kl}(\bphi^*)$, $\bV^* = \sqrt{\nu}\bS_{nq}^{-1}(\bv_{11}^*,\bv_{12}^*,\dots,\bv_{1r}^*,\bv_{21}^*,\dots,\bv_{rr}^*)$, $\bV^*_z = \bV^*_{1:nr,}$, $\bV^*_a = \bV^*_{(nr+1):(nr+qr),}$, and
\begin{equation*}
    \bH^*: = \bH(\bphi^*) = \begin{pmatrix}
        \bH_{zz}^*& \bH_{za}^*\\
        \bH_{az}^* & \bH_{aa}^*
    \end{pmatrix}  =  \begin{pmatrix}
        \bH_{Lzz}^*& \bH_{Lza}^*\\
        \bH_{Laz}^* & \bH_{Laa}^*
    \end{pmatrix} +  \bV^*\bV^*{}^\T 
\end{equation*}
with $\bH_{zz}^* = \bH_{Lzz}^* + \bV_z^*\bV_z^*{}^\T = \bH_{Lzz}(\bphi^*) + \bV_z^*\bV_z^*{}^\T$, $\bH_{aa}^* = \bH_{Laa}^* + \bV_a^*\bV_a^*{}^\T= \bH_{Laa}(\bphi^*)+ \bV_a^*\bV_a^*{}^\T$, $\bH_{za}^* = \bH_{Lza}^* + \bV_z^*\bV_a^*{}^\T= \bH_{Lza}(\bphi^*)+ \bV_z^*\bV_a^*{}^\T $, and $\bH_{az}^* = \bH_{za}^*{}^\T = \bH_{Lza}(\bphi^*) + \bV_a^*\bV_z^*{}^\T$. 
For $i\in[n]$, let $\bH_{Lz_iz_i}^* = (\bH_{Lzz}^*)_{\cR_i,\cR_i}$ and $\bH_{Lz_ia}^* = (\bH_{Lza}^*)_{\cR_i,}$; for $j\in[q]$, let $\bH_{La_ja_j}^* = (\bH_{Laa}^*)_{\cR_j,\cR_j}$ and $\bH_{La_jz}^* = (\bH_{Laz}^*)_{\cR_j,}$; 
for $i\in[n]$, let $\bV_{z_i}^* = (\bV_z^*)_{\cR_i,}$; for $j\in[q]$, let $\bV_{a_j}^* = (\bV_a^*)_{\cR_j,}$.

We have the following bounds:
\begin{subequations}
    \begin{align}
    &\lambda_{\min}(\bH_{Lzz}^*) = \min_{i\in[n]}\lambda_{\min}(\bH_{Lz_iz_i}^*) = \Theta_p(q);\; \lambda_{\max}(\bH_{Lzz}^*) = \max_{i\in[n]}\lambda_{\max}(\bH_{Lz_iz_i}^*) = O_p(q);\label{eq_bound_hzz}\\
    &\lambda_{\min}(\bH_{Laa}^*) = \min_{j\in[q]}\lambda_{\min}(\bH_{La_ja_j}^*) = \Theta_p(n);\; \lambda_{\max}(\bH_{Laa}^*) = \max_{j\in[q]}\lambda_{\max}(\bH_{La_ja_j}^*) = O_p(n);\label{eq_bound_haa}\\
    &\|\bH_{Lza}^*\|_{2\to\infty} = O_p(\sqrt{q\log n});\;\|\bH_{Laz}^*\|_{2\to\infty} = O_p(\sqrt{n\log q});\label{eq_bound_haz_inf}\\
    &\|\bH_{Lz_ia}^*\| = O_p(\sqrt{q})\text{, }i\in[n];\;\|\bH_{La_jz}^*\| = O_p(\sqrt{n})\text{, }j\in[q];\; \|\bH_{Lza}\| = O_p(\sqrt{nq});\label{eq_bound_haz}\\ &\max_{i\in[n]}\|\bV_{z_i}^*\| = O(\sqrt{q/n});\;\max_{j\in[q]}\|\bV_{a_j}^*\| = O(\sqrt{n/q});\|\bV_{z}^*\| = O(\sqrt{q});\;\|\bV_{a}^*\| = O(\sqrt{n}).\label{eq_bound_V}
\end{align}
\label{eq_bound_h_ast}
\end{subequations}
Here, Equations~\eqref{eq_bound_hzz}, \eqref{eq_bound_haa}, and \eqref{eq_bound_haz_inf} follow from Lemma~\ref{lemma_hessian_property}; Equation~\eqref{eq_bound_haz} follows from Lemma~\ref{lemma_concentration_l_mat} and the boundedness of $\|\ell_{ij}^{\prime\prime}(\theta_{ij}^*)\bz_i^*\ba_j^*{}^\T\|$ uniformly over $i\in[n]$ and $j\in[q]$; Equation~\eqref{eq_bound_V} follows from the construct of $\bV^*$, Assumption~\ref{assump_psd_covariance}$(ii)$ and $\|\bA^*\|_{2\to\infty}\le M$.

Now we invert $\bH^*$ into a blockwise form as 
\begin{equation*}
    (\bH^*)^{-1} = \begin{pmatrix}
        \bH_{zz}^{\dagger}& \bH_{za}^{\dagger}\\
        \bH_{az}^{\dagger} & \bH_{aa}^{\dagger}
    \end{pmatrix},
\end{equation*}
where by Sherman–Morrison–Woodbury formula, there is
\begin{align*}
    \bH_{zz}^\dagger =&\, (\bH_{zz}^*)^{-1}+(\bH_{zz}^*)^{-1}\bH_{za}^*\bH_{-a}^*\bH_{az}^*(\bH_{zz}^*)^{-1};\\
    \bH_{za}^\dagger = &\,-\big\{(\bH_{zz}^*)^{-1}+(\bH_{zz}^*)^{-1}\bH_{za}^*\bH_{-a}^*\bH_{az}^*(\bH_{zz}^*)^{-1}\big\}\bH_{ za}^*(\bH_{aa}^*)^{-1};\\
    \bH_{az}^\dagger = &\,-\big\{(\bH_{aa}^*)^{-1}+(\bH_{aa}^*)^{-1}\bH_{za}^*\bH_{-z}^*\bH_{za}^*(\bH_{aa}^*)^{-1}\big\}\bH_{za}^*(\bH_{zz})^{-1};\\
    \bH_{aa}^\dagger =&\, (\bH_{aa}^*)^{-1}+(\bH_{aa}^*)^{-1}\bH_{az}^*\bH_{-z}^*\bH_{za}^*(\bH_{aa}^*)^{-1}.
\end{align*}
Here, $\bH_{-a}^*$ denotes Schur complement of $\bH_{aa}^*$ and $\bH_{-z}^*$ is the Schur complement of $\bH_{zz}^*$. By Lemma~\ref{lemma_hessian_property}, we know that $\lambda_{i}\big(\bS_{nq}\bH^*\bS_{nq}\big)$ is of constant order for any $i\in[nr+qr]$, with probability approaching 1. Therefore $\big\|\bH_{-z}\big\| = O_p(q^{-1})$ (because the scaling constant at the corresponding block is $q$) and $\big\|\bH_{-a}\big\| = O_p(n^{-1})$.

Next, we bound $\bH_{zz}^{\dagger} - \{\partial_{\bphi_z\bphi_z}^2\cL(\bphi^*)\}^{-1} = \bH_{zz}^{\dagger} - (\bH_{Lzz}^*)^{-1}$. By Woodbury identity
\begin{equation}\begin{aligned}
    (\bH_{zz}^* )^{-1}  = &\,(\bH_{Lzz}^* + \bV^*_z\bV^*_z{}^\T)^{-1}  \\=&\,(\bH_{Lzz}^*)^{-1} -  \underbrace{(\bH_{Lzz}^*)^{-1}\bV_z^*\big\{\bI_{r^2} +  (\bV_z^*)^\T(\bH_{Lzz}^*)^{-1}\bV_z^*\big\}^{-1}(\bV_z^*)^\T (\bH_{Lzz}^*)^{-1}}_{\alpha}
\end{aligned}\label{eq_woodbury_expand_Hzz}\end{equation}
By \eqref{eq_bound_hzz} and \eqref{eq_bound_V}, we first note that $\big\|(\bV_z^*)^\T(\bH_{Lzz}^*)^{-1}\bV_z^*\big\| = O_p(q\times q^{-1}) =O_p(1)$. Since $(\bV_z^*)^\T(\bH_{Lzz}^*)^{-1}\bV_z^*$ is positive semi-definite, we note that 
\begin{equation*}
    \big\|\big\{\bI_{r^2} +  (\bV_z^*)^\T(\bH_{Lzz}^*)^{-1}\bV_z^*\big\}^{-1}\big\| = O_p(1).
\end{equation*}
Then for $\alpha$, with \eqref{eq_bound_hzz} and \eqref{eq_bound_V}, we obtain
\begin{equation*}
    \|\alpha\| = O_p(q^{-1}\times \sqrt{q}\times 1\times \sqrt{q}\times q^{-1}) = O_p(q^{-1}),
\end{equation*}
and by the block-diagonal structure of $(\bH_{Lzz}^*)^{-1}$, for each $i\in[n]$
\begin{align*}
    \|\alpha_{\cR_i,}\| \le &\,\big\|(\bH_{Lz_iz_i}^*)^{-1}\bV_{z_i}^*\big\{\bI_{r^2} +  (\bV_z^*)^\T(\bH_{Lzz}^*)^{-1}\bV_z^*\big\}^{-1}(\bV_z^*)^\T (\bH_{Lzz}^*)^{-1}\big\|\\\le &\,O_p(q^{-1}\times \sqrt{q/n}\times 1\times \sqrt{q}\times q^{-1}) = O_p(n^{-1/2}q^{-1}).
\end{align*}
Moreover, with \eqref{eq_bound_haz_inf}, we know
\begin{align*}
    \|\alpha\|_{2\to\infty}\le&\, \big\|(\bH_{Lzz}^*)^{-1}\bV_{z}^*\big\|_{2\to\infty}\big\|\big\{\bI_{r^2} +  (\bV_z^*)^\T(\bH_{Lzz}^*)^{-1}\bV_z^*\big\}^{-1}(\bV_z^*)^\T (\bH_{Lzz}^*)^{-1}\big\|\\\le &\,\max_{i\in[n]}\|(\bH_{Lz_iz_i}^*)^{-1}\|\|\bV_{z_i}^*\|\times O_p(q^{-1/2})\\ = &\,O_p(n^{-1/2}q^{-1}).
\end{align*}
Then with \eqref{eq_bound_hzz}, we arrive at
\begin{equation*}
    \big\|(\bH^*_{zz})^{-1}\big\| \le \|(\bH_{Lzz}^*)^{-1}\| + \|\alpha\| = O_p(q^{-1}),
\end{equation*}
\begin{equation*}
    \big\|\{(\bH^*_{zz})^{-1} - (\bH_{Lzz}^*)^{-1}\}_{\cR_i,}\big\| = \|\alpha_{\cR_i,}\| = O_p(n^{-1/2}q^{-1}),
\end{equation*}
and 
\begin{equation*}
    \big\|(\bH^*_{zz})^{-1} - (\bH_{Lzz}^*)^{-1}\big\|_{2\to\infty} = O_p(n^{-1/2}q^{-1}),
\end{equation*}
Similarly, we can obtain
\begin{equation*}
    \big\|(\bH^*_{aa})^{-1}\big\| = O_p(n^{-1}),\;\big\|(\bH^*_{aa})^{-1} - (\bH_{Laa}^*)^{-1}\big\|_{2\to\infty} = O_p(n^{-1}q^{-1/2}),
\end{equation*}
and for $j\in[q]$,
\begin{equation*}
    \big\|\{(\bH^*_{aa})^{-1} - (\bH_{Laa}^*)^{-1}\}_{\cR_j,}\big\|= O_p(n^{-1}q^{-1/2}).
\end{equation*}
Now for $\bH_{zz}^{\dagger}$, we have with \eqref{eq_bound_haz}, $\|(\bH_{zz}^*)^{-1}\| = O_p(q^{-1})$ and $\big\|\bH_{-a}\big\| = O_p(n^{-1})$ that
\begin{equation*}
    \big\|\bH^{\dagger}_{zz}\big\|\le \|(\bH_{zz}^*)^{-1}\| + \|(\bH_{zz}^*)^{-1}\|\|\bH_{za}^*\|\|\bH_{-a}^*\|\|\bH_{za}^*\|\|(\bH_{zz}^*)^{-1}\| = O_p(q^{-1}),
\end{equation*}
and for each $i\in[n]$,
\begin{equation}\label{eq_bound_compound_produc_1}\begin{aligned}
   &\, \big\|\{\bH^{\dagger}_{zz} - (\bH_{Lzz}^*)^{-1}\}_{\cR_i,}\big\|\\\le&\, \|\{(\bH_{zz}^*)^{-1} - (\bH_{Lzz}^*)^{-1}\}_{\cR_i,}\| + \|(\bH_{Lzz}^*)^{-1}_{\cR_i,}\bH_{za}^*\bH_{-a}^*\bH_{az}^*(\bH_{zz}^*)^{-1}\|\\&\,+ \|\{(\bH_{zz}^*)^{-1} - (\bH_{Lzz}^*)^{-1}\}_{\cR_i,}\bH_{za}^*\bH_{-a}^*\bH_{az}^*(\bH_{zz}^*)^{-1}\| \\\le&\, \|\{(\bH_{zz}^*)^{-1} - (\bH_{Lzz}^*)^{-1}\}_{\cR_i,}\|+ \|(\bH_{Lz_iz_i}^*)^{-1}\|\|\bH_{z_ia}^*\|\|\bH_{-a}^*\|\|\bH_{za}^*\|\|(\bH_{zz}^*)^{-1}\|\\&\,+ \|\{(\bH_{zz}^*)^{-1} - (\bH_{Lzz}^*)^{-1}\}_{\cR_i,}\|\|\bH_{za}^*\|\|\bH_{-a}^*\|\|\bH_{za}^*\|\|(\bH_{zz}^*)^{-1}\| \\=&\,O_p
    \big(n^{-1/2}q^{-1} + q^{-1}q^{1/2}n^{-1}(nq)^{1/2}q^{-1} + n^{-1/2}q^{-1} (nq)^{1/2}n^{-1}(nq)^{1/2}q^{-1}\big)\\=  &\,O_p
    (n^{-1/2}q^{-1}).
\end{aligned}\end{equation}
Furthermore, with $\big\|(\bH^*_{zz})^{-1} - (\bH_{Lzz}^*)^{-1}\big\|_{2\to\infty} = O_p(n^{-1/2}q^{-1})$,
\begin{equation}\label{eq_bound_compound_produc_3}\begin{aligned}
   &\, \big\|\bH^{\dagger}_{zz} - (\bH_{Lzz}^*)^{-1}\big\|_{2\to\infty}\\\le&\, \|(\bH_{zz}^*)^{-1} - (\bH_{Lzz}^*)^{-1}\|_{2\to\infty} + \|(\bH_{Lzz}^*)^{-1}\bH_{za}^*\|_{2\to\infty}\|\bH_{-a}^*\bH_{az}^*(\bH_{zz}^*)^{-1}\|\\&\,+ \|(\bH_{zz}^*)^{-1} - (\bH_{Lzz}^*)^{-1}\|_{2\to\infty}\|\bH_{za}^*\bH_{-a}^*\bH_{az}^*(\bH_{zz}^*)^{-1}\| \\\le&\, \|(\bH_{zz}^*)^{-1} - (\bH_{Lzz}^*)^{-1}\|_{2\to\infty} + \max_{i\in[n]}\|(\bH_{Lz_iz_i}^*)^{-1}\|\|\bH_{z_ia}^*\|\|\bH_{-a}^*\bH_{az}^*(\bH_{zz}^*)^{-1}\|\\&\,+ \|(\bH_{zz}^*)^{-1} - (\bH_{Lzz}^*)^{-1}\|_{2\to\infty}\|\bH_{za}^*\bH_{-a}^*\bH_{az}^*(\bH_{zz}^*)^{-1}\| \\=&\,O_p
    \big(n^{-1/2}q^{-1} + q^{-1}(q\log n)^{1/2}n^{-1}(nq)^{1/2}q^{-1} + n^{-1/2}q^{-1} (nq)^{1/2}n^{-1}(nq)^{1/2}q^{-1}\big)\\=  &\,O_p
    (n^{-1/2}q^{-1}\log^{1/2} n).
\end{aligned}\end{equation}
Here, we use $\max_{i\in[n]}\|\bH^*_{z_ia}\|\le \sqrt{r}\|\bH^*_{za}\|_{2\to\infty}=O_p(\sqrt{q\log n})$ from \eqref{eq_bound_haz_inf}.

For $\bH_{za}^{\dagger}$, invoke the expansion and apply a similar trick to get
\begin{equation}\label{eq_bound_compound_produc_2}\begin{aligned}
    \big\|(\bH_{za}^{\dagger})_{\cR_i,}\big\| =&\, \big\|\big\{(\bH_{zz}^*)^{-1}+(\bH_{zz}^*)^{-1}\bH_{za}^*\bH_{-a}^*\bH_{az}^*(\bH_{zz}^*)^{-1}\big\}_{\cR_i,}\bH_{ za}^*(\bH_{aa}^*)^{-1}\big\|\\\le &\,\big\|(\bH_{Lz_iz_i}^*)^{-1}\bH_{ z_ia}^*(\bH_{aa}^*)^{-1}\big\| + \big\|\big\{(\bH_{zz}^*)^{-1} - (\bH_{Lzz}^*)^{-1}\}_{\cR_i,}\bH_{ za}^*(\bH_{aa}^*)^{-1}\big\| \\&\,+ \big\|(\bH_{Lz_iz_i}^*)^{-1}\bH_{z_ia}^*\bH_{-a}^*\bH_{az}^*(\bH_{zz}^*)^{-1}\bH_{ za}^*(\bH_{aa}^*)^{-1}\big\|\\&\,+ \big\|\big\{(\bH_{zz}^*)^{-1} - (\bH_{Lzz}^*)^{-1}\big\}_{\cR_i,}\bH_{za}^*\bH_{-a}^*\bH_{az}^*(\bH_{zz}^*)^{-1}\bH_{ za}^*(\bH_{aa}^*)^{-1}\big\|\\= &\,O_p(n^{-1}q^{-1/2}),
\end{aligned}\end{equation}
for each $i\in[n]$, and
\begin{equation}\label{eq_bound_compound_produc_4}\begin{aligned}
    \big\|\bH_{za}^{\dagger}\big\|_{2\to\infty} \le &\,\big\|(\bH_{Lzz}^*)^{-1}\bH_{ za}^*\big\|_{2\to\infty}\|(\bH_{aa}^*)^{-1}\|\\&\, + \big\|\big\{(\bH_{zz}^*)^{-1} - (\bH_{Lzz}^*)^{-1}\big\}\big\|_{2\to\infty}\big\|\bH_{ za}^*(\bH_{aa}^*)^{-1}\big\| \\&\,+ \big\|(\bH_{Lzz}^*)^{-1}\bH_{za}^*\big\|_{2\to\infty}\big\|\bH_{-a}^*\bH_{az}^*(\bH_{zz}^*)^{-1}\bH_{ za}^*(\bH_{aa}^*)^{-1}\big\|\\&\,+ \big\|\big\{(\bH_{zz}^*)^{-1} - (\bH_{Lzz}^*)^{-1}\big\}\big\|_{2\to\infty}\big\|\bH_{za}^*\bH_{-a}^*\bH_{az}^*(\bH_{zz}^*)^{-1}\bH_{ za}^*(\bH_{aa}^*)^{-1}\big\|\\\le &\,\max_{i\in[n]}\big\|(\bH_{Lz_iz_i}^*)^{-1}\bH_{ z_ia}^*\big\|\big\|(\bH_{aa}^*)^{-1}\big\| \\&\,+ \max_{i\in[n]}\big\|\big\{(\bH_{zz}^*)^{-1} - (\bH_{Lzz}^*)^{-1}\}_{\cR_i,}\big\|\big\|\bH_{ za}^*(\bH_{aa}^*)^{-1}\big\| \\&\,+ \max_{i\in[n]}\big\|(\bH_{Lz_iz_i}^*)^{-1}\bH_{z_ia}^*\big\|\big\|\bH_{-a}^*\bH_{az}^*(\bH_{zz}^*)^{-1}\bH_{ za}^*(\bH_{aa}^*)^{-1}\big\|\\&\,+ \max_{i\in[n]}\big\|\big\{(\bH_{zz}^*)^{-1} - (\bH_{Lzz}^*)^{-1}\big\}_{\cR_i,}\big\|\big\|\bH_{za}^*\bH_{-a}^*\bH_{az}^*(\bH_{zz}^*)^{-1}\bH_{ za}^*(\bH_{aa}^*)^{-1}\big\|\\= &\,O_p(n^{-1}q^{-1/2}\log^{1/2}n),
\end{aligned}\end{equation}

Similarly, we can show that for each $j\in[q]$,
\begin{equation*}
    \big\|\{\bH_{aa}^{\dagger} - (\bH_{Laa}^*)^{-1}\}_{\cR_j,}\big\| = O_p(n^{-1}q^{-1/2});\;\big\|\bH_{aa}^{\dagger} - (\bH_{Laa}^*)^{-1}\big\|_{2\to\infty}=O_p(n^{-1}q^{-1/2}\log^{1/2} q),
\end{equation*}
and for each $j\in[q]$,
\begin{equation*}
    \big\|(\bH_{az}^{\dagger})_{\cR_j,}\big\|=O_p(n^{-1/2}q^{-1});\;\big\|\bH_{az}^{\dagger}\big\|_{2\to\infty}=O_p(n^{-1/2}q^{-1}\log^{1/2} q).
\end{equation*}
Putting these altogether, we obtain \eqref{lemma_hessian_property_2_eq1} and \eqref{lemma_hessian_property_2_eq3}.
For \eqref{lemma_hessian_property_2_eq5}, we already know $\|\bH_{zz}^{\dagger}\| = O_p(q^{-1})$ and $\|(\bH_{Lzz}^*)^{-1}\| = O_p(q^{-1})$. For $\bH_{za}^{\dagger}$, repeat the calculation of \eqref{eq_bound_compound_produc_4} and we know $\|\bH_{za}^{\dagger}\| = O_p((nq)^{-1/2})$. Similarly, one can check $\|\bH_{aa}^{\dagger}\| = O_p(n^{-1})$ and $\|(\bH_{Laa}^*)^{-1}\| = O_p(n^{-1})$, which proves \eqref{lemma_hessian_property_2_eq5}.

For \eqref{lemma_hessian_property_2_eq2}--\eqref{lemma_hessian_property_2_eq6}, it can be similarly established by repeating the above procedure and the following lemma.
\begin{lemma}\label{lemma_score_hessian_concern}
    Under the setup of Lemma~\ref{lemma_hessian_property_2}, we have 
    \begin{align*}
    (i)\;&\big\|(\bV_z^*)^\T(\bH_{Lzz}^*)^{-1}\partial_{\bphi_z}\cL(\bphi^*)\big\|=O_p(1 + \sqrt{n/q});\;\big\|(\bV_a^*)^\T(\bH_{Laa}^*)^{-1}\partial_{\bphi_a}\cL(\bphi^*)\big\|=O_p(1 + \sqrt{q/n});\\
        (ii)\;& 
        \big\|\bH_{az}(\bH_{zz}^*)^{-1}\partial_{\bphi_z}\cL(\bphi^*)\big\| = O_p(n/\sqrt{q} + \sqrt{n});\\
        (iii)\;& 
        \big\|\bH_{za}(\bH_{aa}^*)^{-1}\partial_{\bphi_a}\cL(\bphi^*)\big\| = O_p(q/\sqrt{n} + \sqrt{q});
    \end{align*}
\end{lemma}

\begin{proof}
See Section~\ref{supp_sec_prove_lemma_score_hessian_concern}.\end{proof}

Then, with $(i)$ of Lemma~\ref{lemma_score_hessian_concern} and the Woodbury identity, we can similarly obtain that for each $i\in[n]$
\begin{align*}
    &\,\big\|\{(\bH^*_{zz})^{-1} - (\bH_{Lzz}^*)^{-1}\}_{\cR_i,}\partial_{\bphi_z}\cL(\bphi^*)\big\| \\=&\, \big\|(\bH_{Lz_iz_i}^*)^{-1}\bV_{z_i}^*\big\{\bI_{r^2} +  (\bV_z^*)^\T(\bH_{Lzz}^*)^{-1}\bV_z^*\big\}^{-1}(\bV_z^*)^\T (\bH_{Lzz}^*)^{-1}\partial_{\bphi_z}\cL(\bphi^*)\big\|\\\le &\,O_p(q^{-1}\times \sqrt{q/n}\times (1+ \sqrt{n/q})) = O_p(q^{-1} + (nq)^{-1/2}),
\end{align*}
and similarly,
\begin{align*}
    &\,\big\|\{(\bH^*_{zz})^{-1} - (\bH_{Lzz}^*)^{-1}\}\partial_{\bphi_z}\cL(\bphi^*)\big\|_{\infty} \\=&\, \max_{i\in[n]}\big\|(\bH_{Lz_iz_i}^*)^{-1}\bV_{z_i}^*\big\{\bI_{r^2} +  (\bV_z^*)^\T(\bH_{Lzz}^*)^{-1}\bV_z^*\big\}^{-1}(\bV_z^*)^\T (\bH_{Lzz}^*)^{-1}\partial_{\bphi_z}\cL(\bphi^*)\big\|\\\le &\,O_p(q^{-1}\times \sqrt{q/n}\times (1+ \sqrt{n/q})) = O_p(q^{-1} + (nq)^{-1/2}),
\end{align*}

Then, for the decomposition
\begin{align*}
    &\, \big\|\{\bH^{\dagger}_{zz} - (\bH_{Lzz}^*)^{-1}\}_{\cR_i,}\partial_{\bphi_z}\cL(\bphi^*)\big\|\\\le&\, \|\{(\bH_{zz}^*)^{-1} - (\bH_{Lzz}^*)^{-1}\}_{\cR_i,}\partial_{\bphi_z}\cL(\bphi^*)\| + \|(\bH_{Lzz}^*)^{-1}_{\cR_i,}\bH_{za}^*\bH_{-a}^*\bH_{az}^*(\bH_{zz}^*)^{-1}\partial_{\bphi_z}\cL(\bphi^*)\|\\&\,+ \|\{(\bH_{zz}^*)^{-1} - (\bH_{Lzz}^*)^{-1}\}_{\cR_i,}\bH_{za}^*\bH_{-a}^*\bH_{az}^*(\bH_{zz}^*)^{-1}\partial_{\bphi_z}\cL(\bphi^*)\|,
\end{align*}
with $(ii)$ of Lemma~\ref{lemma_score_hessian_concern} and $\big\|\{(\bH^*_{zz})^{-1} - (\bH_{Lzz}^*)^{-1}\}_{\cR_i,}\partial_{\bphi_z}\cL(\bphi^*)\big\| = O_p(q^{-1} + (nq)^{-1/2})$, follow \eqref{eq_bound_compound_produc_1} and we can get 
\begin{equation*}
    \big\|\{\bH^{\dagger}_{zz} - (\bH_{Lzz}^*)^{-1}\}_{\cR_i,}\partial_{\bphi_z}\cL(\bphi^*)\big\| = O_p(q^{-1} + (nq)^{-1/2})\text{ for each }i\in[n];
\end{equation*}
with $(ii)$ of Lemma~\ref{lemma_score_hessian_concern} and $\big\|\{(\bH^*_{zz})^{-1} - (\bH_{Lzz}^*)^{-1}\}\partial_{\bphi_z}\cL(\bphi^*)\big\|_{\infty} = O_p(q^{-1} + (nq)^{-1/2})$, follow \eqref{eq_bound_compound_produc_3} and we get
\begin{align*}
    \big\|\{\bH^{\dagger}_{zz} - (\bH_{Lzz}^*)^{-1}\}\partial_{\bphi_z}\cL(\bphi^*)\big\|_{\infty}\le&\,\max_{i\in[n]}\big\|\{\bH^{\dagger}_{zz} - (\bH_{Lzz}^*)^{-1}\}_{\cR_i,}\partial_{\bphi_z}\cL(\bphi^*)\big\|\\ =&\, O_p(\{q^{-1} + (nq)^{-1/2}\}\log^{1/2}n).
\end{align*}
For $\big\|\{\bH^{\dagger}_{zz} - (\bH_{Lzz}^*)^{-1}\}\partial_{\bphi_z}\cL(\bphi^*)\big\|$, with $(ii)$ of Lemma~\ref{lemma_score_hessian_concern}, \eqref{eq_bound_h_ast}, and $\|\bH_{-a}^*\| = O_p(n^{-1})$ there is  
\begin{align*}
    &\, \big\|\{\bH^{\dagger}_{zz} - (\bH_{Lzz}^*)^{-1}\}\partial_{\bphi_z}\cL(\bphi^*)\big\|\\\le&\, \|\{(\bH_{zz}^*)^{-1} - (\bH_{Lzz}^*)^{-1}\}\partial_{\bphi_z}\cL(\bphi^*)\| + \|(\bH_{zz}^*)^{-1}\bH_{za}^*\bH_{-a}^*\bH_{az}^*(\bH_{zz}^*)^{-1}\partial_{\bphi_z}\cL(\bphi^*)\|\\=&\,\big\|(\bH_{Lzz}^*)^{-1}\bV_{z}^*\big\{\bI_{r^2} +  (\bV_z^*)^\T(\bH_{Lzz}^*)^{-1}\bV_z^*\big\}^{-1}(\bV_z^*)^\T (\bH_{Lzz}^*)^{-1}\partial_{\bphi_z}\cL(\bphi^*)\big\| \\&\,+ O_p\big(q^{-1}\times (nq)^{1/2}\times n^{-1}\times (n/\sqrt{q} + \sqrt{n}) \big)\\ = &\,O_p\big(\sqrt{n}(q^{-1} + (nq)^{-1/2})\big).
\end{align*}

Similarly with $(iii)$ of Lemma~\ref{lemma_score_hessian_concern} and follow \eqref{eq_bound_compound_produc_2} to get 
\begin{align*}
    \big\|(\bH_{za}^{\dagger})_{\cR_i,}\partial_{\bphi_a}\cL(\bphi^*)\big\|  = O_p(n^{-1} + (nq)^{-1/2})\text{ for each }i\in[n];
\end{align*}
follow \eqref{eq_bound_compound_produc_4} to get
\begin{align*}
    \big\|\bH_{za}^{\dagger}\partial_{\bphi_a}\cL(\bphi^*)\big\|_{\infty}  = O_p(\{n^{-1} + (nq)^{-1/2}\}\log^{1/2}n),
\end{align*}
and \begin{align*}
    \big\|\bH_{za}^{\dagger}\partial_{\bphi_a}\cL(\bphi^*)\big\|  = O_p\big(\sqrt{n}(n^{-1} + (nq)^{-1/2})\big),
\end{align*}
By symmetry, we can show that for each $j\in[q]$,
\begin{equation*}
    \big\|\{\bH_{aa}^{\dagger} - (\bH_{Laa}^*)^{-1}\}_{\cR_j,}\partial_{\bphi_a}\cL(\bphi^*)\big\| = O_p(n^{-1} + (nq)^{-1/2});
\end{equation*}
\begin{equation*}
    \big\|\{\bH_{aa}^{\dagger} - (\bH_{Laa}^*)^{-1}\}\partial_{\bphi_a}\cL(\bphi^*)\big\|_{\infty} = O_p(\{n^{-1} + (nq)^{-1/2}\}\log^{1/2}q);
\end{equation*}
\begin{equation*}
    \big\|\{\bH_{aa}^{\dagger} - (\bH_{Laa}^*)^{-1}\}\partial_{\bphi_a}\cL(\bphi^*)\big\| = O_p\big(\sqrt{q}(n^{-1} + (nq)^{-1/2})\big),
\end{equation*}
and for each $j\in[q]$,
\begin{equation*}
    \big\|(\bH_{az}^{\dagger})_{\cR_j,}\partial_{\bphi_z}\cL(\bphi^*)\big\|=O_p(q^{-1} + (nq)^{-1/2});
\end{equation*}
\begin{equation*}
    \big\|\bH_{az}^{\dagger}\partial_{\bphi_z}\cL(\bphi^*)\big\|_{\infty}=O_p(\{q^{-1} + (nq)^{-1/2}\}\log^{1/2}q);\;\big\|\bH_{az}^{\dagger}\partial_{\bphi_z}\cL(\bphi^*)\big\|=O_p(\sqrt{q}(n^{-1} + (nq)^{-1/2})).
\end{equation*}
Putting these altogether, we obtain \eqref{lemma_hessian_property_2_eq2}--\eqref{lemma_hessian_property_2_eq6}.

\subsection{Proof of Lemma~\ref{lemma_asymp_expand_res}}\label{supp_sec_prove_lemma_asymp_expand_res}
    Recall that when $i\le n$ and $l\in[r]$, the $\{(i-1)r+l\}$-th entry of $\bR$ is $(\hat\bphi^o - \bphi^*)^\T\partial_{\bphi\bphi z_{il}}^3\cL_{\nu}(\tilde\bphi^{\flat})(\hat\bphi^o - \bphi^*) = (\hat\bphi^o - \bphi^*)^\T\partial_{\bphi\bphi z_{il}}^3\cL(\tilde\bphi^{\flat})(\hat\bphi^o - \bphi^*) + \nu nq(\hat\bphi^o - \bphi^*)^\T\partial_{\bphi\bphi z_{il}}^3\cP^*(\tilde\bphi^{\flat})(\hat\bphi^o - \bphi^*)$. For the first part, recall that the second-order derivative has block form
    \begin{equation*}
    \partial_{\bphi\bphi}^2\cL(\bphi) = \begin{pmatrix}
        \bH_{Lzz}(\bphi)  & \bH_{Lza}(\bphi) \\
        \bH_{Laz}(\bphi) & \bH_{Laa}(\bphi) 
    \end{pmatrix},
\end{equation*}
with $\bH_{Lzz}(\bphi) = \partial_{\bphi_z\bphi_z}^2\cL(\bphi)$, $\bH_{Laa}(\bphi) = \partial_{\bphi_a\bphi_a}^2\cL(\bphi)$, and $\bH_{Lza}(\bphi) = \{\bH_{Laz}(\bphi)\}^\T = \partial_{\bphi_z\bphi_a}^2\cL(\bphi)$; and $\bphi_z$, $\bphi_a$ given in Lemma~\ref{lemma_hessian_property}.
    Then one can calculate
    \begin{align}
        (\hat\bphi^o - \bphi^*)^\T\partial_{\bphi\bphi z_{il}}^3\cL(\tilde\bphi^{\flat})(\hat\bphi^o - \bphi^*) = &\,(\hat\bphi^o_z - \bphi^*_z)^\T\partial_{z_{il}}\bH_{Lzz}(\tilde\bphi^{\flat})(\hat\bphi^o_z - \bphi^*_z)\label{eq_expand_third_order_cl_1} \\&\, + (\hat\bphi^o_a - \bphi^*_a)^\T\partial_{z_{il}}\bH_{Laa}(\tilde\bphi^{\flat})(\hat\bphi^o_a - \bphi^*_a) \label{eq_expand_third_order_cl_2}\\&\,+ 2(\hat\bphi^o_z - \bphi^*_z)^\T\partial_{z_{il}}\bH_{Lza}(\tilde\bphi^{\flat})(\hat\bphi^o_a - \bphi^*_a).\label{eq_expand_third_order_cl_3}
    \end{align}
    For \eqref{eq_expand_third_order_cl_1}, by the block-diagonal structure of $\bH_{Lzz}(\cdot)$, we have
    \begin{align*}
        |\eqref{eq_expand_third_order_cl_1}| &\le\big\|\partial_{z_{il}}\big\{\sum_{j=1}^q\ell_{ij}^{\prime\prime}\big(\ba_j^\T\bz_i\big)\ba_j\ba_j^\T\big\}\big|_{\tilde\bphi^{\flat}}\big\|\big\|\hat\bz_i^o - \bz_i^*\big\|^2\\&\le \big\|\sum_{j=1}^q\ell_{ij}^{\prime\prime\prime}\big(\tilde\ba_j^{\flat}{}^\T\tilde\bz_i^{\flat}\big)\tilde\ba_j^{\flat}\tilde\ba_j^{\flat}{}^\T\tilde a_{jl}^{\flat}\big\|\big\|\hat\bz_i^o - \bz_i^*\big\|^2 = O_p\big(q/(n\wedge q)\big),
    \end{align*}
    where the last inequality follows from $|\ell_{ij}^{\prime\prime\prime}(\theta)|\le b_U$ by Assumption~\ref{assump_limit}, $\|\hat\bz_i^o - \bz_i^*\| = O_p((n\wedge q)^{-1/2})$ by \eqref{eq_ind_consistency_Zo}, and the boundedness of $\tilde\bphi^{\flat}$ as $\bphi^*,\hat\bphi^o\in\Theta^*(\epsilon_{nq})$. Similarly, one has
    \begin{align*}
        \sum_{i=1}^n\sum_{l=1}^r|\eqref{eq_expand_third_order_cl_1}|^2 & \le \sum_{i=1}^n\sum_{l=1}^r\big\|\sum_{j=1}^q\ell_{ij}^{\prime\prime\prime}\big(\tilde\ba_j^{\flat}{}^\T\tilde\bz_i^{\flat}\big)\tilde\ba_j^{\flat}\tilde\ba_j^{\flat}{}^\T\tilde a_{jl}^{\flat}\big\|^2\big\|\hat\bz_i^o - \bz_i^*\big\|^4 \le O\big(q^2\sum_{i=1}^n\|\hat\bz_i^o - \bz_i^*\|^4\big)\\& = O_p\big(nq^2/(n\wedge q)^2\big) 
    \end{align*}
    For \eqref{eq_expand_third_order_cl_2}, we use the block-diagonal structure of $\bH_{Laa}(\cdot) $ to get
    \begin{align*}
        |\eqref{eq_expand_third_order_cl_2}| = &\, \Big|\sum_{j=1}^q(\hat\ba_j^o - \ba_j^*)^\T \partial_{z_{il}}\big\{\ell_{ij}^{\prime\prime}\big(\ba_j^\T\bz_i\big)\bz_i\bz_i^\T\big\}\big|_{\tilde\bphi^\flat}(\hat\ba_j^o - \ba_j^*)\Big|\\=  &\,\Big|\sum_{j=1}^q(\hat\ba_j^o - \ba_j^*)^\T \big\{\ell_{ij}^{\prime\prime\prime}\big(\tilde\ba_j^{\flat}{}^\T\tilde\bz_i^{\flat}\big)\tilde\bz_i^{\flat}\tilde\bz_i^{\flat}{}^\T \tilde a_{jl}^{\flat} + \ell_{ij}^{\prime\prime}\big(\tilde\ba_j^{\flat}{}^\T\tilde\bz_i^{\flat}\big)(\tilde\bz_i^{\flat}\be_l^\T + \be_l\tilde\bz_i^{\flat}{}^\T)\big\}(\hat\ba_j^o - \ba_j^*)\Big|\\\le &\,O\big(\sum_{j=1}^q\|\hat\ba_j^o - \ba_j^*\|^2\big)\\ = &\,O_p\big(q/(n\wedge q)\big).
    \end{align*}
Here, the inequality follows from the boundedness of $\ell_{ij}^{\prime\prime}(\cdot)$, $\ell_{ij}^{\prime\prime\prime}(\cdot)$, and $\tilde\bphi^{\flat}$; the last equality follows from \eqref{eq_ave_consistency_Ao}. Similarly,
\begin{equation*}
    \sum_{i=1}^n\sum_{l=1}^r|\eqref{eq_expand_third_order_cl_2}|^2 = O_p\big(nq^2/(n\wedge q)^2\big).
\end{equation*}

For \eqref{eq_expand_third_order_cl_3}, we calculate
\begin{align*}
    &\,|\eqref{eq_expand_third_order_cl_3}|\\ = &\, 2\Big|\sum_{j=1}^q(\hat\bz_i^o - \bz_i^*)^\T\partial_{z_{il}}\big\{\ell_{ij}^{\prime\prime}(\ba_j^\T\bz_i)\ba_j\bz_i^\T + \ell_{ij}^{\prime}(\ba_j^\T\bz_i)\bI_r\big\}\big|_{\tilde\bphi^{\flat}}(\hat\ba_j^o - \ba_j^*)\Big|\\\le  &\,2\|\hat\bz_i^o - \bz_i^*\|\big\|\sum_{j=1}^q\big\{\ell_{ij}^{\prime\prime\prime}(\tilde\ba_j^{\flat}{}^\T\tilde\bz_i^{\flat})\tilde\ba_j^{\flat}\tilde\bz_i^{\flat}{}^\T \tilde a_{jr}^{\flat} + \ell_{ij}^{\prime\prime}(\tilde\ba_j^{\flat}{}^\T\tilde\bz_i^{\flat})\tilde\ba_j^{\flat}\be_l{}^\T + \ell_{ij}^{\prime\prime}(\tilde\ba_j^{\flat}{}^\T\tilde\bz_i^{\flat})\bI_r\tilde a_{jl}^{\flat}\big\}(\hat\ba_j^o - \ba_j^*)\big\|\\\le  &\,O_p((n\wedge q)^{-1/2})\times \big(\sum_{j=1}^q\|\hat\ba_j^o - \ba_j^*\|^2\big)^{1/2}\\&\qquad\qquad\times \Big(\sum_{j=1}^q\big\|\ell_{ij}^{\prime\prime\prime}(\tilde\ba_j^{\flat}{}^\T\tilde\bz_i^{\flat})\tilde\ba_j^{\flat}\tilde\bz_i^{\flat}{}^\T \tilde a_{jr}^{\flat} + \ell_{ij}^{\prime\prime}(\tilde\ba_j^{\flat}{}^\T\tilde\bz_i^{\flat})\tilde\ba_j^{\flat}\be_l{}^\T + \ell_{ij}^{\prime\prime}(\tilde\ba_j^{\flat}{}^\T\tilde\bz_i^{\flat})\bI_r\tilde a_{jl}^{\flat}\big\|^2\Big\}^{1/2}\\=&\,O_p\big(q/(n\wedge q)\big)
\end{align*}Here, the last inequality follows from {Cauchy–Schwarz inequality} and \eqref{eq_ind_consistency_Zo}; the last equality follows from \eqref{eq_ave_consistency_Ao} and the boundedness of $\ell_{ij}^{\prime\prime}(\cdot)$, $\ell_{ij}^{\prime\prime\prime}(\cdot)$, and $\tilde\bphi^{\flat}$. Similarly, 
\begin{equation*}
    \sum_{i=1}^n\sum_{l=1}^r|\eqref{eq_expand_third_order_cl_3}|^2 = O_p\big(nq^2/(n\wedge q)^2\big).
\end{equation*}

Next, we handle the second part $\nu nq(\hat\bphi^o - \bphi^*)^\T\partial_{\bphi\bphi z_{il}}^3\cP^*(\tilde\bphi^{\flat})(\hat\bphi^o - \bphi^*)$. Recall that 
\begin{equation}\begin{aligned}
    \partial_{\bphi\bphi}^2\cP^*(\bZ,\bA) = &\,\sum_{k=1}^r\sum_{l=1}^r(nq)^{-1}\bS_{nq}^{-1}\{\bv_{kl}(\bphi) + \bu_{kl}(\bphi)\}\{\bv_{kl}(\bphi) + \bu_{kl}(\bphi)\}^\T\bS_{nq}^{-1} \\&\,+ (nq)^{-1}\bS_{nq}^{-1}\sum_{k=1}^r\sum_{l=1}^r\omega_{kl}(\bphi)\bE_{kl}\bS_{nq}^{-1},
    \end{aligned}
\end{equation}
where for $k,l\in[r]$, $\bv_{kl}(\bphi)$, $\bu_{kl}(\bphi)$, and $\bE_{kl}(\bphi)$ are defined in \eqref{eq_define_vkl}, \eqref{eq_define_ukl}, and \eqref{eq_define_Ekl}, respectively. By definition, one has
\begin{equation*}
    \partial_{z_{il}}\{\bv_{kh}(\bphi) + \bu_{kh}(\bphi)\} = n^{-1/2}\{1_{(l=h)}\be_{(i-1)r + k}^{(nr+qr)}+1_{(l=k)}\be_{(i-1)r + h}^{(nr+qr)}\},
\end{equation*}
and $\partial_{z_{il}}\omega_{kh}(\bphi) = 1_{(l=h)}n^{-1}z_{ik} + 1_{(l=k)}n^{-1}(z_{ih}-z_{ih}^*)$. Therefore,
\begin{align}
    &\,nq(\hat\bphi^o - \bphi^*)^\T\partial_{\bphi\bphi z_{il}}^3\cP^*(\tilde\bphi^{\flat})(\hat\bphi^o - \bphi^*)\nonumber\\ = &\,2\sum_{k=1}^rq^{1/2}n^{-1/2}(\hat z_{ik}^o - z_{ik}^*)\{\bv_{kl}(\tilde\bphi^{\flat}) + \bu_{kl}(\tilde\bphi^{\flat})+\bv_{lk}(\tilde\bphi^{\flat}) + \bu_{lk}(\tilde\bphi^{\flat})\}^\T\bS_{nq}^{-1}(\hat\bphi^o-\bphi^*)\label{eq_expand_third_order_cp_1}\\&\,2\sum_{k=1}^rn^{-1}(2\tilde z_{ik}^{\flat}-z_{ik}^*)\nonumber\\&\quad\times \big\{q\sum_{s=1}^n(\hat z_{sk}^o - z_{sk}^*)(\hat z_{sl}^o - z_{sl}^*) -n\sum_{j=1}^q(\hat a_{jk}^o - a_{jk}^*)(\hat a_{jl}^o - a_{jl}^*) \big\}.\label{eq_expand_third_order_cp_2}
\end{align}
For \eqref{eq_expand_third_order_cp_1}, one has
\begin{align*}
    |\eqref{eq_expand_third_order_cp_1}|\le &\,Cn^{-1/2}q^{1/2}\|\hat \bz_{i}^o - \bz_{i}^*\|\max_{k\in[r]}\|\bv_{kl}(\tilde\bphi^{\flat}) + \bu_{kl}(\tilde\bphi^{\flat})+\bv_{lk}(\tilde\bphi^{\flat}) + \bu_{lk}(\tilde\bphi^{\flat})\|\\ &\,\quad\times (nq)^{1/2}\|(nq)^{-1/2}\bS_{nq}^{-1}(\hat\bphi^o - \bphi^*)\|\\ = &\,O_p\Big(n^{-1/2}q^{1/2}\times (n\wedge q)^{-1/2}\times 1 \times (nq)^{1/2}\times (n\wedge q)^{-1/2}\Big) = O_p\big(q/(n\wedge q)\big).
\end{align*}
Here, $\|\bv_{kl}(\tilde\bphi^{\flat}) + \bu_{kl}(\tilde\bphi^{\flat})\| = O_p(1)$ from the boundedness of $\hat\bphi^{\flat}$ and definition of $\bv_{kl}(\bphi)$ and $\bu_{kl}(\bphi)$; and the first inequality follows from \eqref{eq_ave_consistency_Ao}, \eqref{eq_ave_consistency_Zo}, and \eqref{eq_ind_consistency_Zo}. Similarly,
\begin{equation*}
    \sum_{i=1}^n\sum_{l=1}^r|\eqref{eq_expand_third_order_cp_1}|^2 = O_p\big(nq^2/(n\wedge q)^2\big).
\end{equation*}
For \eqref{eq_expand_third_order_cp_2}, we have
\begin{equation*}
    |\eqref{eq_expand_third_order_cp_2}| \le O\big(n^{-1}\times (nq)\big\{n^{-1}\|\hat\bZ^o - \bZ^*\|_{\Fn}^2 + q^{-1}\|\hat\bA^o - \bA^*\|_{\Fn}^2\big\}\big) =O_p\big(q/(n\wedge q)\big).
\end{equation*}Similarly,
\begin{equation*}
    \sum_{i=1}^n\sum_{l=1}^r|\eqref{eq_expand_third_order_cp_2}|^2 = O_p\big(nq^2/(n\wedge q)^2\big).
\end{equation*}

Putting these altogether, we know when $i\in[n]$ $\|(\bS_{nq}^2\bR)_{(i-1)r+l}\| =q^{-1}\|\bR_{(i-1)r+l}\| = O_p\big((n\wedge q)^{-1}\big)$ and $\|(\bS_{nq}\bR)_{1:nr}\| =q^{-1/2}\|\bR_{1:nr}\| = O_p\big(\sqrt{nq}/(n\wedge q)\big)$. Similarly, one can show that when $n<i\le n+q$ $\|(\bS_{nq}^2\bR)_{(i-1)r+l}\| =n^{-1}\|\bR_{(i-1)r+l}\| = O_p\big((n\wedge q)^{-1}\big)$ and $\|(\bS_{nq}\bR)_{(nr+1):(nr+nq)}\| =n^{-1/2}\|\bR_{(nr+1):(nr+qr)}\| = O_p\big(\sqrt{nq}/(n\wedge q)\big)$. Combine these two parts and we finish the proof.

\subsection{Proof of Lemma~\ref{lemma_unif_sub_exp}}\label{sec_prove_lemma_unif_sub_exp}
Without loss of generality, we assume that $\mu_{jk}$ are positive. The proof can be readily extended to the general case of $\mu_{jk}$ by symmetry. Invoke the expression of $\xi_{jk}$ and one has the tail probability at
\begin{align*}
    \PP(|\xi_{jk}| \ge t)\le&\, \PP\left(\big\|\sum_{i=1}^n\ell_{ij}^{\prime}(\theta_{ij}^*)\bz_i^*\big\|\ge n^{-1/2}\min_{j\in[q]}\lambda_{\min}\Big\{-\sum_{i=1}^n\ell_{ij}^{\prime\prime}(\theta_{ij}^*)\bz_i^*\bz_i^*\Big\}t\right)\\\le &\,\PP\left(\big\|\sum_{i=1}^n\ell_{ij}^{\prime}(\theta_{ij}^*)\bz_i^*\big\|\ge t\big\{b_U\lambda_{\min}(\bSigma_f^*)\sqrt{n}/2\big\}\right)\\
    \le &\,2\exp\left(-c\min\big\{t^2,t\sqrt{n}\big\}\right)\quad :=\gamma_t,
\end{align*}
for some absolute constant $c$ independent of $j$ and $k$. Here the last step follows from that $\max_{i\in[n]}\|\bz_i^*\|$ and $\max_{j\in[q]}\|\ba_j^*\|$ are bounded and {$\|\ell_{ij}^{\prime}(\theta_{ij}^*)\|_{\psi_1}\le C$} uniformly over $i,j$.

Next, we note that
\begin{align*}
    &\,\PP\left\{|\xi_{jk} - \mu_{jk}| - \EE\big[|\xi_{jk} - \mu_{jk}|\big] \ge t\right\}
    \\\le&\, \PP\left\{\xi_{jk}  \ge t + \EE\big[|\xi_{jk} - \mu_{jk}|  +\mu_{jk}\big]\right\} +\PP  \left\{\xi_{jk}  \le -t - \EE\big[|\xi_{jk} - \mu_{jk}|  +\mu_{jk}\big]\right\}
    \\\le&\, \PP\left\{\xi_{jk}  \ge t \right\} +\PP  \left\{\xi_{jk}  \le -t - \EE\big[|\xi_{jk} - \mu_{jk}|  +\mu_{jk}\big]\right\}\\
    \le &\, \gamma_t + \PP  \left\{\xi_{jk}  \le -t - \EE\big[|\xi_{jk} - \mu_{jk}|  +\mu_{jk}\big]\right\}.
\end{align*}
To control $\mu_{jk} - \EE\big[|\xi_{jk} - \mu_{jk}|]$, we first select some universal threshold $t_0$. Then since $\EE|\xi_{jk}|$ is bounded, there exists some constant $C_1$ such that $|\mu_{jk} - \EE\big[|\xi_{jk} - \mu_{jk}|]|\le C_1$ whenever $\mu_{jk}\le t_0$. When $\mu_{jk}>t_0$, we compute 
\begin{align*}
    \mu_{jk} - \EE\big[|\xi_{jk} - \mu_{jk}|] =&\, \int \left(\mu_{jk} - |\xi_{jk} - \mu_{jk}|\right) d\PP_{\xi_{jk}}\\=&\,\int_{\xi_{jk}\le \mu_{jk}} \xi_{jk} d\PP_{\xi_{jk}} + \int_{\xi_{jk}> \mu_{jk}}  \left(2\mu_{jk} - \xi_{jk} \right) d\PP_{\xi_{jk}}\\\le &\,\EE|\xi_{jk}| + \mu_{jk}\gamma_{\mu_{jk}}\\
    \le &\,\EE|\xi_{jk}| + 2t_0\gamma_{t_0}.
\end{align*}
The last inequality follows from that, when $t_0$ is large enough, $x\gamma_x$ is non-increasing in $[t_0,\infty)$, as $\gamma_x$ represents the sub-exponential tail. These together implies that $\mu_{jk} - \EE\big[|\xi_{jk} - \mu_{jk}|]$ is upper bounded, denoted by $C_2$. Then one has
\begin{equation*}
    \PP\left\{|\xi_{jk} - \mu_{jk}| - \EE\big[|\xi_{jk} - \mu_{jk}|\big] \ge t\right\} \le \gamma_t + \PP  \left\{\xi_{jk}  \le -t + C_2\right\} = \gamma_t + \gamma_{t-C_2}.
\end{equation*}
Since $t - C_2\ge t/2$ when $t$ is large enough, we conclude that 
\begin{equation}
    \PP\left\{|\xi_{jk} - \mu_{jk}| - \EE\big[|\xi_{jk} - \mu_{jk}|\big] \ge t\right\}\le 4\exp\big(-4c\min\{t^2,t\sqrt{n}\}\big).
\end{equation}
Similar results can be obtained for the left tail, and thus we know {$\big\||\xi_{jk} - \mu_{jk}| - \EE\big[|\xi_{jk} - \mu_{jk}|\big]\big\|_{\psi_1}\le C$} for some constant $C$ uniformly over $j,r,\mu_{jk}$.

\subsection{Proof of Lemma~\ref{lemma_fold_nor_dif}}\label{sec_prove_lemma_fold_nor_dif}
    Without loss of generality, we assume $a>0$. Invoke the mean of the folded normal distribution, and we obtain
     \begin{equation*}
        \EE[|\xi-a|] - \EE[|\xi|] = a\times \mathrm{erf}\Big(\frac{a}{\sqrt{2\phi^2}}\Big) + \phi\sqrt{2/\pi}\Big[\exp\big\{-a^2/(2\phi^2)\big\} - 1\Big].
    \end{equation*}
    where $\mathrm{erf}(x) = 2\pi^{-1/2}\int_0^xe^{-t^2}dt\ge 2\pi^{-1/2} (x - \frac{1}{2}x^2)$. For any $c>0$, $\mathrm{erf}(x/\sqrt{2\phi^2}) > \mathrm{erf}(c/\sqrt{2\phi^2})$ when $x\ge c$. Subsequently, the first term can be lowered bounded as
    \begin{equation*}
        a\times \mathrm{erf}\Big(\frac{a}{\sqrt{2\phi^2}}\Big) \ge \left\{\begin{aligned}
            &\sqrt{2/\pi}\phi^{-1}\big(a^2-a^3/(2\sqrt{2\phi^2})\big)\text{, when }a\le c\\
            &\mathrm{erf}(c/\sqrt{2\phi^2}) a\text{, when }a\ge c
        \end{aligned}\right.
    \end{equation*}
    For $\exp\{-a^2/(2\phi^2)\} - 1$, we take the same threshold and bound it from below as
    \begin{equation*}
        \phi\sqrt{2/\pi}\big[\exp\{-a^2/(2\phi^2)\} - 1\big]\ge \left\{\begin{aligned}
            -(2\pi)^{-1/2}\phi^{-1}a^2\text{, when }a\le c\\
            -\phi\sqrt{2/\pi}\text{, when }a\ge c
        \end{aligned}\right.
\end{equation*}
Combine these, and we arrive at
\begin{equation}\label{eq_sharp_lower_bound_folded_normal_diff}
        \EE[|\xi-a|] - \EE[|\xi|]\ge \left\{\begin{aligned}
            &\sqrt{2/\pi}\phi^{-1}\big(\frac{1}{2}a^2-a^3/(2\sqrt{2\phi^2})\big)\text{, when }a\le c\\
            &\mathrm{erf}(c/\sqrt{2\phi^2}) a - \phi\sqrt{2/\pi}\text{, when }a\ge c
        \end{aligned}\right.
    \end{equation}
     Note that $\sqrt{2\phi^2}\times \mathrm{erf}(1) > 0.75 \phi\sqrt{2}$. Then we can select $c$ such that for some small constant $\epsilon_0$, there are
     \begin{equation*}
        (1-\epsilon_0)c\times \mathrm{erf}(c/\sqrt{2\phi^2})\ge \phi \sqrt{2/\pi} \text{ and }c<(1-\epsilon_0)\sqrt{2\phi^2}.
     \end{equation*}
     For $c$ that obeys these requirements, \eqref{eq_sharp_lower_bound_folded_normal_diff} becomes
     \begin{equation*}
        \EE[|\xi-a|] - \EE[|\xi|]\ge \left\{\begin{aligned}
            &\epsilon_0/\sqrt{2\pi}\phi^{-1}a^2\text{, when }a\le c\\
            &\epsilon_0\mathrm{erf}(c/\sqrt{2\phi^2}) a \text{, when }a\ge c
        \end{aligned}\right.
    \end{equation*}
     immediately implies the result.

\subsection{Proof of Lemma~\ref{lemma_regularize_loadings}}\label{sec_prove_lemma_regularize_loadings}
To start, we bound $\sum_{k=1}^r\sum_{l=1,l\neq k}^r(\bG^{-1})_{kl}^2$. Suppose $\bG - \bI_r = \bDelta$ where $r_0\le\|\bDelta\| \le R_0\ll1$. Taylor expansion gives
\begin{equation*}
        \bG^{-1} = \bI_r - \bDelta + \bDelta^2 + O(\|\bDelta\|^3).
    \end{equation*}
    By the definition of feasible set $\Xi(R_0,\hat\bZ(\hat\bG^*)^\T)$, we know $(\bG\hat\bM^*\bG^\T)_{kk} = 1$ for all $k\in[r]$ with $\hat\bM^*$ defined as $n^{-1}\hat\bG^*\Mffh(\hat\bG^*)^\T$. For $\hat\bM^*$, 
  we note that
    \begin{equation}\label{eq_expan_ff1}
\begin{aligned}\|\hat\bM^* - n^{-1}\bZ^*{}^\T\bZ^*\|=&\,\big\|\hat\bG^*n^{-1}\hat\bZ^\T\hat\bZ(\hat\bG^*)^\T - n^{-1}\bZ^*{}^\T\bZ^*\big\|\\ \le &\,\big\|n^{-1}\big\{\hat\bZ(\hat\bG^*)^\T-\bZ^*\big\}^\T\bZ^*\big\| \\&\,+ \big\|n^{-1}\bZ^*{}^\T\big\{\hat\bZ(\hat\bG^*)^\T-\bZ^*\big\}\big\|\\&\,+\big\|n^{-1}\big\{\hat\bZ(\hat\bG^*)^\T-\bZ^*\big\}^\T\big\{\hat\bZ(\hat\bG^*)^\T-\bZ^*\big\}\big\|.
    \end{aligned}\end{equation}
 Apply Corollary~\ref{coro_asymp_start} and we know \eqref{eq_expan_ff1} can be bounded by $O_p\big((n\wedge q)^{-1}\big)$. 
 Therefore, from $(ii)$ of Assumption~\ref{assump_psd_covariance_prime}, we arrive at
    
    \begin{equation*}
        (\hat\bM^*)_{k,k} = 1 + O_p\big((n\wedge q)^{-1}\big),\text{ and }\Big|(\hat\bM^*)_{k,k} - (\bM^*)_{k,k}\Big| = O_p\big((n\wedge q)^{-1}\big),
    \end{equation*}
which implies with $(\bG\hat\bM^*\bG^\T)_{kk} = 1$ that
    \begin{equation*}
        1 = \big(\hat\bM^* + \hat\bM^*\bDelta^\T + \bDelta\hat\bM^* \big)_{kk} + O(\|\bDelta\|^2) = 1 + 2\sum_{i=1}^r\hat\bM^*_{k,i}\Delta_{ki} + O(\|\bDelta\|^2) + O_p\big((n\wedge q)^{-1}\big).
    \end{equation*}
    Here the second equality is due to $\hat\bM^* = (\hat\bM^*)^\T$. 
    Subsequently, one has
    \begin{equation*}
        \Delta_{kk} = -(\hat\bM_{kk}^*)^{-1}\sum_{l\neq k}\Delta_{kl}\hat\bM^*_{k,l} + O(\|\bDelta\|^2) + O_p\big((n\wedge q)^{-1}\big).
    \end{equation*}
    Now we compute
    \begin{align*}
        \sum_{k=1}^r\sum_{l=1,l\neq k}^{r}(\bG^{-1})_{k,l}^2 \ge &\,\sum_{k=1}^r\sum_{l=1,l\neq k}^{r}(\Delta_{kl})^2 - O\big(\|\bDelta\|^3\big) \\\ge &\,\big\|\bG  -\bI_r\big\|_F^2- \big\|\mathrm{diag}(\bG - \bI_r)\big\|^2_{F}- O\big(\|\bDelta\|^3\big)\\
        \ge &\,r_0^2 - \sum_{k=1}^r\Big(\sum_{l\neq k}\Delta_{kl}\hat\bM^*_{k,l}\Big)^2 - O\big(\|\bDelta\|^3\big)-  O_p\big(\|\bDelta\|(n\wedge q)^{-1}\big)\\
        \ge &\,r_0^2 - \sum_{k=1}^r\Big(\sum_{l=1,l\neq k}^r\Delta_{k,l}^2\Big)\Big\{\sum_{l=1,l\neq k}^r(\hat\bM^*_{k,l})^2\Big\} - O\big(\|\bDelta\|^3\big)-  O_p\big(\|\bDelta\|(n\wedge q)^{-1}\big)
        \\
        \ge &\,r_0^2 - \sum_{k=1}^r\Big\{\sum_{l=1,l\neq k}^r(\bG^{-1})_{k,l}^2\Big\}\Big\{\sum_{l=1,l\neq k}^r(\bM_{k,l}^*)^2\Big\} - O\big(\|\bDelta\|^3\big)\\&\,-  O_p\big(\|\bDelta\|(n\wedge q)^{-1}\big).
    \end{align*}Here, the last inequality follows from $\|\hat\bM^* - \bM^*\big\| = O_p\big((n\wedge q)^{-1}\big)$.
   With $|\bM_{k,l}^*|$ uniformly bounded over $k,l\in[r]$, $\|\bDelta\|\le R_0\ll 1$, we conclude that, when $n$ and $q$ are large enough, with probability approaching 1,
    \begin{equation*}
        \sum_{k=1}^r\sum_{l=1,l\neq k}^{r}(\bG^{-1})_{k,l}^2\ge c_0r_0^2 - C_2^{\prime}R_0\{R_0^2+ (n\wedge q)^{-1}\},
    \end{equation*}
    for some absolute constants $c_0,C_2$. Without loss of generality, we assume that
    \begin{equation*}
        \sum_{l=2}^r(\bG^{-1})_{l,1}^2 \ge \frac{c_0r_0^2 -C_2^{\prime}R_0\{R_0^2+ (n\wedge q)^{-1}\}}{r}.
    \end{equation*}
    Then with $\lambda_{r-1}\{q^{-1}(\bA^*_{\cT_1^*,})^\T\bA^*_{\cT_1^*,}\}\ge \sigma_q^{-1}$, one arrives at
    \begin{align*}
        \sum_{j\in\cT_1^*}\mu_{j1}^2\ge &\,n\big\|\bA^*_{\cT_1,-1}(\bG^{-1})_{-1,1}\big\|^2\\\ge &\, n \big\{\sigma_{r-1}\big(q^{-1/2}\bA^*_{\cT^*_1,}\big)\big\}^2q\big\|(\bG^{-1})_{1,1}\big\|^2 \\\ge&\, C_1^{\prime}nq\sigma_q^{-1}\big[r_0^2 - C_2^{\prime}R_0\{R_0^2+ (n\wedge q)^{-1}\}\big],
    \end{align*}
    for some absolute constant $C_1^{\prime}$. To further lower bound $\sum_{j\in\cT_1^*}\min\big(c_1\mu_{j1}^2,c_2|\mu_{j1}|\big)$, we introduce the following lemma
    \begin{lemma}\label{lemma-cn-cn}
        Let $c_1,c_2>0$ be fixed constants and let $x_1,\dots,x_n$ be real numbers.
Suppose there exist positive sequences $C_n c_n\to \infty$ with $C_n\to \infty$ and $c_n\gtrsim 1$ such that
\[
\sum_{i=1}^n x_i^2 \;\ge\; C_n
\quad\text{and}\quad
|x_i| \;\le\; c_n \quad \text{for all } i=1,\dots,n.
\]
Then 
\begin{equation}
\label{eq_lemma-cn-cn}
\sum_{i=1}^n \min\bigl(c_1 |x_i|,\, c_2 x_i^2\bigr)
\;\ge\;
\Big(c_2\wedge\frac{c_1}{c_n}\Big)\, C_n.
\end{equation}
    \end{lemma}
    \begin{proof}
        See Section~\ref{supp_sec_prove_lemma-cn-cn}
    \end{proof}
    Note that, with $\max_{j\in[q]}\|\ba_j^*\|\le M$, $\mu_{j1}$ can be bounded uniformly as  $|\mu_{j1}|\le \sqrt{n}M\|\bG^{-1} - \bI_r\|\lesssim\sqrt{n}R_0$. Since $\sqrt{n}R_0$ is non-diminishing, when $n,q$ are large enough,
    \begin{equation*}
       \sum_{j\in\cT_1^*}\min\big(c_1\mu_{j1}^2,c_2|\mu_{j1}|\big)\ge \frac{\sqrt{n}q{\big[C_1r_0^2 -C_2R_0\{R_0^2+ (n\wedge q)^{-1}\}\big]}}{R_0\sigma_q},
    \end{equation*}
    for some constant $C_1,C_2>0$.

\subsection{Proof of Lemma~\ref{lemma_score_hessian_concern}}\label{supp_sec_prove_lemma_score_hessian_concern}
For $(i)$, recall that $\bV_z^*$ can be written as $\bV_z^* = \sqrt{\nu q}(\bv_{z,11}^*,\bv_{z,12}^*,\dots,\bv_{z,1r}^*,\bv_{z,21}^*,\dots,\bv_{z,rr}^*)$ with
    \begin{equation*}
    \bv_{z,kl}^* := \big(n^{-1/2}z_{1k}^*\be_l^\T,\dots,n^{-1/2}z_{nk}^*\be_l^\T\big)^\T,
\end{equation*}
for each $k,l\in[r]$. Write $\bH_i=\bH_{Lz_iz_i}^*$, $\bar\bH_i=\EE\bH_i$, and $\bs_i=\sum_{j=1}^q\ell_{ij}^{\prime}(\theta_{ij}^*)\ba_j^*$.  Both $\bH_i$ and $\bar\bH_i$ have eigenvalues of order $q$, so $\|q\bar\bH_i^{-1}\|=O(1)$ uniformly in $i$. One can expand fourth and eighth moments of the independent centred summands in $\bs_i$ and $\bH_i-\bar\bH_i$ to get  $\bH_i^{-1}\bs_i=\bar\bH_i^{-1}\bs_i+\bR_i$, $\EE\|\bR_i\|^2=O(q^{-2})$, and $\EE\|\bR_i\|^4=O(q^{-4})$,
uniformly in $i\in[n]$.
For each $k,l\in[r]$, by the block-diagonal structure of $\bH_{Lzz}^*$,
\begin{align*}
    \big|(\bv_{z,kl}^*)^\T(\bH_{Lzz}^*)^{-1}\partial_{\bphi_z}\cL(\bphi^*)\big| = &\,\big|n^{-1/2}\sum_{i=1}^n\Trc\big\{z_{ik}^*\be_l^\T(\bH_{Lz_iz_i}^*)^{-1}\sum_{j=1}^q\ell_{ij}^{\prime}(\theta_{ij}^*)\ba_j^*\big\}\big|\\\le &\,n^{-1/2}\Big|\sum_{i=1}^n\sum_{j=1}^q
\ell_{ij}^{\prime}(\theta_{ij}^*)z_{ik}^*\be_l^\T \bar\bH_i^{-1}\ba_j^*\Big|
+n^{-1/2}\sum_{i=1}^n|z_{ik}^*|\|\bR_i\|\\
{\color{blue}\le} &\,O_p\big(q^{-1/2}+\sqrt n/q\big).
\end{align*}
Here, $q z_{ik}^*\be_l^\T\bar\bH_i^{-1}\ba_j^*$ are uniformly bounded by Assumption~\ref{assump_psd_covariance}$(ii)$ and $\|\bA^*\|_{2\to\infty}\le M$. Then we can use $(iii)$ of Lemma~\ref{lemma_concentration} to bound the centred leading sum, and use $\EE\|\bR_i\|\lesssim q^{-1}$ and Markov's inequality to bound the remainder.
Then since $\bV_z^*$ has only $r^2$ columns, we arrive at
\begin{align*}
    \big\|(\bV_z^*)^\T(\bH_{Lzz}^*)^{-1}\partial_{\bphi_z}\cL(\bphi^*)\big\|
    \le r^2\times\sqrt{\nu q}\times O_p\big(q^{-1/2}+\sqrt n/q\big)
    =O_p(1+\sqrt{n/q}).
\end{align*}
Similarly, one can show $\big\|(\bV_a^*)^\T(\bH_{Laa}^*)^{-1}\partial_{\bphi_a}\cL(\bphi^*)\big\|=O_p(1+\sqrt{q/n})$.
This proves part $(i)$.

For part $(ii)$, similar to the above derivation, we leverage the block-diagonal structure of $\bH_{Lzz}^*$ to get

\begin{equation}\begin{aligned}
    &\,\sum_{j\in[q]}\|\bH_{La_jz}^*(\bH_{Lzz}^*)^{-1}\partial_{\bphi_z}\cL(\bphi^*)\|^2\\
    = &\,\sum_{j\in[q]}\Big\|\sum_{i=1}^n
    \{\ell_{ij}^{\prime\prime}(\theta_{ij}^*)\bz_i^*\ba_j^*{}^\T+
    \ell_{ij}^{\prime}(\theta_{ij}^*)\bI_r\}
    (\bH_{Lz_iz_i}^*)^{-1}\sum_{t=1}^q\ell_{it}^{\prime}(\theta_{it}^*)\ba_t^*\Big\|^2\\
    \le &\,3q^{-1}\Big\{
    \sum_{j\in[q]}\underbrace{\Big\|\sum_{i=1}^n\sum_{t=1}^q
    \ell_{it}^{\prime}(\theta_{it}^*)\ell_{ij}^{\prime\prime}(\theta_{ij}^*)
    \bz_i^*\ba_j^*{}^\T\bar\bH_i^{-1}\ba_t^*\Big\|^2}_{\alpha_1}
    +\sum_{j\in[q]}\underbrace{\Big\|\sum_{i=1}^n\sum_{t=1}^q
    \ell_{it}^{\prime}(\theta_{it}^*)\ell_{ij}^{\prime}(\theta_{ij}^*)
    \bar\bH_i^{-1}\ba_t^*\Big\|^2}_{\alpha_2}\Big\}\\
    &\,+3\sum_{j\in[q]}\underbrace{\Big\|\sum_{i=1}^n
    \{\ell_{ij}^{\prime\prime}(\theta_{ij}^*)\bz_i^*\ba_j^*{}^\T+
    \ell_{ij}^{\prime}(\theta_{ij}^*)\bI_r\}\bR_i\Big\|^2}_{\alpha_3}.
\end{aligned}\label{eq_bound_Hza_Hlzz_s}\end{equation}
Since $\max_{i\in[n],j\in[q],t\in[q]}q\|\ell_{ij}^{\prime\prime}(\theta_{ij}^*)\bz_i^*\ba_j^*{}^\T\bar\bH_i^{-1}\ba_t^*\|=O(1)$ by Assumptions~\ref{assump_psd_covariance}$(ii)$, $\|\bA^*\|_{2\to\infty}\le M$, and~\ref{assump_limit},
one can separate $t=j$ from $t\ne j$ in $\alpha_1$ to see that its independent row summands have means of order $O(1)$ and second moments of order $O(q)$. Therefore, we have
\begin{equation*}
\EE\sum_{j\in[q]}\alpha_1=O(nq^2+n^2q),\quad\text{and}\quad
\sum_{j\in[q]}\alpha_1=O_p(nq^2+n^2q).
\end{equation*}
For $\alpha_2$, note
\begin{equation*}
\begin{aligned}
&q\sum_{i=1}^n\sum_{t=1}^q\ell_{it}^{\prime}(\theta_{it}^*)\ell_{ij}^{\prime}(\theta_{ij}^*)\bar\bH_i^{-1}\ba_t^*\\
&\quad=
\underbrace{q\sum_{i=1}^n\ell_{ij}^{\prime}(\theta_{ij}^*)\sum_{t=1,t\neq j}^q
\ell_{it}^{\prime}(\theta_{it}^*)\bar\bH_i^{-1}\ba_t^*}_{\beta_1}
+\underbrace{q\sum_{i=1}^n\{\ell_{ij}^{\prime}(\theta_{ij}^*)\}^2
\bar\bH_i^{-1}\ba_j^*}_{\beta_2}.
\end{aligned}
\end{equation*}
Because $\max_{t\in[q]}\|\ba_t^*\|\le M$ by $\|\bA^*\|_{2\to\infty}\le M$, it holds that
\begin{align*}
    \EE\sum_{j=1}^q\alpha_2 \le&\,2\sum_{j=1}^q\EE\|\beta_1\|^2+
    2\sum_{j=1}^q\EE\|\beta_2\|^2\\
    \le&\,2\sum_{j=1}^q\EE\Big[\sum_{i=1}^n
    \{\ell_{ij}^{\prime}(\theta_{ij}^*)\}^2
    \Big\|q\sum_{t=1,t\neq j}^q\ell_{it}^{\prime}(\theta_{it}^*)
    \bar\bH_i^{-1}\ba_t^*\Big\|^2\Big]+O(n^2q)\\
    =&\,O(nq^2+n^2q).
\end{align*}
Here the second inequality follows from independence across rows and the independence between $\ell_{ij}^{\prime}(\theta_{ij}^*)$ and $q^{-1}\sum_{t=1,t\neq j}^q\ell_{it}^{\prime}(\theta_{it}^*)\bar\bH_i^{-1}\ba_t^*$.
By Markov's inequality, we conclude that $\sum_{j=1}^q\alpha_2=O_p(nq^2+n^2q)$.

For $\alpha_3$, the moment bounds and Cauchy--Schwarz inequality give
\[
\EE\big\|\{\ell_{ij}^{\prime\prime}(\theta_{ij}^*)\bz_i^*\ba_j^*{}^\T+
\ell_{ij}^{\prime}(\theta_{ij}^*)\bI_r\}\bR_i\big\|^2=O(q^{-2}).
\]
Applying Cauchy--Schwarz to the sum over $i$ gives
$\EE\sum_{j=1}^q\alpha_3=O(n^2/q)$.
Combining these bounds with \eqref{eq_bound_Hza_Hlzz_s} and Markov's inequality yields
\[
\|\bH_{Laz}^*(\bH_{Lzz}^*)^{-1}\partial_{\bphi_z}\cL(\bphi^*)\|
=O_p(\sqrt n+n/\sqrt q).
\]
The augmentation term $\bV_a^*(\bV_z^*)^\T$ has the same bound by $(i)$ and $\|\bV_a^*\|=O(\sqrt n)$. Finally, the Woodbury identity transfers this result from $(\bH_{Lzz}^*)^{-1}$ to $(\bH_{zz}^*)^{-1}$, as by \eqref{eq_bound_h_ast},
$\|\bH_{az}^*(\bH_{Lzz}^*)^{-1}\bV_z^*\|=O_p(\sqrt n)$, and the middle Woodbury factor has norm at most $1$. Its additional contribution is thus $O_p\{\sqrt n(1+\sqrt{n/q})\}$.
This proves $(ii)$, and $(iii)$ can be proved by symmetry.

\subsection{Proof of Lemma~\ref{lemma-cn-cn}}\label{supp_sec_prove_lemma-cn-cn}

Fix $i \in \{1,\dots,n\}$ and write $s_i := x_i^2$. Then $s_i \in [0,c_n^2]$
because $|x_i|\le c_n$. Define
$f(s) := \min(c_1 \sqrt{s},\, c_2 s), $ for $ s \ge 0.$
We first show that for all $s \in (0,c_n^2]$,
\begin{equation}
\label{eq:pointwise-bound}
f(s) \;\ge\; \alpha_n s
\quad\text{where}\quad
\alpha_n := c_2 \wedge \frac{c_1}{c_n}.
\end{equation}

Consider the ratio $f(s)/s$ for $s>0$.
Note that for $s_0 = (c_1/c_2)^2$
\[
f(s)/s =
\begin{cases}
c_2, & 0 < s \le s_0,\\[3pt]
c_1 s^{-1/2}, & s \ge s_0.
\end{cases}
\]

\noindent\textbf{Case 1: $0 < s \le s_0$.}
Then $f(s) = c_2 s$, so
\[
\frac{f(s)}{s} = c_2 \;\ge\; \alpha_n,
\]
because $\alpha_n \le c_2$ by definition. Thus \eqref{eq:pointwise-bound} holds.

\medskip
\noindent\textbf{Case 2: $s \ge s_0$.}
Then $f(s) = c_1 \sqrt{s}$, and we have
\[
\frac{f(s)}{s}
= \frac{c_1}{\sqrt{s}}
\;\ge\;
\frac{c_1}{c_n},
\]
since $s \le c_n^2$ implies $\sqrt{s} \le c_n$.
Therefore
\[
\frac{f(s)}{s} \;\ge\; \frac{c_1}{c_n} \;\ge\; \alpha_n,
\]
because $\alpha_n \le c_1/c_n$ by definition. This again yields
\eqref{eq:pointwise-bound}.

\medskip
Combining the two cases, we have shown that for all $s \in (0,c_n^2]$,
\[
f(s) \;\ge\; \alpha_n s.
\]
Substituting $s = s_i = x_i^2$ gives, for each $i$,
\[
\min\bigl(c_1 |x_i|,\, c_2 x_i^2\bigr)
= f(x_i^2)
\;\ge\;
\alpha_n x_i^2.
\]
Summing over $i=1,\dots,n$ yields
\[
\sum_{i=1}^n \min\bigl(c_1 |x_i|,\, c_2 x_i^2\bigr)
\;\ge\;
\alpha_n \sum_{i=1}^n x_i^2
\;\ge\;
\alpha_n C_n
=
\Bigl( c_2 \wedge \frac{c_1}{c_n} \Bigr) C_n,
\]
This completes the proof.

\section{Proof of Proposition~\ref{prop_linear_convergence}}\label{supp_sec_prove_prop_linear_convergence}
We start with the following lemma.
\begin{lemma}\label{lemma_8}
    Under the setup of Proposition~\ref{prop_linear_convergence}, when $n$ and $q$ are large enough, with probability approaching $1$, the following events hold
    \begin{enumerate}[label=$(\roman*)$]
        \item $\partial_{\bZ}\cL_G(\hat\bZ^0,\hat\bA^0) = \zero$ and $\partial_{\bA}\cL_G(\hat\bZ^0,\hat\bA^0) = \zero$;
        \item for sufficiently small $\epsilon>0$, there is some absolute constant $c_0$ such that, \begin{equation*}\min_{(\bZ,\bA)\in\Theta^*(\epsilon)}\lambda_{\min}(\bS_{nq}\partial_{\bphi\bphi}^2\cL_{G}(\bZ,\bA)\bS_{nq})\ge c_0b_L\min\{\lambda_{\min}(\bSigma_z^*),\lambda_{\min}(\bSigma_{a}^*)\}/2;\end{equation*}
        \item the bounds in Lemma~\ref{lemma_hessian_property} hold.
    \end{enumerate}
    Here, $\bphi = (\bz_1^\T,\dots,\bz_n^\T,\ba_1^\T,\dots,\ba_q^\T)^\T$ and $\bS_{nq}$ is defined in \eqref{eq_define_snq} as
    \begin{equation*}
        \bS_{nq} = \begin{pmatrix}
            q^{-1/2}\bI_{nr} & \zero \\ \zero & n^{-1/2}\bI_{qr}
        \end{pmatrix}.
    \end{equation*}
\end{lemma}
\begin{proof}
For $(i)$, we note that in Section~\ref{supp_sec_prove_lemma_combine_zero}, we have shown $(\hat\bZ^0\hat\bG^{\star}{}^\T,\hat\bA^0\hat\bG^{\star}{}^{-1}) = (\hat\bZ^o,\hat\bA^o)$ with probability approaching $1$. Then with \eqref{eq_first_order_condition} and the structure of the first order derivative, one can verify that $\partial_{\bphi}\cL(\hat\bZ^0,\hat\bA^0) = 0$. Furthermore, since $\|n^{-1}\hat\bZ^0{}^\T\hat\bZ^0-\bI_r\|_{\Fn}^2 + \|\mathrm{ndiag}(q^{-1}\hat\bA^0{}^\T\hat\bA^0)\|_{\Fn}^2 = 0$, we know $\partial_{\bphi}\big\{\|n^{-1}\bZ^\T\bZ-\bI_r\|_{\Fn}^2 + \|\mathrm{ndiag}(q^{-1}\bA^\T\bA)\|_{\Fn}^2\big\} = \zero$. Together, we conclude that $\partial_{\bphi}\cL_{G}(\hat\bZ^0,\hat\bA^0) = \zero$ with probability approaching $1$.
    Part $(ii)$ follows from Lemma 2 of \citet{cui2025identifiability}; and part $(iii)$ is a consequence of Lemma~\ref{lemma_hessian_property}. 
\end{proof}

We already know that the error contractions hold for $\tilde e_{\ell,2}^{(t)}$ and $\tilde e_{\ell,\infty}^{(t)}$ by $\tilde e_{\ell,\infty}^{(0)} = o(1)$. Then it suffices to show \eqref{eq_l_2_cont_star} for $t+1$, given it holds for $0,\dots,t$.

We first calculate $\tilde e_{\ell,2}^{(t+1)}$. Note that in Algorithm~\ref{alg_gradient_descent}, $\tilde\bZ^{(t+1)}$ and $\tilde\bA^{(t+1)}$ are given as
\begin{equation*}
    \tilde\bZ^{(t+1)} = \cT_{M}\big\{\tilde\bZ^{(t)} - q^{-1}\alpha\partial_{\bZ}\cL_{G}(\tilde\bZ^{(t)},\tilde\bA^{(t)})\big\}\text{ and }\tilde\bA^{(t+1)} = \cT_{M}\big\{\tilde\bA^{(t)} - n^{-1}\alpha\partial_{\bA}\cL_{G}(\tilde\bZ^{(t)},\tilde\bA^{(t)})\big\}.
\end{equation*}
We first show that the projection $\cT_M$ always induces smaller error, i.e., $\|\cT_M(\bZ) - \hat\bZ^0\|_{\Fn}\le \|\bZ - \hat\bZ^0\|_{\Fn}$}. By definition, we know $\|\hat\bZ^0\|_{2\to\infty}\le M$. It suffices  to verify that $\|\cT_M(\bZ_{i,}) - \hat\bZ_{i,}^0\| \le \|\bZ_{i,} - \hat\bZ_{i,}^{0}\|$ for each $i\in[n]$. Note that when $\|\bZ_{i,}\|\le M$, the equality holds naturally. When $\|\bZ_{i,}\| > M$, we note that
\begin{equation*}
    \|\bZ_{i,}\|^2 - 2\|\bZ_{i,}\|M + M^2\ge 0,
\end{equation*}
which leads to 
\begin{equation*}
    \|\bZ_{i,}\|^2 - M^2\ge 2M\|\bZ_{i,}\|\big(1-\frac{M}{\|\bZ_i\|}\big)\ge 2\bZ_{i,}(\hat\bZ_{i,}^0)^\T\big(1-\frac{M}{\|\bZ_i\|}\big).
\end{equation*}Here, the last inequality follows from $\|\bZ_{i,}\|M \ge \bZ_{i,}(\hat\bZ_{i,}^0)^\T$ and $M/\|\bZ_{i,}\|< 1$
Therefore,
\begin{align*}
    \|\cT_M(\bZ_{i,}) - \hat\bZ_{i,}^0\|^2  = &\,\big\|\frac{M\bZ_{i,}}{\|\bZ_{i,}\|} - \hat\bZ_{i,}^0\big\|^2\\= &\, \|\hat\bZ_{i,}^0\|^2 + M^2 - 2\frac{M}{\|\bZ_{i,}\|}\bZ_{i,}(\hat\bZ_{i,}^0)^\T\\\le &\,\|\hat\bZ_{i,}^0\|^2 +M^2 +\|\bZ_{i,}\|^2 - M^2 - 2\bZ_{i,}(\hat\bZ_{i,}^0)^\T\\ = &\,\|\bZ_{i,} - \hat\bZ_{i,}^0\|^2.
\end{align*}
Therefore, we conclude that $\|\cT_M(\bZ) - \hat\bZ^0\|_{\Fn}\le \|\bZ - \hat\bZ^0\|_{\Fn}$, and similarly, one can verify $\|\cT_M(\bA) - \hat\bA^0\|_{\Fn}\le \|\bA - \hat\bA^0\|_{\Fn}$. Then
\begin{align*}
    (nq)\tilde e_{\ell,2}^{(t+1)}\le &\,q\big\|\tilde\bZ^{(t)} - q^{-1}\alpha\partial_{\bZ}\cL_{G}(\tilde\bZ^{(t)},\tilde\bA^{(t)}) - \hat\bZ^0\big\|_{\Fn}^2\\ &\, + n\big\|\tilde\bA^{(t)} - n^{-1}\alpha\partial_{\bA}\cL_{G}(\tilde\bZ^{(t)},\tilde\bA^{(t)}) - \hat\bA^{0}\big\|_{\Fn}^2\\\le &\,q\big\|\tilde\bZ^{(t)}  - \hat\bZ^0- q^{-1}\alpha\big\{\partial_{\bZ}\cL_{G}(\tilde\bZ^{(t)},\tilde\bA^{(t)}) - \partial_{\bZ}\cL_{G}(\hat\bZ^0,\hat\bA^0)\big\}\big\|_{\Fn}^2\\ &\, + n\big\|\tilde\bA^{(t)} - \hat\bA^{0} - n^{-1}\alpha\big\{\partial_{\bA}\cL_{G}(\tilde\bZ^{(t)},\tilde\bA^{(t)}) - \partial_{\bA}\cL_{G}(\hat\bZ^0,\hat\bA^0)\big\}\big\|_{\Fn}^2.
\end{align*}
Here, the second inequality is due to $(i)$ of Lemma~\ref{lemma_8}. Then let $\tilde\bphi^{(t)}$, $\hat\bphi^0$ be the vectorized parameters $(\tilde\bZ^{(t)},\tilde\bA^{(t)})$ and $(\hat\bZ^0,\hat\bA^0)$, and by the fundamental theorem of calculus (Theorem 4.2 in Chapter XIII in \citet{lang2012real}), we obtain
\begin{align*}
    &nq\tilde e_{\ell,2}^{(t+1)}\le \big\|\bS_{nq}^{-1}\big(\tilde\bphi^{(t)} - \hat\bphi^0\big) - \alpha\bS_{nq}\big\{\partial_{\bphi}\cL_{G}(\tilde\bphi^{(t)}) - \partial_{\bphi}\cL_{G}(\hat\bphi^{0})\big\}\big\|^2\\&= \big\|\bS_{nq}^{-1}\big(\tilde\bphi^{(t)} - \hat\bphi^0\big)  - \alpha \big\{\int_0^1\bS_{nq}\partial_{\bphi\bphi}^2\cL_{G}(\hat\bphi^0 + s(\tilde\bphi^{(t)} - \hat\bphi^0))\bS_{nq}ds\big\}\bS_{nq}^{-1}\big(\tilde\bphi^{(t)} - \hat\bphi^0\big) \big\|^2.
\end{align*}
With $\tilde e_{\ell,2}^{(t)} = o(1)$, $\tilde e_{\ell,\infty}^{(t)} = o(1)$ and $(\hat\bZ^0,\hat\bA^0),(\tilde\bZ^{(t)},\tilde\bA^{(t)})\in\Theta$, we know by $(ii)$ of Lemma~\ref{lemma_8} and the continuity of the eigenvalues that 
\begin{equation*}
\lambda_{\min}\big\{\int_0^1\bS_{nq}\partial_{\bphi\bphi}^2\cL_{G}(\hat\bphi^0 + s(\tilde\bphi^{(t)} - \hat\bphi^0))\bS_{nq}ds\big\}\ge  c_0b_L\min\{\lambda_{\min}(\bSigma_z^*),\lambda_{\min}(\bSigma_{a}^*)\}/2 := \gamma.
\end{equation*}
Consequently, one has
\begin{equation*}
    nq\tilde e_{\ell,2}^{(t+1)}\le (1-\alpha\gamma)^2\|\bS_{nq}^{-1}(\tilde\bphi^{(t)} - \hat\bphi^0)\|^2=nq(1-\alpha\gamma)^2\tilde e_{\ell, 2}^{(t)}\le nqC_0(1-C_1\alpha)^{2(t+1)}\tilde  e_{\ell ,2}^{(0)},
\end{equation*}
where the last inequality follows from the error contraction at the $t$-th iteration and that $C_1\le \gamma$. 

Next, we calculate $\tilde e_{\ell,\infty}^{(t+1)}$. We partition $\partial_{\bphi\bphi}^2\cL_{G}(\bphi)$ into three parts:
\begin{equation*}
    \partial_{\bphi\bphi}^2\cL_{G}(\bphi) = \bH_{LD}(\bphi) + \bH_{LOD}(\bphi) + \bH_{G}(\bphi),
\end{equation*}where $\bH_{LD}(\bphi)$ is as defined in \eqref{eq_define_h_ld}, $\bH_{LOD} = \partial_{\bphi\bphi}^2\cL(\bphi) - \bH_{LD}(\bphi)$, and $\bH_{G}(\bphi) = \frac{c nq}{2}\partial_{\bphi\bphi}\big\{\|n^{-1}\bZ^\T\bZ-\bI_r\|_{\Fn}^2 + \|\mathrm{ndiag}(q^{-1}\bA^\T\bA)\|_{\Fn}^2\big\}$.
Since $\|\hat\bZ^{0}\|_{2\to\infty}\le M$, we know $\|\cT_M(\bZ)-\hat\bZ^0\|_{2\to\infty}\le \|\bZ-\hat\bZ^0\|_{2\to\infty}$ for any $\bZ\in\RR^{n\times r}$. Similar inequality holds for $\|\cT_M(\bA)-\hat\bA^0\|_{2\to\infty}$. Let $\cR_i = \{(i-1)r+1,(i-1)r+2,\dots,ir\}$ and write $\tilde\bphi^{(t),s} = \hat\bphi^0 + s(\tilde\bphi^{(t)} - \hat\bphi^0)$. Then use the fundamental theorem of calculus in a similar way to get
\begin{align}
    \tilde e_{\ell,\infty}^{(t+1)}\le&\, \max\Big\{\big\|\tilde\bZ^{(t)} - q^{-1}\alpha\partial_{\bZ}\cL_{G}(\tilde\bZ^{(t)},\tilde\bA^{(t)}) - \hat\bZ^0\big\|_{2\to\infty},\nonumber\\&\,\big\|\tilde\bA^{(t)} - n^{-1}\alpha\partial_{\bA}\cL_{G}(\tilde\bZ^{(t)},\tilde\bA^{(t)}) - \hat\bA^{0}\big\|_{2\to\infty}\Big\}\nonumber\\\le&\, \max_{i\in[n+q]}\Big\|\big[\big\{\bI_{(n+q)r} - \alpha\bS_{nq}^{2}\int_0^1\partial_{\bphi\bphi}^2\cL_G(\hat\bphi^0 + s(\tilde\bphi^{(t)} - \hat\bphi^0))ds\big\}(\tilde\bphi^{(t)} - \hat\bphi^0)\Big]_{\cR_i}\Big\|\nonumber
    \\\le &\,\max\Big\{\max_{i\in[n]}\big\|(\tilde\bz_{i}^{(t)} - \hat\bz_i^0)^\T\big\{\bI_r - q^{-1}\alpha \int_0^1\bH_{Lz_iz_i}(\tilde\bphi^{(t),s})ds\big\}\big\|,\nonumber\\&\quad\qquad \max_{j\in[q]}\big\|(\tilde\ba_{j}^{(t)} - \hat\ba_j^0)^\T\big\{\bI_r - n^{-1}\alpha \int_0^1\bH_{La_ja_j}(\tilde\bphi^{(t),s})ds\big\}\big\|\Big\}\label{eq_bound_inf_error_contr_1}\\ &\,+\alpha\max_{i\in[n+q]}\big\|\big[\bS_{nq}^2\int_0^1\big\{\bH_{LOD}(\tilde\bphi^{(t),s}) + \bH_G(\tilde\bphi^{(t),s})\big\}ds(\tilde\bphi^{(t)}-\hat\bphi^0)\big]_{\cR_i,}\big\|,\label{eq_bound_inf_error_contr_2}
\end{align}Here, the last inequality follows from the block-diagonal structure of $\bH_{LD}(\bphi)$.

For \eqref{eq_bound_inf_error_contr_1}, with $\tilde e_{\ell,2}^{(t)} = o(1)$, $\tilde e_{\ell,\infty}^{(t)} = o(1)$ and $(\hat\bZ^0,\hat\bA^0),(\tilde\bZ^{(t)},\tilde\bA^{(t)})\in\Theta$, we know $\tilde\bphi^{(t),s}\in\Theta^*(\epsilon)$ for any constant $\epsilon>0$. Consequently, by Lemma~\ref{lemma_hessian_property} and the continuity of the eigenvalues, \eqref{eq_bound_inf_error_contr_1} can be bounded as
\begin{align*}
    \eqref{eq_bound_inf_error_contr_1}\le &\,\max\Bigg(\Big[1 - \min_{i\in[n]}\lambda_{\min}\big\{q^{-1}\alpha \int_0^1\bH_{Lz_iz_i}(\tilde\bphi^{(t),s})ds\big\}\Big]\big\|\tilde\bZ^{(t)} - \hat\bZ^0\big\|_{2\to\infty},\\&\quad\qquad \Big[1 - \min_{j\in[q]}\lambda_{\min}\big\{n^{-1}\alpha \int_0^1\bH_{La_ja_j}(\tilde\bphi^{(t),s})ds\big\}\Big]\big\|\tilde\bA^{(t)} - \hat\bA^0\big\|_{2\to\infty}\Bigg)\\\le &\,(1-\alpha\gamma)\max\{\|\tilde\bZ^{(t)} - \hat\bZ^0\|_{2\to\infty},\|\tilde\bA^{(t)} - \hat\bA^0\|_{2\to\infty}\}\\ \le &\,(1-\alpha\gamma)\tilde e_{\ell,\infty}^{(t)}.
\end{align*}
For \eqref{eq_bound_inf_error_contr_2}, by Lemma~\ref{lemma_hessian_property}, we know
\begin{align*}
    &\,\max_{i\in[n+q]}\big\|\big[\bS_{nq}^2\int_0^1\bH_{LOD}(\tilde\bphi^{(t),s})ds(\tilde\bphi^{(t)} - \hat\bphi^0)\big]_{\cR_i}\big\|\\=&\,\max\Big[\max_{i\in[n]}q^{-1}\big\|\int_0^1\bH_{Lza}(\tilde\bphi^{(t),s})ds\big\|_{2\to\infty}\|\tilde\bphi^{(t)}_a-\hat\bphi^0_a\|,\\&\qquad\quad\max_{j\in[q]}n^{-1}\big\|\int_0^1\bH_{Laz}(\tilde\bphi^{(t),s})ds\big\|_{2\to\infty}\|\tilde\bphi^{(t)}_z-\hat\bphi^0_z\|\Big]\\\le&\,C_{\delta}\sqrt{\log(n\vee q)}(nq)^{-1/2}\big\|\bS_{nq}^{-1}(\tilde\bphi^{(t)} - \hat\bphi^0)\big\| \\\le &\,C_{\delta}\sqrt{\log(n\vee q)\tilde e_{\ell,2}^{(t)}}.
\end{align*}
Similar to the above derivation and the calculation of \eqref{eq_integral_expansion_w2} in Section~\ref{supp_sec_prove_ora_o_step1}, we can show that 
\begin{equation*}
    \max_{i\in[n+q]}\big\|\big[\bS_{nq}^2\int_0^1\bH_{G}(\tilde\bphi^{(t),s})ds(\tilde\bphi^{(t)} - \hat\bphi^0)\big]_{\cR_i}\big\|\le C_{\delta}\sqrt{\log(n\vee q)\tilde e_{\ell,2}^{(t)}}.
\end{equation*}
Here, we use the same $C_{\delta}$ to denote a constant depending on $\delta$, while they may have different magnitudes. 
Combine the upper bounds for \eqref{eq_bound_inf_error_contr_1} and \eqref{eq_bound_inf_error_contr_2} to obtain
\begin{equation*}
    \tilde e_{\ell,\infty}^{(t+1)}\le (1-\alpha\gamma)\tilde e_{\ell,\infty}^{(t)} + C_{\delta}\alpha\sqrt{\log (n\vee q)\tilde e_{\ell,2}^{(t)}}.
\end{equation*}
Invoke the error contraction for the $t$-th iteration, and we arrive at
\begin{align*}
    \tilde e_{\ell,\infty}^{(t+1)}\le &\,C_0(1-\gamma\alpha)(1-C_1\alpha)^{t}{\log(n\vee q)}\tilde e_{\ell,\infty}^{(0)} + C_{\delta}\alpha\sqrt{C_0\log (n\vee q)\tilde e_{\ell,2}^{(0)}}(1-C_1\alpha)^{t}\\\le &\,C_0(1-C_1\alpha)^{t+1}{\log(n\vee q)}\tilde e_{\ell,\infty} ^{(0)}
\end{align*}
Here, the second inequality follows from the fact that $\tilde e_{\ell,2}^{(0)}\le 2(\tilde e_{\ell,\infty}^{(0)})^2$, and thus
\begin{equation*}
    C_1 = c_0b_L\min\{\lambda_{\min}(\bSigma_z^*),\lambda_{\min}(\bSigma_a^*)\}/4\le \gamma - \frac{C_{\delta}\sqrt{C_0\tilde e_{\ell,2}^{(0)}}}{C_0\sqrt{\log(n\vee q)}\tilde e_{\ell,\infty}^{(0)}},
\end{equation*} always holds for sufficiently large $n,q$.

\end{document}